\documentclass[11pt,english]{elsarticle}
\usepackage{mathptmx}
\usepackage[T1]{fontenc}
\usepackage[latin9]{inputenc}
\usepackage{array}
\usepackage{verbatim}
\usepackage{varioref}
\usepackage{float}
\usepackage{textcomp}
\usepackage{amsthm}
\usepackage{relsize}
\usepackage{amsbsy}
\usepackage{graphicx}
\usepackage{amssymb}
\usepackage{amsmath}
\usepackage{hyperref}

\hypersetup{
    bookmarks=true,         
    unicode=false,          
    pdftoolbar=true,        
    pdfmenubar=true,        
    pdffitwindow=false,     
    pdfstartview={FitH},    
    pdftitle={My title},    
    pdfauthor={Author},     
    pdfsubject={Subject},   
    pdfcreator={Creator},   
    pdfproducer={Producer}, 
    pdfkeywords={keywords}, 
    pdfnewwindow=true,      
    colorlinks=true,       
    linkcolor=blue,          
    citecolor=blue,        
    filecolor=magenta,      
    urlcolor=blue           
}

\makeatletter

\DeclareRobustCommand{\lyxmathsym}[1]{\ifmmode\begingroup\def\b@ld{bold}
  \def\rmorbf##1{\ifx\math@version\b@ld\textbf{##1}\else\textrm{##1}\fi}
  \mathchoice{\hbox{\rmorbf{#1}}}{\hbox{\rmorbf{#1}}}
  {\hbox{\smaller[2]\rmorbf{#1}}}{\hbox{\smaller[3]\rmorbf{#1}}}
  \endgroup\else#1\fi}

\providecommand{\tabularnewline}{\\}
\newcommand{\lyxdot}{.}

\theoremstyle{plain}

  \theoremstyle{definition}

\usepackage{mathptmx}
\usepackage{array}
\usepackage{geometry}
\usepackage{graphicx}
\usepackage{babel}

\begin{document}

\begin{frontmatter}

\title{Statistical mechanics of two-dimensional and geophysical
flows}

\date{Version 2 - March, 9, 2011}

\author[label1]{Freddy Bouchet}

\address[label1]{ Universit\'e de Lyon, Laboratoire de Physique de l'Ecole Normale Sup\'erieure de Lyon, ENS-Lyon, CNRS, 46 all\'ee d'Italie, 69364 Lyon cedex 07, France. \href{mailto:Freddy.Bouchet@ens-lyon.fr}{\nolinkurl{Freddy.Bouchet@ens-lyon.fr}}}

\author[label2]{ Antoine Venaille}

\address[label2]{GFDL-AOS Princeton University,
Forrestal Campus,  NJ 08542 Princeton, United States.}

\begin{abstract}
The theoretical study of the self-organization of two-dimensional
and geophysical turbulent flows is addressed based on statistical mechanics
methods. This review is a self-contained presentation of classical
and recent works on this subject; from the statistical mechanics basis
of the theory up to applications to Jupiter's troposphere and ocean
vortices and jets. Emphasize has been placed on examples with available
analytical treatment in order to favor better understanding of the
physics and dynamics.

After a brief presentation of the 2D Euler and quasi-geostrophic equations,
the specificity of two-dimensional and geophysical turbulence is emphasized.
The equilibrium microcanonical measure is built from the Liouville
theorem. Important statistical mechanics concepts (large deviations,
mean field approach) and thermodynamic concepts (ensemble inequivalence,
negative heat capacity) are briefly explained and described.

On this theoretical basis, we predict the output of the long time evolution
of complex turbulent flows as statistical equilibria. This is applied
to make quantitative models of two-dimensional turbulence, the Great
Red Spot and other Jovian vortices, ocean jets like the Gulf-Stream,
and ocean vortices. A detailed comparison between these statistical
equilibria and real flow observations is provided.

We also present recent results for non-equilibrium situations, for
the studies of either the relaxation towards equilibrium or non-equilibrium
steady states. In this last case, forces and dissipation are in a
statistical balance; fluxes of conserved quantity characterize the
system and microcanonical or other equilibrium measures no longer describe
the system.
\end{abstract}


\end{frontmatter}

\newpage

\tableofcontents

\newpage

{\large \bf List of figure captions}\\

\begin{figure}[h!]
\begin{center}
\includegraphics[height=0.35\textheight]{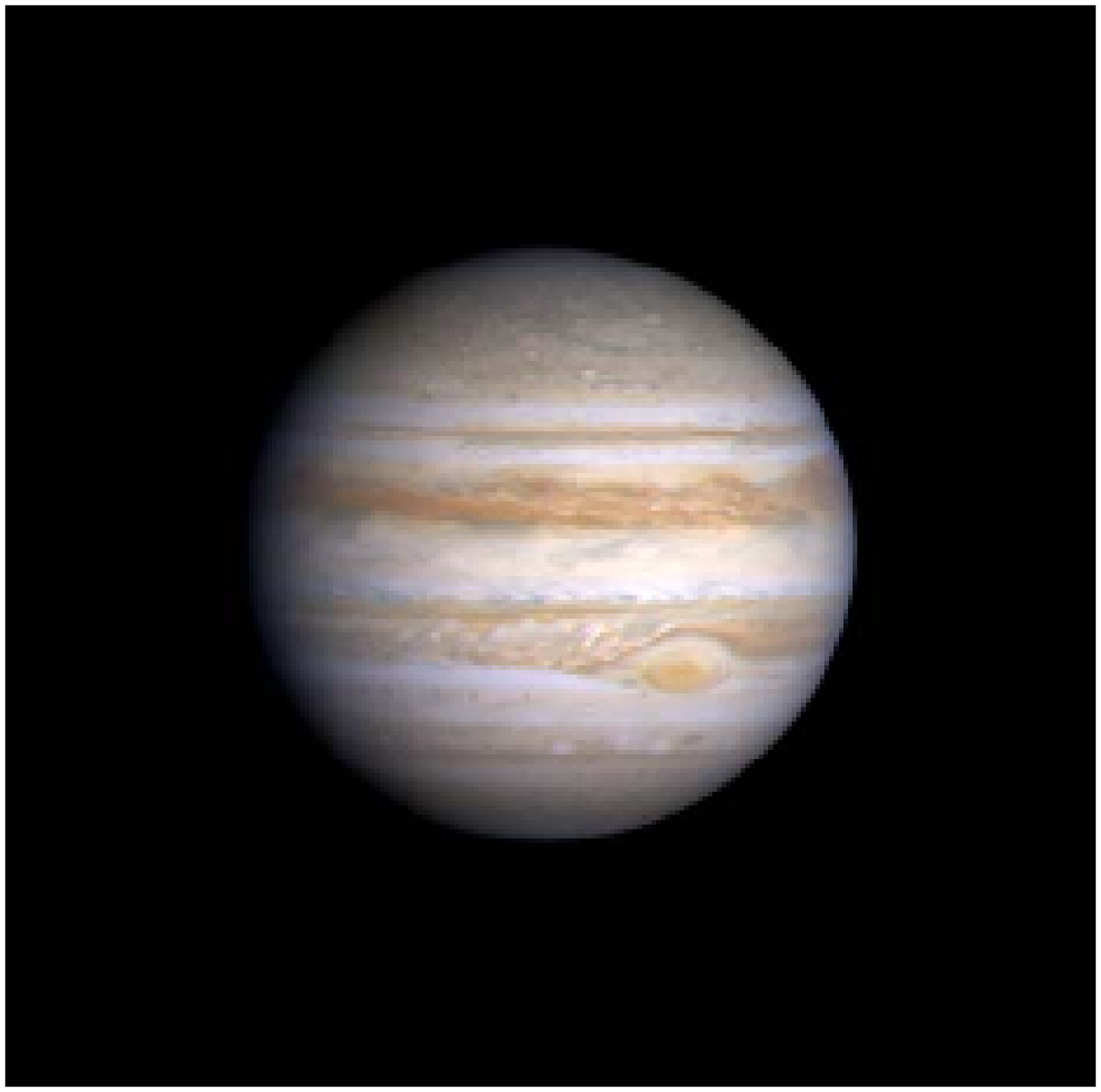}
\end{center}
\caption{{\footnotesize Observation of the Jovian atmosphere from Cassini (Courtesy of NASA/JPL-Caltech). See figure  \ref{Fig:ColorJupiter} page  \pageref{Fig:ColorJupiter}  for more detailed legends.}}
\label{Fig:ColorJupiter_ColorPage}
\end{figure}

Figure \ref{Fig:ColorJupiter_ColorPage} page \pageref{Fig:ColorJupiter_ColorPage}. Observation of the Jovian atmosphere from Cassini (Courtesy of NASA/JPL-Caltech). See figure  \ref{Fig:ColorJupiter} page  \pageref{Fig:ColorJupiter}  for more detailed legends.\\

\begin{figure}[h!]
\begin{center}
\includegraphics[height=0.4\textheight]{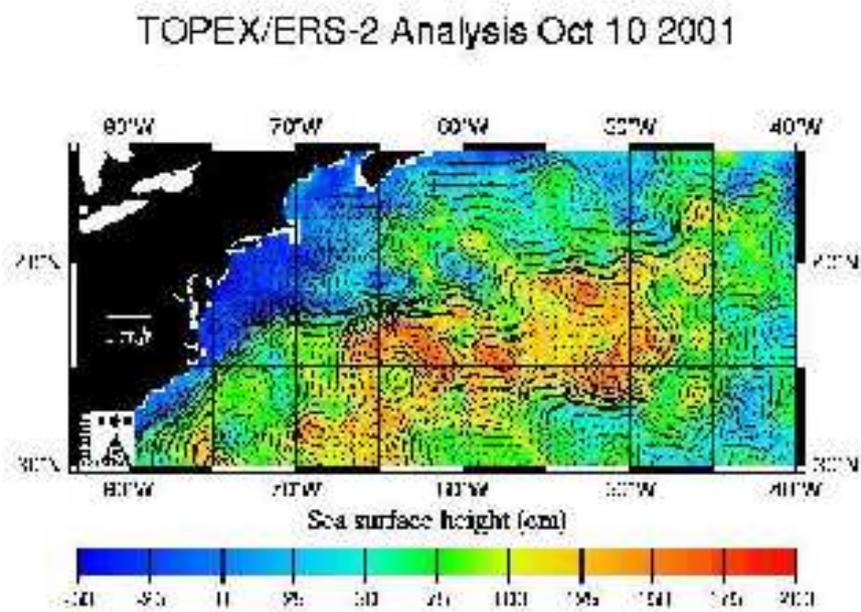}
\end{center}
\caption{{\footnotesize Observation of the north
Atlantic ocean from altimetry. See figure \ref{Fig:gulfstream}  page \pageref{Fig:gulfstream} for more detailed legends.}}

\label{Fig:gulfstream_ColorPage}
\end{figure}

Figure \ref{Fig:gulfstream_ColorPage} page \pageref{Fig:gulfstream_ColorPage}. Observation of the north
Atlantic ocean from altimetry. See figure \ref{Fig:gulfstream}  page \pageref{Fig:gulfstream} for more detailed legends.\\

\begin{figure}[h!]
\begin{center}
\includegraphics[height=0.25\textheight]{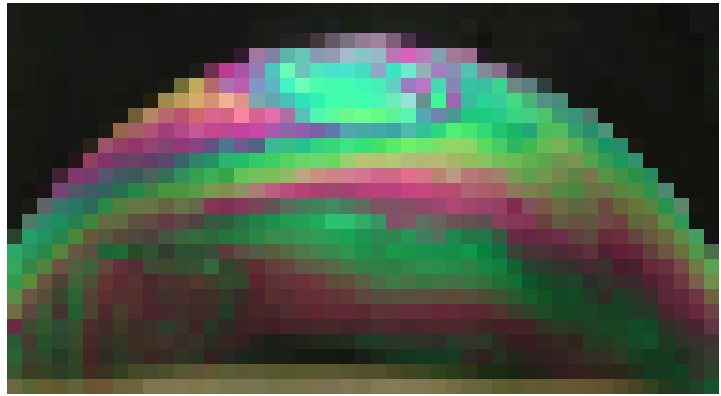}
\end{center}
\caption{{\footnotesize Example of an experimental realization of a 2D flow
in a soap bubble, courtesy of American Physical Society. See \cite{kellayPRL08}
 and \cite{Kellay_Glodburg_2002_Rep_Prog_Phsyics} for further details.}}

\label{Fig:bubble}
\end{figure}

Figure \ref{Fig:bubble} page \pageref{Fig:bubble}. Example of an experimental realization of a 2D flow
in a soap bubble, courtesy of American Physical Society. See \cite{kellayPRL08}
 and \cite{Kellay_Glodburg_2002_Rep_Prog_Phsyics} for further details.\\

\begin{figure}[h!]
\begin{center}
\includegraphics[height=0.45\textheight]{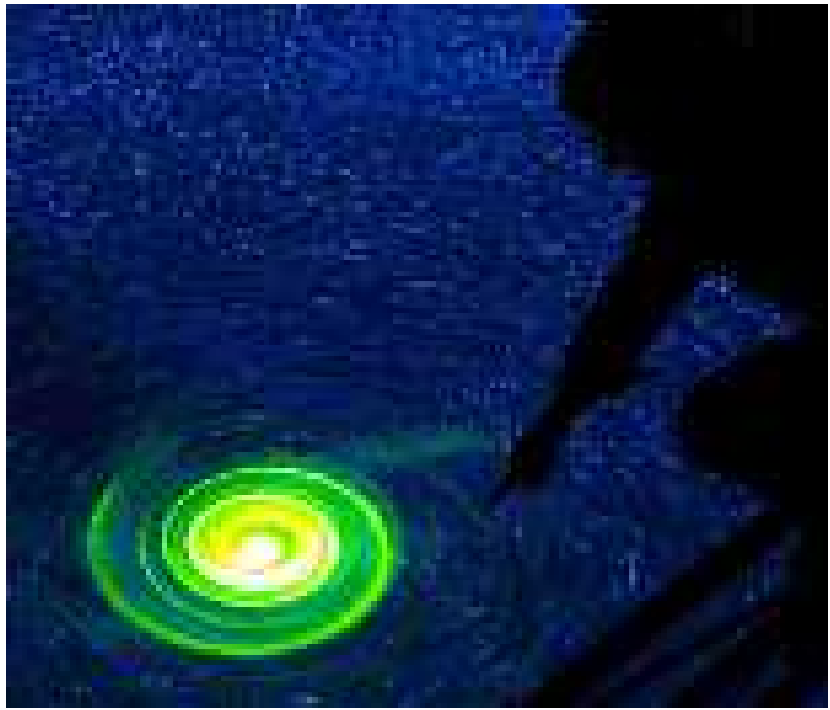}
\end{center}
\caption{{\footnotesize Experimental observation of a 2D long lived coherent
vortex on the $14 \textrm{m}$ diameter Coriolis turntable (photo gamma production).
 }}

\label{Fig:ColorCoriolis}
\end{figure}

Figure \ref{Fig:ColorCoriolis} page \pageref{Fig:ColorCoriolis}. Experimental observation of a 2D long lived coherent
vortex on the $14 \textrm{m}$ diameter Coriolis turntable (photo gamma production).\\

Figure  \ref{Fig:vertical_structure} page \pageref{Fig:vertical_structure}. Vertical structure of the 1.5-layer quasi-geostrophic model: a deep layer of density $\rho+\Delta\rho$ and a lighter upper
layer of thickness $H$ and density $\rho$. Because of the inertia
of the lower layer, the dynamics is limited to the upper layer.\\

Figure \ref{Fig:SelfOrganization} page \pageref{Fig:SelfOrganization}. Snapshot of electron density (analogous to vorticity
field) at successive time from an initial condition with two vortices
to a single large scale coherent structure via turbulent mixing (see
\cite{Scecter_etal_2000_PhysicsFluids,Schecter_Dubin_etc_Vortex_Crystals_2DEuler1999PhFl}).
The best experimental realization of inviscid 2D Euler equations is
probably so far achieved in those magnetized electron plasma experiments
where the electrons are confined in a Penning trap. The dynamics of
both systems are indeed isomorphic, where the electron density plays
the role of vorticity. The major drawback of this experimental setting
comes from its observation, since any measurement requires the destruction
of the plasma itself.\\

Figure \ref{delta_a4} page \pageref{delta_a4}. Bifurcation diagram for the statistical equilibria of the 2D Euler equations in a doubly periodic domain with aspect ratio $\delta$, in the limit where the normal form treatment is valid, in the $g$-$a_{4}$ parameter plane. The geometry parameter $g$ is inversely proportional to the energy and proportional to the difference between the two first eigenvalues of the Laplacian (or equivalently to $\delta-1$ in the limit of small $\delta-1$), the parameter $a_4$ measures the non-quadratic contributions to the Casimir functional. The solid line is a second order phase transition between a dipole (mixed state) and a parallel flow along the $y$ direction (pure state $X=0$). Along the dashed line, a metastable parallel flow (along the $x$ direction, pure state $X=1$) loses its stability.\\

Figure \ref{fig:Equilibre} page \pageref{fig:Equilibre}. Bifurcation diagrams for statistical equilibria of the 2D Euler equations in a doubly periodic domain
a) in the $g$-$a_{4}$ plane (see figure \ref{delta_a4}) b) obtained numerically in
 the $E-a_{4}$ plane, in the case of doubly periodic geometry with
aspect ratio $\delta=1.1$. The colored insets are streamfunction
and the inset curve illustrates good agreement between numerical and
theoretical results in the low energy limit.\\

Figure \ref{fig_f} page \pageref{fig_f}. The double well shape of the specific free energy $f\left(\phi\right)$
(see equation (\ref{eq:Van Der Waals Cahn Hilliard})). The function
$f\left(\phi\right)$ is even and possesses two minima at $\phi=\pm u$.
At equilibrium, at zeroth order in $R$, the physical system will
be described by two phases corresponding to each of these minima.\\

Figure \ref{fig_domaine} page \pageref{fig_domaine}. At zeroth order, $\phi$ takes the two values $\pm u$
on two sub-domains $A_{\pm}$. These sub-domains are separated by
strong jets. The actual shape of the structure, or equivalently the
position of the jets, is given by the first order analysis.\\

Figure \ref{fig_plateau} page \pageref{fig_plateau}. Illustration of the Plateau problem (or minimal area
problem) with soap films: the spherical bubble minimizes its area
for a given volume (Jean Simeon Chardin, \emph{\footnotesize Les bulles
de savon}{\footnotesize , 1734)}.\\

Figure \ref{Fig:ColorJupiter} page \pageref{Fig:ColorJupiter}. Observation of the Jovian atmosphere from Cassini (Courtesy
of NASA/JPL-Caltech). One of the most striking feature of the Jovian
atmosphere is the self organization of the flow into alternating eastward
and westward jets, producing the visible banded structure and the
existence of a huge anticyclonic vortex $\sim20,000\ km$ wide, located
around $20 �$ South: the Great Red Spot (GRS).
The GRS has a ring structure: it is a hollow vortex surrounded by
a jet of typical velocity $\sim100\ m.s^{\{-1\}}\ $ and width $\sim1,000\, km$.
Remarkably, the GRS has been observed to be stable and quasi-steady for
many centuries despite the surrounding turbulent dynamics. The explanation
of the detailed structure of the GRS velocity field and of its stability
is one of the main achievement of the equilibrium statistical mechanics
of two dimensional and geophysical flows (see figure \ref{fig:Emergence_Numerique_tache_rouge}
and section \ref{sec:First Order GRS and rings}).\\

Figure \ref{fig:Emergence_Numerique_tache_rouge} page \pageref{fig:Emergence_Numerique_tache_rouge}. Left: the observed velocity field is from Voyager spacecraft data,
from Dowling and Ingersoll \cite{Dowling_Ingersoll_1988JAtS...45.1380D}
; the length of each line is proportional to the velocity at that
point. Note the strong jet structure of width of order $R$, the Rossby
deformation radius. Right: the velocity field for the statistical
equilibrium model of the Great Red Spot. The actual values of the
jet maximum velocity, jet width, vortex width and length fit with
the observed ones. The jet is interpreted as the interface between
two phases; each of them corresponds to a different mixing level of
the potential vorticity. The jet shape obeys a minimal length variational
problem (an isoperimetrical problem) balanced by the effect of the
deep layer shear.\\

Figure \ref{fig_ellipse} page \pageref{fig_ellipse}. Left panel: typical vortex shapes obtained from the isoperimetrical
problem (curvature radius equation (\ref{Rayon_courbure})), for two
different values of the parameters (arbitrary units). The characteristic
properties of Jupiter's vortex shapes (very elongated, reaching extremal
latitude $y_{m}$ where the curvature radius is extremely large) are
well reproduced by these results. Central panel: the Great Red Spot
and one of the White Ovals. Right panel: one of the Brown Barge cyclones
of Jupiter's north atmosphere. Note the very peculiar cigar shape
of this vortex, in agreement with statistical mechanics predictions
(left panel).\\

Figure \ref{phase_top} page \pageref{phase_top}. Phase diagram of the statistical equilibrium states versus the energy $E$ and a parameter related to the asymmetry between positive
and negative potential vorticity $B$, with a quadratic topography.
The inner solid line corresponds to a phase transition, between vortex
and straight jet solutions. The dash line corresponds to the limit
of validity of the small deformation radius hypothesis. The dot lines
are constant vortex aspect ratio lines with values 2,10,20,30,40,50,70,80
respectively. We have represented only solutions for which anticyclonic
potential vorticity dominate ($B>0$). The opposite situation may
be recovered by symmetry. For a more detailed discussion of this figure,
the precise relation between $E$, $B$ and the results presented
in this review, please see \cite{Bouchet_Dumont_2003_condmat}.\\

Figure \ref{Fig:SouthernOcean} page \pageref{Fig:SouthernOcean}. Snapshot of surface velocity field from a comprehensive
numerical simulation of the southern Oceans \cite{HallbergMESO_06}.
Left: coarse resolution, the effect of mesoscale eddies ($\sim100km$)
is parameterized. Right: higher resolution, without parameterization
of mesoscale eddies. Note the formation of large scale coherent structure
in the high resolution simulation: there is either strong and thin
eastward jets or rings of diameter $\sim200\ km$. Typical velocity
and width of jets (be it eastward or around the rings) are respectively
$\sim1\ m.s^{-1}$ and $\sim20\ km$. The give a statistical mechanics
explanation and model for these rings.\\

Figure \ref{rings_shape} page \pageref{rings_shape}. Vortex statistical equilibria in the quasi-geostrophic model. It is a circular patch of (homogenized) potential vorticity in a background of homogenized potential vorticity, with two different mixing values. The velocity field (right panel) has a very clear ring structure, similarly to the Gulf-Stream rings and to many other ocean vortices. The width of the jet surrounding the ring has the order of magnitude of the Rossby radius of deformation $R$.\\

Figure \ref{fig:Chelton} page \pageref{fig:Chelton}. Altimetry observation of the westward drift of oceanic eddies (including rings) from \cite{Chelton07}, figure 4. The red line is the zonal average (along a latitude circle) of the propagation speeds of all eddies with life time greater than 12 weeks. The black line represents the velocity  $\beta_c R^2$ where $\beta_c$ is  the meridional gradient of the Coriolis parameter and  $R$ the first baroclinic Rossby radius of deformation. This eddy propagation speed is a prediction of statistical mechanics, when the linear momentum conservation, due to translational invariance, is taken into account (see section \ref{sub:The-westward-drift}).\\

Figure \ref{Fig:gulfstream} page \pageref{Fig:gulfstream}. Observation of the sea surface height of the north
Atlantic ocean (Gulf Stream area) from altimetry REF. As explained
in section \ref{sub:Euler-QG-equations}, for geophysical flows, the
surface velocity field can be inferred from the see surface height
(SSH): strong gradient of SSH are related to strong jets. The Gulf
stream appears as a robust eastward jet (in presence of meanders),
flowing along the east coast of north America and then detaching the
coast to enter the Atlantic ocean, with an extension $L\sim2000\ km$.
The jet is surrounded by numerous westward propagating rings of typical
diameters $L\sim200\ km$. Typical velocities and widths of both the
Gulf Stream and its rings jets are respectively $1\ m.s^{-1}$ and
$50\ km$, corresponding to a Reynolds number $Re\sim10^{11}$. Such
rings can be understood as local statistical equilibria, and strong
eastward jets like the Gulf Stream and obtained as marginally unstable
statistical equilibria in simple academic models (see subsections
\ref{sub:Gulf Stream Rings}-\ref{sec:Gulf Stream and Kuroshio}).\\

Figure \ref{Fig:frontBerloff} page \pageref{Fig:frontBerloff}. b) and c) represent respectively a snapshot of the streamfunction
and potential vorticity (red: positive values; blue: negative values)
in the upper layer of a three layers quasi-geostrophic model in a
closed domain, representing a mid-latitude oceanic basin, in presence
of wind forcing. Both figures are taken from numerical simulations
\cite{berloff}, see also \cite{BerloffHogg}. a) Streamfunction predicted
by statistical mechanics, see section \vref{sec:Gulf Stream and Kuroshio}
for further details. Even in an out-equilibrium situation like this one, the equilibrium statistical mechanics predicts correctly the overall qualitative structure of the flow.\\

Figure \ref{fig_pvfronts} page \pageref{fig_pvfronts}. a) Eastward jet: the interface is zonal, with positive
potential vorticity $q=u$ on the northern part of the domain. b)
Westward jet: the interface is zonal, with negative potential vorticity
$q=-u$ in the northern part of the domain. c) Perturbation of the
interface for the eastward jet configuration, to determine when this
solution is a local equilibrium (see subsection \ref{sub:Eastward-jets-with_topography}).
Without topography, both (a) and (b) are entropy maxima. With positive
beta effect (b) is the global entropy maximum; with negative beta
effect (a) is the global entropy maximum.\\

Figure \ref{fig:Fofonoff} page \pageref{fig:Fofonoff}. Phase diagrams of RSM statistical equilibrium states of the 1.5 layer quasi-geostrophic model,  characterized by a linear $q-\psi$ relationship, in a rectangular domain elongated in the $x$ direction.  $S(E,\Gamma)$ is the equilibrium entropy, $E$ is the energy and $\Gamma$ the circulation. Low energy states are the celebrated Fofonoff solutions  \cite{Fofonoff:1954_steady_flow_frictionless}, presenting a weak westward flow in the domain bulk.  High energy states have a very different structure (a dipole). Please note that at high energy the entropy is non-concave. This is related to ensemble inequivalence (see \ref{sub:Long-range-interactions} page \pageref{sub:Long-range-interactions}), which explain why such states were not  computed in previous studies. The method to compute explicitly this phase diagram is the same as the one presented in subsection \ref{sub:The-example-doubly-periodic} page \pageref{sub:The-example-doubly-periodic}.  See \cite{Venaille_Bouchet_PRL_2009} for more details.\\

Figure \ref{Fig:CascadeJoel} page \pageref{Fig:CascadeJoel}. First experimental observation of the inverse energy cascade
and the associated $k^{-5/3}$ spectrum, from \cite{Sommeria_1986_JFM_2Dinverscascade_MHD}.
The 2D turbulent flow is approached here by a thin layer of mercury
and a further ordering from a transverse magnetic field. The flow
is forced by an array of electrodes at the bottom, with an oscillating electric field. The parameter Rh is the ratio between inertial to bottom friction
terms. At low $Rh$ the flow has the structure of the forcing (left
panel). At sufficiently high $Rh$ the prediction of the self similar
cascade theory is well observed (right panel, bottom), and at even
higher $Rh$, the break up of the self similar theory along with the
organization of the flow into a coherent large scale flow is observed (see right
panel above).\\

Figure \ref{fig:Omega-Psi} page \pageref{fig:Omega-Psi}. $\omega-\psi$ scatter-plots (cyan) (see color
figure on the .pdf version). In black the same after time averaging
(averaging windows $1\ll\tau\ll1/\nu$, the drift due to translational
invariance has been removed)\textbf{. }Left: dipole case with $\delta=1.03$.
Right: unidirectional case $\delta=1.10$.\\

Figure \ref{ts_f} page \pageref{ts_f}. Dynamics of the 2D Navier--Stokes equations with stochastic forces in a doubly periodic domain of aspect ratio $\delta$, in a non-equilibrium phase transition regime. The two main plots are the time series and probability density functions (PDFs) of the modulus of the Fourier component $z_{1}=\frac{1}{\left(2\pi\right)^{2}}\int_{\mathcal{D}}\mathrm{d}\mathbf{r}\, \omega(x,y)\exp(iy)$ illustrating random changes between dipoles ($|z_1| \simeq 0.55$) and unidirectional flows ($|z_1| \simeq 0.55$). As discussed in section \ref{sub:Phase Transition 2}, the existence of such a non-equilibrium phase transition can be guessed from equilibrium phase diagrams (see figure \ref{fig:Equilibre}).\\

Figure \ref{fig:Swinney} page \pageref{fig:Swinney}.  Bistability in a rotating tank experiment with topography (shaded area)\cite{Tian_Weeks_etc_Ghil_Swinney_2001_JFM_JetTopography,Weeks_Tian_etc_Swinney_Ghil_Science_1997}. The dynamics in this experiment would be well modeled by a 2D barotropic model with topography (the quasi-geostrophic model with $R=\infty$). The flow is alternatively close to two very distinct states, with random switches from one state to the other. Left: the streamfunction of each of these two states. Right: the time series of the velocity measured at the location of the black square on the left figure, illustrating clearly the bistable behavior. The similar theoretical structures for the 2D Euler equations on one hand and the quasi-geostrophic model on the other hand, suggest that the bistability in this experiment can be explained as a non equilibrium phase transition, as done in section \ref{sub:Phase Transition 2} (see also figure \ref{ts_f}).\\

Figure \ref{fig:SST-Kuroshio} page \pageref{fig:SST-Kuroshio}. Kuroshio: sea surface temperature of the pacific ocean east of Japan,
February 18, 2009, infra-red radiometer from satellite (AVHRR, MODIS)
(New Generation Sea Surface Temperature (NGSST), data from JAXA (Japan
Aerospace Exploration Agency)).\protect \\
 The Kuroshio is a very strong current flowing along the coast,
south of Japan, before penetrating into the Pacific ocean. It is similar
to the Gulf Stream in the North Atlantic. In the picture, The strong
meandering color gradient (transition from yellow to green) delineates
the path of the strong jet (the Kuroshio extension) flowing eastward
from the coast of Japan into the Pacific ocean.\protect \\
 South of Japan, the yellowish area is the sign that, at the time
of this picture, the path of the Kuroshio had detached from the Japan
coast and was in a meandering state, like in the 1959-1962 period
(see figure \ref{fig:Kuroshio-Path}).\\

Figure \ref{fig:Kuroshio-Path} page \pageref{fig:Kuroshio-Path}. Bistability of the paths of the Kuroshio during the 1956-1962 period
: paths of the Kuroshio in (left) its small meander state and (right)
its large meander state. The 1000-m (solid) and 4000-m (dotted) contours
are also shown. (figure from Schmeits and Dijkstraa \cite{Schmeits_Dijkstraa_2001_JPO_BimodaliteGulfStream},
adapted from Taft 1972).\\

Figure \ref{fig:Kuroshio-Time-Series} page \pageref{fig:Kuroshio-Time-Series}. Bistability of the paths of the Kuroshio, from Qiu and Miao \cite{Qiu_Miao_2000JPO....30.2124Q}:
time series of the distance of the Kuroshio jet axes from the coast,
averaged other the part of the coast between 132 degree and 140 degrees East,
from a numerical simulation using a two layer primitive equation model.\\

Figure \ref{fig:shear_vorticity} page \pageref{fig:shear_vorticity}. Evolution of $\omega(x,y,t)$ from an initial vorticity perturbation
$\omega(x,y,0)=\omega_{1}\left(y,0\right)\cos\left(x\right)$, by
the linearized 2D Euler equations close to a shear flow $U\left(y\right)=\sigma y$
 (colors in the .PDF document).\\

Figure \ref{fig:Vorticity_Depletion} page \pageref{fig:Vorticity_Depletion}. Evolution of the vorticity perturbation
$\omega(x,y,t)=\omega\left(y,t\right)\exp\left(ikx\right)$, close
to a parallel flow $\mathbf{v}_{0}(x,y)=U(y)\mathbf{e}_{x}$ with
$U\left(y\right)=\cos\left(y\right)$, in a doubly periodic domain
with aspect ratio $\delta$. The figure shows the modulus of the perturbation
$\left|\omega\left(y,t\right)\right|$ as a function of time and $y$.
One clearly sees that the vorticity perturbation rapidly converges
to zero close to the points where the velocity profile $U\left(y\right)$
has extrema ($U'(y_{0})=0$, with $y_{o}=0$ and $\pi$). This\emph{
depletion of the perturbation vorticity} at the stationary streamlines
$y_{0}$ is a new generic self-consistent mechanism, understood mathematically
as the regularization of the critical layer singularities at the edge
of the continuous spectrum (see \cite{Bouchet_Morita_2010PhyD}).\\

Figure \ref{fig:v1_stseri} page \pageref{fig:v1_stseri}. The space-time series of perturbation velocity components, $|v_{\delta,x}(y,t)|$
(a) and $|v_{\delta,y}(y,t)|$ (b), for the initial perturbation profile
$\cos\left(x/\delta\right)$ in a doubly periodic domain with aspect
ratio $\delta=1.1$. Both the components relax toward zero, showing
the asymptotic stability of the Euler equations (colors in the .pdf
document). \\

Figure \ref{fig:decroissance_vitesse} page \pageref{fig:decroissance_vitesse}. The time series of perturbation velocity components $|v_{\delta,x}(y,t)|$
(a) and $|v_{\delta,y}(y,t)|$ (b) at three locations, $y=0$ (vicinity
of the stationary streamline) (red), $y=\pi/4$ (green), and $y=\pi/2$
(blue), for the initial perturbation profile $A(y)=1$ and the aspect
ratio $\delta=1.1$. We observe the asymptotic forms $|v_{\delta,x}(y,t)|\sim t^{-\alpha}$,
with $\alpha=1$, and $|v_{\delta,y}(y,t)|\sim t^{-\beta}$, with
$\beta=2$, in accordance with the theory for the asymptotic behavior
of the velocity (equations (\ref{eq:Vitesse-x-algebraic-linear})
and (\ref{eq:Vitesse-y-algebraic-linear})). The initial perturbation
profile is $\cos\left(x/\delta\right)$ in a doubly periodic domain
with aspect ratio $\delta=1.1$ (colors in the .pdf document).\\

Figure \ref{fig:w0_stseri} page \pageref{fig:w0_stseri}. The space-time series of the $x$-averaged perturbation vorticity,
$\omega_{0}(y,t)=\omega(y,t$)-$\Omega_{0}\left(y,0\right)$. The
initial condition is $\omega\left(y,t\right)=\Omega_{0}\left(y,0\right)+\epsilon\cos\left(x\right)$,
in a doubly periodic domain with aspect ratio $\delta=1.1$ (colors
in the .pdf document).\\

\newpage

\newpage

\newpage

\section{Introduction}

\subsection{Two-dimensional and geostrophic turbulence}

For many decades, two-dimensional turbulence has been a very active
subject for theoretical investigations, motivated not only by the
conceptual interest in understanding atmosphere and ocean turbulence,
but also by the beauty and precision of the theoretical and mathematical
achievements obtained thereby. For over two decades, two-dimensional
flows have been studied experimentally in many different laboratory
setups, as for instance illustrated in figures \ref{Fig:ColorCoriolis},
\ref{Fig:SelfOrganization}, \ref{Fig:CascadeJoel}, and \ref{fig:Swinney} (see also \cite{Kellay_Glodburg_2002_Rep_Prog_Phsyics,Sommeria_2001_CoursLesHouches}
and references therein for further details).\\

Although they both involve a huge range of temporal and spatial scales,
two-dimensional and three-dimensional turbulent flows are very different
in nature.

The first difference is that whereas in three-dimensional turbulence
energy flows forward (from the largest towards the smallest scales),
it flows backward (from the smallest towards the largest scales) in
two-dimensional turbulence. Three-dimensional turbulence transfers
energy towards the viscous scale where it is dissipated into heat at a finite rate,
no matter how small the viscosity. By contrast, in the absence of
any strong dissipation mechanism at the largest scales, the dissipation
of energy remains weak in two-dimensional turbulence. As a consequence,
the flow dynamics is dominated by large scale coherent structures,
such as vortices or jets. This review is devoted to the understanding
and prediction of these stable and quasi-steady structures in two-dimensional
turbulent flows.

The second fundamental difference between 2D and 3D turbulence is
that the level of fluctuations in two-dimensional turbulence is very
small. The largest scales of three-dimensional turbulent flows are
the place of incessant instabilities, whereas the largest scales of
two-dimensional turbulence are often quasi-stationary and evolve over
a very long time scale, compared for instance to the turnover time
of the large scale coherent structures.

As explained in this review, the above-mentioned peculiarities
of two-dimensional turbulent flows are theoretically understood as
the consequences of dynamical invariants of two-dimensional perfect
flows, which are not invariants of perfect three-dimensional flows.
These invariants, including the enstrophy, make the forward energy
cascade impossible in two-dimensional flows, and explain the existence
of an extremely large number of stable stationary solutions of the
2D Euler equations, playing a major role in the dynamics.\\

Atmospheric and oceanic flows are three-dimensional, but strongly
dominated by the Coriolis force, mainly balanced by pressure gradients
(geostrophic balance). The turbulence that develops in such flows
is called geostrophic turbulence. Models describing it have the same
type of additional invariants as two-dimensional turbulence has. As
a consequence, energy flows backward and the main phenomenon is the
formation of large scale coherent structures (jets, cyclones and anticyclones)
(see figures \ref{Fig:ColorJupiter} and \ref{Fig:gulfstream}). The
analogy between two-dimensional turbulence and geophysical turbulence
is further emphasized by the theoretical similarity between the 2D Euler
equations -- describing 2D flows -- and the layered quasi-geostrophic
or shallow-water models -- describing the largest scales of geostrophic
turbulence --: both are transport equations of a scalar quantity by
a non-divergent flow, conserving an infinity of invariants.

The formation of large scale coherent structures is a fascinating
problem and an essential part of the dynamics of Earth's atmosphere
and oceans. This is the main motivation for setting up a theory for
the formation of the largest scales of geostrophic and two-dimensional
turbulence.

\subsection{Turbulence and statistical mechanics}

Any turbulence problem involves a huge number of degrees of freedom
coupled via complex nonlinear interactions. The aim of any theory
of turbulence is to understand the statistical properties of the velocity
field. It is thus extremely tempting and interesting to attack these
problems from a statistical mechanics point of view. Statistical mechanics
is indeed a very powerful theory that allows us to reduce the complexity
of a system down to a few thermodynamic parameters. As an example,
the concept of phase transition allows us to describe drastic changes
of the whole system when a few external parameters are changed. Statistical
mechanics is the main theoretical approach that we develop in this review, and we show that it succeeds in explaining many
of the phenomena associated with two-dimensional turbulence.

This may seem surprising at first, as it is a common belief that statistical
mechanics is not successful in handling turbulence problems. The reason
for this belief is that most turbulence problems are intrinsically far
from equilibrium. For instance, the forward energy cascade
in three-dimensional turbulence involves a finite energy dissipation
flux no matter how small the viscosity (anomalous dissipation). Because
of this flux, the flow cannot be considered close to some equilibrium
distribution. By contrast, two-dimensional turbulence does not suffer
from this problem (there is no anomalous dissipation of the energy),
so that equilibrium statistical mechanics, or close to equilibrium
statistical mechanics makes sense when small fluxes are present.\\

The first attempt to use equilibrium statistical mechanics ideas to
explain the self-organization of 2D turbulence comes from Onsager
in 1949 \cite{Onsager:1949_Meca_Stat_Points_Vortex} (see \cite{Eyink_Sreenivasan_2006_Rev_Modern_Physics}
for a review of Onsager's contributions to turbulence theory). Onsager
worked with the point-vortex model, a model made of singular point
vortices, first used by Lord Kelvin and which is a special class of
solution of the 2D Euler equations. The equilibrium statistical mechanics
of the point-vortex model has a long and very interesting history,
with wonderful pieces of mathematical achievements \cite{Onsager:1949_Meca_Stat_Points_Vortex,Joyce_Montgommery_1973,CagliotiLMP:1995_CMP_II(Inequivalence),Kiessling_Lebowitz_1997_PointVortex_Inequivalence_LMathPhys,Dubin_ONeil_1988_PhysRevLett_Kinetic_Point_Vortex,Chavanis_houches_2002,Eyink_Spohn_1993_JSP....70..833E,IUTAM_Symposium08}.
In order to treat flows with continuous vorticity fields, another
approach, taking account of the quadratic invariants only, was proposed
by Kraichnan \cite{Kraichnan_Motgommery_1980_Reports_Progress_Physics}.
This last work has inspired a quadratic-invariant statistical theory
for quasi-geostrophic flows over topography: the Salmon--Holloway--Hendershott
theory \cite{SalmonHollowayHendershott:1976_JFM_stat_mech_QG,Salmon_1998_Book}.
Another phenomenological approach based on a minimal enstrophy principle and  leading to similar predictions for the large scale flow as the Salmon--Holloway--Hendershott theory has been independently proposed by Bretherton-Haidvogel \cite{BrethertonHaidvogel}.
The generalization of Onsager's ideas to the 2D Euler equation with continuous
vorticity field, taking into account all invariants, has been proposed
in the beginning of the 1990s \cite{Robert:1990_CRAS,Miller:1990_PRL_Meca_Stat,Robert:1991_JSP_Meca_Stat,SommeriaRobert:1991_JFM_meca_Stat},
leading to the Robert--Sommeria--Miller theory (RSM theory). The
RSM theory includes the previous Onsager, Kraichnan, Salmon--Holloway--Hendershott and Bretherton-Haidvogel theories
and determines the particular limits %
\footnote{Corresponding to special classes of initial conditions%
} within which those give relevant predictions and general results.
The part of this review dealing with equilibrium statistical mechanics
mainly falls within the framework of the RSM theory and presents its
further developments.

Over the last fifteen years, the RSM equilibrium theory has been applied
successfully to a large class of problems, for both the Euler and
quasi-geostrophic equations. We cite and describe all relevant works and contributions to this subject.  The aim of this review is also pedagogical, and as such we have chosen to emphasize
on a class of problems that can be understood using analytical solutions. We give a comprehensive description only of those works. Nevertheless, this includes many interesting applications, such as predictions
of phase transitions in different contexts, a model for the Great
Red Spot and other Jovian vortices, and models of ocean vortices and
jets.\\

Most turbulent flows are forced, and reach a statistically steady state
where forcing is balanced on average by a dissipative mechanism. Such situations
are referred to in statistical mechanics as Non-Equilibrium Steady
States (NESS). One class of such problems in two-dimensional turbulence
are the self-similar inertial cascades first described by Kraichnan
\cite{Kraichnan_1967PhFl...10.1417K}: the backward energy cascade
and the forward enstrophy cascade. These are essential concepts of
two-dimensional turbulence that will be briefly described. However,
in the regime where the flow is dominated by large scale coherent
structures, these self-similar cascades are no longer relevant and Kraichnan's
theory provides no prediction. We will explain in this review how
the vicinity to statistical equilibrium can be invoked in order to
provide partial responses to the description of the non-equilibrium
situations, for instance prediction of non-equilibrium phase transitions.
We will also emphasize why and how such predictions based on equilibrium
statistical mechanics are necessarily limited in scope, and explain
how a non-equilibrium theory can be foreseen based on kinetic theory
approaches.

\subsection{About this review}

The aim of this review is to give a self-contained description
of statistical mechanics of two-dimensional and geophysical turbulence,
and of its applications to real flows. For pedagogical purposes, we will emphasize
analytically solvable cases, so the physics can be easily understood.

The typical audience should be graduate students and researchers from
different fields. One of the difficulty with this review is that
knowledge is required from statistical physics \cite{Landau_Lifshitz_1996_Book},
thermodynamics \cite{Callen_Thermodynamics_1985tait.book.....C},
geophysical fluid dynamics \cite{PedloskyBook,VallisBook,Salmon_1998_Book,GillBook}
and two-dimensional turbulence \cite{Kraichnan_Motgommery_1980_Reports_Progress_Physics,Sommeria_2001_CoursLesHouches,Tabeling02}.
For each of these subjects, the notions needed will be briefly presented,
in a self-contained way, but we refer to classical textbooks or review
papers for more detailed presentations.

There already exist several presentations of the equilibrium statistical
mechanics of two-dimensional and geostrophic turbulent flows \cite{Sommeria_2001_CoursLesHouches,Majda_Wang_Book_Geophysique_Stat},
some emphasizing kinetic approaches of the point-vortex model \cite{Chavanis_houches_2002},
other focusing on the legacy of Onsager \cite{Eyink_Sreenivasan_2006_Rev_Modern_Physics}.
Parts of the introductory sections of this review (two-dimensional
fluid mechanics and the mean-field equilibrium statistical mechanics
theory) are similar to those found in previous reviews or lectures (especially
\cite{Sommeria_2001_CoursLesHouches}). However, the statistical mechanics
foundations of the theory is explained in further details and none of the
applications discussed in this review, with emphasis on analytically
solvable cases, were described in previous books or reviews. For instance,
the present review gives i) a precise explanation of the statistical
mechanics basis of the theory, ii) a detailed discussion of the validity
of the mean-field approximation, iii) an analytic treatment of phase
diagrams for small energy and analytic models for the Great Red Spot
as well as for ocean jets and vortices, iv) a detailed discussion
of the irreversible behavior of the 2D Euler equations despite its
being actually a time reversible equation. In addition, we present new
results on non-equilibrium studies, on the different regime description,
on non-equilibrium phase transitions and kinetic theories. Most of
these new results have been derived over the last few years. Other
important recent developments of the theory such as statistical ensemble
inequivalence \cite{EllisHavenTurkington:2000_Inequivalence,EllisHavenTurkington:2002_Nonlinearity_Stability}
and related phase transitions \cite{Bouchet_Barre:2005_JSP} would
be natural extensions of this review, but were considered too advanced
for such a first introduction. We however always describe the main results and give the appropriate references to the appropriate papers, for an interested reader to be able to understand these more technical points.

We apologize that this review leaves little room for the description
of experiments, for the cascade regimes of two dimensional turbulence
or for the kinetic theory of the point-vortex model. For these we
refer the reader to \cite{Sommeria_2001_CoursLesHouches,Tabeling02}, \cite{Kraichnan_Motgommery_1980_Reports_Progress_Physics,Bernard_Boffetta_Celani_Falkovich_2007PRL,Eyink_Aluie_2009PhFl...21k5107E,Eyink_Aluie_II_2009PhFl...21k5108A}
and \cite{Dubin_ONeil_1988_PhysRevLett_Kinetic_Point_Vortex,Chavanis_houches_2002}
respectively. Interesting related problems insufficiently covered
in this review also include the mathematical works on the point-vortex
model \cite{CagliotiLMP:1995_CMP_II(Inequivalence),Kiessling_Lebowitz_1997_PointVortex_Inequivalence_LMathPhys,Eyink_Spohn_1993_JSP....70..833E}
or on the existence of invariant measures and their properties for
the 2D stochastic Navier-Stokes equation \cite{Kuksin_Penrose_2005_JPhysStat_BalanceRelations,Kuksin_Shirikyan_2000_CMaPh,Mattingly_Sinai_1999math_3042M},
as well as studies of the self-organization of quasi-geostrophic
jets on a beta-plane (see \cite{Held_Larichev_1996JAtS...53..946H,Lapeyre_Held_2003JAtS,Thomson_Young_2007JAts,Farrel_Ioannou_2009JAtS,Dritschel_McIntyre_2008JAtS} and references therein).

\subsection{Detailed outline}

Section \ref{sec:2D-Geostrophic-Turbulence} is a general presentation
of the equations and phenomenology of two-dimensional (2D Euler equations)
and geophysical turbulence. One of the simplest possible models for
geophysical flows, namely the 1.5-layer quasi-geostrophic model (also
called Charney--Hasegawa--Mima model), is presented in section \ref{sub:Euler-QG-equations}.

Section \ref{sub:Theory-General-Euler-QG} deals with important properties
of 2D Euler and quasi-geostrophic equations, and their physical consequences:
the Hamiltonian structure (section \ref{sub:Hamiltonian-structure}),
the existence of an infinite number of conserved quantities (section
\ref{sub:Casimirs-conservation-laws}).

These conservation laws play a central part in the theory. They are
for instance responsible for: i) the existence of multiple stationary solutions of the 2D Euler equations and the stability of some of these states (section
\ref{sub:multiple_equilibria}), ii) the cascade phenomenology, with
energy transferred upscale, and enstrophy downscale (section \ref{sub:Fjortof_Argument}),
iii) the most striking feature of 2D and geophysical flows: their
self-organization into large scale coherent structures (section \ref{sub:Self_Organization}),
iv) the non-trivial predictions of equilibrium statistical mechanics
of two dimensional turbulence, compared to statistical mechanics of
three-dimensional turbulence (sections \ref{sub:3D-Stat-mech-and-Energy-Enstrophy}
and \ref{sub:Mean_Field}). Sections \ref{sub:3D-Stat-mech-and-Energy-Enstrophy}
and \ref{sub:Mean_Field} also explain in details the relations
between the Kraichnan energy-enstrophy equilibrium theory and the
Robert-Sommeria-Miller theory, and justify the validity of a mean
field approach.\\

The self-organization of two-dimensional and geostrophic flows is
the main motivation for a statistical mechanics approach of the problem.
The presentation of the equilibrium theory is the aim of section \ref{sec:Equilibrium-statistical-mechanics}.
A reader more interested in applications than in the statistical mechanics
basis of the theory can start her reading at the beginning of section
\ref{sec:Equilibrium-statistical-mechanics}.

Section \ref{sub:The-mixing-entropy} explains how the microcanonical
mean field variational problem describes statistical equilibria. All
equilibrium results presented afterwards rely on this variational
problem. The ergodicity hypothesis is also discussed in section \ref{sub:The-mixing-entropy}.
Section \ref{sub:Canonical-and-Grand-Canonical} explains the practical
and mathematical interest of canonical ensembles, even if they are
not really relevant from a physical point of view. Section \ref{sub:Long-range-interactions}
explains the relations between the statistical mechanics of two dimensional
flows and the statistical mechanics of other systems with long range
interactions.

An analytically solvable case of phase transitions in a doubly periodic
domain is presented section \ref{sub:The-example-doubly-periodic}.
This example chosen for its pedagogical interest, illustrate the scope
and type of results one can expect from statistical theory of two-dimensional
and geophysical flows. The concepts of bifurcations, phase transitions,
and phase diagrams reducing the complexity of turbulent flows to a
few parameters are emphasized. \\

Section \ref{sec:First Order GRS and rings} is an application of
the equilibrium statistical mechanics theory to the explanation of
the stability and formation, and precise modeling of large scale vortices
in geophysical flows, such as Jupiter's celebrated Great Red Spot
and the ubiquitous oceanic mesoscale rings. The analytical computations
are carried out in the limit of a small Rossby radius (the typical
length scale characterizing geostrophic flows) compared to the
domain size, through an analogy with phase coexistence in classical
thermodynamics (for instance the equilibrium of a gas bubble in a
liquid).

Section \ref{sub:Van Der Waals} gives an account of the Van der Waals--Cahn--Hilliard
model of first order transitions, which is the relevant theoretical
framework for this problem. The link between Van der Waals--Cahn--Hilliard
model and the statistical equilibria of the 1.5-layer quasi-geostrophic
model is clarified in subsection \ref{sub:QG strong jet generql}.
This analogy explains the formation of strong jets in geostrophic
turbulence. All the geophysical applications presented in this review
come from this result.

Subsection \ref{sub:Gulf Stream Rings} deals with the application
to mesoscale ocean vortices. Their self-organization into circular
rings and their observed westward drift are explained as a result
of equilibrium statistical mechanics.

Subsection \ref{sub:Application-to-Jupiter} deals with the application
to Jovian vortices. The stability and shapes of the Red Great Spot,
white ovals and brown barges are explained by equilibrium statistical
mechanics. A detailed comparison of statistical equilibrium predictions
with the observed velocity field is provided. These detailed quantitative
results are one of the main achievements of the application of the
statistical equilibrium theory.\\

Section \ref{sec:Gulf Stream and Kuroshio} gives another application
of the statistical theory, now to the self-organization of ocean currents.
By considering the same analytical limit and theoretical framework
as in the previous section, we investigate the applicability of the
equilibrium statistical theory to the description of strong mid-latitude
eastward jets, such as the Gulf Stream or the Kuroshio (north Pacific
Ocean). These jets are found to be marginally stable. The variations
of the Coriolis parameter (beta effect) or a possible zonal deep current
are found to be key parameters for the stability of these flows.
\\

Section \ref{sec:Out of equilibrium} deals with non-equilibrium situations:
Non-Equilibrium Steady States (NESS), where an average balance between
forces and dissipation imposes fluxes of conserved quantity (sections
\ref{sub:NESS Two regimes} to \ref{sub:Kinetic_Theory}) and relaxation
towards equilibrium (section \ref{sub:Relaxation-towards-equilibrium}).
Section \ref{sub:NESS Two regimes} is a general discussion about
the 2D Navier-Stokes equations and conservation laws. The two regimes
of two-dimensional turbulence, the inverse energy cascade and direct
enstrophy cascade on one hand, and the regime dominated by large scale
coherent structures on the other hand, are clearly delimited in sections
\ref{sub:Kraichnan} and \ref{sub:Second-regime:Large-Scales}. Section
\ref{sub:Phase Transition-1} delineates what can be learned from
equilibrium statistical mechanics, and what cannot, in a non-equilibrium
context. We also present predictions of non-equilibrium phase transitions
using equilibrium phase diagrams and compare these predictions with
direct numerical simulations. Section \ref{sub:Kinetic_Theory} comments
progresses and challenges for a non-equilibrium theory based on kinetic
theory approach. Section \ref{sub:Relaxation-towards-equilibrium}
presents recent results on the asymptotic behavior of the linearized
2D Euler equations and relaxation towards equilibrium of the 2D Euler
equations.

\newpage

\section{Two-dimensional and geostrophic turbulence\label{sec:2D-Geostrophic-Turbulence}}

In this section, we present the 2D Euler equations and the quasi-geostrophic
equations, the simplest model of geophysical flows such as ocean
or atmosphere flows. We also describe the Hamiltonian structure of these
equations, the related dynamical invariants. The consequences of these
invariants are explained: i) for the inverse energy cascade, ii) for
the existence of multiple (stable and unstable) steady states for
the equations.

\subsection{2D Euler and quasi-geostrophic equations \label{sub:Euler-QG-equations}}

\subsubsection{2D Euler equations}

The incompressible 3D Euler equations describe the momentum transport
of a perfect and non-divergent flow. They read

\begin{equation}
\partial_{t}\mathbf{u}+\mathbf{u}\cdot\nabla\mathbf{u}=-\frac{1}{\rho}\nabla P\quad\text{with}\quad\nabla \cdot \mathbf{u}=0\ .\label{eq:Euler 2D Incompressible Vitesse}\end{equation}
 where $\mathbf{u}=(\mathbf{v},w)=\mathbf{v}+w\mathbf{e}_{z}$ ($\mathbf{v}$
is the projection of $\mathbf{u}$ in the plane ($\mathbf{e}_{x}$,$\mathbf{e}_{y}$)). The density $\rho$ is assumed to be  constant.  If we assume the flow to be two-dimensional ($w=0$ and $\mathbf{v}=\mathbf{v}\left(\mathbf{r}\right)$
with $\mathbf{r}=(x,y)$), then it is easily verified that the vorticity
is a scalar quantity: $\nabla\times\mathbf{v}$ is along $\mathbf{e}_{z}$.
Defining the vorticity as $\omega=\left(\nabla\times\mathbf{v}\right)\centerdot\mathbf{e}_{z}$,
the 2D Euler equations take the simple form of a conservation law
for the vorticity. Indeed, taking the curl of (\ref{eq:Euler 2D Incompressible Vitesse})
gives

\begin{equation}
\partial_t \omega + \mathbf{v}\boldsymbol{\centerdot\nabla}\omega=0\,\,\mathbf{;\,\, v}=\mathbf{e}_{z}\times\boldsymbol{\nabla}\psi\,\,;\,\,\omega=\Delta\psi,\label{eq:Euler_2D_Vorticity}\end{equation}
where we have expressed the non divergent velocity as the curl of
a streamfunction $\psi$. We complement the equation $\omega=\Delta\psi$
with boundary conditions: if the flow takes place in a simply connected
domain $\mathcal{D}$, then the condition that $\mathbf{v}$ has no
component along the normal to the interface (impenetrability condition)
imposes $\psi$ to be constant on the interface. This constant being
arbitrary, we impose $\psi=0$ on the interface. We may also consider
flows on a doubly periodic domain $(0,\ 2\pi\delta)\times(0,\ 2\pi)$ of aspect ratio $\delta$,
in which case $\psi(x+2\pi\delta,y)=\psi(x,y)$ and $\psi(x,y+2\pi)=\psi(x,y)$.\\

The (purely kinetic) energy of the flow reads \begin{equation}
\mathcal{E}\left[\omega\right]=\frac{1}{2}\int_{\mathcal{D}}\mathrm{d}\mathbf{r}\,\mathbf{v}^{2}=\frac{1}{2}\int_{\mathcal{D}}\mathrm{d}\mathbf{r}\,\left(\nabla\psi\right)^{2}=-\frac{1}{2}\int_{\mathcal{D}}\mathrm{d}\mathbf{r}\,\omega\psi,\label{eq:Energy}\end{equation}
 where the last equality has been obtained with an integration by parts. This quantity is conserved by the dynamics ( $d_{t}\mathcal{E}=0$).  As will be seen in section \ref{sub:Theory-General-Euler-QG}, the 2D Euler
equations have an infinity of other conserved quantities.

Given the strong analogies between the 2D Euler and quasi-geostrophic
equations, we further present the theoretical properties of both equations
in section \ref{sub:Theory-General-Euler-QG}.\\

In the preceding paragraph, we started from the 3D Euler equation
and assumed that the flow is two-dimensional. A natural question to
raise is whether such two-dimensional flows actually exist. Over the
last decades, a number of experimental realizations of two-dimensional
flows have been performed. Two-dimensionality can be achieved using
strong geometrical constraints, for instance soap film flows \cite{Bruneau_Kellay_2005PhRvE,Kellay_Glodburg_2002_Rep_Prog_Phsyics}
(see figure \ref{Fig:bubble}, page \pageref{Fig:bubble}) or very
thin fluid layers over denser fluids \cite{Marteau_Cardoso_Tabeling_1995PhRvE,Paret_Tabeling_1998_PhysFluids}
(figure \ref{Fig:CascadeTabeling} page \pageref{Fig:CascadeTabeling}). Another way to achieve two-dimensionality
is to use a very strong transverse ordering field: a strong transverse
magnetic field in a metal liquid setup \cite{Sommeria_1986_JFM_2Dinverscascade_MHD}
(see figure \ref{Fig:CascadeJoel}, page \pageref{Fig:CascadeJoel}),
or the Coriolis force on a rapidly rotating fluids (see figure \ref{Fig:ColorCoriolis},
page \pageref{Fig:ColorCoriolis}). Another original way to mimic
the 2D Euler equations (\ref{eq:Euler_2D_Vorticity}) is to look at
the dynamics of electrons in a Penning trap \cite{Schecter_Dubin_etc_Vortex_Crystals_2DEuler1999PhFl,Scecter_etal_2000_PhysicsFluids}
(see figure \ref{Fig:SelfOrganization}, page \pageref{Fig:SelfOrganization}).

\subsubsection{Large scale geophysical flows: the geostrophic balance\label{sub:Geostrophic_Equilibrium}}

The quasi-geostrophic equations are the simplest relevant model to
describe mid- and high-latitude atmosphere and ocean flows. The model
itself will be presented in section \ref{sub:Quasi-Geostrophic Model}.
To understand its physics, we need to introduce four fundamental concepts
of geophysical fluid dynamics: beta-plane approximation, hydrostatic
balance, geostrophic balance and Rossby radius of deformation.
This section gives a basic introduction to these concepts, that is
sufficient for understanding the discussions in the following sections
; a more precise and detailed presentation can be found in geophysical
fluid dynamics textbooks \cite{PedloskyBook,VallisBook,Salmon_1998_Book,GillBook}.\\

To begin with, we write the momentum equations in a rotating frame
($\mathbf{\boldsymbol{\Omega}}$ being the Earth's rotation vector),
with gravity $\mathbf{g}$, in Cartesian coordinates, calling $\mathbf{e}_{z}$
the vertical direction (upward) along $\mathbf{g}$, $\mathbf{e}_{y}$
the meridional direction (northward), and $\mathbf{e}_{x}$ the zonal
direction (eastward) \begin{equation}
\partial_{t}\mathbf{u}+\mathbf{u}\cdot\nabla\mathbf{u}+2\boldsymbol{\Omega}\times\mathbf{u}=-\frac{1}{\rho}\nabla P+\mathbf{g}\,.\label{eq:Euler_Rotation}\end{equation}

\paragraph*{Beta-plane approximation}

One can show that for mid-latitude oceanic basin of typical meridional
extension $L\sim5000\ km$, the lowest order effect of Earth's sphericity
appears only through the projection of Earth's rotation vector on
the local vertical axis: $f=2\boldsymbol{\Omega}\cdot\mathbf{e}_{z}=f_{0}+\beta_{c}y$,
with $f_{0}=2\Omega\sin\theta_{0}$, where $\theta_{0}$ is the mean
latitude where the flow takes place, and $\beta_{c}=2\Omega\cos\theta_{0}/r_{e}$,
where $r_{e}$ is the Earth's radius \cite{VallisBook,PedloskyBook}.
At mid latitudes $\theta_{0}\sim45^0$, so
that $f_{0}\sim10^{-4}\ s^{-1}$ and $\beta_{c}\sim10^{-11}\ m^{-1}s^{-1}$.

\paragraph*{Hydrostatic balance}

Recalling $\mathbf{u}=\left(\mathbf{v},w\right)$, the momentum equations
(\ref{eq:Euler_Rotation}) along the vertical axis read

\[
\partial_{t}w+\mathbf{u}\cdot\nabla w+2\left(\boldsymbol{\Omega}\times\mathbf{\mathbf{u}}\right)\cdot\mathbf{e}_{z}=-\frac{1}{\rho}\partial_{z}P+g.\]
In the ocean or atmosphere context, an estimation of the order of
magnitude of each term \cite{VallisBook,PedloskyBook} lead to the
conclusion that the dominant terms are the vertical pressure gradients
and gravitation. Neglecting the others gives the hydrostatic balance:
\begin{equation}
\partial_{z}P=-\rho g\,.\label{eq:Hydrostatic_Balance}\end{equation}

\paragraph*{Geostrophic balance}

In the plane $(x,y)$ perpendicular to the gravity direction, the
momentum equations read

\begin{equation}
\partial_{t}\mathbf{v}+\mathbf{v}\cdot\nabla\mathbf{v}+w\partial_{z}\mathbf{v}+f\mathbf{e}_{z}\times\mathbf{v}=-\frac{1}{\rho}\nabla_{h}P,\label{eq:Euler 2D Incompressible Vitesse Rotation}\end{equation}
where $\nabla_{h}P$ denotes the horizontal pressure gradient.

The Rossby number $\epsilon$ is defined by the ratio of the order
of magnitude of the advection term $\mathbf{v}\cdot\nabla\mathbf{v}$
over that of the Coriolis term $f\mathbf{e}_{z}\times\mathbf{v}$.
Introducing typical velocity $U$ and length $L$ for the flow, $\epsilon=U/fL$.
In mid-latitude atmosphere, $L\sim10^{4}\ km$ (size of cyclones and
anticyclones), $U\sim10\ m.s^{-1}$, so that $\epsilon\approx0.01$.
In the ocean $L\approx10^{2}\ km$ (width of ocean currents), $U\approx1\ m.s^{-1}$,
so that $\epsilon\approx0.1$. In both cases, this number is small:
$\epsilon\ll1$. In the limit of small Rossby numbers, the advection
term becomes negligible in (\ref{eq:Euler 2D Incompressible Vitesse Rotation}),
and at leading order there is a balance between the Coriolis term
and pressure gradients. This is called the geostrophic balance:

\begin{equation}
f\mathbf{e}_{z}\times\mathbf{v}_{g}=-\frac{1}{\rho}\nabla_{h}P,\label{eq:Geostrophic equilibrium}\end{equation}
where $\mathbf{v}_{g}$ is the geostrophic velocity. From (\ref{eq:Geostrophic equilibrium}),
we see that the geostrophic velocity is orthogonal to horizontal pressure
gradients. Taking the curl of the geostrophic balance (\ref{eq:Geostrophic equilibrium}),
and noting that horizontal variations of $\rho$ and $f$ are much
weaker than variations of $\mathbf{v}_{g}$, we see that the two-dimensional
velocity field $\mathbf{v}_{g}$ is at leading order non-divergent:
$\nabla\centerdot\mathbf{v}_{g}=0$.

Let us consider the case of a flow with constant density $\rho=\rho_{0}$.
Then the combination of the vertical derivative of (\ref{eq:Geostrophic equilibrium})
and of the hydrostatic equilibrium gives $f\mathbf{e}_{z}\times\partial_{z}\mathbf{v}_{g}=\frac{g}{\rho_{0}}\nabla_{h}P=0$:
the geostrophic flow does not vary with depth. This is the Taylor-Proudman
theorem. However geophysical flows have slightly variable densities
and display therefore vertical variations. But the Taylor-Proudman
theorem shows that such vertical variations are strongly constrained
and explain the tendency of geostrophic flows towards two-dimensionality.
Please see \cite{VallisBook,PedloskyBook,Salmon_1998_Book} for further
discussions of the geostrophic balance.\\

\paragraph*{The Rossby radius of deformation}

\begin{figure}[t!]
\begin{center}
\includegraphics[width=7cm]{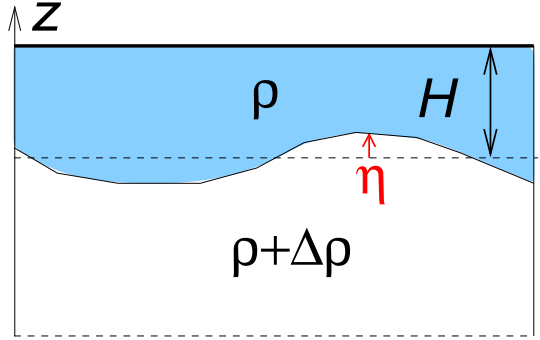}
\end{center}
\caption{{\footnotesize Vertical structure of the 1.5-layer quasi-geostrophic
model: a deep layer of density $\rho+\Delta\rho$ and a lighter upper
layer of thickness $H$ and density $\rho$. Because of the inertia
of the lower layer, the dynamics is limited to the upper layer.}}

\label{Fig:vertical_structure}
\end{figure}

A consequence of the combined effects of geostrophic and hydrostatic
balances is the existence of density and pressure fronts, whose typical
width is called the Rossby radius of deformation. This length plays
a central role in geostrophic dynamics. In order to give a physical
understanding of the Rossby radius of deformation, we consider a situation
where a light fluid of density $\rho$ lies above a denser fluid of
density $\rho+\Delta\rho$ with $\Delta\rho\ll\rho$. We also assume
here that the bottom layer is much thicker than the upper one. Then,
because of the inertia of the deep layer, the dynamics will be limited
to the upper layer of depth $H$ (see figure \ref{Fig:vertical_structure}).

We consider an initial condition where the interface has a steep slope
of amplitude $\eta$, and study the relaxation of the interface slope.
This classical problem is called the Rossby adjustment problem \cite{GillBook,VallisBook}.

Without rotation, the only equilibrium is a horizontal interface.
If the interface is not horizontal, pressure gradients induce dynamics,
for instance gravity waves, that transport potential energy and mass
in order to restore the horizontal equilibrium. A typical velocity
for this dynamics is the velocity of gravity waves $c$. Recalling
that the top layer has a thickness $H$ much smaller than the other
one, and considering waves with wavelengths much longer than $H$
(this is the classic shallow-water approximation), the velocity of
the gravity waves is $c=\sqrt{Hg^{\prime}}$ ($m.s^{-1}$) where $g^{\prime}=g\Delta\rho/\rho$
is called the reduced gravity.

With rotation, we see from (\ref{eq:Euler 2D Incompressible Vitesse Rotation})
and (\ref{eq:Geostrophic equilibrium}) that horizontal pressure gradients
can be balanced by the Coriolis force. It is then possible to maintain
a stationary non-horizontal slope for the interface (or front) in
this case.

The dynamical processes leading from an unstable front to a stable
one is called the Rossby adjustment. Initially, the dynamics is dominated
by gravity waves of typical velocity $c=\sqrt{Hg^{\prime}}$. This
initial process reduces the front slope until Coriolis forces become
as important as pressure terms (related to gravity through hydrostatic
balance). The typical time of the adjustment process is $\tau=f^{-1}$
, where $f$ is the planetary vorticity (also called Coriolis parameter).
This time can be estimated by considering that it is the time scale
at which velocity variations $\partial_{t}\mathbf{u}$ become of the
order of the Coriolis force $-f\mathbf{e}_{z}\times\mathbf{u}$. These
typical time $\tau$ and velocity $c$ are the two important physical
parameters of the adjustment. Then, the typical horizontal width of
the front at geostrophic equilibrium can be estimated by a simple
dimensional analysis: $R\sim c\tau$, which finally gives \[
R\sim\frac{\sqrt{g^{\prime}H}}{f}\ .\]

This length is the Rossby radius of deformation. It depends on the
stratification $\Delta\rho/\rho$, on $g$, on the Coriolis parameter
$f$, and on a typical thickness of the fluid $H$. This is the typical
length at which many fronts form in geophysical flows, resulting from
a balance between Coriolis force and pressure gradients which are
related to the stratification via the hydrostatic balance.

For the sake of simplicity, we have introduced the Rossby radius of
deformation with a simple dimensional analysis. The Rossby adjustment
is a very interesting physical problem in itself. Please see \cite{GillBook,VallisBook}
for more detailed analysis and discussions of this process. \\

In mid-latitude oceans, $\Delta \rho/\rho \sim3 \ 10^{-3}$ so  $g^{\prime}=0.03\ m.s^{-2}$ and $H=500\ m$,
then the Rossby radius of deformation is $R\sim60\ km$ and tends
to a few kilometers closer to the poles. This length is easily observable
on snapshots of oceanic currents, as shown in figures \ref{Fig:SouthernOcean}
and \ref{Fig:gulfstream}. It corresponds for example to the jet width,
either when jets are organized into either rings or zonal (eastward)
flows. In the Earth's atmosphere $R\sim1000\ km$, which is also the
typical size of cyclones responsible for mid-latitude weather features.
In the Jovian atmosphere $R\sim2000\ km$, which corresponds to the
typical width of the jet around the Great Red Spot. It is remarkable
that in this latter case, the large scale flow, i.e. the Great Red
Spot itself, has a much bigger length scale $\sim20000\ km$.

\subsubsection{The quasi-geostrophic model \label{sub:Quasi-Geostrophic Model}}

We now present the quasi-geostrophic equations, a model for the dynamics
of mid and high latitude flows, where the geostrophic balance (\ref{sub:Geostrophic_Equilibrium})
holds at leading order.

On the previous section \ref{sub:Geostrophic_Equilibrium}, we have
seen that for geophysical flows, the Rossby number $\epsilon=U/fL$
is small, leading to the geostrophic balance (\ref{eq:Geostrophic equilibrium})
at leading order. In order to capture the dynamics, the quasi-geostrophic
model is obtained through an asymptotic expansion of the Euler equations
in the limit of small Rossby number $\epsilon$, together with a Burger
$R/L$ of order one (where $R$ is the Rossby radius of deformation
introduced in the previous paragraph, and $L$ a typical length for
horizontal variations of the fields). We refer to \cite{VallisBook}
and \cite{PedloskyBook} for a comprehensive derivation. Here we give
only the resulting model and the physical interpretation.

We consider in this review the simplest possible model for the
vertical structure of the ocean, that takes into account the stable
stratification: an upper active layer where the flow takes place,
and a lower denser layer either at rest or characterized by a prescribed
stationary current (see figure \ref{Fig:vertical_structure}). This
is called the 1.5 layer quasi-geostrophic model. The full dynamical
system reads

\begin{equation}
\partial_t q + \mathbf{v}\cdot\nabla q=0,\label{QG}\end{equation}
 \begin{equation}
\mathrm{with}\,\,\, q=\Delta\psi-\frac{\psi}{R^{2}}+\eta_{d}(y),\label{dir}\end{equation}
 \begin{equation}
\mathrm{and}\,\,\,\mathbf{v}=\mathbf{e}_{z}\times\nabla\psi,\label{u}\end{equation}
with the impenetrability boundary condition, equivalent to $\psi$
being constant on the domain boundary $\partial\mathcal{D}$.

The complete derivation shows that the streamfunction gradient $\nabla\psi$
is proportional to the pressure gradient along the interface between
the two layers; then relation (\ref{u}) is actually the geostrophic
balance (\ref{eq:Geostrophic equilibrium}). The dynamics (\ref{QG})
is a non-linear transport equation for a scalar quantity, the potential
vorticity $q$ given by (\ref{dir}). The potential vorticity is a
central quantity for geostrophic flows \cite{GillBook,VallisBook,Salmon_1998_Book,PedloskyBook}.
The term $\Delta\psi=\omega$ is the relative vorticity%
\footnote{ The term {}``relative'' refers to the vorticity $\omega$ in the
rotating frame.%
}. The term $\psi/R^{2}$ is related to the interface pressure gradient
and thus to the interface height variations through the hydrostatic
balance (see section \ref{sub:Geostrophic_Equilibrium}). $R$ is
the Rossby radius of deformation introduced in section \ref{sub:Geostrophic_Equilibrium}.
Physically, an increase of $-\psi/R^{2}$ implies a stretching of
the upper layer thickness. Since the potential vorticity is conserved,
a stretching of the fluid column in the upper layer (i.e. an increase
of $-\psi/R^{2}$) is associated with a decrease of the relative vorticity
$\omega=\Delta\psi$, i.e. a tendency toward an anticyclonic rotation
of the fluid column \cite{Salmon_1998_Book}. The term $\eta_{d}$
represents the combined effects of the planetary vorticity gradient
(remember that $f=f_{0}+\beta y$) and of a given stationary flow
in the deep layer. We assume that this deep flow is known and unaffected
by the dynamics of the upper layer. It is described by the streamfunction
$\psi_{d}$ which induces a permanent deformation of the interface
with respect to its horizontal position at rest%
\footnote{\label{noteh}A real topography $h(y)$ would correspond to $h(y)=-f_{0}\eta_{d}(y)/H$
where $f_{0}$ is the reference planetary vorticity at the latitude
under consideration and $H$ is the mean upper layer thickness. Due
to the sign of $f_{0}$, the signs of $h$ and $\eta_{d}$ would be
the same in the south hemisphere and opposite in the north hemisphere.
As we will discuss extensively the Jovian south hemisphere vortices,
we have chosen this sign convention for $\eta_{d}$. %
}. This is why the deep flow acts as a topography on the active layer.
The detailed derivation gives \[
\eta_{d}=\beta_{c}y+\psi_{d}/R^{2}\ .\]

Starting from (\ref{QG}) and (\ref{dir}) and assuming $\psi=0$
at boundaries it is possible to prove that quasi-geostrophic flows
conserve the energy:
\begin{equation}
\mathcal{E}\left[q\right]=\frac{1}{2}\int_{\mathcal{D}}\mathrm{d}\mathbf{r}\,\left[(\nabla\psi)^{2}+\frac{\psi^{2}}{R^{2}}\right]=-\frac{1}{2}\int_{\mathcal{D}}\mathrm{d}\mathbf{r}\,\left(q-\eta_{d}\right)\psi.\label{ene}
\end{equation}
Two contributions are distinguished in (\ref{ene}): a kinetic energy
term $\frac{1}{2}\int_{\mathcal{D}}\mathrm{d}\mathbf{r}\,(\nabla\psi)^{2}$,
as in the Euler equations, and a (gravitational) available potential
energy term $\frac{1}{2}\int_{\mathcal{D}}\mathrm{d}\mathbf{r}\,\frac{\psi^{2}}{R^{2}}$.

\subsection{Hamiltonian structure, Casimir's invariants and microcanonical measures
\label{sub:Theory-General-Euler-QG}}

This subsection deals with theoretical properties of the 2D Euler
and quasi-geostrophic equations. As already noticed, these properties
are very similar because both dynamics are the non-linear advection
of a scalar quantity, the vorticity for the 2D Euler case or the potential
vorticity for the quasi-geostrophic case. In the following, we
discuss these properties in terms of the potential vorticity $q$,
but they are also valid for the 2D Euler equation. Indeed the 2D Euler
equation is included in the 1/2-layer quasi-geostrophic equation,
as can be seen by considering the limit $R\rightarrow+\infty$, $\eta_{d}=0$,
in the expression (\ref{dir}) of the potential vorticity.

\subsubsection{The theoretical foundations of equilibrium statistical mechanics \label{sub:Hamiltonian-structure}}

Let us consider a canonical Hamiltonian system: $\{q_{i}\}_{1\leq i\leq N}$
denote the generalized coordinates, $\{p_{i}\}_{1\leq i\leq N}$ their
conjugate momenta, and $H(\{q_{i},p_{i}\})$ the Hamiltonian. The
variables $\{q_{i},p_{i}\}_{1\leq i\leq N}$  belong to a
$2N$-dimensional space $\Omega$ called the phase space. Each point
$(\{q_{i},p_{i}\})$ is called a microstate. The equilibrium statistical
mechanics of such a canonical Hamiltonian system is based on the Liouville
theorem, which states that the non-normalized measure \[ \mu=\prod_{i=1}^N\mathrm{d}p_{i}\mathrm{d}q_{i} \]  is dynamically invariant.
The invariance of $\mu$ is equivalent to
\begin{equation}
\sum_{i}\left(\frac{\partial\dot{q}_{i}}{\partial q_{i}}+\frac{\partial\dot{p}_{i}}{\partial p_{i}}\right)=0,
\label{eq:Liouville}
\end{equation}
 which is a direct consequence of the Hamiltonian equations of motion
\[
\left\{ \begin{aligned}\dot{q_{i}} & =\frac{\partial H}{\partial p_{i}}\,,\\
\dot{p}_{i} & =-\frac{\partial H}{\partial q_{i}}\,.\end{aligned}
\right.\]
 Note that the equations of motion can also be written in a Poisson
bracket form: \begin{equation}
\left\{ \begin{aligned}\dot{q_{i}} & =\left\{ q_{i},H\right\} ,\\
\dot{p_{i}} & =\left\{ p_{i},H\right\} .\end{aligned}
\right.\label{eq:Crochets_Poisson_Canonique}\end{equation}

The terms in the sum (\ref{eq:Liouville}) actually vanish independently:
\[
\forall i,\quad\frac{\partial\dot{q}_{i}}{\partial q_{i}}+\frac{\partial\dot{p}_{i}}{\partial p_{i}}=0.\]
 This is called a detailed Liouville theorem.

For any conserved quantities $\left\{ \mathcal{I}_{1}\left(p,q\right),...,\mathcal{I}_{n}\left(p,q\right)\right\} $
of the Hamiltonian dynamics, the measures \[
\mu_{F}=\frac{1}{Z_{F}}\prod_{i}\mathrm{d}p_{i}\mathrm{d}q_{i}F\left(\mathcal{I}_{1},...,\mathcal{I}_{n}\right)\]
where $Z_{F}$ is a normalization constant, are also invariant measures.
An important question is to know which of these is relevant for describing
the statistics of the physical system.

In the case of an isolated system, the dynamics is Hamiltonian and
there is no exchange of energy or other conserved quantities with
the environment. It is therefore natural to consider a measure that
takes into account all these dynamical invariants as constraints.
This justifies the definition of the microcanonical measure\textbf{
}(for a given set of the values  $\left\{ I_{1}^0(q,p),\ldots,I_{n}^0(q,p)\right\} $ of the invariants $\left\{ \mathcal{I}_{1}(q,p),\ldots,\mathcal{I}_{n}(q,p)\right\} $):
\begin{equation}
\mu_{m}\left(I_{1}^0,...,I_{n}^0\right)=\frac{1}{\Omega\left(I_{1}^0,...,I_{n}^0\right)}\prod_{i}\mathrm{d}p_{i}\mathrm{d}q_{i}\prod_{k=1,n}\delta\left(\mathcal{I}_{k}\left(p,q\right)-I_{k}^0\right),
\label{eq:Microcanonical_Measure_H}
\end{equation}
 where $n$ is the number of constraints and  $\Omega\left(I_{1}^0,...,I_{n}^0\right)$ is a normalization
constant
\footnote{A more natural definition of the microcanonical measure would be as
the uniform measure on the submanifold defined by $\mathcal{I}_{k}=I_{k}^0$
for all $k$. This would request adding determinants in the formula
(\ref{eq:Microcanonical_Measure_H}), and imply further technical
difficulties. In most cases, however, in the limit of a large number
of degrees of freedom $N$, these two definitions of the microcanonical
measure become equivalent because the measures have large deviations
properties (saddle points evaluations) where $N$ is the large parameter,
and such determinants become irrelevant. We note that in the original works of Boltzmann and Gibbs, the microcanonical
measure refers to a measure where only the energy constraint is considered.}.
For small variations of the constraints $\left\{ \Delta \mathcal{I}_{k}\right\}_{1 \le k \le n} $,  the volume of the phase space with the constraint $I_{k}^0\leq \mathcal{I}_{k}\leq I_{k}^0+\Delta \mathcal{I}_{k}$ is given by $\Omega\left(I_{1}^0,...,I_{n}^0\right)\prod_{k=1,n}\Delta \mathcal{I}_{k}$.

Then the Boltzmann entropy of the Hamiltonian system is \[
S=k_{B}\log\Omega.\]

When the system considered is not isolated, but coupled with an external
thermal bath of conserved quantities, other measures need to be used to describe
properly the system by equilibrium statistical mechanics. Such measures
are usually referred to as canonical or grand-canonical.
A classical statistical mechanics result then proves that the relevant functions
$F$ are exponential (Boltzmann factors):
\begin{equation}
\mu_{c}=\frac{1}{Z_{c}}\prod_{i}\mathrm{d}p_{i}\mathrm{d}q_{i}exp\left(-\beta_1I_{1}-...-\beta_nI_{n}\right)
\label{eq:Canonical_Measure_H}
\end{equation}
where $Z_{c}$ is a normalization constant. When coupled to a thermal bath, a system can receive from and give energy to the thermal bath, the resulting balance leading to the Boltzmann factor, as explained in statistical mechanics textbooks. Flows are forced and stirred by mechanisms that do not allow for this two-way exchange of energy characteristic of thermal baths. It is then hard to imagine the
coupling of flows described by the Euler or quasi-geostrophic dynamics, with baths of energy, vorticity or potential vorticity.
Then the relevant statistical ensemble for these models is the microcanonical one, and we will work
in the following only starting from microcanonical measures. See subsection \ref{sub:Physical-interpretation-canonical}, page \pageref{sub:Physical-interpretation-canonical} on the physical interpretation of the microcanonical ensemble.

In statistical mechanics studies, it is sometimes argued  that, in the
limit of an infinite number of degrees of freedom, canonical and microcanonical
measures are equivalent. Then as canonical measures are more easily
handled, they are preferred in many works. However, whereas the equivalence
of canonical and microcanonical ensembles is very natural and usually
true in systems with short range interactions, common in condensed
matter theory, it is often wrong in systems like the Euler equations.
As a consequence, we will avoid the use of canonical measures in the
following (see for instance \cite{Bouchet_Barre:2005_JSP,Dauxois_Ruffo_Arimondo_Wilkens_2002LNP...602....1D,Campa_Dauxois_Ruffo_Revues_2009_PhR...480...57C,Bouchet_Gupta_Mukamel_PRL_2009,Chavanis_2006IJMPB_Revue_Auto_Gravitant,Bouchet:2008_Physica_D,EllisHavenTurkington:2000_Inequivalence}
and references therein). \\

In statistical mechanics, a macrostate $M$ is a set of microstates
verifying some conditions. The conditions are usually chosen such
that they describe conveniently the macroscopic behavior of the physical
systems through a reduced number of variables. For instance, in a
magnetic system, a macrostate $M$ could be the ensemble of microstates
with a given value of the total magnetization; in the case of a gas,
a macrostate could be the ensemble of microstates corresponding\textbf{
}to a given local density $f\left(\mathbf{x},\mathbf{p}\right)$ in
the six dimensional space $(\mathbf{x},\mathbf{p}$) ($\mu$ space),
where $f$ is defined  for instance through some coarse-graining.
In our fluid problem, an interesting macrostate will be the local
probability distribution $\rho\left(\mathbf{x},\sigma\right)d\sigma$
to observe vorticity values $\omega\left(\mathbf{x}\right)=\sigma$
at $\mathbf{x}$ with precision $\mathrm{d}\sigma$.

If we identify the macrostate $M$ with the values of the constraints
that define it, we can define the probability of a macrostate $P\left(M\right)dM$.
If the microstates are distributed according to the microcanonical
measure, $P\left(M\right)$ is proportional to the volume of the subset
$\Omega_{M}$ of phase space where microstates $\left\{ q_{i},p_{i}\right\} _{1\leq i\leq N}$
realize the state $M$. The Boltzmann entropy of a macrostate $M$
is then defined to be proportional to the logarithm of the phase space
volume of the subset $\Omega_{M}$ of all microstates $\left\{ q_{i},p_{i}\right\} _{1\leq i\leq N}$
that realize the state $M$.

In systems with a large number of degrees of freedom, it is customary
to observe that the probability of some macrostates is concentrated
close to a unique macrostate. There exist also cases where the probability
of macrostates concentrates close to larger set of macrostates (see
for instance \cite{Kiessling_2008AIPC}). Such a concentration is
a very important information about the macroscopic behavior of the
system. The aim of statistical physics is then to identify the physically
relevant macrostates, and to determine their probability and where
this probability is concentrated. This is the program we will follow
in the next sections, for the 2D Euler equations.

In the preceding discussion, we have explained that the microcanonical
measure is a natural invariant measure with given values of the invariants.
An important issue is to know if this measure describes also the statistics
of the temporal averages of the Hamiltonian system. This issue, called
ergodicity will be discussed in section \ref{sec:ergodicity}.

The first step to define the microcanonical measure is to identify
the equivalent of a Liouville theorem and the invariants. The Euler
and quasi-geostrophic equations describe a conservative dynamics.
They can be derived from a least action principle \cite{Salmon_1998_Book,Holm_Marsden_Ratiu_1998_EulerPoincare},
like canonical Hamiltonian systems. It is thus natural to expect
Hamiltonian structure. There are however fundamental differences between
infinite dimensional systems like the Euler equations and canonical
Hamiltonian systems:
\begin{enumerate}
\item The Euler equation is a dynamical system of infinite dimension. The
notion of the volume of an infinite dimensional space is meaningless.
Then the microcanonical measure can not be defined straightforwardly.
\item For such infinite dimensional systems, we can not in general find
a canonical structure (pair of canonically conjugated variables $\left\{ q_{i},p_{i}\right\} $
describing all degrees of freedom). There exists however a Poisson structure:
one can define a Poisson bracket $\left\{ .,.\right\} $, like in
canonical Hamiltonian systems (\ref{eq:Crochets_Poisson_Canonique})
and the dynamics reads \begin{equation}
\partial_{t}q=\left\{ q,\mathcal{H}[q]\right\} ,\label{eq:Poisson Bracket}\end{equation}
 where $\mathcal{H}$ is the Hamiltonian.
\end{enumerate}
For infinite dimensional Hamiltonian systems like the 2D Euler equations
or quasi-geostrophic model, the Poisson bracket in (\ref{eq:Poisson Bracket})
is degenerate \cite{Holm_etal_PhysRep_1985,Morrison_1998_HamiltonianFluid_RvMP}, leading
to the existence of an infinite number of conserved quantities, the
Casimir's functionals. These conservations laws have very important
dynamical consequences, as explained in the next section. A detailed description
of the Hamiltonian structure of infinite dimensional systems is beyond
the scope of this review. We refer to \cite{Holm_etal_PhysRep_1985,Morrison_1998_HamiltonianFluid_RvMP}
for the description of the Poisson structure for many fluid systems.
The conservation laws and the Liouville theorem are however essential
consequences and we discuss them in the next two sections.

\subsubsection{Casimir's conservation laws \label{sub:Casimirs-conservation-laws}}

Both Euler (\ref{eq:Euler_2D_Vorticity}) and quasi-geostrophic (\ref{QG})
equations conserve an infinite number of functionals, named Casimirs.
They are all functional of the form:\begin{equation}
\mathcal{C}_{s}[q]=\int_{\mathcal{D}}\mathrm{d}\mathbf{r}\, s(q),\label{eq:casimir}\end{equation}
 where $s$ is any function sufficiently smooth. Here and in the following, $q$ is the transported field, either the potential vorticity (\ref{dir}) for the quasi-geostrophic model or the vorticity in the case of the Euler equations for which $q=\omega$. As said in section \ref{sub:Hamiltonian-structure}, Casimir conserved quantities are related to the degenerate structure of infinite dimensional Hamiltonian systems. They can be also understood as the invariants
arising from the Noether's theorem, as a consequence of the relabeling
symmetry of fluid mechanics (see for instance \cite{Salmon_1998_Book}).

Let us define $A\left(\sigma\right)$ the area of $\mathcal{D}$ with
potential vorticity values lower than $\sigma$, and $\gamma\left(\sigma\right)$
the potential vorticity distribution
\begin{equation}
\gamma\left(\sigma\right)=\frac{1}{\left|\mathcal{D}\right|}\frac{dA}{d\sigma}\,\,\,\mbox{with\ensuremath{\,\,\,}}A\left(\sigma\right)=\int_{\mathcal{D}}\mathrm{d}\mathbf{r}\,\chi_{\left\{ q\left({\bf x}\right)\leq\sigma\right\} },
\label{eq:distribution_vorticite}
\end{equation}
where $\chi_{\mathcal{B}}$ is the characteristic function of the set
$\mathcal{B}\subset\mathcal{D}$ ($\chi_{\mathcal{B}}(x)=1$ for $x \in \mathcal{B}$),
and $\left|\mathcal{D}\right|$ is the area of $\mathcal{D}$.
As quasi-geostrophic (\ref{QG}) and 2D Euler equations (\ref{eq:Euler_2D_Vorticity}) are transport equations
by an incompressible flow, the area $\gamma\left(\sigma\right)$ occupied
by a given vorticity level $\sigma$ (or equivalently $A\left(\sigma\right)$)
is a dynamical invariant.

The conservation of the distribution $\gamma\left(\sigma\right)$
is equivalent to the conservation of all Casimir's functionals (\ref{eq:casimir}).
The domain averaged potential vorticity $\Gamma$, the enstrophy $\mathcal{G}_{2}$
and the other moments of the potential vorticity $\mathcal{G}_{n}$ are
Casimirs of a particular interest \begin{equation}
\mathcal{G}\left[q\right]=\mathcal{G}_{1}\left[q\right]=\int_{\mathcal{D}}\mathrm{d}\mathbf{r}\, q\mbox{\ensuremath{\,\,\,} and}\,\,\,\mathcal{G}_{n}\left[q\right]=\int_{\mathcal{D}}\mathrm{d}\mathbf{r}\, q^{n}.\label{eq:Enstrophy}\end{equation}
 For the 2D Euler equations in a bounded domain, $\mathcal{G}$ is also
the circulation $\mathcal{G}=\int_{\partial\mathcal{D}}\mathbf{v}\cdot\ d\mathbf{l}$.
\\

In any Hamiltonian systems, symmetries are associated with conservation
laws, as a consequence of Noether's theorem (see e.g. \cite{Salmon_1998_Book} and references therein). Then if the flow domain
$\mathcal{D}$ is invariant under rotations or translations, it will
be associated with angular momentum and momentum conservation. For domains with symmetries, these conservation laws have to be taken into account in a statistical mechanics analysis.

\subsubsection{Detailed Liouville theorem and microcanonical measure for the dynamics
of conservative flows \label{sub:microcanonical-measure}}

In order to discuss the detailed Liouville theorem, and build microcanonical
measure, in the following we decompose the potential vorticity field
on the eigenmodes of the Laplacian on $\mathcal{D}$; where $\mathcal{D}$
is the domain on which the flow takes place. We could have decomposed
the field on any other orthogonal basis. Whereas the Laplacian and
Fourier basis are simpler for the following discussion, finite elements
basis are much more natural to justify mean field approximation and
to obtain large deviation results for the measures, as discussed in
section \ref{sub:Mean_Field}.\\

We call $\{e_{i}\}_{i\geq1}$ the orthonormal family of eigenfunctions
of the Laplacian on the domain $\mathcal{D}$, with Dirichlet boundary conditions (see subsection  \ref{sub:Euler-QG-equations}, page \pageref{sub:Euler-QG-equations}):
\begin{equation}
-\Delta e_{i}=\lambda_{i}e_{i},\quad\int_{\mathcal{D}}\mathrm{d}\mathbf{r}\ e_{i}e_{j}=\delta_{ij}.
\label{eq:LaplacianEigenmodes}
\end{equation}
The eigenvalues $\lambda_{i}$ are arranged in increasing order. For
instance for a doubly periodic domain or infinite domain, $e_{i}\left(\mathbf{r}\right)$
are Fourier modes. Any function $g$ defined on the domain can be
decomposed into $g=\sum_{k}g_{k}(t)e_{k}(\mathbf{r})$ with $g_{k}=\int d\mathbf{r}\ ge_{k}$.
Then \[
q\left(\mathbf{r},t\right)=\sum_{i=1}^{+\infty}q_{i}\left(t\right)e_{i}\left(\mathbf{r}\right).\]
 From (\ref{QG}), the quasi-geostrophic equations are
\begin{equation}
\dot{q}_{i}=\sum_{j=1}^{+\infty}\sum_{k=1}^{+\infty} A_{ijk}q_{j}q_{k},\label{eq:Euler-Fourrier}\end{equation}
where the explicit expression for $A_{ijk}$ will not be needed in
the following discussion. For (\ref{eq:Euler-Fourrier}), a detailed
Liouville theorem holds: \begin{equation}
\forall i,\quad\frac{\partial\dot{q_{i}}}{\partial q_{i}}=0 \, ,\label{eq:Detailed_Liouville_Euler}
\end{equation}
see \cite{Lee52}, \cite{Kraichnan_Motgommery_1980_Reports_Progress_Physics}.
Note that while we have discussed here the detailed Liouville theorem
in the context of mode decomposition, more general results exist \cite{Robert_2000_CommMathPhys-TruncationEuler,Zeitlin_1991_HamiltonianTruncations}
\footnote{A direct consequence of the detailed Liouville theorem (\ref{eq:Detailed_Liouville_Euler})
is that any truncation of the 2D Euler or quasi-geostrophic equations
also verifies a Liouville theorem \cite{Kraichnan_Motgommery_1980_Reports_Progress_Physics}.
This result is actually much more general: any approximation of the
Euler equation obtained by an $L_{2}$ projection on a finite dimensional
basis verify a Liouville theorem, see \cite{Robert_2000_CommMathPhys-TruncationEuler}).
For truncations preserving the Hamiltonian structure and a finite
number of Casimir invariants, see \cite{Zeitlin_1991_HamiltonianTruncations}.%
}.\\

From the detailed Liouville theorem, we can define the microcanonical
measure. First the $n$ moment microcanonical measure (which, by including the energy, makes $n+1$ constraints) is defined as

\begin{equation}
\mu_{m,n}\left(E,\Gamma_{1},...,\Gamma_{n}\right)=\frac{1}{\Omega_{n}\left(E,\Gamma_{1},...,\Gamma_{n}\right)}\prod_i \mathrm{d}q_{i}\ \delta\left(\mathcal{E}\left[q\right]-E\right)\prod_{k=1,n}\delta\left(\mathcal{G}_{k}\left[q\right]-\Gamma_{k}\right),
\label{eq:microcanonical_measure_n}
\end{equation}
where $\mathcal{E}$ (\ref{ene}) is the energy, $\Gamma_{n}$
(\ref{eq:Enstrophy}) the vorticity moments and $\delta(\cdot)$ the Dirac delta function. A precise definition of $\mu_{m,n}$ goes
through the definition of approximate finite dimensional measures:
for any observable $\phi_{K}$ depending on $K$ components $\left\{ q_{i}\right\} _{1\leq i\leq K}$
of $q$ , we define \[
<\mu_{m,n}^{N},\phi_{K}>=\frac{\int\prod_{_{i=1,N}}\mathrm{d}q_{i}\ \delta\left(\mathcal{E}_{N}\left[q\right]-E\right)\prod_{k=1,n}\delta\left(\mathcal{G}_{N,k}\left[q\right]-\Gamma_{k}\right)\phi_{K}}{\Omega_{n,N}\left(E,\Gamma_{1},...,\Gamma_{n}\right)},\]
 where $\mathcal{E}_{N}$ and $\mathcal{G}_{N,n}$ are finite dimensional
approximations of $\mathcal{E}$ (\ref{ene}) and $\mathcal{G}_{n}$ (\ref{eq:Enstrophy}),
and $\Omega_{N}$ is a normalization factor. Then we define $<\mu_{m,n},\phi_{K}>=\lim_{N\rightarrow\infty}<\mu_{m,n}^{N},\phi_{K}>$.
Usually $\Omega_{n,N}$ has no finite limit when $N$ goes to infinity,
and the definition of $\Omega_{n}\left(E,\Gamma_{1},...,\Gamma_{n}\right)$
in the formal notation (\ref{eq:microcanonical_measure_n}) implies
a proper rescaling.

$\mu_{m,n}$ are ensembles of invariant measures. The microcanonical
measure corresponding to the infinite set of invariants $\left\{ \Gamma_{i}\right\} $
is then defined as \[
\mu_{m}\left(E,\left\{ \Gamma_{i}\right\} \right)=\lim_{n\rightarrow\infty}\mu_{m,n}\left(E,\Gamma_{1},...,\Gamma_{n}\right),\]
 and is denoted \begin{equation}
\mu_{m}\left(E,\left\{ \Gamma_{i}\right\} \right)=\frac{1}{\Omega\left(E,\left\{ \Gamma_{i}\right\} \right)}\prod_{i=1..\infty}\mathrm{d}q_{i}\ \delta\left(\mathcal{E}\left[q\right]-E\right)\prod_{k=1..\infty}\delta\left(\mathcal{G}_{k}\left[q\right]-\Gamma_{k}\right).\label{eq:microcanonical_measure}\end{equation}

\subsection{Specificity of 2D and geostrophic turbulence as a consequence of
Casimir's invariants}

We discuss in this section the consequences of the conservation laws
presented above. These consequences are important physical properties:
i) the existence of an infinite number of stationary solutions to the 2D Euler equations, and
the stability of some of these flows (section \ref{sub:multiple_equilibria}),
ii) the existence of an inverse (or upscale) energy cascade and of
a direct (or downscale) cascade of enstrophy (section \ref{sub:Fjortof_Argument}),
iii) the self organization of the large scale flow (section \ref{sub:Self_Organization}),
iv) non-trivial results from the equilibrium statistical mechanics
of two-dimensional flows by contrast with three dimensional flows
(section \ref{sub:3D-Stat-mech-and-Energy-Enstrophy}), and v) the
validity of a mean-field treatment of equilibrium statistical mechanics
(section \ref{sub:Mean_Field}).

\subsubsection{First physical consequence of 2D invariants: multiple stationary flows \label{sub:multiple_equilibria} }

Let us consider a dynamical system $\mathcal{G}$: $\dot{x}=G\left(x\right)$,
where $\dot{x}$ is the temporal derivative of $x$,  with conserved quantity $F\left(x\right)$ ($\dot{F}$$\left(x\right)=0$).
It can be proved easily that any non-degenerate extrema $x_{0}$ of
$F$ ($F'\left(x_{0}\right)=0$) is a stationary solution ($\dot{x}=0$)  of $\mathcal{G}$
($G\left(x_{0}\right)=0$) and if, in addition, the second variations
of $G$ are either positive-definite or negative-definite, then this
stationary solution is stable \cite{Holm_etal_PhysRep_1985}. This general
result seems natural when one considers the examples of energy and
angular momentum extrema encountered in classical mechanics. This
simple idea, coupled to convexity estimates, was used for instance
by Arnold \cite{Arnold_1966} to prove the stability of stationary solutions of the 2D Euler equations. Generalizations of these ideas
to larger classes of stationary flows of the 2D Euler equations
can be found in \cite{Wolansky_Ghil_1998_CMaPh,EllisHavenTurkington:2002_Nonlinearity_Stability,Caglioti_Rousset_2007_JStatPhys_QSS}.
Generalizations of these ideas to many other fluid mechanics equations
can be found in \cite{Holm_etal_PhysRep_1985}.

If we apply this idea to the 2D Euler and quasi-geostrophic equations,
as a consequence of the infinite number of Casimir's invariants (\ref{sub:Casimirs-conservation-laws}),
there exists an infinite number of stationary flows, a large number
of them being stable. In any dynamical system, fixed points play a
major role. In the case of the 2D Euler equations, moreover they turn
out to be attractive, as discussed in section \ref{eq:Euler_Relaxation}.\\

We discuss now the case of the quasi-geostrophic equations, but the
case of the 2D Euler equations is exactly similar. The conserved quantities
we use are the so called Energy-Casimir functionals \begin{equation}
\mathcal{F}=\mathcal{E}[q]+\mathcal{C}_{s}[q]=-\frac{1}{2}\int_{\mathcal{D}}\mathrm{d}\mathbf{r}\, q\psi+\int_{\mathcal{D}}\mathrm{d}\mathbf{r}\, s\left(q\right)\,,\label{eq:Energy_Casimir}\end{equation}
where $s$ is an arbitrary function. They are the sum of the energy
(\ref{ene}) and a Casimir invariant (\ref{eq:casimir}). The critical
points $q_{e}$ of this functional (satisfying $\delta\mathcal{F}=\int_{\mathcal{D}}\mathrm{d}\mathbf{r}\,\left(\psi_{e}-s^{\prime}\left(q_{e}\right)\right)\delta q=0$
for any perturbation $\delta q$) verify the equation

\begin{equation}
\psi=s^{\prime}\left(q\right).\label{eq:equilibres_energy_Casimir}\end{equation}
As expected from the general argument above, these critical points
should be stationary solutions of the quasi-geostrophic equation (\ref{QG}).
From (\ref{QG}), we see that any dynamical invariant verifies $\nabla\psi \times \nabla q = 0 $ (we recall that $q$, $\psi$ and the velocity $\mathbf{v}$ are related by (\ref{dir}-\ref{u})). Then the dynamical invariants of the quasi-geostrophic
equation (\ref{QG}), are all potential vorticity fields $q$, such
that the isocontour lines for $q$ and for $\psi$ are the same.
A special class of dynamical invariant are the potential functional
vorticity fields $q$, such that a relation $q=g\left(\psi\right)$
between $q$ and $\psi$ exist, with $g$ an arbitrary function. Then
solutions to (\ref{eq:equilibres_energy_Casimir}) are indeed stationary flows.

The case when $s$ is either strictly convex or concave is very interesting.
Indeed, then $s'$ is monotonous and (\ref{eq:equilibres_energy_Casimir})
can be inverted: $q=\left(s'\right)^{-1}\left(\psi\right)$. Moreover
if the functional (\ref{eq:Energy_Casimir}) is either strictly convex
or concave, then we expect the critical points to exist and to be
unique, and we expect them to be non degenerate (the second variations
are either positive-definite or negative-definite). Then according
to the general argument above, in this case we expect the stationary flows (\ref{eq:Energy_Casimir}) to be dynamically stable.

In the case of fluid dynamics, there are further difficulties with
the general argument above, because the potential vorticity field $q$ lies in an infinite dimensional
space variable. Roughly speaking, these difficulties are related to
continuity properties of the functionals, which may depend on the
chosen norm for the  potential vorticity field. One then has to defines carefully the norm for the perturbation
and a norm with respect to which the dynamics is stable. In the case
of the Euler equations, these difficulties have first been dealt by
Arnold \cite{Arnold_1966}, proving that when $\mathcal{F}$ is either
strictly convex or strictly concave, the stationary flows (\ref{eq:Energy_Casimir}) are indeed stable. Among Arnold's results, we learn that a sufficient
condition for $\mathcal{F}$ to be strictly convex is $s$ convex
and a sufficient condition for $\mathcal{F}$ to be strictly concave
is $s$ concave with $s^{\prime\prime}(q_{e})\geq c>\lambda_{1}$,
where $\lambda_{1}$ is the smallest eigenvalue of the Laplacian on
the domain $\mathcal{D}$, with Dirichlet boundary conditions (these
two condition can be easily worked out). These results have found
to be valid for weaker hypothesis and generalized to the quasi-geostrophic
model and a number of other models in fluid dynamics and plasma physics
(see for instance \cite{Holm_etal_PhysRep_1985,Wolansky_Ghil_1998_CMaPh,EllisHavenTurkington:2002_Nonlinearity_Stability}).\\

In the preceding paragraphs, we have applied the property that nondegenerate extremum of conserved quantities are stable equilibria, to the minimization of Energy-Casimir functionals (\ref{eq:Energy_Casimir}), following Arnold \cite{Arnold_1966}. The same idea and property could be applied to other conserved functionals or conserved functionals with constrains. For instance, we may use that the dynamics conserve all Casimirs (\ref{eq:casimir}). Then the extremum of the energy (\ref{ene}) for fixed values of the Casimirs (\ref{eq:casimir}) (a constrained variational problem) should be a stable equilibria\footnote{See \cite{VallisYoungCarnevaleJFM,1983_Turkington_CommMathPhys1,1983_Turkington_CommMathPhys2} for interesting algorithms that allow to compute energy maxima while preserving the Casimir functionals.}. Such an extremization is called a Kelvin energy principle as Lord Kelvin was the first to realize this property \cite{Kelvin_1887_bis}. We note that critical points of a Kelvin energy principle, like critical points of Energy-Casimir functionals (\ref{eq:Energy_Casimir}), are stationary solutions of the quasi-geostrophic (or 2D Euler) equations. On the one hand, the class of stable solutions obtained through Kelvin energy principle is larger than the class obtained from Energy-Casimir variational problem, and in this sense, the Kelvin energy principle is less restrictive. On the other hand the stability of Kelvin energy minimizers is expected to be weaker compared to the stability of Energy-Casimir minimizer, as perturbations modifying the value of the Casimirs may destabilize the flow (for instance we know no counterparts of the Arnold theorems for Kelvin energy minimizers). We refer to \cite{ChavanisEPJB2009} for a recent comprehensive review of these different variational problems and other related ones, and for a discussion of the conditions for second variations of these variational problems to be definite positive or definite negative.\\

We conclude that due to the infinite number of their invariants,
the 2D Euler and quasi-geostrophic equations have an infinite number
of stationary flows. Moreover an infinite class of these stationary flows can be proved to be stable. As in any dynamical system,
we expect these stationary flows to play a very important role
in the dynamics. We will see in section \ref{sub:The-mixing-entropy} that
the microcanonical measure of statistical mechanics is concentrated
close to some of these stationary flows. Moreover, the arguments
of this section, or some generalizations, can be used to prove the
dynamical stability of classes of statistical equilibria \cite{Michel_Robert_1994_JSP_GRS,EllisHavenTurkington:2002_Nonlinearity_Stability}.

\subsubsection{Second physical consequence of 2D invariants: the inverse energy
cascade \label{sub:Fjortof_Argument}}

We saw that the infinite number of steady states of the
2D Euler and quasi-geostrophic equations, and the stability of some
of these states, can be understood as a consequence of the conservation
of Casimir invariants. We now look at another consequence of these
conservation laws: the direction of the energy fluxes in spectral
space is upscale. The argument developed in this section is a very classical
one for physical systems with multiple invariants.

We treat here the case of decaying two-dimensional turbulence (for which the potential vorticity $q$ is simply the relative vorticity $\omega$), following \cite{Nazarenko_Quinn_PRL_2009}.
A discussion of the direction of the energy and enstrophy fluxes was
originally given by Fjortoft \cite{Fjortoft53}. The case of statistically
stationary cascade \cite{Kraichnan_Phys_Fluid_1967_2Dturbulence}
is treated in section \ref{sub:Kraichnan}. Let us consider an infinite
2D domain and decompose the vorticity into Fourier eigenmodes. The energy
spectrum $E(k)$ is defined such that $E\left(k\right)\mathrm{d}k$ is the energy
contained in modes with wave numbers $\mathbf{k}'$ with $k\leq k'\leq k+\Delta k$, and such that the total energy is $E=\int\mathrm{d}k\, E\left(k\right)$.
We define similarly the enstrophy spectrum $\Gamma_{2}\left(k\right)$:
$\Gamma_{2}=\int\mathrm{d}k\,\Gamma_{2}\left(k\right)$ (see \ref{eq:Enstrophy}
for $\Gamma_{2}$). It is easy to show that $\Gamma_{2}\left(k\right)=2k^{2}E\left(k\right)$.

A question of interest is to determine whether the energy goes towards
large scales or small scales. To answer this, we look at rigorous
bounds on the $k$-centroids $k_{E}$ (and $l$-centroids $l_{E}$)
for the energy: \[
k_{E}=\frac{1}{E}\int\mathrm{d}k\, kE\left(k\right)\,\,\,\mathrm{and}\,\,\, l_{E}=\frac{1}{E}\int\mathrm{d}k\, k^{-1}E\left(k\right),\]
and for the enstrophy
\[ k_{\Gamma_{2}}=\frac{1}{\Gamma_{2}}\int\mathrm{d}k\, k\Gamma_{2}\left(k\right)dk\,\,\,\mathrm{and}\,\,\, l_{\Gamma_{2}}=\frac{1}{\Gamma_{2}}\int\mathrm{d}k\, k^{-1}\Gamma_{2}\left(k\right).\]
A transfer of energy toward large scales during the flow evolution
is equivalent to an increase of the $k$- or the $l$-centroid.

Using Cauchy-Schwartz inequalities $\left(\int\mathrm{d}k\, f(k)g(k)\le\sqrt{\int\mathrm{d}k\, f^{2}(k)\int\mathrm{d}k\, g^{2}(k)}\right)$,
one can easily show that \begin{equation}
k_{E}\le\sqrt{\frac{\Gamma_{2}}{2E}},\qquad k_{\Gamma_{2}}\ge\sqrt{\frac{\Gamma_{2}}{2E}},\qquad k_{E}k_{\Gamma_{2}}\ge\frac{\Gamma_{2}}{2E},\label{eq:centroids-k}\end{equation}
 \begin{equation}
l_{E} \ge \sqrt{\frac{2E}{\Gamma_{2}}},\qquad l_{\Gamma_{2}} \le \sqrt{\frac{E}{\Gamma_2}},\qquad l_{E}l_{\Gamma_{2}} \ge \frac{2E}{\Gamma_{2}}\ .\label{eq:centroids-l}\end{equation}
The first inequalities of (\ref{eq:centroids-k}) and (\ref{eq:centroids-l})
imply that the energy cannot be transferred to scales smaller
than $\sqrt{2E/\Gamma_{2}}$ , and enstrophy cannot be transferred
to scales larger than $\sqrt{2E/\Gamma_{2}}$.

The last inequality of (\ref{eq:centroids-k}) implies that if the
energy goes to larger and larger scales ($k_{E}\rightarrow0$), then
the enstrophy goes to smaller and smaller scales ($k_{\Gamma_{2}}\rightarrow+\infty$):
an evolving state presenting an inverse flux of energy implies
a simultaneous direct flux of enstrophy. Similarly, the last inequality of (\ref{eq:centroids-l})
implies than if the enstrophy goes to smaller and smaller scales ($l_{\Gamma_{2}}\rightarrow0$)
, then the energy goes to larger and larger scales ($l_{E}\rightarrow+\infty$):
a direct flux of enstrophy implies an inverse flux of energy. A sufficient
and necessary condition for the existence of a forward enstrophy flux
is then the existence of an inverse energy flux.

\subsubsection{Third physical consequence: the phenomenon of large scale self-organization of the flow\label{sub:Self_Organization}}

\begin{figure}[t!]
\begin{center}
\includegraphics[width=1\textwidth]{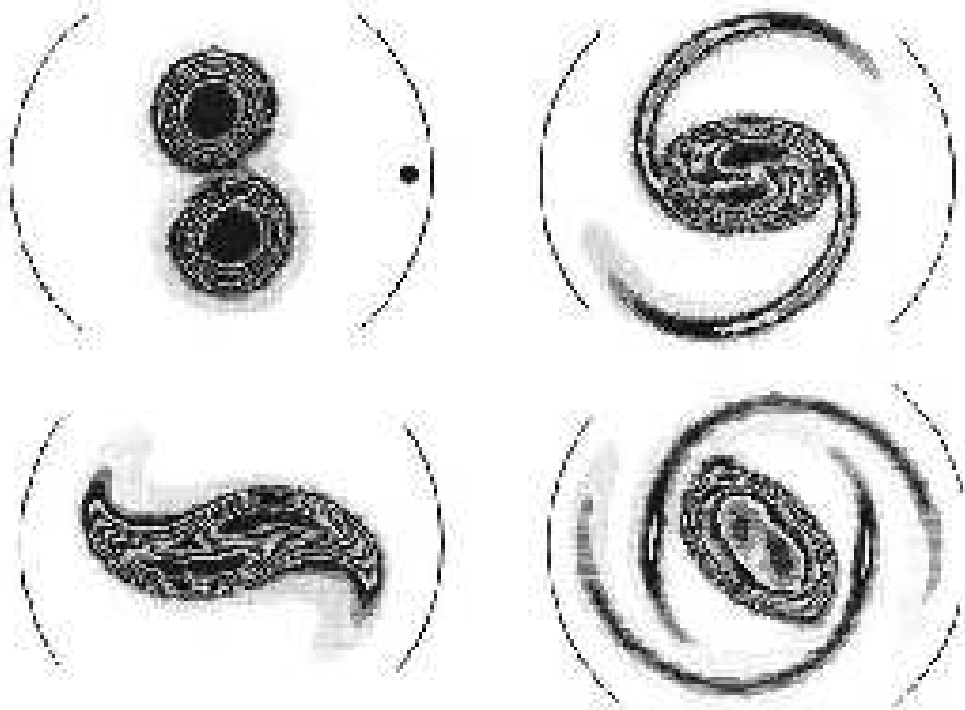}
\end{center}

\caption{\footnotesize Snapshot of electron density (analogous to vorticity
field) at successive time from an initial condition with two vortices
to a single large scale coherent structure via turbulent mixing (see
\cite{Scecter_etal_2000_PhysicsFluids,Schecter_Dubin_etc_Vortex_Crystals_2DEuler1999PhFl}).
The best experimental realization of inviscid 2D Euler equations is
probably so far achieved in those magnetized electron plasma experiments
where the electrons are confined in a Penning trap. The dynamics of
both systems are indeed isomorphic, where the electron density plays
the role of vorticity. The major drawback of this experimental setting
comes from its observation, since any measurement requires the destruction
of the plasma itself.}

\label{Fig:SelfOrganization}
\end{figure}

The most striking feature of 2D and geostrophic flows, and by far the
most important phenomenon for applications, is their tendency to organize
into large scale coherent structures. Be it in laboratory experiments
(with the formation of long lived and robust 2D vortices, see for instance figure \ref{Fig:SelfOrganization}), in the ocean
(with the formation of jets and rings), in the Jovian atmosphere
(with the Great Red Spot and other vortices), or in numerical simulations,
these coherent structures are yet ubiquitous, and represent the main
qualitative feature of turbulent 2D flows. Understanding their formation
is thus a major challenge in geophysical fluid dynamics.

In the previous section, we proved that an upscale energy flux
is always accompanied by a downscale enstrophy flux, and that there is a lower bound for the energy centroid. Using
heuristic statistical mechanics arguments, we know that the dynamics will tend to partition
as much as possible energy and enstrophy among the modes.
The combination of these two arguments and the preceding are sufficient to conclude that the complex non-linear dynamics of the flow will tend to a transfer
of energy toward largest scales and a transfer of enstrophy towards
 smallest scales.

We also explained in section \ref{sub:multiple_equilibria} why
the equations have infinitely many multiple stable stationary flows. This,
together with the energy fluxes towards the largest scales is already
sufficient to explain qualitatively the self-organization of the flow.
On an inertial time scale (given for instance by the turnover time
of the large scales of the turbulent flow), these large scale structures
can be considered stationary solutions, in contrast to the complicated
dynamics of the small scale turbulent flow.

The aim of this review is to present predictive theories for these
large scale structures. We need to explain the physical mechanism at work
and describe theoretically the dynamical mechanisms that will select
some states among all the possible stationary flows. This is is
where statistical mechanics will be very useful. The equilibrium theory
predicts that turbulent mixing\footnote{Here mixing does not refers to the effect of molecular viscosity, but rather to the stirring by the flow dynamics.} will drive the flow toward a stationary
state that maximizes a Boltzmann-Gibbs entropy formulae, while satisfying all the constraints
of the dynamics presented in the previous subsections. This mixing
entropy, derived from the Liouville theorem, will allow us to build theoretically
 natural invariant measures for the dynamics. We will also consider
non-equilibrium theories for forced and dissipated flows.

Statistical mechanics is an extremely powerful tool that allows to
reduce a  complicated problem (the description of a fine-grained
turbulent flow, our microscopic state with a huge number of degrees
of freedom) to the study of a few parameters, which describes the
large scale structures of the flow, our macroscopic state.

\subsubsection{Fourth consequence: about Jeans paradox, why can we get a non-trivial equilibrium statistical mechanics
for 2D flows by contrast with the 3D Euler ultraviolet divergence\label{sub:3D-Stat-mech-and-Energy-Enstrophy} }

In this section, for pedagogical reasons, we  try to apply equilibrium statistical mechanics ideas to the three-dimensional Euler equations.
This very simple discussion illustrates, with the example of the 3D Euler equations, that in most of Hamiltonian field equation a straight application of equilibrium statistical mechanics fails because of the so called Jeans paradox \cite{Pomeau_Cargese_1995}. The reason is that in the simplest cases, for instance when energy is the only conserved quantity, at the statistical equilibria each degree of freedom has on average the same energy. Then because there are infinitely many degrees of freedom for a field, either the total energy is infinite, or if the energy is kept constant the average energy per modes is zero.

Then equilibrium statistical mechanics predicts that all the energy
flows towards the smallest scales, for the three-dimensional Euler equations, in accordance with basic observations.
But the microcanonical measure obtained in the limit of an infinite
number of degrees of freedom is a trivial one, with no more energy
in the largest scales. This argument proves that, because of Jeans paradox, three-dimensional
turbulence is intrinsically a non-equilibrium process.

The main interest of this discussion is to show that, by contrast, two dimensional flows
are different from this point of view and give non-trivial microcanonical
measures. Basically thanks to the presence of more invariants, the Jeans paradox is avoided for the energy.
Since the discussions of this section involves technical
computations, it can be skipped at first reading. However, this
discussion is essential for a physically relevant interpretation
of the theory.\\

In section \ref{sub:microcanonical-measure}, we defined a microcanonical
measure for the 2D Euler equations. We proceed here similarly. First,
we note that for the three-dimensional Euler equations, for instance on
a periodic cube, the velocity can be decomposed in Fourier modes $\mathbf{u}_{\mathbf{k}}=(u_{1,\mathbf{k}},u_{2,\mathbf{k}},u_{3,\mathbf{k}})$.
The 3D Euler equations then read\[
\dot{u}_{i,\mathbf{k}}=\sum_{\mathbf{p},\mathbf{q}\,;\, j,l=1,2,3}A_{j,l,\mathbf{k},\mathbf{p},\mathbf{q}}u_{j,\mathbf{p}}u_{l,\mathbf{q}}.\]
The explicit expression of $A$ is not important for this discussion.
The important point is that a detailed Liouville theorem holds:

\[
\forall \mathbf{k},\quad\sum_{i}\left(\frac{\partial\dot{u}_{i,\mathbf{k}}}{\partial u_{i,\mathbf{k}}}+\frac{\partial\dot{u}_{i,-\mathbf{k}}}{\partial u_{i,-\mathbf{k}}}\right)=0, \]
see for instance \cite{Lee52} for more details. As the kinetic energy $\mathcal{E}=\frac{1}{2}\sum_{\mathbf{k}}\left|\mathbf{u}{}_{\mathbf{k}}\right|^{2}=\frac{1}{2}\int d\mathbf{r}\ \mathbf{u}^{2}$
is the only invariant in this case, following discussion in section
\ref{sub:microcanonical-measure}, the microcanonical measure is defined
as
\begin{equation}
\mu_{m}=\lim_{K\rightarrow\infty}\mu_{m}^{K}\,\,\,\mbox{with}\,\,\,\mu_{m}^{K}=\frac{1}{\Omega_{K}\left(E\right)}\prod_{\left|\mathbf{k}\right|\leq K}\mathrm{d}\mathbf{v}{}_{\mathbf{k}}\delta\left(\mathcal{E}_{K}-E\right),\label{eq:microcanonical_Euler3D}\end{equation}
where $2\mathcal{E}_{K}=\sum_{\left|\mathbf{k}\right|\leq K}\left|\mathbf{v}{}_{\mathbf{k}}\right|^{2}$
and $m$ stand for ``microcanonical''.

Let us compute the average of $E_{\mathbf{k}}=\left|\mathbf{v}{}_{\mathbf{k}}\right|^{2}/2$,
the energy of mode $\mathbf{k}$ for the measure $\mu_{m}^{K}$.
We note this average $\left\langle E_{\mathbf{k}}\right\rangle _{m,K}=\int\mu_{m}^{K}E_{\mathbf{k}}$. In (\ref{eq:microcanonical_Euler3D}), all
degrees of freedoms $\mathbf{v}{}_{\mathbf{k'}}$ readily play a symmetric
role. Henceforth we have energy equipartition $\left\langle E_{\mathbf{k}}\right\rangle _{m,K}=\left\langle E_{\mathbf{k'}}\right\rangle _{m,K}$
and $\left\langle E_{\mathbf{k}}\right\rangle _{m,K}=E/N\left(K\right)$
where $N\left(K\right)$ is the total number of modes such that $\left|\mathbf{k}\right|\leq K$.
This equipartition for the finite-dimensional measure $\mu_{m}^{K}$
leads to an energy spectrum $E\left(k\right)$
proportional to $k^{2}$ (see section \ref{sub:Fjortof_Argument} for definition of $E(k)$). This has been described by Kraichnan
\cite{Kraichnan_1967PhFl...10.1417K}. Recent applications to Galerkin
truncations of the Euler equations and bottlenecks in turbulence can be
found in Frisch \textit{et al} \cite{FrishKurien08}.

We are now interested in invariant measures for the Euler equations
themselves, not in invariant measures for truncated dynamics.
We thus take the limit $K\rightarrow\infty$ (which then implies $N\left(K\right)\rightarrow+\infty$), with fixed energy because each trajectory has a finite energy. We
obtain
\[ \left\langle E_{\mathbf{k}}\right\rangle _{m}=\lim_{K\rightarrow\infty}\left\langle E_{\mathbf{k}}\right\rangle _{m,K}=0.\]
Then, at statistical equilibrium, the average energy of each mode
is exactly zero. Due to the infinite number of degrees of freedom,
the tendency towards equipartition spreads energy on modes corresponding to smaller and smaller scales. This phenomenon associated with
the existence of an infinite number of degrees of freedom is called the Jeans paradox \cite{Pomeau_Cargese_1995}. It is a form
of ultraviolet divergence.

We thus conclude that the equilibrium statistical mechanics of the 3D Euler equations explains the tendency for the energy to flow towards smaller and smaller scales. However, because in the microcanonical distribution, the energy per
degree of freedom is zero, the microcanonical measure is trivial
and thus useless. We see that for 3D turbulence, the statistically
stationary flux of energy towards the smallest scales is intrinsically
a far from equilibrium process. The equilibrium statistical mechanics
is of no help to understand these fluxes, and the associated energy spectrum
and velocity increment statistics.\\

Because of the additional Casimir invariants, the situation is quite
different in 2D turbulence. All configurations of the microcanonical
ensemble described in section \ref{sub:microcanonical-measure} have
the same energy $E$ and enstrophy $\Gamma_{2}$. Then, for any configurations
the $l_{E}$ centroid inequality (\ref{eq:centroids-l}, page \pageref{eq:centroids-l})
holds: $l_{E}>\sqrt{\frac{2E}{\Gamma_{2}}}$. Henceforth, this is also true
for the average over the microcanonical measure: \[
\left\langle l_{E}\right\rangle _{m}>\sqrt{\frac{2E}{\Gamma_{2}}}.\]
This simple argument shows that for the microcanonical measure, the
energy cannot flow to the smallest scales. The microcanonical measure
is thus non-trivial for two-dimensional conservative flows. It
will describe large scale features that cannot be guessed straightforwardly.
A detailed understanding of these equilibrium structures is the aim
of the next sections. \\

\paragraph*{The energy-enstrophy microcanonical measure of two dimensional flows}

In order to illustrate that statistical mechanics gives non-trivial
predictions for the 2D Euler case, we now consider the energy-enstrophy
canonical measure \[
\mu_{m,2}\left(E,\Gamma_{2}\right)=\frac{1}{\Omega\left(E,\Gamma_{2}\right)}\prod_{i=1}^{\infty}\mathrm{d}\omega_{i}\,\delta\left(\mathcal{E}\left[\omega\right]-E\right)\delta\left(\mathcal{G}_{2}\left[\omega\right]-\Gamma_{2}\right)\]
This is the measure where we take into account only the quadratic
invariants. There is  no physical reason to exclude the other
invariants; however the energy-enstrophy measure can be interesting
because it may be in some cases a good approximation of the complete
microcanonical measure.  The interest and limitation of the energy-enstrophy measures are further discussed at the end of this section.

Our real motivation here is more pedagogical:
it will be very useful to introduce mean-field
treatment, and to explain on a simple example the relation between
microcanonical measures defined on section \ref{sub:microcanonical-measure}
and mixing entropy used in section \ref{sub:The-mixing-entropy} and
the following. The following discussion and the computation performed in the following paragraphs are adapted from the original presentation in \cite{Bouchet_Corvellec_JSTAT_2010}.

An energy-enstrophy ensemble has been treated and discussed in length
 by many authors, including Kraichnan  (see \cite{Kraichnan_Motgommery_1980_Reports_Progress_Physics}), but in the canonical ensembles (that is using the canonical measure \ref{eq:Canonical_Measure_H}, page \pageref{eq:Canonical_Measure_H} rather than the microcanonical one). See also \cite{Majda_Wang_Book_Geophysique_Stat} for a precise discussion for the energy-enstrophy-circulation ensemble. It has also been proven, through computations of explicit inequalities, that the statistics of the small scales of the velocity field, in the energy-enstrophy ensemble, are incompatible with 2D Navier-Stokes invariant measures \cite{Biryuk06}).

 The following discussion gives the first derivation of the microcanonical ensemble, its relation to the mean field variational problem, and the first observation
of ensemble inequivalence for the energy-enstrophy ensemble. We come
back to discuss Kraichnan type results in the end of this section.\\

Following the discussion in section \ref{sub:microcanonical-measure},
the microcanonical measure is defined through $N$ dimensional approximations:
\begin{equation}
\mu_{m,2}=\lim_{N\rightarrow\infty}\mu_{m,2}^{N}\,\,\,\mbox{with}
\nonumber \end{equation}
\begin{equation}
\mu_{m,2}^{N}=\frac{1}{\Omega_{N}\left(E,\Gamma_{2}\right)}\prod_{i=1..N}\mathrm{d}\omega_{i}\,\delta\left(\mathcal{E}_{N}\left[\omega\right]-E\right)\delta\left(\mathcal{G}_{2,N}\left[\omega\right]-\Gamma_{2}\right)\label{eq:microcanonical_energy_enstrophy_N}\end{equation}
where, $\omega_{i}$ are components of $\omega$ on the base of eigenmodes
$e_{i}$ (see \ref{eq:LaplacianEigenmodes}), $\mathcal{E}_{N}$ and
$\Gamma_{2,N}$ are $N$ dimensional approximations of the
energy (\ref{eq:Energy}) and enstrophy (\ref{eq:Enstrophy}): $2\mathcal{E}_{N}=\sum_{n=1..N}\omega_{n}^{2}/\lambda_{n}$
and $\mathcal{G}_{2,N}\left[\omega\right]=\sum_{n=1..N}\omega_{n}^{2}$. In the following, to simplify the argument, we assume that the first eigenvalue is non degenerate: $\lambda_{1}\neq\lambda_{2}$, which is generically the case\footnote{This is always true for simply connected bounded Lipshitz domains. An example of geometry for which the first eigenvalue is degenerate is a doubly periodic domain with aspect ratio $\delta=1$.}.

The main technical difficulty is to compute
\begin{equation}
\Omega_{N}\left(E,\Gamma_{2}\right)=\int\prod_{i=1..N}\mathrm{d}\omega_{i}\,\delta\left(\mathcal{E}_{N}\left[\omega\right]-E\right)\delta\left(\mathcal{G}_{2,N}\left[\omega\right]-\Gamma_{2}\right).
\label{eq:Omega_N}
\end{equation}
The computation of this result, using representation of the delta functions as integral
in the complex plane, is given in \cite{Bouchet_Corvellec_JSTAT_2010}, where it is shown that
\begin{equation}
\Omega_{N}\left(E,\mathcal{G}_{2}\right)\underset{N\rightarrow\infty}{=}C_{3}\left(N,\left\{ \lambda_{i}\right\} \right)C_{4}\left(\left\{ \lambda_{i}\right\} ,\Gamma_{2},N\right)\frac{\exp\left[NS\left(E,\Gamma_{2}\right)\right]}{\sqrt{2E}}+o\left(N\right) \nonumber
\end{equation}
\begin{equation}
\mbox{with}\,\,\, S\left(E,\Gamma_{2}\right)=\frac{1}{2}\log\left(\Gamma_{2}-2\lambda_{1}E\right)+\frac{\log2}{2},
\label{eq:Entropie-energie-enstrophie-texte}
\end{equation}
where $C_{3}\left(N,\left\{ \lambda_{i}\right\} \right)$ depends
only on $N$ and $\left\{ \lambda_{i}\right\} $ (i.e. does not depend
on the physical parameters $E$ and $\Gamma_2$) and $C_{4}$ has no exponentially large contribution ($\lim_{N\rightarrow\infty} \left( \log C_{2} \right ) / N = 0$). The notation $o\left(N\right)$ refers to corrections that are negligible with respect to  $N$ when $N$ becomes large enough.
From (\ref{eq:Entropie-energie-enstrophie-texte}) we have
\begin{equation}
S\left(E,\Gamma_{2}\right)=\lim_{N\rightarrow\infty}\frac{1}{N}\log\left(\Omega_{N}\left(E,\Gamma_{2}\right)\right)-C\left(N,\left\{ \lambda_{i}\right\} \right),\label{eq:Definition-Entropy}
\end{equation}
where $C$ can be computed from $C_{3}$, and depends only on $N$
and on the geometric factors $\left\{ \lambda_{i}\right\} $ (the
entropy is defined up to an arbitrary constant).

We see that the quantity $S(E,\Gamma_2)$ is  the Boltzmann entropy  rescaled by $1/N$ with an unimportant additional constant. It counts the number of microstates $\Omega_N$ satisfying the constraints of the problem, i.e. characterized by energy $E$ and enstrophy $\Gamma_2$.

From the entropy, we can compute the temperature $\beta=\partial S/\partial E=-\lambda_{1}/\left(\Gamma_{2}-2\lambda_{1}E\right)\leq0$
and chemical potential $\alpha=\partial S/\partial\Gamma_{2}=1/\left[2\left(\Gamma_{2}-2\lambda_{1}E\right)\right]$.
These thermodynamic potentials are related by $\beta=-2\lambda_{1}\alpha$.
Then some couples of thermodynamic parameter are not obtained in
the microcanonical ensemble, by contrast with what would be expected
in the thermodynamics of classical condensed matter systems, which are most of the time short range interacting systems. Moreover
the determinant of the Hessian of $S$ ($\partial^{2}S/\partial E^{2}.\partial^{2}S/\partial\Gamma_{2}^{2}-\left(\partial^{2}S/\partial E\partial\Gamma_{2}\right)^{2}$)
is zero, showing that $S$ is not strictly concave as one would expect
for an entropy in the case of short range interacting systems. Both
of these properties are signs of non-equivalence between the microcanonical
and the canonical ensembles.
(see for instance \cite{Bouchet_Barre:2005_JSP,EllisHavenTurkington:2000_Inequivalence,Dauxois_Ruffo_Arimondo_Wilkens_2002LNP...602....1D}).
This case of the energy-enstrophy ensemble is actually a case of partial
equivalence (see \cite{EllisHavenTurkington:2000_Inequivalence} for
a definition). \\

From (\ref{eq:microcanonical_energy_enstrophy_N}) we see that for
the finite $N$ dimensional measure $\mu_{N}$, the distribution function
for $\omega_{n}$ the amplitude of mode $e_{n}$ is \[
P_{N,n}\left(\omega_{n}\right)=\frac{\Omega_{N-1;\lambda_{n}}\left(E-\omega_{n}^{2}/2\lambda_{n},\Gamma_{2}-\omega_{n}^{2}\right)}{\Omega_{N}\left(E,\Gamma_{2}\right)},\]
where $\Omega_{N-1;\lambda_{n}}$ is defined as $\Omega_{N}$ (equation.
(\ref{eq:Omega_N})), but with the integration over $\omega_{n}$
excluded, and with the constraint $\omega_{n}^{2}\leq\max\left\{ 2\lambda_{n}E,\Gamma_{2}\right\} $.
The distribution function for $E_{n}=\omega_{n}^{2}/2\lambda_{n}$,
the energy of the mode $e_{n}$, is obtained by the change of variable
$P_{N,n}\left(E_{n}\right)\mathrm{d}E_{n}=P_{N,n}\left(\omega_{n}\right)\mathrm{d}\omega_{n}$.
Using result (\ref{eq:Entropie-energie-enstrophie-texte}) for both
$\Omega_{N-1;\lambda_{1}}$ (then $\lambda_{1}$ has to be replaced
by $\lambda_{2}$) and $\Omega_{N}$, we obtain \[
P_{N,1}\left(E_{1}\right)\underset{N\rightarrow\infty}{\sim}C\frac{\exp\left[N\log\left(\Gamma_{2}-2\lambda_{2}E+2\left(\lambda_{2}-\lambda_{1}\right)E_{1}\right)/2\right]}{\sqrt{E_{1}\left(E-E_{1}\right)}}\]
\[ \mbox{for}\,\,\,0\leq E_{1}\leq E,\]
and $P_{N,1}\left(E_{1}\right)=0$ otherwise, where $C$ does not
depend on $E_{1}$ (normalization constant). From this expression,
we see that the most probable energy is $E_{1}=E$. Moreover, the
distribution is exponentially picked close to $E_{1}=E$, such that
in the infinite $N$ limit (the microcanonical distribution) we have
\[
P_{1}\left(E_{1}\right)=\delta\left(E-E_{1}\right).\]
All the energy condenses to the first mode.

If one disregards large deviations for $E-E_{1}$, a good approximation
for large $N$ of the finite $N$ distribution is the exponential
distribution \begin{equation}
P_{N,1}\left(E_{1}\right)=\underset{N\rightarrow\infty}{\sim}C\frac{\exp\left[-N\frac{\lambda_{2}-\lambda_{1}}{\Gamma_{2}-2\lambda_{1}E}\left(E-E_{1}\right)\right]}{\sqrt{E-E_{1}}}\,\,\,\mbox{for}\,\,\,0\leq E_{1}\leq E \nonumber \end{equation} \begin{equation}\mbox{and}\,\,\, N^{1/2}\left(E_{1}-E\right)\ll1\,;\label{eq:PE1}\end{equation}
the distribution for $\omega_{1}$ being also exponential. The amplitude
of the departure of $E_{1}$ from the value $E$ is thus proportional
to $1/N$ and to $\left(\Gamma_{2}-2\lambda_{1}E\right)/\left(\lambda_{2}-\lambda_{1}\right).$

The distribution of the energy $E_{n}$ of mode $n$ is obtained similarly
as \begin{equation}
P_{N,n}\left(E_{n}\right)\underset{N\rightarrow\infty}{\sim}C\frac{\exp\left[N\log\left(\Gamma_{2}-2\lambda_{1}E-2\left(\lambda_{n}-\lambda_{1}\right)E_{n}\right)\right]}{\sqrt{E_{n}}}\end{equation}
\begin{equation}\mbox{for}\,\,\,0\leq E_{n}\leq E.\label{eq:PEn}\end{equation}
For infinite $N$, the microcanonical distribution are thus a delta
function with zero energy: \[
P_{n}\left(E_{n}\right)=\delta\left(E_{n}\right).\]
Disregarding large deviations, finite $N$ distributions is also well
approximated by an exponential distribution (a Gaussian distribution
this time for $\omega_{n}$) with typical energy departure from $0$
of order $1/N$ for the energy and a variance of order $1/\sqrt{N}$
for $\omega_{n}$. One may also check that for large $n$ ($\lambda_{n}\gg\lambda_{1}$,
the variance for the enstrophy becomes independent of $n$ (asymptotic
equipartition of the enstrophy). \\

Even if we have described finite $N$ effects for finite $N$ approximations
of the microcanonical measure $\mu_{m}^{N}$, the only invariant measure
for the Euler equation is the limit one $\mu_{m}$. From the preceding
discussion we see that all the energy is concentrated on the first
mode and that the excess enstrophy $\Gamma_{2}-2\lambda_{1}E$ goes
to smaller and smaller scales, leading to a zero energy or zero enstrophy
per mode in the infinite $N$ limit. This condensation of the energy
in the first mode is the main physical result of this energy-enstrophy
ensemble. This is a non trivial prediction of equilibrium statistical
mechanics of two-dimensional flows, by contrast with the triviality
of the results for three-dimensional flows.

\paragraph*{The Kraichnan energy-enstrophy theory}

The term of condensation has been proposed by Kraichnan from the analysis
of the energy-enstrophy canonical ensembles \cite{Kraichnan_Motgommery_1980_Reports_Progress_Physics}.
As explained in section \ref{sub:Hamiltonian-structure}, canonical
measure are not relevant for fluid systems and they may be useful
only when given equivalent results to microcanonical measures. Kraichnan
noticed this and worked nevertheless with the canonical ensemble, maybe because he didn't know how to make microcanonical computations,
and most probably because at that time the possibilities of ensemble
inequivalence were nearly unknown%
\footnote{The first observation of ensemble inequivalence have been made in
the astrophysical context \cite{LyndenBell:1968_MNRAS,Hertel_Thirring_1971_AnnPhys},
and then observed for
two dimensional flows \cite{Smith_ONeil_Physics_Fluids_1990_Inequivalence,Kiessling_Lebowitz_1997_PointVortex_Inequivalence_LMathPhys,Eyink_Spohn_1993_JSP....70..833E,EllisHavenTurkington:2000_Inequivalence,Venaille_Bouchet_PRL_2009}. Thorough study of ensemble inequivalence in the broad class of systems with long range interactions has been addressed during the last decade by many others, see for instance  \cite{Dauxois_Ruffo_Arimondo_Wilkens_2002LNP...602....1D,Chavanis_2006IJMPB_Revue_Auto_Gravitant,Campa_Giansanti_Morigi_Labini_2008AIPC..970.....C,Bouchet_Barre_Venaille_2008_Proceeding_Assise,Campa_Dauxois_Ruffo_Revues_2009_PhR...480...57C,Bouchet_Barre:2005_JSP} and references therein.%
}. Unfortunately, as explained above the energy-enstrophy ensemble
is an example of partial ensemble inequivalence. These remarks explain
the difficulties encountered by Kraichnan by analyzing the canonical
measure and why he wrongly concluded that a statistical mechanics
approach would work only for truncated systems. Working in the microcanonical
ensemble actually allows to build invariant measures of the real Euler
equation. If one is however interested in truncated systems, then
Kraichnan's work remains very useful.

More importantly, when looking closely at Kraichnan's
works (see for instance \cite{Kraichnan_Motgommery_1980_Reports_Progress_Physics}
page 565), one sees that in the canonical ensemble, a complete condensation
of the energy on the gravest mode occurs only for specific values
of the thermodynamical parameters. For most values of the thermodynamical
parameters, an important part of the energy remains on the other modes.
Still Kraichnan argued, probably from numerical observations available
at the time and from physical insight, that these cases leading to
a condensation were the most interesting ones. The microcanonical
treatment we propose here proves that a complete condensation occurs
whatever the values of the energy and of the enstrophy, in the microcanonical
ensemble. A complete condensation is actually observed in many numerical
simulations. We thus conclude that the physical insight of Kraichnan
and his concept of condensation describes the relevant physical mechanism,
but that a treatment in the microcanonical ensemble provides a much
better understanding, and overcomes the preceding contradictions.

\paragraph*{Limitations and interest of the energy-enstrophy approach}

There is no reason to consider only the energy and enstrophy invariants, except for being able to solve easily the mathematics. Here we used this property for instance to illustrate the equivalence of the mean field variational problems with a direct definition of the microcanonical measure. Another class of statistical equilibria with easily solvable solutions is the one for which only energy, enstrophy and circulations are taken into account \cite{Venaille_Bouchet_PRL_2009,NasoChavanisDubrulle} \footnote{\cite{Venaille_Bouchet_PRL_2009} proves relations between phase transitions on one hand, and ensemble equivalence and inequivalence results on the other hand. \cite{NasoChavanisDubrulle} proves specifically the equivalence between entropy maximization at fixed energy, circulation and enstrophy on one hand, and macroscopic enstrophy minimization at fixed energy and circulation on the other hand (see also \cite{Bouchet:2008_Physica_D} for equivalence results in a more general context)}. Moreover, several studies, among which \cite{Abramov_Majda_2003_PNAS,Dubinkina_Frank_2010JCoPh}, have specifically addressed the importance of higher potential vorticity moments, showing that they may be indeed essential in some case.

From the following studies we will see that taking into account all
invariants, it will be wrong that the energy is limited to the first
mode $e_{1}$. However the energy-enstrophy measure may be in some
cases a good approximation: for instance in the limit of small energy,
most of the energy will remain in the first few modes. The notion
of condensation will thus be valid only roughly speaking.

By contrast, in some cases like for instance for doubly periodic domains
with aspect ratio close to one but not exactly one (see section \ref{sub:The-example-doubly-periodic}),
the notion of condensation would lead to completely wrong predictions.

\subsubsection{Validity of a mean field approach to the microcanonical measures\label{sub:Mean_Field}}

For pedagogical reasons, we have considered in the previous section
the energy-enstrophy microcanonical ensemble. It is shown in \cite{Bouchet_Corvellec_JSTAT_2010} that within this ensemble the correlation coefficient
between vorticity at point $\mathbf{r}$ and vorticity at point $\mathbf{r'}$
is zero. It would be possible to prove without much difficulties that
vorticity at points $\mathbf{r}$ and $\mathbf{r}'$ are actually
independent variables. Such a result is extremely important and does
apply to a much wider context than the energy-enstrophy measure, for
instance it will remain true for all the microcanonical invariant
measures, whatever the number of invariants. We will explain why vorticity
fields are independent for microcanonical measures below. Let us first
analyze an extremely important implication: the possibility to quantify
the volume of the phase space through the Boltzmann--Gibbs entropy formula.
\\

A classical example where degrees of freedom can be considered independent
is an ensemble of particles undergoing collisions (for instance hard
spheres) in the dilute limit (the Boltzmann-Grad limit). Microscopically,
particles travel at a typical velocity $\bar{v}$ and collide with
each other after traveling a typical distance $l$ called the mean
free path. Let $\sigma$ be the diffusion cross-section for these
collisions. One has $\sigma=\pi a^{2}$ where the parameter $a$ is
of the order of the particle radius. The mean free path is defined
as $l=1/\left(\pi a^{2}n\right)$, where $n$ is the typical particle
density. The Boltzmann equation applies when the ratio $\Gamma=a/l$
is small (the Boltzmann-Grad limit \cite{Spohn_1991}). In the limit
$\Gamma\rightarrow0$, any two colliding particles can be considered
as independent (uncorrelated) as they come from very distant areas.
This is the basis of Boltzmann hypothesis of molecular chaos (Stosszahl
Ansatz). It explains why the evolution of the $\mu$-space distribution
function $f(\mathbf{x},\mathbf{p},t)$ may be described by an autonomous
equation, the Boltzmann equation ($\mathbf{x}$,$\mathbf{p}$ refers respectively to position and momentum, the $\mu$-space
is the six dimensional space of spatial variable $\mathbf{x}$ and
momentum $\mathbf{p}$).

There is a classical argument by Boltzmann (that one can found in any
good textbook in statistical mechanics) to prove that the Boltzmann
entropy of the distribution $f$ is, up to a multiplicative constant,
given by the Boltzmann--Gibbs formula:
\begin{equation}
\mathcal{S}=-\int\mathrm{d}\mathbf{x}\mathrm{d}\mathbf{p}\, f\log f.\label{eq:Maxwell-Boltzmann-Entropy}
\end{equation}
We stress that this formula for the Boltzmann entropy is not a Gibbs
entropy %
\footnote{The Gibbs entropy $S=-k\int\rho(p_{i},q_{i})\log_{2}(\rho(p_{i},q_{i}))\, dp_{i}dq_{i}$
is an ensemble entropy, a weight on the phase space, whereas the Boltzmann--Gibbs
formula for the entropy is an integral over the $\mu$-space. In the case of dilute
gases, the Boltzmann--Gibbs formulae for the entropy is just the opposite of the $H$
function of Boltzmann. We avoid this terminology here since our discussion
is not related to relaxation towards equilibrium, and because the
equivalent of an $H$ theorem has never been proved for the 2D Euler
equations.%
}.
The essential point is that this formula is a valid counting of the
volume of the accessible part of the phase space only when particles
can be considered as independent. For instance, for particles with
short range interactions studied by Boltzmann, this is valid only
in the Boltzmann-Grad limit.

As discussed above, in the energy-enstrophy ensemble, vorticity field
values are independent. As we will explain bellow, the reason is completely
different from the Boltzmann case, there is here no dilute-gas (Boltzmann-Grad)
limit. However the consequences will be the same: if we define $\rho\left(\mathbf{r},\sigma\right)$
as $\rho\left(\mathbf{r},\sigma\right)\mathrm{d\mathbf{r}}\mathrm{d}\sigma$
being the probability to have values of the vorticity $\omega$ between
$\sigma$ and $\sigma+\mathrm{d}\sigma$ in the area element $\mathrm{d}\mathbf{r}$
around $\mathbf{r}$, then the entropy \begin{equation}
\mathcal{S}=-\int_{\mathcal{D}}\mathrm{d}\mathbf{r}\int_{-\infty}^{+\infty}\mathrm{d}\sigma\,\rho\ln\rho,\label{eq:Maxwell-Boltzmann-Entropy-Euler}\end{equation}
actually quantifies the volume of the phase space. Let us explain
the meaning of this last sentence, for instance in the case of the
energy-enstrophy ensemble. The probability $\rho$ is normalized (
$N\left[\rho\right](\mathbf{r})\equiv\int_{-\infty}^{+\infty}\hspace{-0.3cm}\mathrm{d}\sigma\,\rho\left(\sigma,\mathbf{r}\right)=1$)
and we define the average vorticity as $\bar{\omega}\left(\mathbf{r}\right)=\int_{-\infty}^{+\infty}  \mathrm{d}\sigma\,\sigma\rho\left(\sigma,\mathbf{r}\right)$.
Then the equilibrium entropy
\begin{equation}
S\left(E,\Gamma_{2}\right)=\sup_{\left\{ \rho|N\left[\rho\right]=1\right\} }\hspace{-0.1cm}\left\{ \frac{1}{\mathcal{\left|D\right|}}\mathcal{S}[\rho]\ |\ \mathcal{E}\left[\overline{\omega}\right]=E\ ,\,\int\mathrm{d}\mathbf{r}\mathrm{d}\sigma\,\sigma^{2}\rho=\Gamma_{2}\right\}
\label{eq:Entropie_Equilibre_MeanField_Energie_Enstrophie}
\end{equation}
is exactly the same as the Boltzmann entropy defined from the rescaled
logarithm of volume of the phase space defined by equation (\ref{eq:Definition-Entropy}). This variational problem (\ref{eq:Entropie_Equilibre_MeanField_Energie_Enstrophie}) means that the equilibrium entropy $S(E,\Gamma)$ is the supremum  of the mixing entropy $\mathcal{S}[\rho]$ defined above, among all the normalized probability $\rho(\sigma,\mathbf{r})$ that are characterized by a given value of the energy $E$ and enstrophy $\Gamma_2$.

The definition of the entropy (\ref{eq:Entropie-energie-enstrophie-texte}-\ref{eq:Definition-Entropy})
and the variational problem (\ref{eq:Entropie_Equilibre_MeanField_Energie_Enstrophie})
are so different, that the fact that they express the same concept
seems astonishing. These types of results are indeed one of the great
achievements of statistical mechanics. It is shown in \cite{Bouchet_Corvellec_JSTAT_2010} that starting from (\ref{eq:Definition-Entropy}) and computing $S(E,\Gamma_{2})$ gives the same result as (\ref{eq:Entropie-energie-enstrophie-texte}).\\

The deep reason why vorticity fields are independent for microcanonical
measures, and henceforth why entropy can be expressed by (\ref{eq:Entropie_Equilibre_MeanField_Energie_Enstrophie})
can be explained rather easily on a heuristic level. Correlations
between variables could appear through the invariants constrains only.
For instance the 2D Euler equation energy can be expressed in the form
where interactions between vorticity appear explicitly, using the
Laplacian Green function $H\left(\mathbf{r},\mathbf{r}^{\prime}\right)$
($\Delta H\left(\mathbf{r},\mathbf{r}^{\prime}\right)=\delta\left(\mathbf{r},\mathbf{r}^{\prime}\right)$
with Dirichlet boundary conditions):

\begin{equation}
\mathcal{E}[\omega]=-\frac{1}{2}\int_{\mathcal{D}}\mathrm{d}\mathbf{r}\,\omega\Delta^{-1}\omega=-\frac{1}{2}\int_{\mathcal{D}}\int_{\mathcal{D}}\mathrm{d}\mathbf{r}\mathrm{d}\mathbf{r}'\,\omega\left(\mathbf{r}\right)H(\mathbf{r},\mathbf{r}^{\prime})\omega\left(\mathbf{r}'\right).\label{eq:EnergyGreen}\end{equation}
In the formula above, $H\left(\mathbf{r},\mathbf{r}^{\prime}\right)$
appears as the coupling between vorticity at point $\mathbf{r}$ and
vorticity at point $\mathbf{r}'$. The Green function of the Laplacian
in a two-dimensional space is logarithmic, which is not integrable
in the whole plane, hence lead to a non-local interaction. Then the vorticity at point $\mathbf{r}$ is coupled
to the vorticity at any other points of the domains and not only close
ones.

For people trained in statistical mechanics, it is natural that in
systems where degrees of freedom are coupled to many other, these
degrees of freedom can be considered as statistically independent
and a mean-field approach will be valid. For instance in systems with
nearest neighbor interactions, a mean field approach becomes exact
for large dimensions, when the effective number of degrees of freedom
to which one degree of freedom is coupled becomes infinite. For people
not trained in statistical mechanics, this can be understood simply,
as one increases the number of coupling, the interaction felt by
one degree of freedom is no more sensitive to the fluctuations of the
others but just to their average value, due to an effect similar to
what happens for the law of large numbers. Then a mean field treatment
becomes exact, which is equivalent to saying that different degrees of
freedom can be considered as statistically independent.

Because of the non locality of the Green function, the vorticity field
is virtually coupled to an infinite number of degrees of freedom,
so that a mean-field is actually exact. This also explains why the
energy computed from the average vorticity field appears in the variational
problem (\ref{eq:Entropie_Equilibre_MeanField_Energie_Enstrophie}).
\\

To formalize the preceding heuristic explanation, in order to prove
that the mean-field approximation is exact and to prove that the Boltzmann--Gibbs formulae
for the entropy (\ref{eq:Maxwell-Boltzmann-Entropy}) is relevant, we need a rather
technical discussion. We will not explain this in details. This has
been for instance justified by theoretical physicists for the point
vortex model in the 1970s (assumed to be valid by Joyce and Montgomery
\cite{Joyce_Montgommery_1973} and then proved to be self-consistent, for instance
in a Cramer Moyal expansions). In
the 1980s, rigorous mathematical proofs have been given also for
the point vortex model (see \cite{Eyink_Spohn_1993_JSP....70..833E,Kiessling_Lebowitz_1997_PointVortex_Inequivalence_LMathPhys,CagliotiLMP:1995_CMP_II(Inequivalence)}
and references therein). In the modern formulation of statistical
mechanics, the entropy appears as a large deviation rate function
for an ensemble of measures, justifying (\ref{eq:Maxwell-Boltzmann-Entropy})
and the variational problem (\ref{eq:Entropie_Equilibre_MeanField_Energie_Enstrophie}).
The proof of such large deviation results leading to the microcanonical
measures for the Euler and quasi-geostrophic equations, justifying
the mean field approach, can be found in \cite{Michel_Robert_LargeDeviations1994CMaPh.159..195M}
(see also \cite{Boucher_Ellis_1999_AP} and references therein).\\

We thus conclude that a mean field approach to the microcanonical
measures of the Euler and quasi-geostrophic equations is valid. This
justifies the use of the entropy (\ref{eq:Maxwell-Boltzmann-Entropy})
and of variational problems similar to (\ref{eq:Entropie_Equilibre_MeanField_Energie_Enstrophie})
but with all invariants of the Euler equations. This is a drastic
simplification compared to direct approaches as the one presented
in section \ref{sub:3D-Stat-mech-and-Energy-Enstrophy} for the energy-enstrophy statistical
mechanics. The first presentation of the equilibrium statistical mechanics
of the 2D Euler and quasi-geostrophic equations on this form dates from
the the beginning of the 1990s with the works of Robert, Sommeria
and Miller \cite{Robert:1990_CRAS,Miller:1990_PRL_Meca_Stat,Robert:1991_JSP_Meca_Stat,RobertSommeria:1992_PRL_Relaxation_Meca_Stat}.
We thus call this theory the Robert-Sommeria-Miller (RSM) theory.

\section{Equilibrium statistical mechanics of two dimensional and geophysical
flows\label{sec:Equilibrium-statistical-mechanics}}

In sections \ref{sub:Hamiltonian-structure} and \ref{sub:microcanonical-measure}
we have recalled the basis of equilibrium statistical mechanics of
Hamiltonian systems: building invariant measures based on the Liouville
theorem, especially the microcanonical measure that takes into account
all of the dynamical invariants of the equations. We have explained
in sections \ref{sub:3D-Stat-mech-and-Energy-Enstrophy} how this
program can be applied to fluid dynamics equations and in section
\ref{sub:Mean_Field} why for the 2D Euler and quasi-geostrophic equations
the microcanonical measure is described by a mean field variational
problem.

The aim of this section is to describe this mean field variational
problem and the tools used to actually compute the equilibrium states.
We consider the limit of small energy, as a simple example were an
analytic treatment is possible in order to illustrate the theory and
especially the notion of phase transition. Phase transition is a key
concept of thermodynamics and statistical physics, where the physical system undergoes drastic qualitative changes
as external parameters are tuned. In the statistical mechanics of
hydrodynamic problems, the flow undergoes continuous or discontinuous
transitions of the topology of the flow streamlines. We discuss applications
of this equilibrium theory to real flows, for instance in the geophysical
context, in sections \ref{sec:First Order GRS and rings} and \ref{sec:Gulf Stream and Kuroshio}
and discussion of out of equilibrium statistical mechanics in section
\ref{sec:Out of equilibrium}.

\subsection{Mixing entropy and equilibrium states\label{sub:The-mixing-entropy}}

\subsubsection{Equilibrium entropy and  microcanonical equilibrium states}

We describe in this section the microcanonical variational problem
and microcanonical entropy, the Robert-Sommeria-Miller theory, following
these first papers \cite{Robert:1990_CRAS,Miller:1990_PRL_Meca_Stat,Robert:1991_JSP_Meca_Stat,RobertSommeria:1992_PRL_Relaxation_Meca_Stat}.

We explained is section \ref{sub:Mean_Field} that for the microcanonical
measure, the vorticity field at different locations are statistically
independent. We denote $\rho\left(\mathbf{r},\sigma\right)\mathrm{d\mathbf{r}}\mathrm{d}\sigma$
the probability for the vorticity $\omega$ to take values between
$\sigma$ and $\sigma+\mathrm{d}\sigma$ in the area element $\mathrm{d}\mathbf{r}$
around $\mathbf{r}$. Then the Boltzmann-Gibbs entropy \begin{equation}
\mathcal{S}\left[\rho\right]\equiv-\int_{\mathcal{D}}d^{2}\mathbf{r}\int_{-\infty}^{+\infty}\hspace{-0.3cm}d\sigma\,\rho\log\rho.\label{eq:Entropie_Maxwell_Boltzmann}\end{equation}
is a quantification of the number of microscopic states (vorticity
fields) corresponding to a macroscopic states (probability density
$\rho$). A more precise meaning of this will be given with the variational
problem (\ref{eq:MVP}) below. The most probable state, close to
which most of the other states will be concentrated, will thus be
the maximizer of the entropy (\ref{eq:Entropie_Maxwell_Boltzmann})
with constraints associated with each dynamical invariant.\\

We now list the constraints. As $\rho$ is a local probability, it
 satisfies a local normalization\begin{equation}
N\left[\rho\right](\mathbf{r})\equiv\int_{-\infty}^{+\infty}\hspace{-0.3cm}\mathrm{d}\sigma\,\,\rho\left(\sigma,\mathbf{r}\right)=1.\label{eq:normalisation}\end{equation}
The conservation of all the Casimir functionals (\ref{eq:casimir}),
or equivalently the known potential vorticity distribution (\ref{eq:distribution_vorticite})
imposes \begin{equation}
D\left[\rho\right](\sigma)\equiv\int_{\mathcal{D}}d\mathbf{r}\,\rho\left(\sigma,{\bf \mathbf{r}}\right)=\gamma\left(\sigma\right).\label{eq:distribution_ro}\end{equation}
 The averaged potential vorticity is \begin{equation}
\overline{q}\left(\mathbf{r}\right)=\int_{-\infty}^{+\infty}\hspace{-0.3cm}\mathrm{d}\sigma\,\,\sigma\rho\left(\sigma,\mathbf{r}\right).\label{eq:vorticite_coarse_grained}\end{equation}
 with the average streamfunction $\bar{\psi}$, defined by $\overline{q}=\Delta\bar{\psi}-\frac{\bar{\psi}}{R^{2}}+h(y)$.
As explained in section \ref{sub:Mean_Field}, because the interactions
are long range and the energy is a sum over infinite contributions,
the energy of the mean field will be equal to the initial energy \begin{equation}
\mathcal{E}\left[\overline{q}\right]\equiv-\frac{1}{2}\int_{\mathcal{D}}\mathrm{d}\mathbf{r}\,\overline{\psi}\overline{q}=E.\label{eq:energie_coarse_grained}\end{equation}

Then the entropy of the system is given by the variational problem

\begin{equation}
S(E,\gamma)= \sup_{\left\{ \rho|N\left[\rho\right]=1\right\} }\hspace{-0.1cm}\left\{ \mathcal{S}[\rho]\ |\ \mathcal{E}\left[\overline{q}\right]=E\ ,D\left[\rho\right]=\gamma\ \right\} \,\,\mbox{(MVP).}\label{eq:MVP}\end{equation}
and, thanks to the large deviation property \cite{Michel_Robert_LargeDeviations1994CMaPh.159..195M}
(see section \ref{sub:Mean_Field}) an overwhelming number of potential vorticity
fields of the microcanonical ensemble will be close to the maximizer
$\rho$ of the variational problem (\ref{eq:MVP}).

Here MPV refers to "microcanonical variational problem". It says that the equilibrium entropy $S(E,\gamma)$  is the supremum of the mixing entropy $\mathcal{S}[\rho]$ among all the normalized probability density field $\rho(\sigma,\mathbf{r})$ that are characterized by a given value of energy $E$ and of the global potential vorticity distribution $\gamma(\sigma)$.\\

Two routes are now possible. The classical one is to look for the critical points of the variational problem (\ref{eq:MVP}). For this, we introduce Lagrange multipliers $\beta$, $\alpha(\sigma)$ and $\zeta(\mathbf{r})$ associated with the conservation of energy $\mathcal{E}$ (\ref{eq:energie_coarse_grained}), vorticity distribution $D(\sigma)$ \ref{eq:distribution_ro}), and normalization constraint $N$ (\ref{eq:normalisation}), respectively. Then critical points $\rho$ are solutions of
\[ \forall\ \delta\rho\quad\delta\mathcal{S}-\beta\delta\mathcal{E}-\int_{-\infty}^{+\infty}\mathrm{d}\sigma\ \alpha(\sigma)\delta D_{\sigma}-\int_{\mathcal{D}}\mathrm{d}\mathbf{r}\ \delta N=0,\] where the entropy is given by  (\ref{eq:Entropie_Maxwell_Boltzmann}). Solving this equation for $\rho$ and using the normalization constraint, we obtain that the critical points $\rho$ verifies the Gibbs state equation  \cite{Robert:1990_CRAS,Miller:1990_PRL_Meca_Stat,Robert:1991_JSP_Meca_Stat,SommeriaRobert:1991_JFM_meca_Stat}:
\begin{equation}
\rho\left(\sigma,\mathbf{r}\right)=\frac{e^{\beta\sigma\psi\left(\mathbf{r}\right)-\alpha(\sigma)}}{Z_{\alpha}\left(\beta\psi\left(\mathbf{r}\right)\right)}\,\,\,\mbox{with}\,\,\, Z_{\alpha}\left(u\right)=\int_{-\infty}^{\infty}\mathrm{d}\sigma\,\exp\left(\sigma u-\alpha\left(\sigma\right)\right),
\label{eq:Gibbs-States}
\end{equation}
We see that $\rho$ depends on $\mathbf{r}$ through the average stream function $\bar{\psi}$. From (\ref{eq:vorticite_coarse_grained}) and (\ref{eq:Gibbs-States}) there is a functional relation between the equilibrium average potential vorticity and the stream functions
\begin{equation}
\bar{q}=g\left(\beta\bar{\psi}\right)\,\,\,\mbox{with}\,\,\, g\left(u\right)=\frac{d}{du}\log Z_{\alpha}.\label{eq:q-psi equilibre}\end{equation}
This last equation characterizes the statistical equilibrium.  One can prove that $g$ is a monotonously increasing function, such that the relation between potential vorticity $\overline{q}$  and streamfunction $\overline{\psi}$ is increasing for positive temperatures ($\beta>0$), and decreasing for negative temperatures ($\beta<0$). This equation has to be solved for any values of the Lagrange parameters $\beta$ and $\alpha\left(\sigma\right)$. Then one has to compute the energy $E$ and potential vorticity distribution $\gamma(\sigma)$ as a function of $\beta$ and $\alpha$.

For a given energy $E$ and distribution $\gamma(\sigma)$, among all the
possible values of $\beta$, $\alpha$ and distribution $\rho$ solving
(\ref{eq:Gibbs-States}-\ref{eq:q-psi equilibre}), the one actually
maximizing the entropy (\ref{eq:MVP}) is selected. In the general
case, this program is not an easy one, and the aim of part of the
following discussion will be to solve this in simple cases and to
describe methods that will possibly make this program simpler.

The second route is to try to work directly with the variational problem
(\ref{eq:MVP}) and to simplify it.
In review we will try to rely as much as possible on variational
problems only. This route will prove to be often physically more enlightening,
at least for the specific examples treated in this review.\\

In his original papers, in the beginning of the 1990s, Miller \cite{Miller:1990_PRL_Meca_Stat,Miller_Weichman_Cross_1992PhRvA} justified formally the mean field variational problem from a formal discretization of the microcanonical measure and its solution through direct computations using the Hubbard-Stratonovich transformation. Robert and Sommeria \cite{Robert:1990_CRAS,Robert:1991_JSP_Meca_Stat,SommeriaRobert:1991_JFM_meca_Stat} were assuming directly and phenomenologically the mean-field variational problem. Only latter on, did the work of Michel and Robert \cite{Michel_Robert_1994_JSP_GRS,Michel_Robert_LargeDeviations1994CMaPh.159..195M} and Boucher, Ellis and Turkington \cite{Boucher_Ellis_1999_AP} justify the mean-field variational problem more rigorously through expliciting the relation with large deviation theory.

We note that a similar mean field variational problem also describes the statistical mechanics of the violent relaxation of the Vlasov equation, both for self-gravitating systems and plasma physics. The mean field variational problem for the Vlasov equation has first been proposed on a phenomenological way by Lynden Bell in the end of the 1960s \cite{LyndenBell:1968_MNRAS}. It can be justified following the same route as for the 2D Euler equations (see for instance \cite{Robert_2000_CommMathPhys-TruncationEuler}). There is indeed a deep analogy, noticed from the 1950s, between the Vlasov and the 2D-Euler equation: both are non-linear conservation laws with an infinite number of Casimir conservation laws, leading to similar properties and analogies, at the level of both dynamics and statistical mechanics. This analogy and its consequences from a statistical mechanics point of view have been used and described in details for instance in \cite{Chavanis_etal_APJ_1996}. A larger class of systems have the same properties: systems with long range interactions. The analogies between the dynamics and statistical mechanics of these systems have been systematically studied during the last decade, see for instance \cite{Dauxois_Ruffo_Arimondo_Wilkens_2002LNP...602....1D,Chavanis_2006IJMPB_Revue_Auto_Gravitant,Campa_Dauxois_Ruffo_Revues_2009_PhR...480...57C,Campa_Giansanti_Morigi_Labini_2008AIPC..970.....C,Bouchet_Gupta_Mukamel_PRL_2009}.

\subsubsection{Ergodicity}

\label{sec:ergodicity}

Section \ref{sec:Equilibrium-statistical-mechanics} describes the
statistical equilibria through the variational problem (\ref{eq:MVP}).
The solution of this variational problem is the most probable state
and also, thanks to the large-deviation property, the state around
which an overwhelming majority of states do concentrate, for the microcanonical
measure. Besides, the microcanonical measure is the most natural invariant
measure of the 2D Euler equations with the dynamical constraints.

Having described a natural invariant measure of the equations is an
important theoretical step. Another important point would be to know
if this invariant measure is the only one having the right values
for the dynamical invariants. The evolution of one trajectory of the
dynamical system also defines a measure (through time averaging).
If we knew the invariant measure were unique, then it would mean that
averaging over the microcanonical measure is equivalent to averaging
over time. When this uniqueness property holds, we call the dynamical
system ergodic.

Generally speaking, the ergodicity of a dynamical system is a property
that is usually extremely difficult to prove. Such proofs exist only
for very few extremely simple systems. Ergodicity is actually thought
to be wrong in general. For instance, in Hamiltonian systems with
a finite number of degrees of freedom, there often exist islands in
phase space in which trajectories are trapped. The common belief in
the statistical mechanics community is that those parts of phase space
where the motion is trapped exist, but occupy an extremely small relative
volume of the phase space, for generic systems with a large number
of degrees of freedom. Apart from a few systems which were proved
to be integrable, this common wisdom has successfully passed empirical
tests of a century of statistical mechanics studies.\\

There is no reason to suspect that this general picture should be
different in the case of the 2D Euler equations, in general. It is
thus thought that an overwhelming number of initial conditions will
have a dynamics consistent with the microcanonical measure predictions.
However, similarly to most other Hamiltonian systems, the 2D Euler
equations are actually non-ergodic.  The proof is quite simple.

Indeed, it is proved in  \cite{Bouchet_Corvellec_JSTAT_2010} that any Young measure for which $\bar{\omega}\left(\mathbf{r}\right)$
is a stationary solution of the 2D Euler equations is either an invariant
or a quasi-invariant measure. The class of invariant measures corresponding
to ensemble of trajectories with given values of the invariants, is
then much, much larger than the class of statistical equilibrium invariant
measures with the same invariants. This proves that nontrivial sets
of vorticity fields are dynamically invariant. In this restricted
sense, this proves that the 2D Euler equations are not ergodic.

This theoretical argument proving non-ergodicity is in accordance
with previous remarks about the phenomenology of the 2D Euler or quasi-geostrophic
equations. For instance, it was observed numerically that initial
conditions with localized vorticity, in large domains, remain localized
(see \cite{Chavanis_Sommeria_1998JFM_LocalizedEquilibria...356..259C}
and references therein; \cite{Chavanis_Sommeria_1998JFM_LocalizedEquilibria...356..259C}
actually proposes an interesting phenomenological modification of
the microcanonical measure approach to cope with this localized dynamics
problem). Another example of possible non-ergodicity is the dynamics
close to stable dynamical equilibria of the equations. When trajectories
come close to such equilibria, they can be trapped (frozen) as was
seen in some numerical simulations. A classical argument by Isichenko
\cite{Isichenko_1997_PhRvL} is that for initial conditions close
to parallel flows, {}``displacement in certain directions is uniformly
small, implying that decaying Vlasov and 2D fluid turbulence are not
ergodic''. Even if the predicted algebraic laws by Isichenko are
most probably wrong, the fact that displacement in directions normal
to the streamlines is uniformly small is probably right, thus being
another argument for non-ergodicity.\\

An important point to be noted, is that the Navier-Stokes equation
with stochastic forces can be proved to be ergodic \cite{Bricmont_Kupianen_2001_Comm_Math_Phys_Ergodicity2DNavierStokes}.
This ergodicity refers to invariant measures of the Navier-Stokes
equations, which are non-equilibrium invariant measures with fluxes
of conserved quantity. A very important point is to understand the
limit of weak forces and dissipation for such invariant measures and
to study their relations with the invariant measures of the 2D Euler
equations. Some very interesting results can be found in \cite{Kuksin_2004_JStatPhys_EulerianLimit}.

\subsubsection{Canonical and Grand Canonical ensembles \label{sub:Canonical-and-Grand-Canonical}}

The microcanonical equilibrium describes the most probable state,
resulting from the microcanonical measure with a given potential vorticity
distribution and energy. From a mathematical point of view, we have
to solve the variational problem (\ref{eq:MVP}). This is a tricky
task, one of the main difficulty being due to the vorticity distribution
and energy constraints. We here define canonical, grand canonical
ensembles, and corresponding equilibrium, that will help us a lot in simplifying
the description of the equilibrium states.\\

It is customary in statistical mechanics to consider statistical ensembles
where the constraints coming from the dynamical invariants are relaxed
(a phrase that will be clarified soon). For instance, in classical
statistical mechanics, the canonical ensemble is obtained by relaxing
the energy constraint, and the grand canonical ensemble by relaxing the
constraint corresponding to the number of particles. We
follow the same paths here. We call grand canonical ensemble, any
ensemble where some or all of the constraints related to the potential
vorticity distribution are relaxed. The meaning of this procedure is
discussed in section \ref{sub:Physical-interpretation-canonical}.

Whereas the microcanonical ensemble is built on the assumption that
all microstates with a given energy are equiprobable (microcanonical
distribution), the canonical ensemble assumes a Boltzmann distribution
of the microstates (distribution weighted by the Boltzmann factor $\exp\left(-\beta E\right)$)). In section \ref{sub:Mean_Field}, we explained
why a mean field description of the microcanonical measure is valid,
and why it leads to the microcanonical variational problem (\ref{eq:MVP}).
The same arguments allow to conclude that, for the canonical distribution,
the most probable state and the Helmholtz free energy can be computed from the
canonical variational problem

\begin{equation}
F(\beta,\gamma)=\inf_{\left\{ \rho|N\left[\rho\right]=1\right\} }\hspace{-0.1cm}\left\{ \mathcal{F}[\rho]=-\mathcal{S}[\rho]+\ \beta\ \mathcal{E}\left[\overline{q}\right]\ |D\left[\rho\right]=\gamma\ \right\} \,\,\mbox{(CVP),}
\label{eq:CVP}
\end{equation}

Here CVP refers to "canonical variational problem". The equilibrium free energy $F(\beta,\gamma)$ is the infimum of the free energy functional $\mathcal{F}[\rho]$ for any normalized field of probability density $\rho(\sigma,\mathbf{r})$ that are characterized by a given global potential vorticity distribution $\gamma(\sigma)$.

 The canonical equilibrium states $\bar{q}=\int d\sigma \sigma \rho $
are the average potential vorticity fields where the free energy minima
are achieved. Comparing (\ref{eq:MVP}) and (\ref{eq:CVP}), the canonical
variational problem (\ref{eq:CVP}) appears to be similar to the variational problem associated with the microcanonical one, but with the energy constraint
 relaxed. This explains the expression ``relaxation''.
Thanks to the Lagrange multiplier rule, the canonical and microcanonical
variational problems have the same critical points (\ref{eq:Gibbs-States}-\ref{eq:q-psi equilibre}),
but the stability properties of the two variational problems may be different
(free energy minima and entropy maxima may be different).
We discuss this last point in more detail in sections \ref{sub:Physical-interpretation-canonical}
and \ref{sub:Long-range-interactions}.

A similar relaxation, this time of the potential vorticity constraint,
leads to fixed-energy grand canonical ensembles. This gives a new
class of variational problems: the minimization of Casimir functionals
\begin{equation}
C(E_{0},s)=\inf_{q}\left\{ \mathcal{C}_{s}[q]=\int_{\mathcal{D}}d^{2}\mathbf{r}\, s(q)\ |\ \mathcal{E}\left[q\right]=E_{0}\ \right\} \,\,\mbox{(EC-VP)},\label{eq:EC-VP}\end{equation}
where $\mathcal{C}_{s}$ is a Casimir functionals, and $s$ a convex
function. Here EC-VP refers to "energy-Casimir variational problem". The fixed-energy grand canonical equilibria are the minimizers
of this variational problem. The relation of this last variational
problem to the microcanonical one is not obvious at a first sight.
The mathematics of such a relation can be skipped, at first reading.
For a detailed account of the derivation of (\ref{eq:EC-VP}) from
the grand canonical distribution, we refer to \cite{Bouchet:2008_Physica_D}.
The convex function $s$ is related to the Lagrange parameters associated
with the conservation of the vorticity distribution (the grand canonical
thermodynamic variables)\footnote{These relations are given by formulas (12) and (16) of \cite{Bouchet:2008_Physica_D}}.

Critical points of (\ref{eq:EC-VP}) are solutions of
\[ \forall\ \delta q\ \ \delta\mathcal{C}_{s}-\beta\delta\mathcal{E}=0,\]
where $\beta$ is the Lagrange parameter associated with the energy
constraint. This yields to the following relation for the critical points:
\[ q=\left(s^{\prime}\right)^{-1}(-\beta \psi).\]
We conclude that if
\begin{equation}
\left(s^{\prime}\right)^{-1}(-u) = g(u),
\label{s_Z}
\end{equation}
where $g(u)$ is defined by equation (\ref{eq:q-psi equilibre}), then the microcanonical variational problem (\ref{eq:MVP}) and  canonical one (\ref{eq:EC-VP}) have the same critical points. Moreover, it is proven in \cite{Bouchet:2008_Physica_D} that if (\ref{s_Z}) holds, then any stable canonical equilibrium (\ref{eq:EC-VP}) is a stable microcanonical equilibrium (\ref{eq:MVP}).

Notice that the variational problem EC-VP (\ref{eq:EC-VP}) was classically
used before this statistical mechanics theory, and independently of
it. It is called the Energy-Casimir variational problem, and was used in
classical works on nonlinear stability of Euler stationary flows \cite{Arnold_1966,Holm_etal_PhysRep_1985}, also allowing to prove the nonlinear stability of some of the statistical equilibrium states \cite{SommeriaRobert:1991_JFM_meca_Stat,Michel_Robert_1994_JSP_GRS}), as discussed in section \ref{sub:multiple_equilibria}. In addition, a minimum enstrophy principle has been previously proposed by  Bretherton--Haidvogel  \cite{BrethertonHaidvogel} to predict the large scale organization of freely evolving 2D and geophysical flows. This approach led to the resolution of the variational problem (\ref{eq:EC-VP}) in the particular case $s(q)=-q^2$ (then the enstrophy of the flow is minimized at fixed energy). Although there is \textit{a priori} no clear physical reason to motivate such a  choice on a general context, the results of this subsection show that the Bretherton--Haidvogel theory is a particular case of the RSM statistical theory, since any solution of (\ref{eq:EC-VP})  is an RSM statistical equilibrium state.

\subsection{Physical interpretation of the canonical and grand canonical ensembles\label{sub:Physical-interpretation-canonical}}

We have just presented canonical and grand canonical ensembles, and
the relaxed variational problems. It is essential to understand their
physical interpretation.

In classical statistical mechanics, two types of interpretations of
canonical ensembles may be relevant, depending on the physical problem
under consideration. The first physical interpretation is that the statistics
of a system in contact with a thermal bath is actually described
by the canonical distribution. The canonical distribution is thus
the natural distribution in many cases, in condensed matter for instance.
But when the physical system can be considered  isolated
(this is usually a matter of comparing the characteristic time
for energy exchanges with the environment with the characteristic time for
relaxation toward equilibrium), then the microcanonical distribution
and ensemble are the relevant ones. In this case of an isolated system,
the canonical distribution can still be considered, based on the fact
that microcanonical and canonical distributions are usually equivalent
in the thermodynamic limit: they give the same predictions for the
average of macroscopic variables. In this second interpretation, the
canonical distribution and ensemble appears as a very useful mathematical
way to avoid some tricky technical difficulties related to the
energy constraint in the microcanonical distribution.\\

Let us discuss more specifically the case of fluid dynamics, and the
2D Euler and quasi-geostrophic equations. First, there
seems to be no way so far to couple such flows with a thermal bath. Also,
for the grand canonical ensemble, it is hard to imagine what a potential vorticity
bath could be. Then, only the second interpretation of the canonical
ensemble can be a relevant interpretation of the relaxed ensembles:
it is a very useful mathematical trick, nothing more. We are thus
led to follow this second interpretation only. CVP (\ref{eq:CVP})
and EC-VP (\ref{eq:EC-VP}) are far more simple variational problems
than the microcanonical ones MVP (\ref{eq:MVP}). Besides all solutions
of CVP and EC-VP are solutions of MVP for some energies and some potential
vorticity distributions (see \cite{Bouchet:2008_Physica_D} for a
proof). The relaxed ensembles are thus very useful.

There is still a crucial difference between usual statistical mechanics
and the statistical mechanics of two-dimensional and geophysical flows:
microcanonical and relaxed ensembles are often non-equivalent. This
is reflected by the fact that there may exist microcanonical equilibria
that are not equilibria of the relaxed ensemble (the converse is
not possible, as just stated above). An affirmative point however, is
that such a situation of ensemble inequivalence can be detected from
the analysis of relaxed ensembles only (see \cite{Bouchet_Barre:2005_JSP}
for a thorough discussion). Therefore, it is always a good choice to begin
with the study of the least constrained ensemble. We show in section
\ref{sub:The-example-doubly-periodic} how to use this general idea
in specific examples.

\subsection{Long range interactions and possible statistical ensemble non-equivalence\label{sub:Long-range-interactions}}

We explained in section \ref{sub:Mean_Field} that the energy
of the 2D Euler equations can be expressed in a form where interactions
between vorticity values appear explicitly, using the Laplacian's Green function
\begin{equation}
\mathcal{E}[\omega]=-\frac{1}{2}\int_{\mathcal{D}}\mathrm{d}\mathbf{r}\,\omega\Delta^{-1}\omega=-\frac{1}{2}\int_{\mathcal{D}}\int_{\mathcal{D}}\mathrm{d}\mathbf{r}\mathrm{d}\mathbf{r}'\,\omega\left(\mathbf{r}\right)H(\mathbf{r},\mathbf{r}^{\prime})\omega\left(\mathbf{r}'\right).\label{eq:EnergyGreen2}\end{equation}
In formula (\ref{eq:EnergyGreen2}), $H\left(\mathbf{r},\mathbf{r}^{\prime}\right)$
appears as the coupling between vorticity at point $\mathbf{r}$ and
vorticity at point $\mathbf{r}'$. The Green function of the Laplacian
in a two-dimensional space is logarithmic, which is not integrable
over the whole plane. The interaction between vorticity at different
points is thus a long-range interaction. This is the main reason why
for statistical equilibria, the vorticity values at different points
can be considered uncorrelated, and why mean-field variational
problems (\ref{eq:MVP}) describe the statistical equilibria.

In physics, there is a large set of systems with long-range interactions
(in the sense of a non-integrable potential). Self-gravitating stars
\cite{Binney_Tremaine_1987_Galactic_Dynamics,Spitzer_1991}, plasma
\cite{Nicholson_1991,Lifshitz_Pitaevskii_1981_Physical_Kinetics},
particles in accelerators, free-electron lasers, magnetic systems
are examples among others of systems with long range interactions.
For the same reasons as the ones presented in section \ref{sub:Mean_Field},
the equilibrium statistical mechanics of these systems will be described
by mean-field variational problems similar to (\ref{eq:MVP}) or (\ref{eq:CVP}).
Moreover, unlike systems with short range interactions,
systems with long range interactions are not additive
(in the limit of a large number of degrees of freedom, if the system
is divided into two macroscopic sub-parts, the total energy is not approximately
equal to the sum of the energies of the two sub-parts). This non-additivity
has drastic physical consequences. For instance, the usual proof for
the concavity of the entropy (in the context of short range interacting
systems) relies on the additivity of the energy. It is then possible
to observe non-concave entropies, and henceforth negative
heat capacities (temperature decreases when energy increases!) for systems
with long-range interactions, as first observed in self-gravitating
systems \cite{LyndenBell:1968_MNRAS}.

The study of the statistical mechanics of systems with long-range
interactions has been a very active branch of statistical mechanics
over the past ten years (see articles in proceedings and reviews
\cite{Dauxois_Ruffo_Arimondo_Wilkens_2002LNP...602....1D,Chavanis_2006IJMPB_Revue_Auto_Gravitant,Campa_Giansanti_Morigi_Labini_2008AIPC..970.....C,Bouchet_Barre_Venaille_2008_Proceeding_Assise,Campa_Dauxois_Ruffo_Revues_2009_PhR...480...57C},
among others). In two-dimensional and geophysical flows, unusual
thermodynamic properties related to long range interactions have
also been observed \cite{Smith_ONeil_Physics_Fluids_1990_Inequivalence,Kiessling_Lebowitz_1997_PointVortex_Inequivalence_LMathPhys,CagliotiLMP:1995_CMP_II(Inequivalence),EllisHavenTurkington:2000_Inequivalence,EllisHavenTurkington:2002_Nonlinearity_Stability,Venaille_Bouchet_PRL_2009}
and their consequence for the stability theory \cite{EllisHavenTurkington:2002_Nonlinearity_Stability}
and related phase transitions \cite{Bouchet_Barre:2005_JSP} has been
discussed.\\

The study of these thermodynamic peculiarities would be a natural
extension of this review. Beyond their theoretical interest, these
studies give important practical outcomes, such as simple characterization
of equivalence between the variational problems (\ref{eq:MVP}) and
(\ref{eq:CVP}) from the entropy curve \cite{EllisHavenTurkington:2000_Inequivalence},
or actually from the free energy curve \cite{Bouchet_Barre:2005_JSP,Venaille_Bouchet_PRL_2009},
classification of all possible phase transitions \cite{Bouchet_Barre:2005_JSP},
which is a very useful guide in any particular study, or new proofs
of flow stability \cite{EllisHavenTurkington:2002_Nonlinearity_Stability}.
A detailed presentation of these results is however beyond the scope
of this review.

\subsection{Statistical equilibria in the limit of affine relations between (potential)
vorticity and streamfunction \label{sub:strong_mixing}}

In section \ref{sub:The-mixing-entropy} we have described the variational
problem which describe statistical equilibria (\ref{eq:MVP}). We
have seen that the critical points of this variational problems give
a nonlinear relation between (potential) vorticity and streamfunction
(Eq. (\ref{eq:q-psi equilibre})), which we write again using the
relation between streamfunction and potential vorticity (\ref{dir})\begin{equation}
q=\Delta\psi-\frac{\psi}{R^{2}}=g\left(\beta\psi\right)\label{eq:q-psi equilibre-1}\end{equation}
(in order to simplify the discussion of this section we assume $\eta_{d}=0$,
the generalization to the case $\eta_{d}\neq0$ would be easy). Equation
(\ref{eq:q-psi equilibre-1}) is a nonlinear elliptic partial differential
equation whose general solution is not easily found. Some algorithms
to solve numerically such an equation (or directly the variational
problem (\ref{eq:MVP})) will be described in section \ref{sub:Numerical-methods-equilibrium}.
Explicit analytic solution can be found only in some specific limit.
A first limit, for which solutions are known is the limit when $g$
is an affine function ($g(x)=ax+b$), called affine $q-\psi$ limit,
that we describe in this section. This limit and normal forms due
to small non-linear effects close phase transitions will also be used
in the treatment of the example discussed in section \ref{sub:The-example-doubly-periodic}.
Another limit which can be treated analytically is a limit of very
strong non-linearity which will be the method used in sections \ref{sec:First Order GRS and rings}
and \ref{sec:Gulf Stream and Kuroshio}.\\

Chavanis and Sommeria first solved statistical equilibria for an affine $g$ \cite{ChavanisSommeria:1996_JFM_Classification}. This work describe for instance phase transitions related to the domain geometry; for instance in a rectangular box a phase transition occurs between monopoles and dipoles when the aspect ratio is changed. We refer to \cite{ChavanisSommeria:1996_JFM_Classification} for a more detailed discussion. Subsequent works \cite{Venaille_Bouchet_PRL_2009,VenailleBouchetJSP}, taking a different perspective by solving directly the variational problems, and describing wider classes of solutions have shown that the affine limit give examples
of statistical ensemble inequivalence (see section \ref{sub:Long-range-interactions})
and of bicritical points and second order azeotropy, two phase transition
types that were not observed before as statistical equilibria for
turbulence problems. Applications to ocean model flows, like Fofonoff
flows have also been discussed recently \cite{Venaille_Bouchet_PRL_2009,VenailleBouchetJSP,NasoChavanisDubrulle,NasoChavanisDubrulle2}.
All these results show that the affine $q-\psi$ limit is extremely rich from a physical point
of view. The reference cited above contain much more results. The
affine $q-\psi$ limit is also extremely interesting from a pedagogical
point of view.

An essential point is to understand the physical circumstances for
which the affine $q-\psi$ limit is relevant. Two different and complementary
types of justification exist. In section \ref{sub:3D-Stat-mech-and-Energy-Enstrophy},
studying the energy-enstrophy ensemble, we have seen that taking into
account enstrophy only as a Casimir invariant leads to an affine $q-\psi$
relation. We however stressed that there is no a priori reason to
take into account only quadratic invariants. As first noticed in \cite{ChavanisSommeria:1996_JFM_Classification},
when $\beta$ is very small (very large temperatures) the energy constraint
has less effect than in other situations and the system has a nearly
homogenized potential vorticity %
\footnote{For a given global distribution $\gamma(\sigma)$, the macroscopic
field $\rho$ (\ref{eq:Gibbs-States}) does not depend on the spatial
coordinates $\mathbf{r}$ when $\beta=0$ (infinite temperatures).
The corresponding potential vorticity field $\bar{q}$, given by equation
(\ref{eq:q-psi equilibre}), is therefore fully homogenized.%
}. Such states correspond to peculiar values of the energy. In this
{}``strong mixing'' limit $\beta\rightarrow0$, an asymptotic expansion
of (\ref{eq:Gibbs-States}) can be performed, by considering $\sigma\beta\psi$
as the small parameter. In this limit, statistical equilibrium states
are characterized by a affine $q-\psi$ relations, whose properties
depend on the energy and the circulation only, even if the infinite
number of constraints are \textit{a priori} considered (see \cite{ChavanisSommeria:1996_JFM_Classification,NasoChavanisDubrulle}
for further discussions). This strong mixing limit is the first type
of justification of the affine $q-\psi$ limit.

A second type of justification for the affine $q-\psi$ limit relies
on the general results about equivalence between variational problems
discussed in section \ref{sub:Canonical-and-Grand-Canonical} (please
see also \cite{Bouchet:2008_Physica_D,Venaille_Bouchet_PRL_2009}).
We explained in the last paragraph of section \ref{sub:Canonical-and-Grand-Canonical}
that the resolution of the variational problem (\ref{eq:EC-VP}) in
the particular case $s(q)=-q^{2}$ , leading to affine $q-\psi$ relation
give access to an admissible class of statistical equilibria. We note,
that actual potential vorticity distribution leading to affine or
close to affine relations may be different in the two cases of the
strong mixing limit, or using the mathematical properties of ensemble
between variational problems. However, the coarse-grained potential
vorticity fields will be the same as they are both described by the
same class of affine $q-\psi$ relations. The phase diagram structure
and phase transitions will also be the same in the two cases

\subsection{Example of doubly-periodic flows\label{sub:The-example-doubly-periodic}}

As an example of application of the equilibrium theory, we treat the
case of the 2D Euler equations ($q=\omega$) on a doubly-periodic domain (torus)
$(x,y)\in\mathcal{D}=(0,2\pi\delta)\times(0,2\pi)$; where $\delta$
is the aspect ratio. We believe this is a very good pedagogical example
because of its simplicity. We will easily solve the problem analytically
in the linearized limit, and make a nonlinear bifurcation analysis,
leading to an interesting phase transition diagram.

This problem is also an interesting one from an academic point of
view, as many direct numerical simulations (DNS) of the Euler or Navier--Stokes
equations are performed in this geometry. The reason is that Fourier
pseudo-spectral codes in periodic domains allow for much more efficient
simulations than any other methods in bounded domains. This geometry
is also advantageous because it does not involve the physics associated
to boundary layers that make the situation more complex. This can
also be considered a drawback, as it is hardly conceivable to make
experimental realizations of this geometry in the lab.

Several generic properties of the theory should emerge through the
study of this example. The first essential point is that the domain
geometry always plays a crucial role. The second point is that the
energy constraint is the one that prevents a complete mixing of the
potential vorticity, it is thus also a key parameter. The last thing
is that, like in usual thermodynamics, phase transitions also play
an essential role for fluid dynamics applications. Indeed, they correspond
to specific values of the physical parameters where drastic changes
to the system occur. As such, these points are particularly interesting
and any theoretical study should emphasize phase transitions.

This study in doubly periodic domains describes phase transitions
between dipoles and parallel flows. The existence of these two types
of statistical equilibria (dipole and parallel flows) were observed
in direct numerical computations and numerical computations of statistical
equilibria in \cite{Yin_Montgomery_Clercx_2003PhFluids}. The bifurcation
theory we present here, predicting the phase diagram, the type of
phase transitions, and the relevant physical control parameter (a
balance between nonlinearity $a_{4}$ and domain geometry $g$) was
first presented in \cite{Bouchet_Simonnet_2008,Morita_Simonnet_Bouchet_2010_NS_Stochastic_Long}.

\subsubsection{Variational problem}

We consider the solutions of the variational problem: \begin{equation}
C(E,s)=\inf_{\omega}\left\{ \mathcal{C}_{s}[\omega]=\int_{\mathcal{D}}d^{2}\mathbf{r}\, s(\omega)\ |\ \mathcal{E}\left[\omega\right]=E\ \right\} \,\,\mbox{(EC-VP)}\label{SE}\end{equation}
 where $E$ is the energy and $s$ a convex function of the vorticity
$\omega$ (the second derivative of $s$ with respect to $\omega$
are positive). This variational problem is the grand canonical variational
problem (\ref{eq:EC-VP}) of section \ref{sub:Canonical-and-Grand-Canonical},
in the case of the 2D Euler equations. We recall that any solution
to (\ref{SE}) is a microcanonical statistical equilibrium, but that
all microcanonical equilibria may not be solutions to (\ref{SE}),
as discussed in section \ref{sub:Physical-interpretation-canonical}.
We also recall (see equation (\ref{s_Z}) page \pageref{s_Z}) that the relation between $s$ and the function
$g$ appearing in the solution of the microcanonical variational problem
(\ref{eq:q-psi equilibre}), is given by \begin{equation}
\overline{\omega}=g(\overline{\psi})=\left(s'\right)^{-1}(-\beta\overline{\psi})\,,\label{faste}\end{equation}
 where $\beta$ is the Lagrange parameter associated with the energy
constraint.

\subsubsection{Quadratic Casimir functionals\label{sub:Quadratic-Casimir}}

We first study the case of a quadratic Casimir functional $s\left(\omega\right)=\omega^{2}/2$. This leads to a linear relation between $\omega$ and $\psi$ (see
(\ref{faste})). As discussed in section \ref{sub:strong_mixing},
a detailed study of microcanonical equilibria with linear $\omega-\psi$
relation was carried out in \cite{ChavanisSommeria:1996_JFM_Classification}
(see also subsection \ref{sub:strong_mixing}), and a detailed study
of variational problems with quadratic functional in relation with
ensemble inequivalence is presented in \cite{Venaille_Bouchet_PRL_2009,VenailleBouchetJSP}. We note that in the case of the doubly periodic geometry, the strong
mixing limit coincide with the weak energy limit.

We want to solve the variational problem
\begin{equation}
C_{2}(E)=\inf_{\omega}\left\{ \mathcal{C}_{2}[\omega]\equiv\frac{1}{2}\int_{\mathcal{D}}d^{2}\mathbf{r}\,\omega^{2}~|~\mathcal{E}[\omega]=E\right\} .\label{SE2}\end{equation}
It is convenient to decompose the fields on the Laplacian eigenmodes.
Let us call $\{e_{i}\}_{i\geq1}$ the orthonormal family of eigenfunctions
of the Laplacian on the domain $\mathcal{D}$: \begin{equation}
-\Delta e_{i}=\lambda_{i}e_{i},\,\,\,\mbox{with}\,\,\,\int_{\mathcal{D}}\mathrm{d}\mathbf{r}\, e_{i}e_{j}=\delta_{ij}.\end{equation}
The eigenvalues ${\lambda_{i}}$ are arranged in increasing order. For
a doubly periodic domain $(x,y)\in(0,2\pi\delta)\times(0,2\pi)$, $e_{i}\left(x,y\right)$
are sines and cosines. For instance, for $\delta>1$: $e_{1}\left(x,y\right)=\sin\left(x/\delta\right)/2\pi\sqrt{\delta},$
$\lambda_{1}=1/\delta^{2}$, $e_{2}\left(x,y\right)=\sin\left(y\right)/2\pi\sqrt{\delta}$
and $\lambda_{2}=1$. We note that $\mbox{cosines}$ are also eigenmodes,
with the same eigenvalues $\lambda_{1}$ and $\lambda_{2}$. This
degeneracy is due to translational invariance. In the following, we
do not take them into account: this amounts to fixing two arbitrary phases
associated with the translational invariance in the directions of $\mathbf{e}_{x}$ and
$\mathbf{e}_{y}$.

We decompose the vorticity on the eigenbasis: $\omega=\sum_{i\geq1}\omega_{i}e_{i}$.
The energy (\ref{eq:Energy}) is then \begin{equation}
\mathcal{E}\left[\omega\right]=\frac{1}{2}\sum_{i\geq1}\lambda_{i}^{-1}\omega_{i}^{2}.\label{eq:Energy_Laplacian_Eigenmodes}\end{equation}
The energy constraint is absorbed into $\omega_{1}$, giving $\omega_{1}^{2}=2\lambda_{1}E-\sum_{i\geq2}(\lambda_{1}/\lambda_{i})\omega_{i}^{2}$.
The condition $\omega_{1}^{2}\geq0$ imposes that the vector $\sum_{i\geq2}\omega_{i}e_i$
 be inside a volume $V_{E}$ defined by \begin{equation}
V_{E}=\left\{ \sum_{i\geq2}\omega_{i}e_i~|~\sum_{i\geq2}\lambda_{i}^{-1}\omega_{i}^{2}\leq2E\right\} .\end{equation}
Substituting the expression for $\omega_{1}^{2}$ in (\ref{SE2}), the
variational problem becomes\begin{equation}
C_{2}(E)=\lambda_{1}E+\inf_{\{\omega_{i}\}_{i\geq2}\in V_{E}}\left\{ \frac{1}{2}\sum_{i\geq2}\frac{\lambda_{i}-\lambda_{1}}{\lambda_{i}}\omega_{i}^{2}\right\} .\label{SE3}\end{equation}
Since for all $i \geq 2$, $\lambda_{i}-\lambda_{1} > 0$, one concludes
that the minimizer of (\ref{SE3}) verifies $\omega_{i}=0$ for all
$i\geq2$: \begin{equation}
\omega=\left(2\lambda_{1}E\right)^{1/2}e_{1}\,\,\,\mbox{and}\,\,\, C_{2}(E)=\lambda_{1}E.\end{equation}
We thus conclude that the equilibrium for a quadratic Casimir functional
with an energy constraint is proportional to the first eigenmode of
the domain.
From the relation between vorticity and streamfunction, we see that
the vorticity field $e_{1}$ corresponds to a parallel flow along
the direction of maximum elongation of the domain \begin{equation}
\mathbf{v}=\frac{\left(2E\right)^{1/2}}{2\pi\delta^{1/2}}\cos\left({x+\phi}{\delta}\right)\mathbf{e}_{y}.\label{eq:paralell_flow_velocity}\end{equation}
\\

This example illustrates general properties of
statistical equilibria:
\begin{enumerate}
\item The equilibrium structure is most of the times at the largest scales
of the domain. This result is in agreement with the widely accepted
empirical rule that the energy piles up to the largest scales of the
domain.
\item The geometry of the domain plays a crucial role for the structure
of the equilibria (in this case, we observe a transition from
flows along the $x$ direction towards flows along the
$y$ direction, as the aspect ratio crosses through the
critical value $\delta_{c}=1$).\\

\end{enumerate}

\paragraph*{The degenerate case. }

A very interesting case is that of the periodic square domain $\mathcal{D}=(0,2\pi)\times(0,2\pi)$.
In this case, the first eigenvalue $\lambda_{1}$ is degenerate, i.e.
$\lambda_{1}=\lambda_{2}$ (this degeneracy is not the degeneracy
due to the translational invariance: there are actually four eigenmodes
for the Laplacian corresponding to eigenvalues $\lambda_{1}=\lambda_{2}$). Then,
for a quadratic functional, a whole family of extrema exists: $\omega=\omega_{1}e_{1}+\omega_{2}e_{2}$
with $\omega_{1}^{2}+\omega_{2}^{2}=2\lambda_{1}E$, and with
entropy $C_{2}(E)=\lambda_{1}E$.

This family of flows includes the parallel flows described previously,
but also dipole flows. For instance, the symmetric dipoles of vorticity
\begin{equation}
\omega\left(x,y\right)=\frac{\left(2E\right)^{1/2}}{2\pi}\left[\sin\left(x+\phi\right)+\sin\left(y+\phi'\right)\right].\label{eq:dipole}\end{equation}
We thus conclude that the maximization of a quadratic Casimir functional
with energy constraint does not select the flow topology in
a square domain; because of the degeneracy, it can be the topology of either parallel
flows or dipoles.

\subsubsection{Weak-energy limit for the maximization of symmetric Casimir functionals }

We now consider the more general case of a Casimir functional \begin{equation}
C_{s}(E)=\inf_{\omega}\left\{ \mathcal{C}_{s}[\omega]\equiv\int_{\mathcal{D}}d^{2}\mathbf{r}\, s\left(\omega\right)\,\,\,|\,\,\,\mathcal{E}[\omega]=E\right\} ,\label{eq:CEs}\end{equation}
where $s$ is a convex function. We suppose that $s$ is even: $s(-\omega)=s(\omega)$. This
is the case for any even initial distribution of the vorticity (an extension of the following discussion to the more general case
would be easy).

It is clear that the multiplication of $s$ by a constant will not
change the minimizers of (\ref{eq:CEs}). Then we can assume, without
loss of generality that $s''\left(0\right)=1$.

We consider the weak-energy limit of the variational problem (\ref{eq:CEs}).
Because the energy is positive-definite, it is clear that for weak
energies, $\omega$ is small. Then $s\left(\omega\right)\sim\omega^{2}/2$.
Then at leading order, in the weak-energy limit, the equilibrium structures
are given by the minimization of a quadratic functional. They are
thus close to the first eigenmodes of the Laplacian of the domain.

In the previous section, we saw that in a doubly periodic square
domain, the minimization of a quadratic functional does not determine
the flow topology, due to the degeneracy of the first Laplacian eigenvalues.
An interesting issue is to understand how this degeneracy is removed
by the next order contribution of a non quadratic functional. We thus
consider\begin{equation}
s\left(\omega\right)=\frac{1}{2}\omega^{2}-\frac{a_{4}}{4}\omega^{4}+o\left(\omega^{4}\right).\label{eq:so}\end{equation}

\paragraph*{Parameter $a_{4}$}

The parameter $a_{4}$ appearing in (\ref{eq:so}) plays a crucial
role. It determines the first correction to a quadratic entropy. Moreover,
it is intimately related to the shape of the relation $\omega=f(\psi)=\left(s'\right)^{-1}(\beta\psi)$.
Indeed, $s'(x)=x-a_{4}x^{3}+o(x^{3})$ and thus $\left(s'\right)^{-1}(x)=x+a_{4}x^{3}+o(x^{3})$.
For instance, when $a_{4}>0$, the curve $\left(s'\right)^{-1}(x)$
bends upward for positive $x$, similarly to an hyperbolic sine; recalling
that $\beta<0$, the curve $f\left(\psi\right)$ is decreasing and
similar to the opposite of an hyperbolic sine. When $a_{4}<0$, $\left(s'\right)^{-1}(x)$
will bend downward, similar to an hyperbolic tangent. We will refer later on
 the case $a_{4}>0$ as the $\sinh$-like case and the case $a_{4}<0$
as the $\tanh$-like case.

\subsubsection{Normal forms and selection between degenerate states in the weak
energy limit}

We now study how the degeneracy between eigenstates for a doubly periodic
square is lifted. We saw in section \ref{sub:Quadratic-Casimir}
that the modification of the domain geometry (aspect ratio) removes
the degeneracy. We suggested in the previous section that the
contribution $a_{4}\omega^{4}$ of the Casimir functional may also
remove the degeneracy. We study how these two effects compete, by making
a quasi-linear study of the variational problem (\ref{eq:CEs}) in
the weak energy limit.

We first evaluate the range of parameters for these two effects to
be of the same order. We have seen that at leading order (maximization
of a quadratic Casimir), the vorticity scales like $(\lambda_{1}E)^{1/2}$.
The fourth order term $a_{4}\omega^{4}$ is thus of order $a_{4}\lambda_{1}^{2}E^{2}$.
The leading order correction due to the geometry in (\ref{SE2}) is
of order $(\lambda_{2}-\lambda_{1})E$. Therefore, one may omit non-quadratic
corrections provided $a_{4}\lambda_{1}^{2}E^{2}\ll(\lambda_{2}-\lambda_{1})E$
(case dominated by the geometry), and one could omit geometry effect
for $a_{4}\lambda_{1}^{2}E^{2}\gg(\lambda_{2}-\lambda_{1})E$ (case
dominated by non quadratic contributions to the Casimir functional).
This suggests that interesting phenomena may occur in the weak-energy
limit when $\lambda_{2}-\lambda_{1}=O(a_{4}E)$ (we assume $\lambda_{1}$
of order one, which is the case if the domain area is of order one).\\

Given the preceding discussion, it is natural to define a geometry
parameter $g$ by \begin{equation}
g=\frac{\lambda_{2}-\lambda_{1}}{E}.\label{resca}\end{equation}

$g>0$ is a measure of the degeneracy removal by the domain geometry
($\lambda_{2}-\lambda_{1}$), scaled by $E$, the scale of relative
non quadratic corrections in the small energy limit%
\footnote{As may be noticed, the actual small parameter in the low energy expansion
is $a_{4}E$. One could have defined $g$ by rescaling $\lambda_{2}-\lambda_{1}$
by $a_{4}E$ rather than by $E$ only. This would however be
inconvenient in the following discussion, as the sign of $a_{4}$
plays an essential role. %
}. For a doubly periodic rectangular domain $(x,y)\in(0,2\pi\delta)X(0,2\pi)$,
with aspect ratio $\delta$, we have $\lambda_{1}=1/\delta^{2}$ and
$\lambda_{2}=1$ (see section \ref{sub:Quadratic-Casimir}). Then
$g=(\delta^{2}-1)/(\delta^{2}E)$.

We now maximize the Casimir (\ref{SE}) using (\ref{resca}) and consider
the limit $a_{4}E\to0$ for fixed values of $g$.

At leading order, the flow is dominated by the degenerate eigenmodes
of the Laplacian. We thus have\begin{equation}
\omega=\left(\omega_{1}e_{1}+\omega_{2}e_{2}\right)\left(1+o\left(a_{4}E\right)\right),\label{eq:omega1-omega2}\end{equation}
The energy constraint (\ref{eq:Energy_Laplacian_Eigenmodes}) can
be expressed as

\begin{equation}
\omega_{2}^{2}=2E\lambda_{2}X+o\left(a_{4}E\right)\,\,\,\mbox{and}\,\,\,\omega_{1}^{2}=2\lambda_{1}E(1-X)+o\left(a_{4}E\right)\label{eq:X}\end{equation}
with $0\le X\leq1$. The expression for the quadratic part (\ref{SE3})
of the Casimir functional is \begin{equation}
C_{2}(E)=E\left(\lambda_{1}+gXE+o\left(gE\right)\right).\label{eq:C2-ordre1}\end{equation}
Similarly, we compute the fourth-order contribution to the entropy.
Let define the structure coefficients by \begin{equation}
\gamma_{n,k}=\int_{\mathcal{D}}d^{2}\mathbf{r}\, e_{1}^{k}e_{2}^{n-k}.\label{eq:Definitions gamma}\end{equation}
Given the symmetric role played by $x$ and $y$, we have $\gamma_{4,0}=\gamma_{4,4}$
and $\gamma_{4,1}=\gamma_{4,3}=0$ for the doubly-periodic square
domain. These equalities will be used in the following for slightly
rectangular domains, which is correct at leading order in $\lambda_{2}-\lambda_{1}$
(or $\delta-1$). We also define $\gamma=3\gamma_{2,2}-\gamma_{4,0}$.
We note that $\gamma=3/8\pi^{2}>0$, which can be verified by a direct
computation.

Straightforward computations then give \begin{equation}
a_{4}\int_{\mathcal{D}}d^{2}\mathbf{r}\,\frac{\omega^{4}}{4}=E\left[\left(\lambda_{1}^{2}\gamma_{4,0}+2\gamma\lambda_{1}^{2}X(1-X)\right)a_{4}E+o\left(a_{4}E\right)\right].\label{eq:C4}\end{equation}
From (\ref{eq:CEs}), (\ref{eq:so}), (\ref{eq:C2-ordre1}) and (\ref{eq:C4})
we conclude, that at leading order, the minimum of the Casimir functional
(\ref{eq:CEs}) is given by \begin{equation}
C(E)=\lambda_{1}E-\gamma_{4,0}\lambda_{1}^{2}a_{4}E^{2}+E^{2}\max_{0\leq X\leq1}h(X),\label{SE5}\end{equation}
 with \begin{equation}
h(X)=-gX+2\gamma\lambda_{1}^{2}a_{4}X(1-X).\label{gx}\end{equation}

\paragraph*{Square geometry.}

In order to understand the effects of the non-quadratic part only,
let us first consider the case of the doubly periodic square domain,
where the geometry parameter $g=0$. The function $X(1-X)$ in $h$
has a single maximum at $X=1/2$. This solution where both eigenmodes
$e_{1}$ and $e_{2}$ coexist equally is called a \textit{mixed state}.
From (\ref{eq:omega1-omega2}-\ref{eq:X}), we see that this mixed
state is the vorticity of a symmetric dipole (see the description
of (\ref{eq:dipole})).

For the square geometry $g=0$, there are also two global minima to
$X(1-X)$: $X=0$ ($\omega_{2}=0$ corresponding to $e_{1}$, see
(\ref{eq:omega1-omega2}-\ref{eq:X})), and $X=1$ ($\omega_{1}=0$,
corresponding to $e_{2}$). We call $e_{1}$ and $e_{2}$ \textit{pure
states}; we recall that $e_{1}$ and $e_{2}$ corresponds to parallel
flows (see (\ref{eq:paralell_flow_velocity})).

From (\ref{gx}), we see that the maximization of the Casimir functional
(\ref{SE5}) depends crucially on the sign of $a_{4}$. For the square
geometry $g=0$, for the $\sinh$-like case $a_{4}>0$, the non quadratic
contribution selects the dipole (mixed state), whereas for the $\tanh$-like
case $a_{4}<0$, the non quadratic contribution selects the parallel
flows (pure states).

\paragraph*{Equilibria in the $\sinh$-like case ($a_{4}>0$) in a rectangular
geometry.}

In a rectangular doubly periodic geometry $g\neq0$, when $a_{4}>0$,
the function $h(X)$ is a concave parabola. It thus has a single maximum
for \begin{equation}
X^{\star}=\frac{1}{2}-\frac{g}{4\gamma\lambda_{1}^{2}a_{4}}.\label{extremum}\end{equation}
 Clearly $X^{\star}\leq1/2$: the dipole is stretched in the same
direction as the domain. The constraint $X^{\star}\geq0$ must be
verified (see \ref{eq:X}). This is the case only if $g\leq g^{\star}$
with \begin{equation}
g^{\star}=2\gamma\lambda_{1}^{2}a_{4}.\label{eq:phase transition g}\end{equation}
 The discussion follows:
\begin{itemize}
\item [1.] For $g>g^{\star}$ the effect of the geometry dominates and
only the parallel flow associated with $e_{1}$ $(X=0$) is observed.
The Casimir minima is \begin{equation}
C_{s}(E)=\lambda_{1}E-\gamma_{4,0}a_{4}\lambda_{1}^{2}E^{2}+o\left(a_{4}E^{2}\right).\label{eq:Cs apositif g grand}\end{equation}

\item [2.] For $g<g^{\star}$ the effect of the non-quadratic term dominates,
we then observe a mixed state corresponding to $X=X^{\star}$. The
entropy is \begin{equation}
C_{s}(E)=\lambda_{1}E+\left[-\gamma_{4,0}\lambda_{1}^{2}a_{4}+\frac{1}{8\gamma\lambda_{1}^{2}a_{4}}\left(g-g^{*}\right)^{2}\right]E^{2}.\label{eq:Cs apositif g petit}\end{equation}
The solution $X=0$ is a local maximizer of (\ref{eq:CEs}) (unstable
state).
\end{itemize}
We thus conclude that in the sinh-like case ($a_{4}>0$), there exists
a second order phase transition (i.e. a discontinuity in the second order derivative of the equilibrium entropy with respect to the energy ) where the flow bifurcates from a dipole
when the non quadratic part dominates ($g<2\gamma\lambda_{1}^{2}a_{4}$)
to a parallel flow when the geometry dominates ($g>2\gamma\lambda_{1}^{2}a_{4}$).

\paragraph*{Equilibria in the $\tanh$-like case ($a_{4}<0$) in a rectangular
domain.}

In a rectangular doubly periodic geometry $g\neq0$, when $a_{4}>0$,
$h(X)$ is a convex parabola. Since $-gX$ favors the state $e_{1}$,
the global statistical equilibrium is always the pure state $e_{1}$
($X=0$). The equilibrium value of the Casimir functional is \begin{equation}
C_{s}(E)=\lambda_{1}E-\gamma_{4,0}\lambda_{1}^{2}a_{4}E^{2}.\label{eq:Cs anegatif}\end{equation}

\begin{figure}[htpb]
\centerline{\includegraphics[height=8cm]{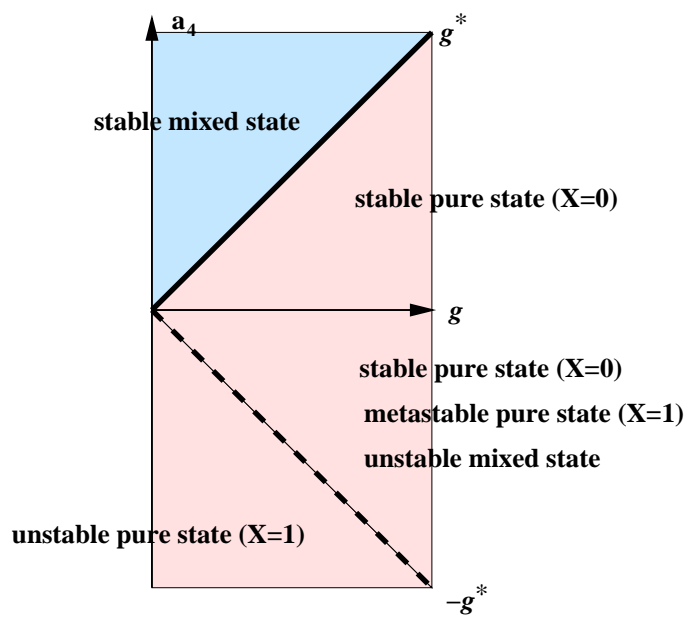}}

\caption{\footnotesize Bifurcation diagram for the statistical equilibria of the 2D Euler equations in a doubly periodic domain with aspect ratio $\delta$, in the limit where the normal form treatment is valid, in the $g$-$a_{4}$ parameter plane. The geometry parameter $g$ is inversely proportional to the energy and proportional to the difference between the two first eigenvalues of the Laplacian (or equivalently to $\delta-1$ in the limit of small$\delta-1$), the parameter $a_4$ measures the non-quadratic contributions to the Casimir functional. The solid line is a second order phase transition between a dipole (mixed state) and a parallel flow along the $y$ direction (pure state $X=0$). Along the dashed line, a metastable parallel flow (along the $x$ direction, pure state $X=1$) loses its stability.}
\label{delta_a4}
\end{figure}

We now study the metastable and unstable equilibria. The function
$h(X)$ has a single minimum for $X=X^{\star}$ (see (\ref{extremum})),
but because now $a_{4}<0$, $X^{\star}\geq1/2$. Depending on the
position of $X^{\star}$ with respect to 1, two cases occur:
\begin{itemize}
\item [1.] For $g>-g^{\star}$, then $X^{\star}<1$, the mixed state exists
as an unstable state. The pure state $e_{2}$ ($X=1)$ is a local
maximum (metastable state).
\item [2.] For $g<-g^{\star}$, then the mixed state is no more a critical
point. The pure state $e_{2}$ corresponding to $X=1$ is a local
minimum (unstable).
\end{itemize}
The results for the equilibrium structures are summarized on figure
\ref{delta_a4}. There is thus a second order phase transition along
the line $g=g^{\star}=2\gamma\lambda_{1}^{2}a_{4}$

\paragraph*{The second order phase transition from the energy point of view.}

In the preceding computations, we have worked with a rescaled geometry
parameter $g$ (\ref{resca}), because this is the correct scaling
for studying the phase transition (balance between the effect of quartic
part of the Casimir functional and the effect of the geometry). From
a physical point of view, in many situations it is more natural to
think in terms of energy, for a fixed geometry configurations.

We now consider fixed aspect ratio $\delta$ and $a_{4}$ parameters.
Using (\ref{resca}), from the phase transition criteria (\ref{eq:phase transition g}),
we deduce that a phase transition occurs for a critical  energy
$E^{*}$ given by \begin{equation}
E^{*}=\frac{4\pi^{2}\delta^{2}\left(\delta^{2}-1\right)}{3a_{4}};\label{eq:Critical Energy}\end{equation}
 we have used $\gamma=3/8\pi^{2}$, $\lambda_{1}=\delta^{-2}$ and
$\lambda_{2}=1$. The phase transition line is thus a hyperbola in
the ($E-a_{4}$ plane). For energies $E<E^{*}$, we have $g>g^{*}$
and equilibria are dipoles, while for $E>E^{*}$ equilibria are parallel
flows.

The computations of last sections are obtained as an expansion in
powers of $a_{4}E$. The result (\ref{eq:Critical Energy}) is thus
valid for small $a_{4}E^{*}$ or equivalently for small values of
$\delta-1$. The transition lines for larger values of the parameter
$\delta-1$ is discussed in next section.

\subsubsection{Larger energy phase diagram}

\begin{figure}
\includegraphics[width=12cm]{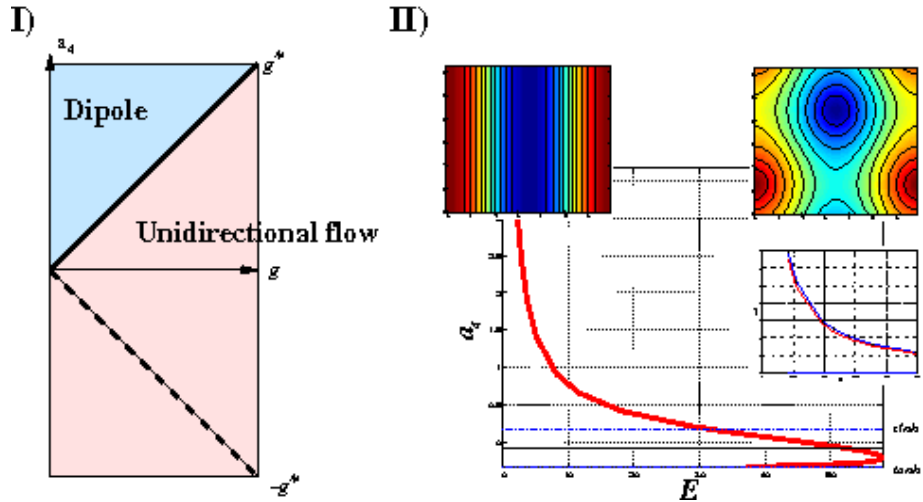}
\caption{\footnotesize Bifurcation diagrams for statistical equilibria of the 2D Euler equations in a doubly periodic domain
a) in the $g$-$a_{4}$ plane (see figure \ref{delta_a4}) b) obtained numerically in
 the $E-a_{4}$ plane, in the case of doubly periodic geometry with
aspect ratio $\delta=1.1$. The colored insets are streamfunction
and the inset curve illustrates good agreement between numerical and
theoretical results in the low energy limit.}
\label{fig:Equilibre}
\end{figure}

In order to look at the phase diagram for larger energies, we use
a continuation algorithm to numerically compute solution to (\ref{faste})
corresponding to $f_{a_{4}}(x)=\left(1/3-2a_{4}\right)\tanh x+\left(2/3+2a_{4}\right)\sinh x$.
Using $f=\left(s'\right)^{-1}$, one can check that (\ref{eq:so})
is verified. The results are shown in figure \ref{fig:Equilibre}.
The inset of Fig \ref{fig:Equilibre} \textbf{a)} shows good agreement
for transition lines obtained either with the continuation algorithm
or the low-energy limit theoretical result, for $\delta=1.01$. Figure
\ref{fig:Equilibre} \textbf{b)} shows the bifurcation diagram for
$\delta=1.1$; in such a case the transition line is still very close
to a hyperbola provided energy is small. \\

In this section, we have computed the phase diagrams for the statistical
equilibria of the 2D Euler equations in a doubly periodic geometry.
This is an illustration of the type of results provided by a statistical
mechanics approach: prediction of large scale flow pattern, of phase
transitions between these, explanation from statistical mechanics
ideas the stability of these flows, and description of the few key
parameters that characterize these flows.

We will come back to the doubly periodic geometry in section \ref{sub:Phase Transition-1},
and show how prediction of equilibrium phase transitions can be useful
also for out of equilibrium situations, when dissipation and forcing
are present.

\subsection{Numerical methods to compute statistical equilibria \label{sub:Numerical-methods-equilibrium}}

We have seen previously that it is possible to compute analytically
equilibrium states of the RSM theory in some limit cases, and to get
important insights on their physical properties through these computations.
However, one might in practice want to be able to compute these equilibrium
states for more general situations. One can distinguish three different numerical
algorithms to find equilibrium states:
\begin{enumerate}
\item The use of an iterative algorithm proposed by Turkington and Whitaker,
that computes local entropy maxima by linearizing the constraints
of the variational problem \cite{1994PhFl....6.3963W,244830}.
\item The use of relaxation equations that maximize the entropy production
of the system while keeping constant the constraints of the problem
(potential vorticity distribution and energy) \cite{RobertSommeria:1992_PRL_Relaxation_Meca_Stat}. This method also drives the system toward a local entropy maximum. Generalization of relaxation equations to large classes of variational problems, with or without constrains have been devised, see for instance \cite{ChavanisEPJB2009} for a recent review.
\item The use of continuation methods to compute the critical points of
the variational problems (\ref{eq:q-psi equilibre}). This method
is very useful to follow a branch of stationary states by changing
one parameter, and to detect bifurcations \cite{thess94,Bouchet_Simonnet_2008}.
\end{enumerate}
One has to be aware that non-linear optimization, with or without
constraint is not an easy task. One has to be able to follow several
bifurcate branches of solutions, and actually be able to track the
good one!

Each of these three methods have its own advantages and drawbacks.
The advantage of methods 1 and 2 is that they actually deal with constrained
variational problems. For instance if we speak about the energy constraint
only, the control parameter of methods 1 and 2 will be the energy
$E$ of the flow and not the inverse temperature. The associated drawback
is that they have to be initialized with fields having either the
energy of interest (method 2) or an energy larger than the energy
of interest (method 1, then one has to be able to compute energy extrema).
Another advantage of methods 1 and 2 is that they actually compute
local extrema. The associated drawback is that it is not possible
to compute saddle or local minima, which is often necessary.

The advantage of method 3 is that it allows to actually follow branches
of solution, whereas methods 1 and 2 lead to jump from one branch
to an other in rather uncontrolled way. It also allows to precisely
tracks bifurcations, and thus finds all the branches of solutions
connected to the initial one. It thus gives a precise and complete
view of the ensemble of critical points. It is however more difficult
to master.

The main drawback of all of these methods is that there is never any
insurance to have caught the actual extrema.

\subsection{Past studies of statistical equilibria and relaxation towards equilibrium}

Most of theoretical contributions are described along this review. As far as applications are concerned, we have described only few examples of statistical equilibrium studies. Our choice was based on their pedagogical interest or on their interest for modeling natural phenomena. There have been however lots of other studies of statistical equilibria, comparisons with direct numerical simulations or experiments, see e.g. \cite{Sommeria_2001_CoursLesHouches,Majda_Wang_Book_Geophysique_Stat} and references therein. We give in this section a brief overview of these works. We also discus briefly phenomenological approaches based on statistical mechanics ideas, discussing relaxation towards equilibrium, the closure problem in turbulence.\\

During the first stage following the appearance of RSM theory, there have been attempt to consider its application to classical fluid mechanics problems, like  shear layer problems  \cite{Staquet}, or von-Karman vortex streets  \cite{thess94}, and to check the prediction of statistical mechanics, as well as to describe symmetry breaking phenomena during the self-organization of initial conditions containing negative and positive vortex patches with equal strength and area, in various domain geometries \cite{Jutner_Thess_Sommeria_1995PhFl}. Statistical equilibria computations were mainly done numerically. It was found that in any of these situation, statistical equilibrium predict that the most probable flow is a self organized large scale structure, qualitatively very similar to the numerically observed one. Quantitative agreement has to be discussed on a case by case basis \cite{Staquet,thess94,Jutner_Thess_Sommeria_1995PhFl}. Similarly, phase diagrams of statistical equilibrium states in a disk (for asymmetric vorticity distribution), and comparison with numerical simulations are provided in \cite{Chen_Cross_1996PhRv,Chen_Cross_1996_PhRvL,Chen_Cross_1994PhRvE}. In the case of a doubly periodic domain, \cite{Yin_Montgomery_Clercx_2003PhFluids} found that freely evolving turbulent flow may for some classes of initial conditions be self-organized into ``bar'' (parallel flows) equilibrium states, different from  the dipole associated with the gravest horizontal mode. An analytical understanding of this phenomenon is given in \cite{Bouchet_Simonnet_2008}, see also section \ref{sub:The-example-doubly-periodic}.

Note that they may also exist some class of initial conditions for which the final state may be unsteady (presenting quasi-periodic movements), which is not described by the statistical mechanics approach \cite{Segre_Kida_1997chao.dyn..9020S}.

Some numerical studies of decaying 2D turbulence have specifically addressed the temporal evolution of the microscopic and macroscopic vorticity distribution \cite{Brands_Stulemeyer_Pasmanter_1997PhFl....9.2465B,Capel_Pasmanter_2000PhFl...12.2514C}, or the effect of boundaries in closed domains \cite{Clercx_Massen_VanHeijst_1999PhFl...11..611C}.\\

The first analytical computations of RSM equilibrium states have been performed in the framework of the 2D Euler equations, for states characterized by a linear $q-\psi$ relation \cite{ChavanisSommeria:1996_JFM_Classification}, which is justified in a strong mixing limit, see section \ref{sub:strong_mixing}.  Generalization to a larger class of flow models (including quasi-geostrophic equations with topography), and relation with possible inequivalence between ensembles is given in \cite{Venaille_Bouchet_PRL_2009,VenailleBouchetJSP}.

As explained in section  \ref{sub:strong_mixing}, some of the states characterized by a linear $q-\psi$ relation had been previously described in the framework of the energy-enstrophy theory (and all are RSM equilibrium states). This includes the original description of ``Fofonoff flows'' as statistical equilibria \cite{SalmonHollowayHendershott:1976_JFM_stat_mech_QG}, see also \cite{Carnevale_Frederiksen_NLstab_statmech_topog_1987JFM} for further discussions and results, and  \cite{Zou_Holloway_1994JFM...263..361Z,WangVallis} for a comparison with direct numerical simulations. Generalization to continuously stratified quasi-geostrophic turbulence in doubly periodic domains (with bottom topography, but without beta effect) is discussed in \cite{Merryfield98JFM}. Generalization to barotropic flows above finite topography is discussed in \cite{Merryfield_Cummins_Holloway_2001JPO....31.1880M}.

Energy enstrophy equilibrium states on a sphere have also been computed for a two-layers quasi-geostrophic model  \cite{fredericksen91GAFDa}, and for a spin-lattice model of fluid vorticity \cite{Lim_2001PhFl...13.1961L,Lim_SinghMavi_2007PhyA}.

Because linear $q-\psi$ relations were also predicted by a phenomenological minimum enstrophy principle  \cite{BrethertonHaidvogel}, the relation between such states and RSM theory has been widely discussed in earlier studies on the RSM theory, see e.g. \cite{ChavanisSommeria:1996_JFM_Classification,Brands_Chavanis_etc_1999PhFl...11.3465B}. It is now understood that minimum enstrophy states are only a particular class of RSM equilibriums states \cite{ChavanisSommeria:1996_JFM_Classification,Bouchet:2008_Physica_D,NasoChavanisDubrulle}.\\

In some cases, the final state flow organization observed in laboratory or numerical experiments is different from the one predicted by the RSM theory. This has lead to several phenomenological approaches inspired by the RSM theory. In most of these approaches, some additional constraints (different from the dynamical invariants) are imposed to the system.

In order to describe the self-organization of turbulent flow in unbounded domains, \cite{Chavanis_Sommeria_1998JFM_LocalizedEquilibria...356..259C} proposed to impose an additional kinetic constraint of entropy maximization in a prescribed ``bubble''.

Another phenomenological approach, assuming \textit{a priori} the existence of different ``mixing regions'' has been proposed to describe the self-organization following the equilibration of an unstable baroclinic jet in a two-layer quasi-geostrophic model in a channel \cite{Esler08}.

It has been observed experimentally and numerically that in some cases, two slightly different initial conditions can lead to very different final states, one being predicted by the RSM theory, the other being a quasi-stationary state in which several vortex are organized into a long-lived crystal configuration, which persists during the time of the experiments, see e.g. \cite{Schecter_Dubin_etc_Vortex_Crystals_2DEuler1999PhFl}, and \cite{KossinSchubert} for an application to mesoscale vortices in cyclones. A phenomenological ``regional entropy maximization'' approach, assuming a priori the existence of several vortex, has been proposed to describe these vortex crystals \cite{JinDubin98,Jin_Dubin_2000PhRvL}.\\

The  idea of an application of equilibrium statistical mechanics to the description of Jovian vortices  was mentioned in the early development of the theory (see for instance \cite{Sommeria_Nore_Dumont_Robert_1991CRASB.312..999S,Miller_Weichman_Cross_1992PhRvA,Michel_Robert_1994_JSP_GRS}). The fact that the ring shape of the Great Red Spot velocity field is related to the small value of the Rossby deformation radius in a Quasi-Geostrophic model has been  understood in \cite{Sommeria_Nore_Dumont_Robert_1991CRASB.312..999S}.  The first theoretical modeling and quantitative predictions are given in \cite{Bouchet_Sommeria:2002_JFM,Bouchet_Dumont_2003_condmat} in the framework of $1.5$ layer quasi-geostrophic equation. In particular, analytical result were obtained by considering the limit of small Rossby radius of deformation (and then strongly non-linear $q-\psi$ relations), see \cite{Bouchet_Sommeria:2002_JFM} and section \ref{sec:First Order GRS and rings}.  At the same time, the conditions for the appearance of the spot on the south hemisphere rather than on the south one, have been discussed in \cite{TurkingtonMHD:2001_PNAS_GRS}. The small Rossby radius of deformation theory has been further developed in the oceanic context to describe rings and jets, see  \cite{Weichman_2006PhRvE,VenailleBouchetJPO} and sections  \ref{sec:First Order GRS and rings} and \ref{sec:Gulf Stream and Kuroshio}.

Another attempt to apply equilibrium statistical mechanics to oceanic flows had been performed by \cite{DibattistaMajda00,DibattistaMajda02} in the framework of the Heton model of \cite{HoggStommel85} for the self-organization phenomena following deep convection events, by  numerically computing statistical equilibrium states of a two-layer quasi-geostrophic model.

In the atmospheric context, the equilibrium statistical theory has been applied by \cite{Prieto_Schubert_2001JAtS} to predict final state organization of the stratospheric polar vortex.\\

In the ocean context, there has been many attempts to propose subgrid-scale parameterizations inspired by the equilibrium statistical mechanics\footnote{We emphasize that such parameterization are phenomenological approaches, contrary to the equilibrium statistical mechanics.}, as first advocated by Holloway in the framework of the energy-enstrophy approach (see e.g. \cite{HollowayReview04,FrederiksenOKane08} for a review and  further references) , and further developed  by \cite{Kazantsev_Sommeria_Verron_1998JPO....28.1017K} to take into account higher order invariants.

The parameterization of  \cite{Kazantsev_Sommeria_Verron_1998JPO....28.1017K} is actually a direct application to the oceans of the relaxation equations proposed by Robert and Sommeria in the case of the Euler equations \cite{RobertSommeria:1992_PRL_Relaxation_Meca_Stat}. The relaxation equations are obtained through an interesting systematic approach based on a maximum entropy production principle (MEPP) in order to obtain equations, preserving the invariant structure of the initial equation but converging towards the equilibrium states. At a phenomenological level, they can be considered as a turbulent closure for the parameterization of small scale mixing. It has been shown empirically that they do not describe the actual turbulent fluxes \cite{Bouchet_2003_condmat}. This is however a drawback shared by most of existing parameterizations, and the essential fact that they preserve the mathematical properties (conservation laws, and so on) make relaxation equation better model candidates than most of other parameterizations. The relaxation equations were further developed in a number of works, see for instance \cite{Robert_Rosier_1997JSP....86..481R,Robert_Rosier_2001NPGeo...8...55R,ChavanisPRL97,Bouchet_2003_condmat,Chavanis_Naso_Dubrulle_2010EPJB_Relaxation} and references therein.

Other closures based on statistical mechanics ideas have been proposed by the group of Majda, sometime at a phenomenological level \cite{Grote_Majda_2000Nonli,Grote_Majda_CrudeClosure_1997PhFl}, sometimes at a more fundamental level for specific problems \cite{Majda_Wang_Bombardement_2006_Comm}.


\section{Statistical equilibria and jet solutions, application to ocean rings
and to the Great Red Spot of Jupiter\label{sec:First Order GRS and rings}}

In section \ref{sub:The-example-doubly-periodic}, we have described
analytically the equilibrium flows with a normal form study close
to a linear relation between potential vorticity and stream function
(or equivalently in the limit of a quadratic Energy-Casimir functional).
We have pointed out that more general solutions are  very
difficult to find analytically, and may require numerical computations,
for instance using continuation algorithms.

There are however other limits where an analytical description becomes
possible. This is for instance the case in the limit of large energies
\cite{1983_Turkington_CommMathPhys1}.
This is a very interesting, nontrivial and subtle limit; we do not
describe it this review. The second interesting limit applies to
the quasi-geostrophic model with 1.5 layers. It is the limit of Rossby
deformation radius $R$ much smaller than the size of the domain%
\footnote{The study of equilibria of the quasi-geostrophic model is a first
step before studying equilibria of the shallow water model, for which
taking the limit $R\ll L$ give similar results.%
} ($R\ll L$), where the nonlinearity of the vorticity-stream function
relation becomes essential. This limit case and its applications to
the description of coherent structures in geostrophic turbulence is
the subject of this section.\\

In the limit $R\ll L$, the variational problems of the statistical
theory are analogous to the Van-Der-Waals Cahn Hilliard model that
describes phase separation and phase coexistence in usual thermodynamics.
The Van-Der-Waals Cahn Hilliard model describes for instance the equilibrium
of a bubble of a gas phase in a liquid phase, or the equilibria of
 soap films in air. For these classical problems, the essential concepts
are the free energy per unit area, the related spherical shape of
the bubbles, the Laplace equation relating the radius of curvature
of the bubble with the difference in pressure inside and outside the
bubble (see section \ref{sub:Van Der Waals}), or properties of minimal
surfaces (the Plateau problem). We will present an analogy
between those concepts and the structures of quasi-geostrophic statistical
equilibrium flows.

For these flows, the limit $R\ll L$ leads to interfaces separating
phases of different free energies. In our case, each phase is characterized
by a different value of average potential vorticity, and corresponds
to sub-domains in which the potential vorticity is homogenized. The
interfaces correspond to strong localized jets of typical width $R$.
This limit is relevant for applications showing such strong jet structures.\\

From a geophysical point of view, this limit $R\ll L$ is relevant
for describing some of Jupiter's features, like for instance the Great
Red Spot of Jupiter (a giant anticyclone) (here $R\simeq500-2000\, km$
and the length of the spot is $L\simeq20,000\, km$) (see section
\ref{sub:Application-to-Jupiter}).

This limit is also relevant to ocean applications, where $R$ is the
internal Rossby deformation radius ($R\simeq50\, km$ at mid-latitude).
We will apply the results of statistical mechanics to the description
of robust (over months or years) vortices such as ocean rings, which
are observed around mid-latitude jets such as the Kuroshio or the Gulf
Stream, and more generally in any eddying regions (mostly localized
near western boundary currents) as the Aghulas current, the confluence
region in the Argentinian basin or the Antarctic Circumpolar circulation
(see section \ref{sub:Gulf Stream Rings}). The length $L$ can be
considered as the diameters of those rings ($L\simeq200\, km$).

We will also apply statistical mechanics ideas in the limit $R\ll L$
to the description of the large scale organization of oceanic currents
(in inertial region, dominated by turbulence), such as the eastward
jets like the Gulf Stream or the Kuroshio extension (the analogue
of the Gulf Stream in the Pacific ocean). In that case the length
$L$ could be thought as the ocean basin scale $L\simeq5,000\, km$
(see section \ref{sec:Gulf Stream and Kuroshio}).

\subsection{The Van der Waals--Cahn Hilliard model of first order phase transitions
\label{sub:Van Der Waals}}

We first describe the Van der Waals--Cahn Hilliard model. We give in
the following subsections a heuristic description based on physical
arguments. Some comments and references on the mathematics of the
problem are provided in section \ref{sec:maths}.

This classical model of thermodynamics and statistical physics describes
the coexistence of phase in usual thermodynamics. It involves the
minimization of a free energy with a linear constraint:\begin{equation}
\left\{ \begin{array}{c}
F=\min\left\{ \mathcal{F}\left[\phi\right]\,\,\left|\,\,\mathcal{A}\left[\phi\right]=-B\right.\right\} \\
{\rm \mbox{with}\,\,\,\,}\mathcal{F}=\int_{\mathcal{D}}\mathrm{d}{\bf \mathbf{r}}\,\left[\frac{R^{2}\left(\nabla\phi\right)^{2}}{2}+f(\phi)\right]\,\,\,\,\mbox{and}\,\,\,\,\mathcal{A}\left[\phi\right]=\int_{\mathcal{D}}\mathrm{d}{\bf r}\,\phi\end{array}\right.\label{eq:Van Der Waals Cahn Hilliard}\end{equation}
 where $\phi$ is the non-dimensional order parameter (for instance
the non-dimensionality local density), and $f\left(\phi\right)$ is
the non-dimensional free energy per unit volume. We consider the limit
$R\ll L$ where $L$ is a typical size of the domain. We assume that
the specific free energy $f$ has a double well shape (see figure
\ref{fig_f}), characteristic of a phase coexistence related to a
first order phase transition. For a simpler discussion, we also assume
$f$ to be even; this does not affect the properties of the solutions
discussed bellow.

\subsubsection{First order phase transition and phase separation \label{sub:phase separation} }

\begin{figure}[t!]
\begin{center}
\includegraphics[width=8cm]{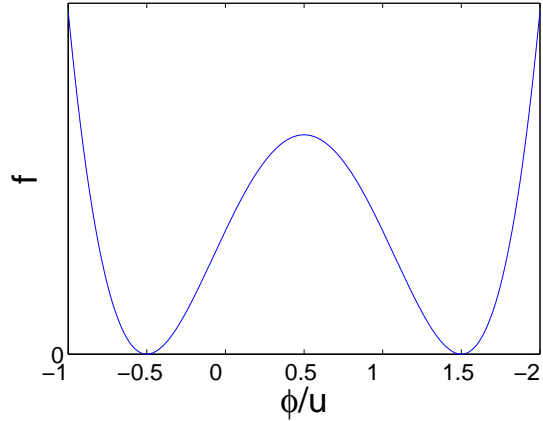}
\end{center}
\caption{{\footnotesize The double well shape of the specific free energy $f\left(\phi\right)$
(see equation (\ref{eq:Van Der Waals Cahn Hilliard})). The function
$f\left(\phi\right)$ is even and possesses two minima at $\phi=\pm u$.
At equilibrium, at zeroth order in $R$, the physical system will
be described by two phases corresponding to each of these minima. }}

\label{fig_f}
\end{figure}

At equilibrium, in the limit of small $R$, the function $f\left(\phi\right)$
plays the dominant role. In order to minimize the free energy, the
system will tend to reach one of its two minima (see figure \ref{fig_f}).
These two minima correspond to the value of the order parameters for
the two coexisting phases, the two phases have thus the same free energy.

\begin{figure}[t!]
\begin{centering}
\includegraphics[width=8cm]{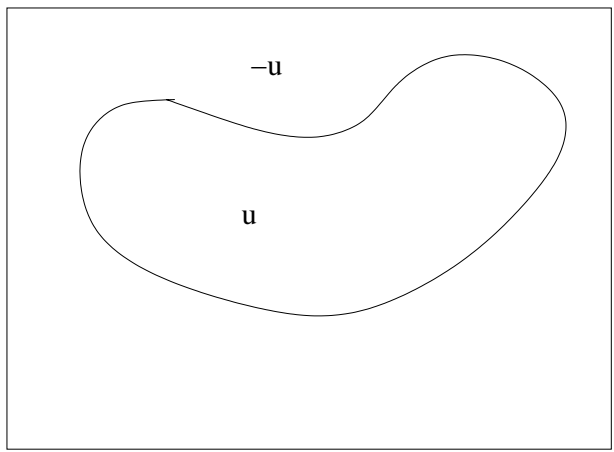}
\par\end{centering}

\caption{{\footnotesize At zeroth order, $\phi$ takes the two values $\pm u$
on two sub-domains $A_{\pm}$. These sub-domains are separated by
strong jets. The actual shape of the structure, or equivalently the
position of the jets, is given by the first order analysis. }\label{fig_domaine}}

\end{figure}

\begin{figure}[t!]
\begin{centering}
\includegraphics[height=8cm]{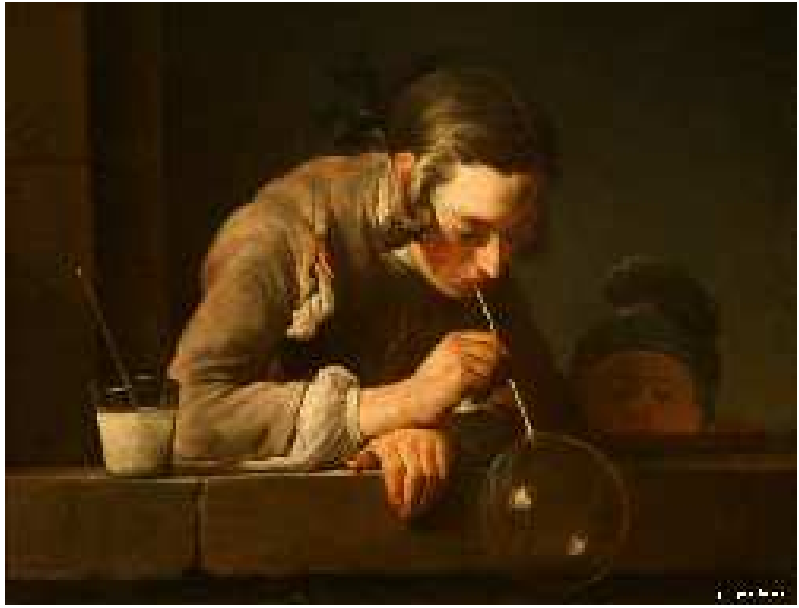}
\par\end{centering}


\caption{{\footnotesize Illustration of the Plateau problem (or minimal area
problem) with soap films: the spherical bubble minimizes its area
for a given volume (Jean Simeon Chardin, }\emph{\footnotesize Les bulles
de savon}{\footnotesize , 1734)}}

\label{fig_plateau}
\end{figure}

\begin{figure}[t!]
\begin{center}
\includegraphics[width=\textwidth]{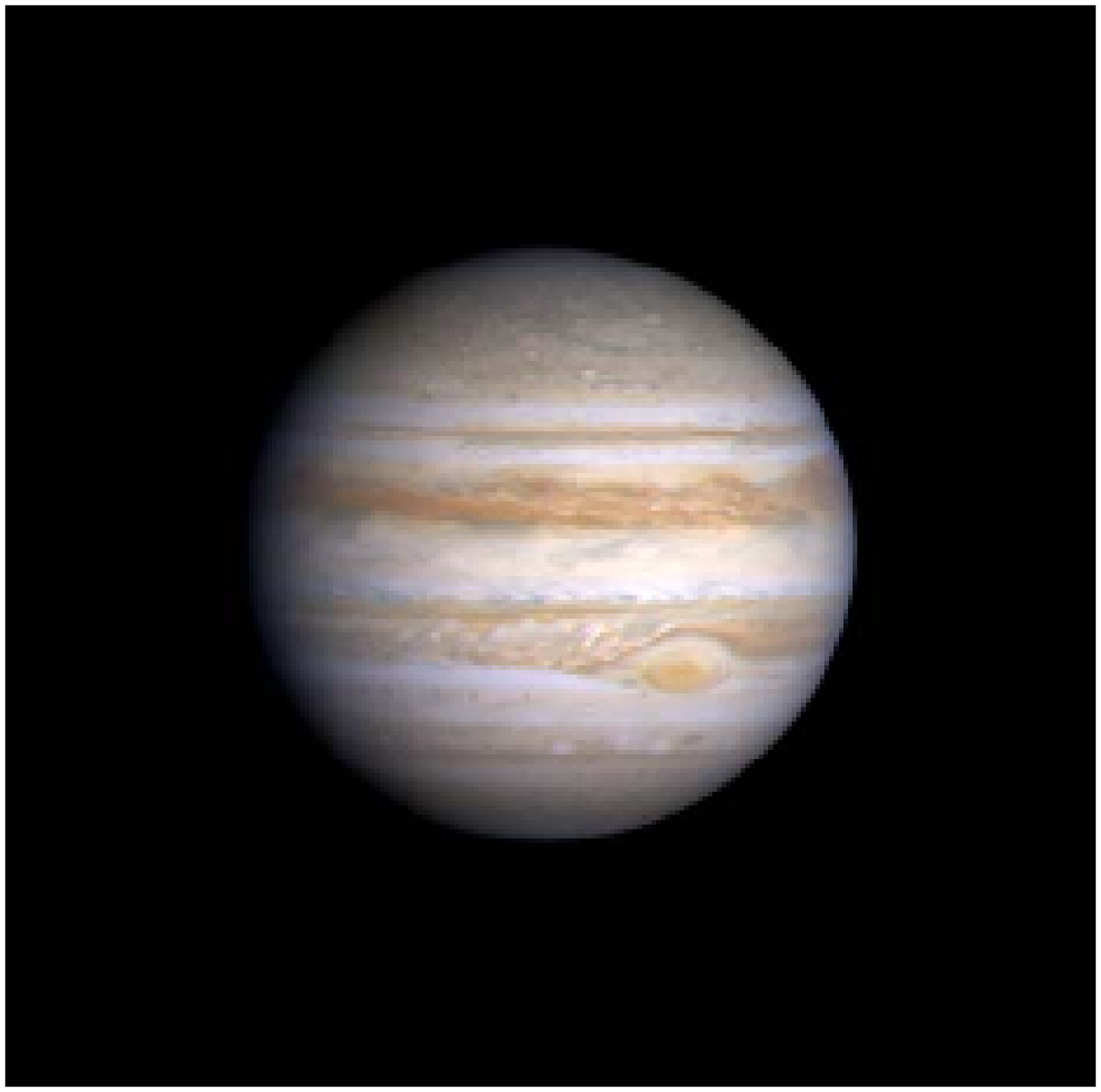}
\end{center}
\caption{{\footnotesize Observation of the Jovian atmosphere from Cassini (Courtesy
of NASA/JPL-Caltech). One of the most striking feature of the Jovian
atmosphere is the self organization of the flow into alternating eastward
and westward jets, producing the visible banded structure and the
existence of a huge anticyclonic vortex $\sim20,000\ km$ wide, located
around $20 �$ South: the Great Red Spot (GRS).
The GRS has a ring structure: it is a hollow vortex surrounded by
a jet of typical velocity $\sim100\ m.s^{\{-1\}}\ $ and width $\sim1,000\, km$.
Remarkably, the GRS has been observed to be stable and quasi-steady for
many centuries despite the surrounding turbulent dynamics. The explanation
of the detailed structure of the GRS velocity field and of its stability
is one of the main achievement of the equilibrium statistical mechanics
of two dimensional and geophysical flows (see figure \ref{fig:Emergence_Numerique_tache_rouge}
and section \ref{sec:First Order GRS and rings}).}}

\label{Fig:ColorJupiter}
\end{figure}

The constraint $\mathcal{A}$ (see equation. \ref{eq:Van Der Waals Cahn Hilliard})
is related to the total mass (due to the translation on $\phi$ to
make $f$ even, it can take both positive and negative values). Without
the constraint $\mathcal{A}$, the two uniform solutions $\phi=u$
or $\phi=-u$ would clearly minimize $\mathcal{F}$: the system would
have only one phase. Because of the constraint $\mathcal{A}$, the
system has to split into sub-domains: part of it with phase $\phi=u$
and part of it with phase $\phi=-u$. In a two dimensional space,
the area occupied by each of the phases are denoted $A_{+}$ and $A_{-}$
respectively. They are fixed by the constraint $\mathcal{A}$ by the
relations $uA_{+}-uA_{-}=-B$ and by $A_{+}+A_{-}=1$ (where $1$
is the total area). A sketch of a situation with two sub-domains each
occupied by one of the two phases is provided in figure \ref{fig_domaine}.

Up to now, we have neglected the term $R^{2}\left(\nabla\phi\right)^{2}$
in the functional (\ref{eq:Van Der Waals Cahn Hilliard}). In classical
thermodynamics, this term is related to non-local contributions to
the free energy (proportional to the gradient rather than to only
point-wise contributions). Moreover the microscopic interactions fix
a length scale $R$ above which such non-local interactions become
negligible. Usually for a macroscopic system such non-local interactions
become negligible in the thermodynamic limit. Indeed as will soon
become clear, this term gives finite volume or interface effects.

We know from observations of the associated physical phenomena (coarsening,
phase separations, and so on) that the system has a tendency to form
larger and larger sub-domains. We thus assume that such sub-domains
are delimited by interfaces, with typical radius of curvature $r$
much larger than $R$%
\footnote{This can indeed be proved mathematically, see section \ref{sec:maths} %
}. Actually the term $R^{2}\left(\nabla\phi\right)^{2}$ is negligible
except on an interface of width $R$ separating the sub-domains.
The scale separation $r\gg R$ allows to consider independently what
happens in the transverse direction to the interface on the one hand
and in the along interface direction on the other hand. As described
in next sections, this explains the interface structure and interface
shape respectively.

\subsubsection{The interface structure}

At the interface, the value of $\phi$ changes rapidly, on a scale
of order $R$, with $R\ll r$. What happens in the direction along
the interface can thus be neglected at leading order. To minimize
the free energy (\ref{eq:Van Der Waals Cahn Hilliard}), the interface
structure $\phi(\zeta)$ needs thus to minimize a one dimensional
variational problem along the normal to the interface coordinate $\zeta$\begin{equation}
F_{int}=\min\left\{ \int\mathrm{d}\zeta\,\left[\frac{R^{2}}{2}\left(\frac{d\phi}{d\zeta}\right)^{2}+f(\phi)\right]\right\} .\label{eq:Variational Free Energy Unit lenght}\end{equation}
 Dimensionally, $F_{int}$ is a free energy $F$ divided by a length.
It is the free energy per unit length of the interface.

We see that the two terms in (\ref{eq:Variational Free Energy Unit lenght})
are of the same order only if the interface has a typical width of
order $R$. We rescale the length by $R$: $\zeta=R\tau$. The Euler-Lagrange
equation of (\ref{eq:Variational Free Energy Unit lenght}) gives
\begin{equation}
\frac{d^{2}\phi}{d\tau^{2}}=\frac{df}{d\phi}.\label{jet}\end{equation}
This equation is a very classical one. For instance making an analogy
with mechanics, if $\phi$ would be a particle position, $\tau$ would
be the time, equation (\ref{jet}) would describe the conservative
motion of the particle in a potential $V=-f$. From the shape of $f$
(see figure \ref{fig_f}) we see that the potential has two bumps
(two unstable fixed points) and decays to $-\infty$ for large distances.
In order to connect the two different phases in the bulk, on each
side of the interface, we are looking for solutions with boundary
conditions $\phi\rightarrow\pm u$ for $\tau\rightarrow\pm\infty$.
It exists a unique trajectory with such limit conditions: in the particle
analogy, it is the trajectory connecting the two unstable fixed points
(homoclinic orbit).

This analysis shows that the interface width scales like $R$. Moreover,
after rescaling the length, one clearly sees that the free energy
per length unit (\ref{eq:Variational Free Energy Unit lenght}) is
proportional to $R$: $F_{int}=eR,$ where $e>0$ could be computed
as a function of $f$ (see e.g. \cite{Bouchet_Sommeria:2002_JFM,VenailleBouchetJPO}).

\subsubsection{The interface shape: an isoperimetrical problem \label{sub:The-interface-shape}}

In order to determine the interface shape, we come back to the free
energy variational problem (\ref{eq:Van Der Waals Cahn Hilliard}).
In the previous section, we have determined the transverse structure
of the interface, by maximizing the one dimensional variational problem
(\ref{eq:Variational Free Energy Unit lenght}). We have discussed
the quantity $F_{int}=Re$, a free energy per unit length, which is
the unit length contribution of the interface to the free energy.
The total free energy is thus \begin{equation}
\mathcal{F}=eRL,\label{eq:Free energy length}\end{equation}
 where we have implicitly neglected contributions of relative order
$R/r$, where $r$ is the curvature radius of the interface.

In order to minimize the free energy (\ref{eq:Free energy length}),
we thus have to minimize the length $L$. We must also take into account
that the areas occupied by the two phases, $A_{+}$ and $A_{-}$ are
fixed, as discussed in section \ref{sub:phase separation}. We thus
look for the curve with the minimal length, that bounds a surface
with area $A_{+}$\begin{equation}
\min\left\{ eRL\left|\mbox{Area}=A_{+}\right.\right\} .\label{eq:Isoperimetric}\end{equation}
This type of problem is called an isoperimetrical problem. In three
dimensions, the minimization of the area for a fixed volume leads
to spherical bubbles or plane surface if the boundaries does not come
into play. When boundaries are involved, the interface shape is more
complex (it is a minimal surface -or Plateau- problem). This can be
illustrated by nice soap films experiments, as may be seen in very
simple experiments or in many science museums. Here, for our two dimensional
problem, it leads to circle or straight lines, as we now prove.

It is a classical exercice of variational calculus to prove that the
first variations of the length of a curve is proportional to the inverse
of its curvature radius $r$.
The solution of the problem (\ref{eq:Isoperimetric}) then leads to

\begin{equation}
\frac{eR}{r}=\alpha,\label{Rayon_courbure}\end{equation}
where $\alpha$ is a Lagrange parameter associated with the conservation
of the area. This proves that $r$ is constant along the interface:
solutions are either circles or straight lines. The law (\ref{Rayon_courbure})
is the equivalent of the Laplace law in classical thermodynamics,
relating the radius of curvature of the interface to the difference
of pressure inside and outside of the bubble%
\footnote{Indeed, at next order, the Lagrange parameter $\alpha$ leads to a
slight imbalance between the two phase free energy, which is related
to a pressure difference for the two phases. This thus gives the relation
between pressure imbalance, radius of curvature and free energy per
unit length (or unit surface in the 3D case).%
}.\\

We have thus shown that the minimization of the Van-Der-Waals Cahn
Hilliard functional, aimed at describing statistical equilibria for
first order phase transitions, predicts phase separation (formation
of sub-domains with each of the two phases corresponding to the two
minima of the free energy). It predicts the interface structure and
that its shape is described by an isoperimetrical problem: the minimization
of the length for a fixed enclosed area. Thus equilibrium structures
are either bubbles (circles) or straight lines. In the following sections,
we see how this applies to the description of statistical equilibria
for quasi-geostrophic flows, describing vortices and jets.

\subsubsection{The mathematics of the Van-Der-Waals Cahn Hilliard problem \label{sec:maths}}

The mathematical study of the Van-Der-Waals Cahn Hilliard functional
(\ref{eq:Van Der Waals Cahn Hilliard}) was a mathematical challenge
during the 1980s. It's solution has followed from the analysis in the
framework of spaces of functions with bounded variations, and on results
from semi-local analysis. One of the main contributions to this problem
was achieved by Modica, in 1987 \cite{1987_Modica_ArchRatMechAna}.
This functional analysis study proves the assumptions of the heuristic
presentation given in the previous subsections: $\phi$ takes the two
values $\pm u$ in sub-domains separated by transition area of width
scaling with $R$.

As a complement to these mathematical works, a more precise asymptotic
expansion based on the heuristic description above, generalizable
at all order in $R$, with mathematical justification of the existence
of the solutions for the interface equation at all order in $R$,
is provided in \cite{Bouchet_These}. Higher order effects are also
discussed in this work.

\subsection{Quasi-geostrophic statistical equilibria and first order phase transitions
\label{sub:QG strong jet generql}}

The first discussion of the analogy between statistical equilibria
in the limit $R\ll L$ and phase coexistence in usual thermodynamics,
in relation with the Van-Der-Waals Cahn Hilliard model is given in
\cite{Bouchet_These,Bouchet_Sommeria:2002_JFM}. This analogy has
been recently put on a more precise mathematical ground, by proving
that the variational problems of the RSM statistical mechanics and
the variational problem are indeed related \cite{Bouchet:2008_Physica_D}.
More precisely, any solution to the variational problem: \begin{equation}
\left\{ \begin{array}{c}
F=\min\left\{ \mathcal{F}\left[\phi\right]\,\,\left|\,\,\mathcal{A}\left[\phi\right]=-B\right.\right\} \\
{\rm \mbox{with}\,\,\,\,}\mathcal{F}=\int_{\mathcal{D}}\mathrm{d}{\bf \mathbf{r}}\,\left[\frac{R^{2}\left(\nabla\phi\right)^{2}}{2}+f\left(\phi\right)-R\phi h\right]\,\,\,\,\mbox{and}\,\,\,\,\mathcal{A}\left[\phi\right]=\int_{\mathcal{D}}\mathrm{d}{\bf \mathbf{r}}\,\phi\end{array}\right.\label{eq:Variational Van-Der-Waals Topography}\end{equation}
 where $\psi=R^{2}\phi$ ($\psi$ is the stream function defined by
equation (\ref{u}) on page \pageref{u}), is a RSM equilibria of
the quasi-geostrophic equations (\vref{QG}).

It is easy to prove that any critical point to (\ref{eq:Variational Van-Der-Waals Topography}) is a critical point to the grand canonical Energy-Casimir functional (\ref{eq:EC-VP}), and is a critical point of the entropy maximization. Considering  the problem (\ref{eq:Variational Van-Der-Waals Topography}), using a part integration and the relation $q=R^{2}\Delta\phi-\phi+Rh$ yields

\[\delta\mathcal{F}=\int\mathbf{\mathrm{d}r}\ \left(f^{\prime}(\phi)-\phi-q\right)\delta\phi \quad \text{and} \quad \delta\mathcal{A}=\int\mathrm{d}\mathbf{r}\ \delta\phi .\]

Critical points of (\ref{eq:Variational Van-Der-Waals Topography})
are therefore solutions of $\delta\mathcal{F}-\alpha\delta\mathcal{A}=0$, for
all $\delta\phi$, where $\alpha$ is the Lagrange multiplier associated
with the constraint $\mathcal{A}$. These critical points satisfy
\[ q=f^{\prime}\left(\frac{\psi}{R^{2}}\right)-\frac{\psi}{R^{2}}-\alpha.\]
We conclude that this equation is the same as (\ref{eq:q-psi equilibre}),
on page \pageref{eq:q-psi equilibre}, provided that $f^{\prime}\left(\frac{\psi}{R^{2}}\right)=g(\beta\psi)+\frac{\psi}{R^{2}}-\alpha$.

The proof that any solution to (\ref{eq:Variational Van-Der-Waals Topography})
is a RSM equilibria involves more complicated mathematical considerations; we assume this in the following and refer the interested readers to \cite{Bouchet:2008_Physica_D} for more details.\\

In the case of an initial distribution $\gamma$ (\ref{eq:distribution_ro})
with only two values of the potential vorticity: $\gamma(\sigma)=\left|\mathcal{D}\right|\left(a\delta(\sigma_{1})+(1-a)\delta(\sigma_{2})\right)$, only two Lagrange multipliers $\alpha_{1}$ and $\alpha_{2}$ are needed, associated with $\sigma_{1}$ and $\sigma_{2}$ respectively, in order to compute $g$, equation  (\ref{eq:q-psi equilibre}), on page \pageref{eq:q-psi equilibre})). In that case, the function $g$ is exactly $\tanh$ function. There exists in practice a much larger class of initial conditions for which the function $g$ would be an increasing function with a single inflexion point, similar to a $\tanh$ function, especially when one considers the limit of small Rossby radius of deformation. The works \cite{Bouchet_Sommeria:2002_JFM,VenailleBouchetJPO} give physical arguments to explain why it is the case for Jupiter's troposphere or oceanic rings and jets.

When $g$ is a $\tanh$-like function, the specific free energy $f$ has a double well shape, provided that the inverse temperature $\beta$ is negative, with sufficiently large values.

\subsubsection{Topography and anisotropy}

The topography term $\eta_{d}=Rh\left(y\right)$ in (\ref{eq:Variational Van-Der-Waals Topography})
is the main difference between the Van-Der-Waals Cahn Hilliard functional
(\ref{eq:Van Der Waals Cahn Hilliard}) and the quasi-geostrophic
variational problem (\ref{eq:Variational Van-Der-Waals Topography}).
We recall that this term is due to the beta plane approximation and
a prescribed motion in a lower layer of fluid (see section \ref{sub:Quasi-Geostrophic Model}).
This topographic term provides an anisotropy in the free energy. Its
effect will be the subject of most of the theoretical discussion in
the following sections.

Since we suppose that this term scales with $R$, the topography term
will not change the overall structure at leading order: there will
still be phase separations in sub-domains, separated by an interface
of typical width $R$, as discussed in section \ref{sub:Van Der Waals}.
We now discuss the dynamical meaning of this overall structure for
the quasi-geostrophic model.

\subsubsection{Potential vorticity mixing and phase separation}

In the case of the quasi-geostrophic equations, the order parameter
$\phi$ is proportional to the stream function $\psi$: $\psi=R^{2}\phi$.
At equilibrium, there is also a functional relation between the stream
function $\psi$ and the coarse-grained potential vorticity $q$ (\ref{eq:q-psi equilibre}).
Then the sub-domains of constant $\phi$ are domains where the (coarse
grained) potential vorticity $q$ is also constant. It means that
the level of mixing of the different fine grained potential vorticity
levels are constant in those sub-domains. We thus conclude that the
coarse grained potential vorticity is homogenized in sub-domains that
corresponds to different phases (with different values of potential
vorticity), the equilibrium being controlled by an equality for the
associated mixing free energy.

\subsubsection{Strong jets and interfaces}

In section \ref{sub:The-interface-shape}, we have described the interface
structure. The order parameter $\phi$ varies on a scale of order
$R$ mostly in the normal to the interface direction, reaching constant
values far from the interface. Recalling that $\phi$ is proportional
to $\psi$, and that $\mathbf{v}=\mathbf{e}_{z}\wedge\nabla\psi$
(\ref{u}), we conclude that:
\begin{enumerate}
\item The velocity field is nearly zero far from the interface (at distances
much larger than the Rossby deformation radius $R$). Non zero velocities
are limited to the interface areas.
\item The velocity is mainly directed along the interface.
\end{enumerate}
These two properties characterize strong jets. In the limit $R\ll L$,
the velocity field is thus mainly composed of strong jets of width
$R$, whose path is determined from an isoperimetrical variational
problem.

\subsection{Application to Jupiter's Great Red Spot and other Jovian features
\label{sub:Application-to-Jupiter}}

Most of Jupiter's volume is gas. The visible features on this atmosphere,
cyclones, anticyclones and jets, are concentrated on a thin outer
shell, the troposphere, where the dynamics is described by similar
equations to the ones describing the Earth atmosphere \cite{Dowling_Review_1995AnRFM..27..293D,Ingersoll_Vasavada_1998IAUSS...1.1042I}.
The inner part of the atmosphere is a conducting fluid, and the dynamics
is described by Magneto-hydrodynamics (MHD) equations.

The most simple model describing the troposphere is the 1-1/2 quasi-geostrophic model, described in section \ref{sub:Quasi-Geostrophic Model}.
This simple model is a good one for localized mid latitude dynamics.
Many classical work have used it to model Jupiter's features, taking
into account the effect of a prescribed steady flow in a deep layer
acting like an equivalent topography $h\left(y\right)$ (see section
\ref{sub:Quasi-Geostrophic Model}). We emphasize that there is no
real bottom topography on Jupiter.

Some works based on soliton theory aimed at explaining the structure
and stability of the Great Red Spot. However, none of these obtained
a velocity field qualitatively similar to the observed one, which
is actually a strongly non-linear structure. Structures similar to
the Great Red Spot have been observed in a number of numerical simulations,
but without reproducing in a convincing way both the characteristic
annular jet structure of the velocity field and the shape of the spot.
Detailed observations and fluid mechanics analysis described convincingly
the potential vorticity structure and the dynamical aspects of the
Great Red Spot (see \cite{Dowling_Review_1995AnRFM..27..293D,Ingersoll_Vasavada_1998IAUSS...1.1042I,1993ARA&A..31..523M}
and references therein). The potential vorticity structure is a constant
vorticity inside the spot surrounded by a gentle shear outside, which
gives a good fluid mechanics theory \cite{1993ARA&A..31..523M}. In
this section we explain this potential vorticity structure thanks
to statistical mechanics. Statistical mechanics provides also more
detailed, and analytical theory of the shape of Jupiter vortices.

The explanation of the stability of the Great Red Spot of Jupiter
using the statistical mechanics of the quasi-geostrophic model is
cited by nearly all the papers from the beginning of the Robert-Sommeria-Miller
theory. Some equilibria having qualitative similarities with the observed
velocity field have been computed in \cite{Sommeria_Nore_Dumont_Robert_1991CRASB.312..999S}.
The theoretical study in the limit of small Rossby deformation radius,
especially the analogy with first order phase transitions \cite{Bouchet_Sommeria:2002_JFM,Bouchet_Dumont_2003_condmat}
gave the theory presented below: an explanation of the detailed shape
and structure and a quantitative model. These results have been extended
to the shallow-water model \cite{Bouchet_Chavanis_Sommeria_2010_SW}.
The work \cite{TurkingtonMHD:2001_PNAS_GRS} argued on the explanation
of the position of the Great Red Spot based on statistical mechanics
equilibria.\\

We describe in the following the prediction of equilibrium statistical
mechanics for the quasi-geostrophic model with topography. The start
from the Van-Der-Waals Cahn Hilliard variational problem in presence
of small topography (\ref{eq:Variational Van-Der-Waals Topography}),
recalling that its minima are statistical equilibria of the quasi-geostrophic
model (see section \ref{sub:QG strong jet generql}).

The Rossby deformation radius at the Great Red Spot latitude is evaluated
to be of order of $500-2000$ km, which has to be compared with the
size of the spot: $10,000\, X\,20,000\,\mbox{km}$. This is thus
consistent with the limit $R\ll L$ considered in the description
of phase coexistence within the Van-Der-Waals Cahn Hilliard model
(section \ref{sub:Van Der Waals}), even if the criteria $r\ll R$
is only marginally verified where the curvature radius $r$ of the
jet is the larger.

\begin{figure}[t!]
\begin{tabular}{cc}
\begin{minipage}[c]{0.45\textwidth}%
\begin{center}
\includegraphics[width=\textwidth]{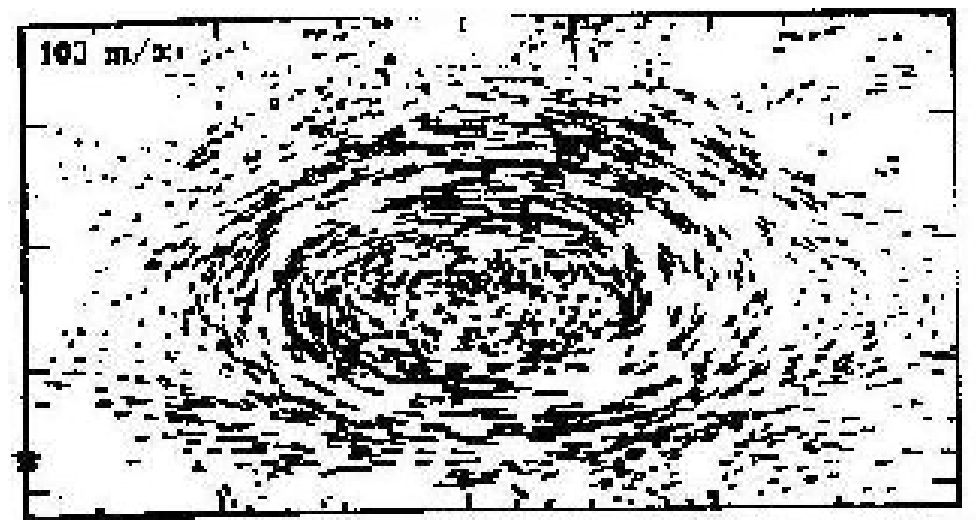} \\
 { Observation (Voyager)}
\par\end{center}%
\end{minipage} & %
\begin{minipage}[c]{0.45\textwidth}%
\begin{center}
\includegraphics[width=\textwidth]{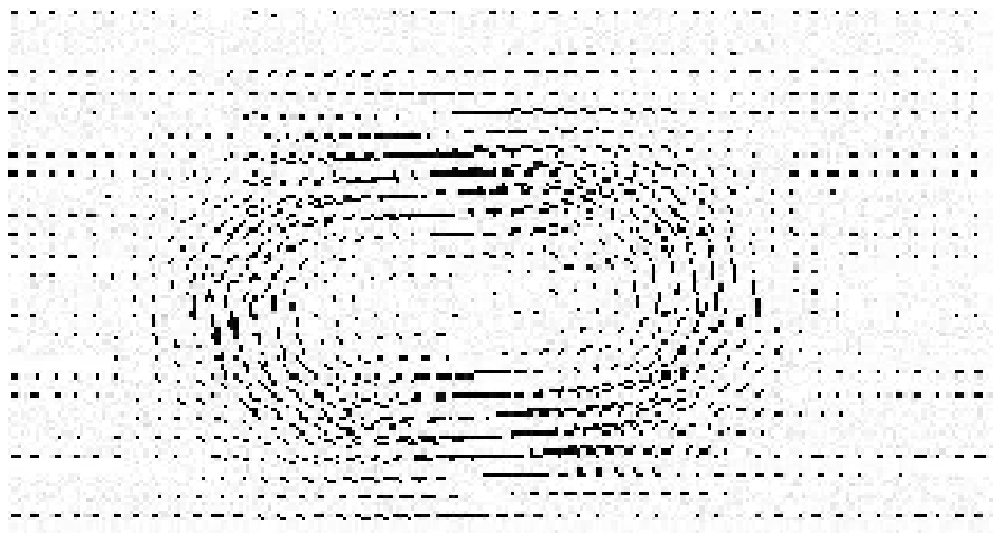} \\
 { Statistical Equilibrium}
\par\end{center}%
\end{minipage}\tabularnewline
\end{tabular}

\caption{\footnotesize Left: the observed velocity field is from Voyager spacecraft data,
from Dowling and Ingersoll \cite{Dowling_Ingersoll_1988JAtS...45.1380D}
; the length of each line is proportional to the velocity at that
point. Note the strong jet structure of width of order $R$, the Rossby
deformation radius. Right: the velocity field for the statistical
equilibrium model of the Great Red Spot. The actual values of the
jet maximum velocity, jet width, vortex width and length fit with
the observed ones. The jet is interpreted as the interface between
two phases; each of them corresponds to a different mixing level of
the potential vorticity. The jet shape obeys a minimal length variational
problem (an isoperimetrical problem) balanced by the effect of the
deep layer shear. \label{fig:Emergence_Numerique_tache_rouge} }

\end{figure}

In the limit of small Rossby deformation radius, the entropy maxima
for a given potential vorticity distribution and energy, are formed
by strong jets, bounding areas where the velocity is much smaller.
Figure \ref{fig:Emergence_Numerique_tache_rouge} shows the observation
of the Great Red Spot velocity field, analyzed from cloud tracking
on spacecraft pictures. The strong jet structure (the interface) and
phase separation (much smaller velocity inside and outside the interface)
is readily visible. The main difference with the structure described
in the previous section is the shape of the vortex: it is not circular
as was predicted in the case without topography or with a linear topography.
We consider the effect of a more general topography in the next section.

\subsubsection{Determination of the vortex shape: the typical elongated shape of
Jupiter's features \label{sec_Vortex_Shape}}

\begin{figure}[t!]

\begin{centering}
\begin{tabular}{ccc}
\begin{minipage}[c]{0.25\textwidth}%
\begin{center}
\includegraphics[width=\textwidth]{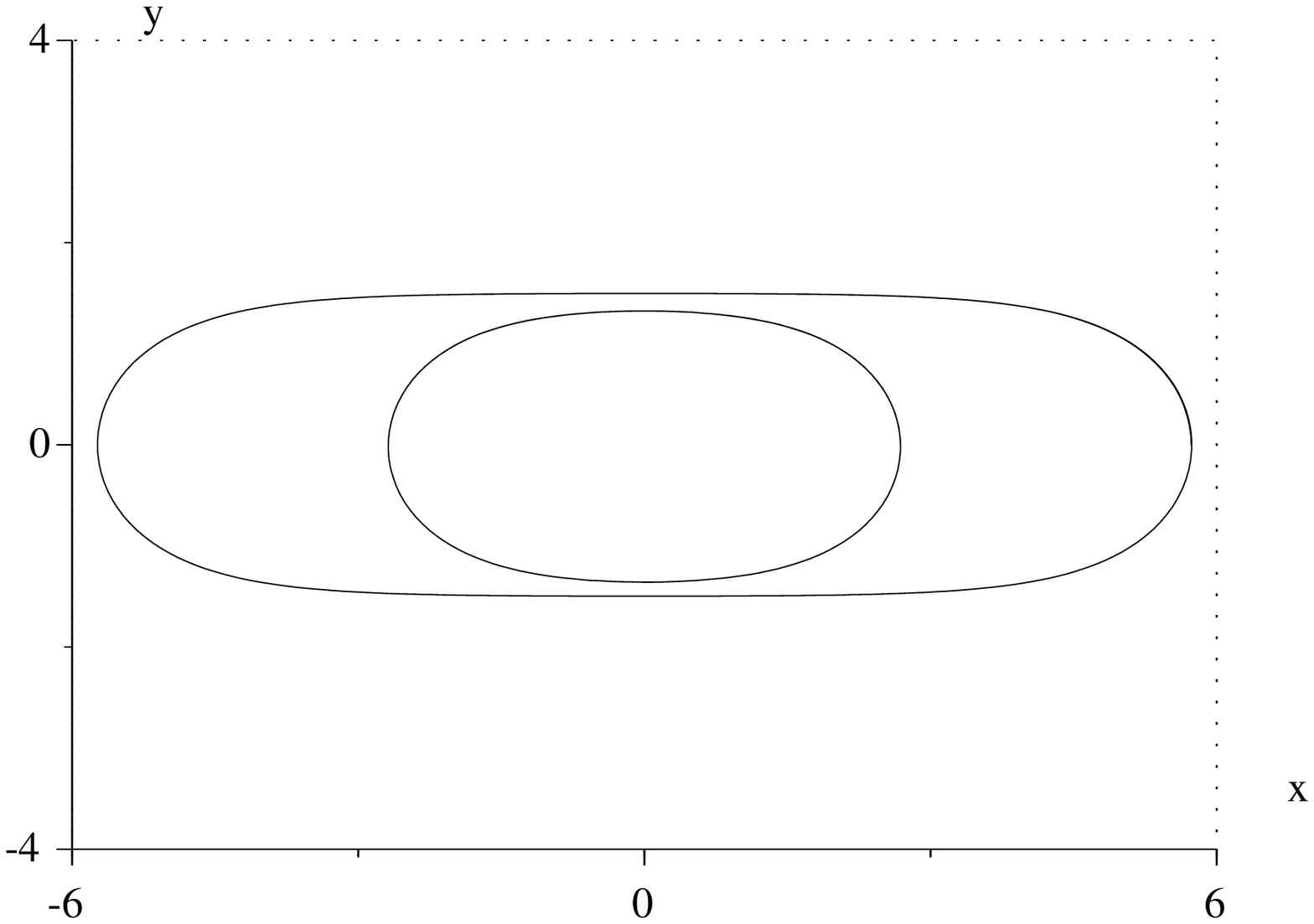} \\
 { Statistical Equilibria}
\par\end{center}%
\end{minipage} & %
\begin{minipage}[c]{0.25\textwidth}%
\begin{center}
\includegraphics[width=\textwidth]{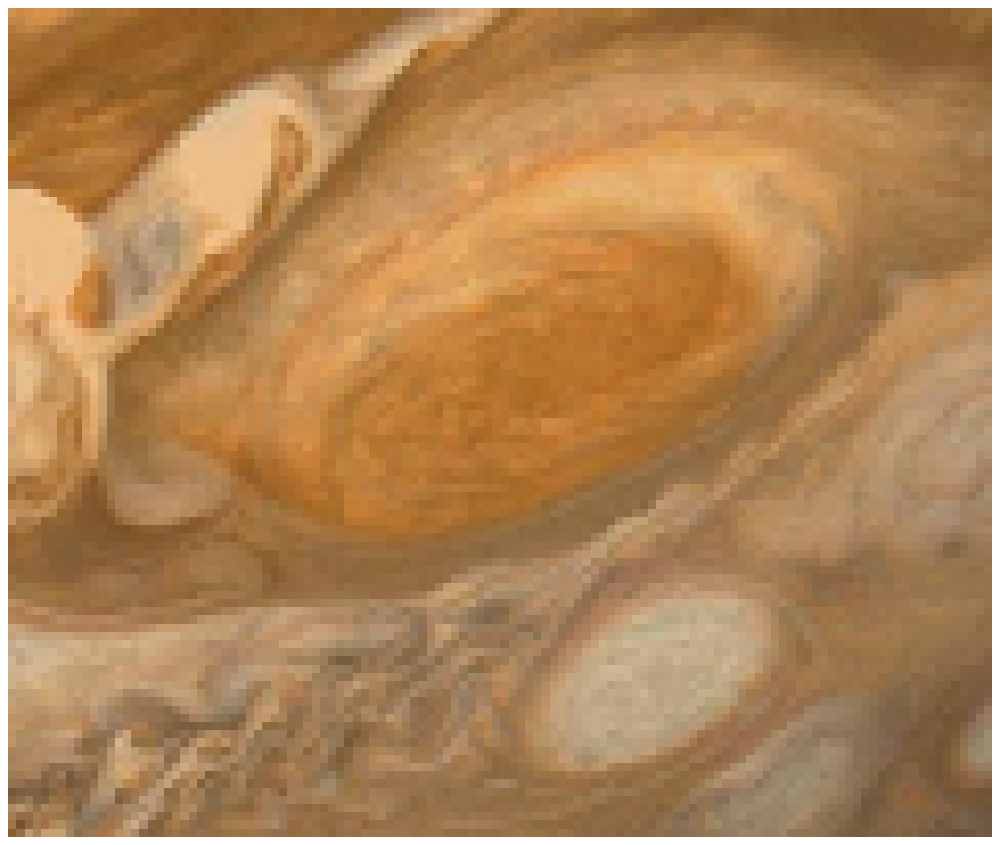} \\
 {Great Red Spot and White Oval BC}
\par\end{center}%
\end{minipage} & %
\begin{minipage}[c]{0.35\textwidth}%
\begin{center}
\includegraphics[width=0.9\textwidth]{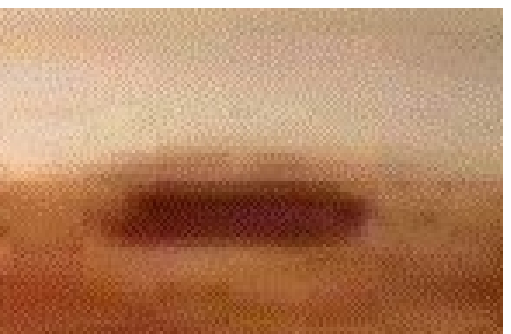} \\
 {A Brown Barge}
\par\end{center}%
\end{minipage}\tabularnewline
\end{tabular}
\par\end{centering}

\caption{{\footnotesize Left panel: typical vortex shapes obtained from the isoperimetrical
problem (curvature radius equation (\ref{Rayon_courbure})), for two
different values of the parameters (arbitrary units). The characteristic
properties of Jupiter's vortex shapes (very elongated, reaching extremal
latitude $y_{m}$ where the curvature radius is extremely large) are
well reproduced by these results. Central panel: the Great Red Spot
and one of the White Ovals. Right panel: one of the Brown Barge cyclones
of Jupiter's north atmosphere. Note the very peculiar cigar shape
of this vortex, in agreement with statistical mechanics predictions
(left panel)..}}
\label{fig_ellipse}
\end{figure}

In order to determine the effect of topography on the jet shape, we
consider again the variational problem (\ref{eq:Variational Van-Der-Waals Topography}).
We note that the topography $\eta_{d}=Rh$ has been rescaled by $R$
in the term $Rh(y)\phi$ appearing in the variational problem. This
corresponds to a regime where the effect of the topography is of the
same order as the effect of the jet free energy. Two other regimes
exist: one for which topography would have a negligible impact (this
would lead to circular vortices, as treated in section \ref{sub:QG strong jet generql})
and another regime where topography would play the dominant role.
This last regime may be interesting in some cases, but we do not treat
it in this review.

Due to the scaling $Rh\phi$, the topography does not play any role
at zeroth order. We thus still conclude that phase separation occurs,
with sub-domains of areas $A_{+}$ and $A_{-}$ fixed by the potential
vorticity constraint (see section \ref{sub:phase separation}), separated
by jets whose transverse structure is described in section \ref{sub:The-interface-shape}.
The jet shape is however given by minimization of the free energy
contributions of order $R$. Let us thus compute the first order contribution
of the topography term $RH=\int_{\mathcal{D}}\mathrm{d}{\bf \mathbf{r}}\,\left(-R\phi h(y)\right)$.
For this we use the zeroth order result $\phi=\pm u$. We then obtain
$H=-u\int_{A_{+}}\mathrm{d}\mathbf{r}\, h+u\int_{A_{-}}\mathrm{d}{\bf \mathbf{r}}\, h=H_{0}-2u\int_{A_{+}}\mathrm{d}{\bf \mathbf{r}}\, h$,
where $H_{0}\equiv u\int_{\mathcal{D}}\mathrm{d}{\bf \mathbf{r}}\, h$.
We note that $H_{0}$ does not depend on the jet shape.

Adding the contribution of the topography to the jet free energy (\ref{eq:Free energy length}),
we obtain the first order expression for the modified free energy
functional

\begin{equation}
\mathcal{F}=RH_{0}+R\left(eL-2u\int_{A_{+}}\mathrm{d}{\bf \mathbf{r}}\, h(y)\right),\label{Energy_libre_ordre1}\end{equation}
 which is valid up to correction of order $e\left(R/r\right)$ and
of order $R^{2}H$. We recall that the total area $A_{+}$ is fixed.
We see that, in order to minimize the free energy, the new term tends
to favor as much as possible the phase $A_{+}$ with positive values
of stream function $\phi=u$ (and then negative values of potential
vorticity $q=-u$) to be placed on topography maxima. This effect
is balanced by the length minimization.

In order to study in more details the shape of the jet, we look at
the critical points of the minimization of (\ref{Energy_libre_ordre1}),
with fixed area $A_{+}$. Recalling that first variations of the length
are proportional to the inverse of the curvature radius,
we obtain \begin{equation}
2uRh(y)+\alpha=\frac{eR}{r},\label{Rayon_courbure_h}\end{equation}
 where $\alpha$ is a Lagrange parameter associated with the conservation
of the area $A_{+}$. This relates the vortex shape to the topography
and parameters $u$ and $e$. From this equation, one can write the
equations for $X$ and $Y$, the coordinates of the jet curve. These
equations derive from a Hamiltonian, and a detailed analysis allows
to specify the initial conditions leading to closed curves and thus
to numerically compute the vortex shape (see \cite{Bouchet_Sommeria:2002_JFM}
for more details)%

Figure \ref{fig_ellipse} compares the numerically obtained vortex
shapes, with the Jovian ones. This shows that the solution to equation
(\ref{Rayon_courbure_h}) has the typical elongated shape of Jovian
vortices, as clearly illustrated by the peculiar cigar shape of Brown
Barges, which are cyclones of Jupiter's north troposphere. We thus
conclude that statistical mechanics and the associated Van-Der-Waals
Cahn Hilliard functional with topography explain well the shape of
Jovian vortices.\\

\begin{figure}[t!]
\begin{centering}
\includegraphics[width=0.85\textwidth]{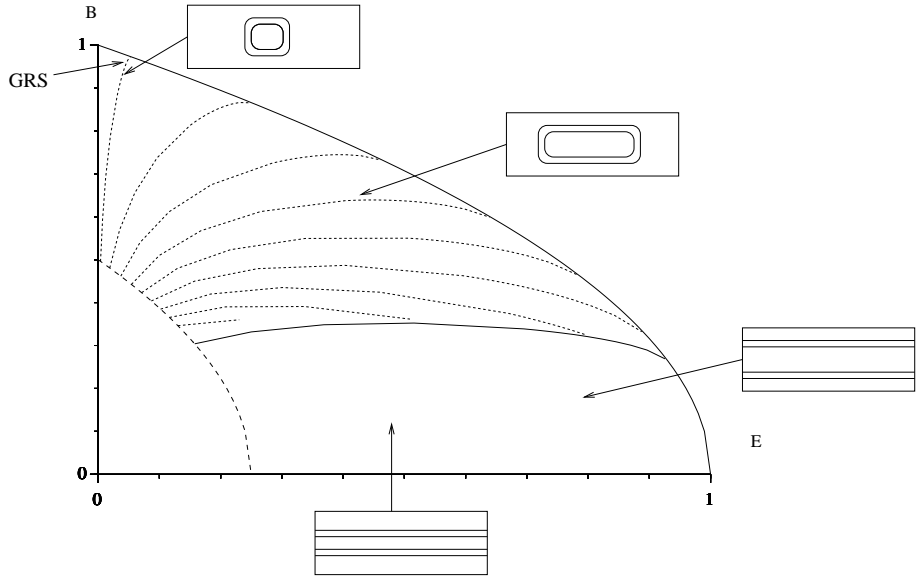}
\par\end{centering}

\caption{{\footnotesize Phase diagram of the statistical equilibrium states versus
the energy $E$ and a parameter related to the asymmetry between positive
and negative potential vorticity $B$, with a quadratic topography.
The inner solid line corresponds to a phase transition, between vortex
and straight jet solutions. The dash line corresponds to the limit
of validity of the small deformation radius hypothesis. The dot lines
are constant vortex aspect ratio lines with values 2,10,20,30,40,50,70,80
respectively. We have represented only solutions for which anticyclonic
potential vorticity dominate ($B>0$). The opposite situation may
be recovered by symmetry. For a more detailed discussion of this figure,
the precise relation between $E$, $B$ and the results presented
in this review, please see \cite{Bouchet_Dumont_2003_condmat}.}}
\label{phase_top}
\end{figure}

Figure \ref{phase_top} shows a phase diagram for the statistical
equilibria, with Jupiter like topography and Rossby deformation radius.
This illustrates the power of statistical mechanics: with only few
parameters characterizing statistical equilibria (here the energy
$E$ and a parameter related to the asymmetry between positive and
negative potential vorticity $B$), we are able to reproduce all the
features of Jupiter's troposphere, from circular white ovals, to the
GRS and cigar shaped Brown Barges. The reduction of the complexity
of turbulent flow to only a few order parameters is the main interest
and achievement of a statistical mechanics theory.

Moreover, as seen on figure \ref{phase_top}, statistical mechanics
predicts a phase transition from vortices towards straight jets. The
concept of phase transition is an essential one in complex systems,
as the qualitative physical properties of the system drastically change
at a given value of the control parameters. This is also an essential
point, to be bring such a concept in turbulent problems. This will
be further emphasized in section \ref{sub:Phase Transition 2}.

\subsubsection{Quantitative comparisons with Jupiter's Great Red Spot \label{sub:Quantitative-comparisons-Jupiter}}

In the preceding section, we have made a rapid description of the
effect of a topography to first order phase transitions. We have obtained
and compared the vortex shape with Jupiter's vortices. A much more
detailed treatment of the applications to Jupiter and to the Great
Red Spot can be found in \cite{Bouchet_Sommeria:2002_JFM,Bouchet_Dumont_2003_condmat}.
The theory can be extended in order to describe the small shear outside
of the spot (first order effect on $\phi$ outside of the interface),
on the Great Red Spot zonal velocity with respect to the ambient shear,
on the typical latitudinal extension of these vortices. A more detailed
description of physical considerations on the relations between potential
vorticity distribution and forcing is also provided in \cite{Bouchet_Sommeria:2002_JFM,Bouchet_Dumont_2003_condmat}.

\subsection{Application to ocean rings \label{sub:Gulf Stream Rings}}

Application of equilibrium statistical mechanics to the description of oceanic flows is a long-standing problem, starting with the work of  Salmon--Holloway--Hendershott \cite{SalmonHollowayHendershott:1976_JFM_stat_mech_QG} in the framework of energy-enstrophy theory.

Another attempt to apply equilibrium statistical mechanics to oceanic flows had been performed by \cite{DibattistaMajda00,DibattistaMajda02} in the framework of the Heton model of \cite{HoggStommel85} for the self-organization phenomena following deep convection events, by  numerically computing statistical equilibrium states of a two-layer quasi-geostrophic model.

None of these previous approaches have explained the ubiquity of oceanic rings. We show in the following that such rings can actually be understood as statistical equilibria by similar arguments that explain the formation of Jovian vortices (see \cite{VenailleBouchetJPO} for more details).

\subsubsection{Rings in the oceans}

\begin{figure}[t!]
\begin{center}
\includegraphics[width=0.8\textwidth]{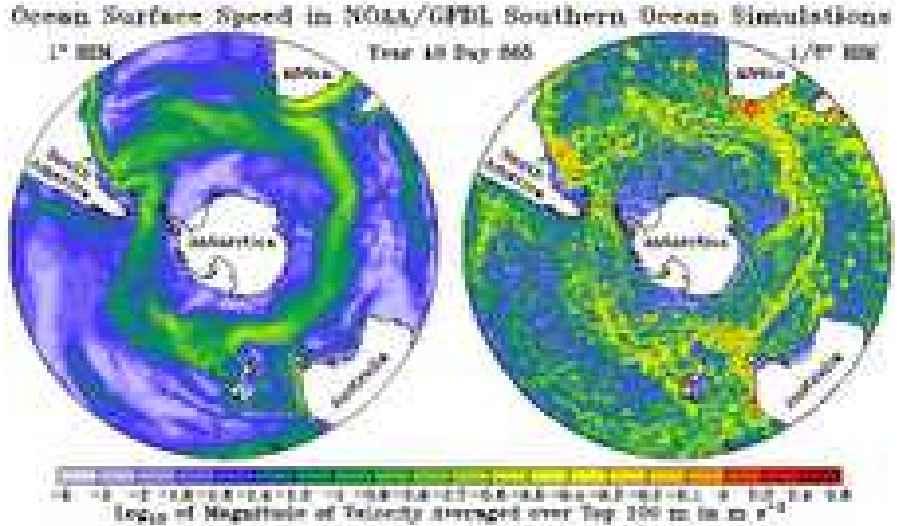}
\end{center}
\caption{{\footnotesize Snapshot of surface velocity field from a comprehensive
numerical simulation of the southern Oceans \cite{HallbergMESO_06}.
Left: coarse resolution, the effect of mesoscale eddies ($\sim100km$)
is parameterized. Right: higher resolution, without parameterization
of mesoscale eddies. Note the formation of large scale coherent structure
in the high resolution simulation: there is either strong and thin
eastward jets or rings of diameter $\sim200\ km$. Typical velocity
and width of jets (be it eastward or around the rings) are respectively
$\sim1\ m.s^{-1}$ and $\sim20\ km$. The give a statistical mechanics
explanation and model for these rings.}}

\label{Fig:SouthernOcean}
\end{figure}

The ocean has long been recognized as a sea of eddies. This has been
first inferred from \emph{in situ} data by Gill, Green and Simmons
in the early 1970s \cite{GGS74}. During the last two decades, the concomitant
development of altimetry \cite{Chelton07,Stammer97} and realistic
ocean modeling \cite{HallbergMESO_06,Barnier06} has made possible
a quantitative description of those eddies. The most striking observation
is probably their organization into westward propagating rings of
diameters $(L_{e}\sim200\ km)$, as for instance seen in figure \ref{Fig:SouthernOcean}.
In that respect, they look like small Jovian Great Red Spots.

Those eddies plays a crucial role for the general ocean circulation
and its energy cycle, since their total energy is one order of magnitude
above the kinetic energy of the mean flow.\\

Those rings are mostly located around western boundary currents, which
are regions characterized by strong baroclinic instabilities%
\footnote{When the mean flow present a sufficiently strong vertical shear, baroclinic
instabilities \cite{PedloskyBook,VallisBook} release part of the
available potential energy associated with this mean flow, which is
generally assumed to be maintained by a large scale, low frequency
forcing mechanism such as surface wind stress or heating \cite{VallisBook}%
}, such as the Gulf Stream, the Kuroshio, the Aghulas currents below
South Africa or the confluence region of the Argentinian basin, as
seen on figures \ref{Fig:SouthernOcean} and \ref{Fig:gulfstream}.
The rings can also propagate far away from the regions where they
are created. \\

Most of those rings have a baroclinic structure, i.e. a velocity field
intensified in the upper layer ($H\sim1\ km$) of the oceans. This
baroclinic structure suggest that the 1.5 layer quasi-geostrophic
model introduced in the previous sections is relevant to this problem.
The horizontal scale of the rings $(L_{e}\sim200\ km)$ are larger
than the width $R\sim50\ km$ of the surrounding jet, of typical velocities
$U=1\ m.s^{-1}$.

The organization of those eddies into coherent rings can be understood
by the same statistical mechanics arguments that have just been presented
in the case of Jupiter's Great Red Spot. The rings correspond to one
phase containing most of the potential vorticity extracted from the mean flow by baroclinic
instability, while the surrounding quiescent flow corresponds to the
other phase. This statistical mechanics approach, the only one to
our knowledge to describe the formation of large scale coherent structures,
might then be extremely fruitful to account for the formation of such
rings. It remains an important open question concerning the criteria
that select the size of such coherent structures. This is an ongoing
subject of investigation.

\begin{figure}[t!]
\begin{center}
\includegraphics[width=0.8\textwidth ]{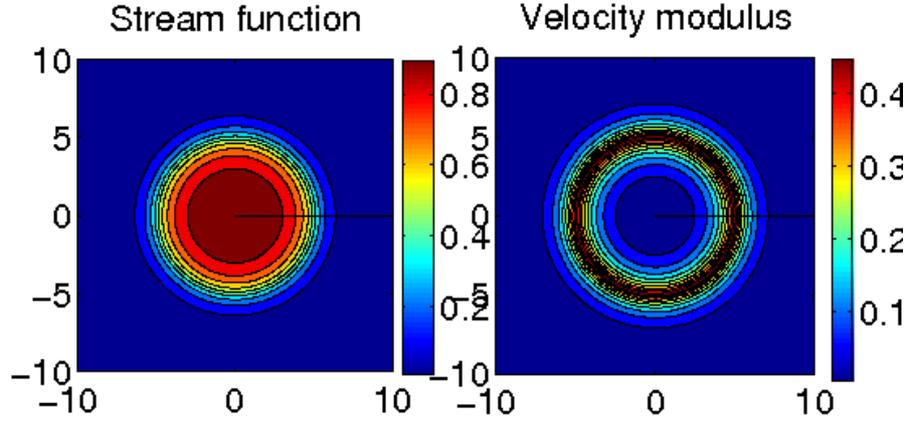}
\end{center}

\caption{\footnotesize Vortex statistical equilibria in the quasi-geostrophic model. It is a circular patch of (homogenized) potential vorticity in a background of homogenized potential vorticity, with two different mixing values. The velocity field (right panel) has a very clear ring structure, similarly to the Gulf-Stream rings and to many other ocean vortices. The width of the jet surrounding the ring has the order of magnitude of the Rossby radius of deformation $R$.}
\label{rings_shape}
\end{figure}

\subsubsection{The westward drift of the rings \label{sub:The-westward-drift}}

In this section, we consider the consequences of the beta effect (see
section \ref{sub:Quasi-Geostrophic Model}), which corresponds to
linear topography $\eta_{d}=\beta_{c}y$ in (\ref{dir}). We prove
that this term can be easily handled and that it actually explains
the westward drift of oceanic rings with respect to the mean surrounding
flow.

We consider the quasi-geostrophic equations on a domain which is invariant
upon a translation along the $x$ direction (either an infinite or
a periodic channel, for instance). Then the quasi-geostrophic equations
are invariant over a Galilean transformation in the $x$ direction.
We consider the transformation \[
\mathbf{v'}=\mathbf{v}+V\mathbf{e}_{x},\]
 where $\mathbf{v}$ is the velocity in the original frame of reference
and $\mathbf{v}'$ is the velocity in the new Galilean frame of reference.

From the relation $\mathbf{v}=\mathbf{e}_{z}\wedge\nabla\psi$ (\ref{u}),
we obtain the transformation law for $\psi$: $\psi'=\psi-Vy$ and
from the expression $q=\Delta\psi-\psi/R^{2}+\beta_{c}y$ (\ref{dir})
we obtain the transformation law for $q$: $q'=q+Vy/R^{2}$. Thus
the expression for the potential vorticity in the new reference frame
is \[
q=\Delta\psi-\frac{\psi}{R^{2}}+\left(\beta_{c}+\frac{V}{R^{2}}\right)y.\]
 From this last expression, we see that a change of Galilean reference
frame translates as a beta effect in the potential vorticity. Moreover,
in a reference frame moving at velocity $-\beta_{c}R^{2}\mathbf{e}_{x}$,
the $\beta_{c}$ effect is exactly canceled out.

From this remark, we conclude that taking into account the beta
effect, the equilibrium structures should be the one described by
the minimization of the Van-Der-Waals Cahn Hilliard variational problem,
but moving at a constant westward speed $V=\beta_{c}R^{2}$.  A more rigorous treatment of the statistical mechanics for the quasi-geostrophic model with translational invariance would require to take into account an additional conserved quantity, the linear momentum, which would lead to the same conclusion: statistical equilibria are rings with a constant westward speed $V=\beta_{c}R^{2}$. See also \cite{VenailleBouchetJPO} for more details and discussions on the physical consequences of this additional constraint.

\begin{figure}[t!]
\begin{center}
\includegraphics[width=6 cm]{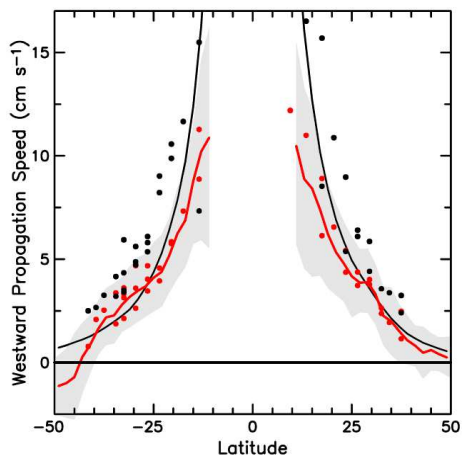}
\end{center}

\caption{\footnotesize Altimetry observation of the westward drift of oceanic eddies (including rings) from \cite{Chelton07}, figure 4. The red line is the zonal average (along a latitude circle) of the propagation speeds of all eddies with life time greater than 12 weeks. The black line represents the velocity  $\beta_c R^2$ where $\beta_c$ is  the meridional gradient of the Coriolis parameter and  $R$ the first baroclinic Rossby radius of deformation. This eddy propagation speed is a prediction of statistical mechanics (see section \ref{sub:The-westward-drift}) }
\label{fig:Chelton}
\end{figure}

\begin{figure}[t!]
\begin{center}
\includegraphics[height=0.8\textwidth]{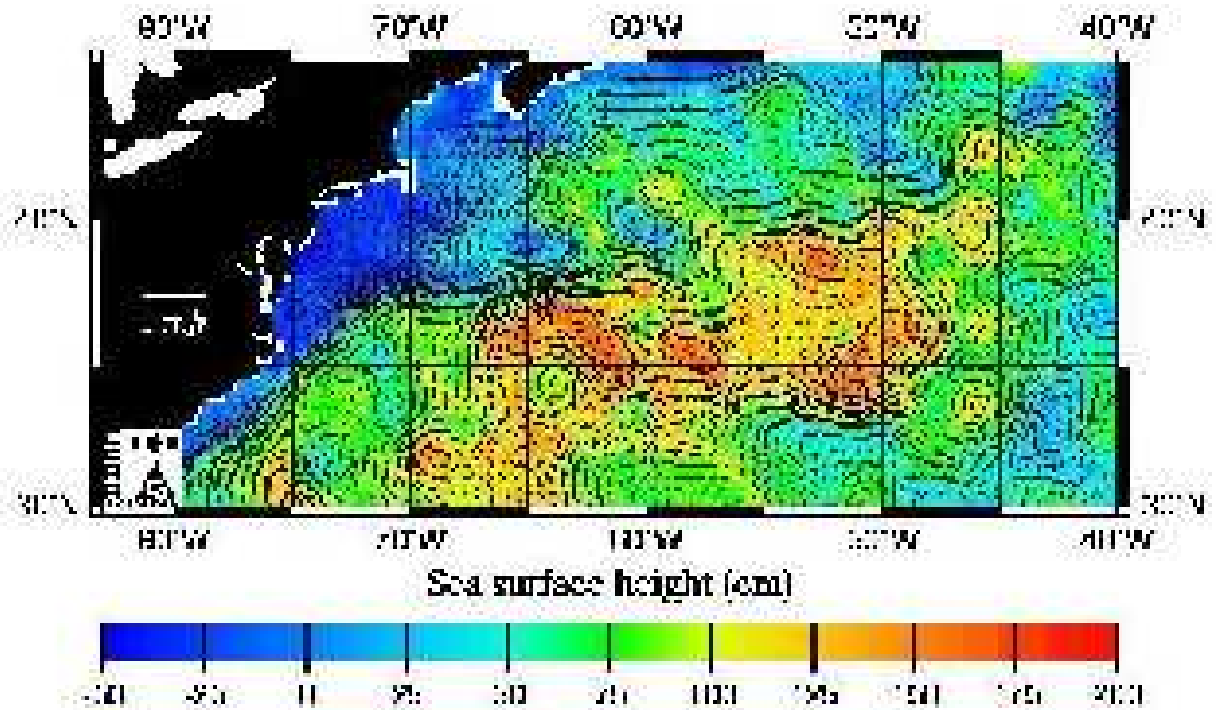}
\end{center}
\caption{{\footnotesize Observation of the sea surface height of the north
Atlantic ocean (Gulf Stream area) from altimetry REF. As explained
in section \ref{sub:Euler-QG-equations}, for geophysical flows, the
surface velocity field can be inferred from the see surface height
(SSH): strong gradient of SSH are related to strong jets. The Gulf
stream appears as a robust eastward jet (in presence of meanders),
flowing along the east coast of north America and then detaching the
coast to enter the Atlantic ocean, with an extension $L\sim2000\ km$.
The jet is surrounded by numerous westward propagating rings of typical
diameters $L\sim200\ km$. Typical velocities and widths of both the
Gulf Stream and its rings jets are respectively $1\ m.s^{-1}$ and
$50\ km$, corresponding to a Reynolds number $Re\sim10^{11}$. Such
rings can be understood as local statistical equilibria, and strong
eastward jets like the Gulf Stream and obtained as marginally unstable
statistical equilibria in simple academic models (see subsections
\ref{sub:Gulf Stream Rings}-\ref{sec:Gulf Stream and Kuroshio}).}}

\label{Fig:gulfstream}
\end{figure}

This drift is actually observed for the oceanic rings, see for instance
figure \ref{fig:Chelton}.

\newpage

\section{Are the Gulf-Stream and the Kuroshio currents close to statistical
equilibria? \label{sec:Gulf Stream and Kuroshio}}

In section \ref{sub:Gulf Stream Rings}, we have discussed applications
of statistical mechanics ideas to the description of ocean vortices,
like the Gulf-Stream rings. We have also mentioned that statistical
equilibria, starting from the Van-Der-Waals Cahn Hilliard functional
(\ref{eq:Variational Van-Der-Waals Topography}), may model physical
situations where strong jets, with a width of order $R$, bound domains
of nearly constant potential vorticity.

\begin{figure}[t!]
\begin{center}
\includegraphics[width=0.9\textwidth]{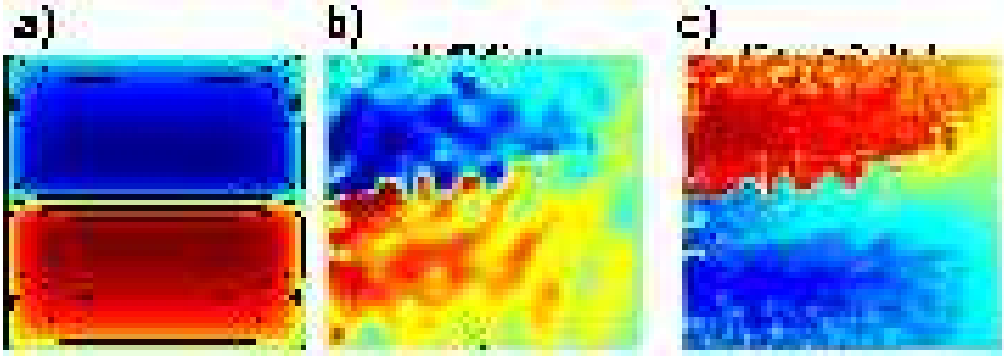}
\end{center}
\caption{ \footnotesize b) and c) represent respectively a snapshot of the streamfunction
and potential vorticity (red: positive values; blue: negative values)
in the upper layer of a three layers quasi-geostrophic model in a
closed domain, representing a mid-latitude oceanic basin, in presence
of wind forcing. Both figures are taken from numerical simulations
\cite{berloff}, see also \cite{BerloffHogg}. a) Streamfunction predicted
by statistical mechanics, see section \vref{sec:Gulf Stream and Kuroshio}
for further details. Even in an out-equilibrium situation like this one, the equilibrium statistical mechanics predicts correctly the overall qualitative structure of the flow.}

\label{Fig:frontBerloff}
\end{figure}

This is actually the case of the Gulf Stream in the North Atlantic
ocean or of the Kuroshio extension in the North Pacific ocean. This
can be inferred from observations, or this is observed in high resolution
numerical simulations of idealized wind driven mid-latitude ocean,
see for instance figure \vref{Fig:frontBerloff} (and ref. \cite{BerloffHogg}
for more details).

It is thus very tempting to interpret the Gulf Stream and the Kuroshio
as interfaces between two phases corresponding to different levels
of potential vorticity mixing, just like the Great Red Spot and ocean
rings in the previous section. The aim of this chapter is to answer
this natural question: are the Gulf-Stream and Kuroshio currents close
to statistical equilibria?\\

{}{}More precisely, we address the following problem: is it possible
to find a class of statistical equilibria with a strong mid-basin
eastward jet similar to the Gulf Stream of the Kuroshio, in a closed
domain? The 1-1/2 layer quasi-geostrophic model (see section \ref{sub:Quasi-Geostrophic Model})
is the simplest model taking into account density stratification for
mid-latitude ocean circulation (in the upper first $1000\, m$) \cite{Pedlosky:1998_OceanCirculationTheory,VallisBook}.
We analyze therefore the class of statistical equilibria which are
minima of the Van-Der-Waals Cahn Hilliard variational problem (\ref{eq:Variational Van-Der-Waals Topography}),
as explained in section \ref{sub:QG strong jet generql}. We ask whether
it exists solutions to \begin{equation}
\left\{ \begin{array}{c}
F=\min\left\{ \mathcal{F}\left[\phi\right]\,\,\left|\,\, A\left[\phi\right]=-B\right.\right\} \\
{\rm \mbox{with}\,\,\,\,}\mathcal{F}=\int_{\mathcal{D}}d{\bf r}\,\left[\frac{R^{2}\left(\nabla\phi\right)^{2}}{2}+f\left(\phi\right)-R\tilde{\beta}_{c}y\phi\right]\,\,\,\,\mbox{and}\,\,\,\,\mathcal{A}\left[\phi\right]=\int_{\mathcal{D}}d{\bf r}\,\phi\end{array}\right.\label{eq:Variational Van-Der-Waals Effet Beta}\end{equation}
in a bounded domain (let say a rectangular basin) with strong mid-basin
eastward jets. At the domain boundary, we fix $\phi=0$ (which using
$\phi=R^{2}\psi$, and (\ref{u}) turns out to be an impermeability
condition). We note that the understanding of the following discussion
requires the reading of sections 4.1 to 4.3. \\

The term $R\tilde{\beta}_{c}y$ is an effective topography including
the beta effect and the effect of a deep zonal flow (see section \ref{sub:Quasi-Geostrophic Model}).
Its significance and effects will be discussed in section 5.2. As
in the previous section, we consider the limit $R\ll L$ and assume
$f$ be a double well function.\\

As discussed in chapter \ref{sub:Van Der Waals}, with these hypothesis,
there is phase separation in two subdomains with two different levels
of potential vorticity mixing. These domains are bounded by interfaces
(jets) of width $R$. In view of the applications to mid-basin ocean
jets, we assume that the area $A_{+}$ occupied by the value $\phi=u$
is half of the total area of the domain (this amounts to fix the total
potential vorticity constraint $\Gamma_{1}$ (\ref{sub:Casimirs-conservation-laws})).
The question is to determine the position and shape of this interface.
The main difference with the cases treated in subsection \ref{sub:Van Der Waals}
is due to the effect of boundaries and of the linear effective topography
$R\tilde{\beta}_{c}y$.

\subsection{Eastward jets are statistical equilibria of the quasi-geostrophic
model without topography\label{sub:Eastward-jets-without_topography}}

\begin{figure}[t!]
\begin{centering}
\includegraphics[width=0.85\textwidth]{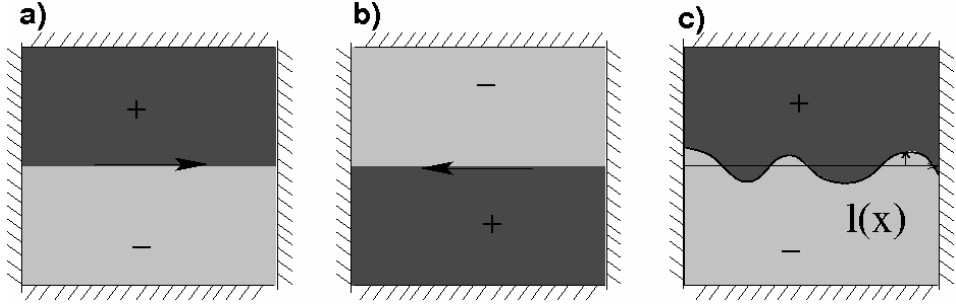}
\par\end{centering}

\caption{{\footnotesize a) Eastward jet: the interface is zonal, with positive
potential vorticity $q=u$ on the northern part of the domain. b)
Westward jet: the interface is zonal, with negative potential vorticity
$q=-u$ in the northern part of the domain. c) Perturbation of the
interface for the eastward jet configuration, to determine when this
solution is a local equilibrium (see subsection \ref{sub:Eastward-jets-with_topography}).
Without topography, both (a) and (b) are entropy maxima. With positive
beta effect (b) is the global entropy maximum; with negative beta
effect (a) is the global entropy maximum.}}
\label{fig_pvfronts}
\end{figure}

The value $\phi=\pm u$ for the two coexisting phases is not compatible
with the boundary condition $\phi=0$. As a consequence, there exists
a boundary jet (or boundary layer) in order to match a uniform phase
$\phi=\pm u$ to the boundary conditions. Just like inner jets, treated
in section \ref{sec:First Order GRS and rings}, these jets contribute to the first order free energy,
which gives the jet position and shape. We now treat the effect of
boundary layer for the case $h=0$ ($\widetilde{\beta}_{c}=0$ in
this case). As explained in section \ref{sub:The-interface-shape},
the jet free energy is the only contribution to the total free energy.

We first quantify the unit length free energy, $F_{b}$, for the boundary
jets. Following the reasoning of section \ref{sub:The-interface-shape},
we have \[
F_{b}=\min\left\{ \int d\zeta\,\left[\frac{R^{2}}{2}\frac{d^{2}\phi}{d\zeta^{2}}+f(\phi)\right]\right\} .\]
 This expression is the same as (\ref{eq:Variational Free Energy Unit lenght}),
the only difference is the different boundary conditions: it was $\phi\rightarrow_{\zeta\rightarrow+\infty}u$
and $\phi\rightarrow_{\zeta\rightarrow-\infty}-u$, it is now $\phi\rightarrow_{\zeta\rightarrow+\infty}u$
and $\phi\left(0\right)=0$. Because $f$ is even, one easily see
that a boundary jet is nothing else than half of a interior domain
jet. Then \[
F_{b}=\frac{1}{2}F_{int}=\frac{e}{2}R,\]
where $F_{int}$ and $e$ are the unit length free energies for the
interior jets, as defined in section \ref{sub:The-interface-shape}.
By symmetry, a boundary jet matching the value $\phi=-u$ to $\phi=0$
gives the same contribution%
\footnote{We have treated the symmetric case when $f$ is even. The asymmetric
case could be also easily treated%
}. Finally, the first order free energy is given by \[
\mathcal{F}=eR\left(L+\frac{L_{b}}{2}\right),\]
where $L_{b}$ is the boundary length. Because the boundary length
$L_{b}$ is a fixed quantity, the free energy minimization amounts
to the minimization of the interior jet length. The interior jet position
and shape is thus given by the minimization of the interior jet length
with fixed area $A_{+}.$ We recall that the solutions to this variational
problem are interior jets which are either straight lines or circles
(see section \ref{sub:The-interface-shape}).\\

In order to simplify the discussion, we consider the case of a rectangular
domain of aspect ratio $\tau=L_{x}/L_{y}$. Generalization to an arbitrary
closed domain could also be discussed. We recall that the two phases
occupy the same area $A_{+}=A_{-}=\frac{1}{2}L_{x}L_{y}$. We consider
three possible interface configurations with straight or circular
jets:
\begin{enumerate}
\item the zonal jet configuration (jet along the $x$ axis) with $L=L_{x}$,
\item the meridional jet configuration (jet along the $y$ axis with $L=L_{y}$,
\item and an interior circular vortex, with $L=2\sqrt{\pi A_{+}}=\sqrt{2\pi L_{x}L_{y}}$
.
\end{enumerate}
The minimization of $L$ for these three configurations shows that
the zonal jet is a global minimum if and only if $\tau<1$. The criterion
for the zonal jet to be a global RSM equilibrium state is then $L_{x}<L_{y}$.
We have thus found zonal jet as statistical equilibria in the case
$h=0$.\\

An essential point is that both the Kuroshio and the Gulf Stream are
flowing eastward (from west to east). From the relation $\mathbf{v}=\mathbf{e}_{z}\times\nabla\psi$
(\ref{u}), we see that the jet flows eastward ($v_{x}>0$) when $\partial_{y}\psi<0$.
Recalling that $\phi=R^{2}\psi$, the previous condition means that
the negative phase $\phi=-u$ has to be on the northern part of the
domain, and the phase $\phi=u$ on the southern part. From (\ref{dir}),
we see that this corresponds to a phase with positive potential vorticity
$q=u$ on the northern sub-domains and negative potential vorticity
$q=-u$ on the southern sub-domain, as illustrated in the panel (a)
of figure (\ref{fig_pvfronts}).

Looking at the variational problems (\ref{eq:Variational Van-Der-Waals Effet Beta}),
it is clear that in the case $\widetilde{\beta}_{c}=0$, the minimization
of $\phi$ is invariant over the symmetry$\phi\rightarrow-\phi$.
Then solutions with eastward or westward jets are completely equivalent.
Actually there are two equivalent solutions for each of the case 1,
2 and 3 above. However, adding a beta effect $h=R\widetilde{\beta}_{c}y$
will break this symmetry. This is the subject of next section.\\

We conclude that in a closed domain with aspect ratio $L_{x}/L_{y}<1$,
without topography, equilibrium states exist with an eastward jet
at the center of the domain, recirculating jets along the domain boundary
and a quiescent interior. For $L_{x}/L_{y}>1$, these solutions become
metastable states (local entropy maximum). This equilibrium is degenerated,
since the symmetric solution with a westward jet is always possible.

\subsection{Addition of a topography \label{sub:Eastward-jets-with_topography}}

For ocean dynamics, the beta effect plays a crucial role. Let us now
consider the case where the topography is $\eta_{d}=\beta_{c}y+\frac{\psi_{d}}{R^{2}}$.
The first contribution comes from the beta-effect (the variation of
the Coriolis parameter with latitude). The second contribution is
a permanent deviation of the interface between the upper layer and
the lower layer. For simplicity, we consider the case where this permanent
interface elevation is driven by a constant zonal flow in the lower
layer: $\psi_{d}=-U_{d}y$ , which gives $\eta_{d}=\left(\beta_{c}-\frac{U_{d}}{R^{2}}\right)y=R\widetilde{\beta}_{c}y$.
Then the combined effect of a deep constant zonal flow and of the
variation of the Coriolis parameter with latitude is an effective
linear beta effect. \\

In the definition of $\widetilde{\beta}_{c}$ above, we use a rescaling
with $R$. This choice is considered in order to treat the case where
the contribution of the effective beta effect appears at the same
order as the jet length contribution. This allows to easily study
how the beta effect breaks the symmetry $\phi\rightarrow-\phi$ between
eastward and westward jets. Following the arguments of section \ref{sec_Vortex_Shape},
we minimize \begin{equation}
\mathcal{F}=RH_{0}+R\left(eL-2u\int_{A_{+}}\mathrm{d}{\bf \mathbf{r}}\,\tilde{\beta}_{c}y\right),\label{eq:Energy_Libre_Jet_Effet_Beta}\end{equation}
(see equation (\ref{Energy_libre_ordre1})), with a fixed area $A_{+}$.
The jet position is a critical point of this functional: $e/r-2u\tilde{\beta}_{c}y_{jet}=\alpha$
(see equation (\ref{Rayon_courbure_h})), where $\alpha$ is a Lagrange
parameter and $y_{jet}$ the latitude of the jet. We conclude that
zonal jets (curves with constant $y_{jet}$ and $r=+\infty$) are
solutions to this equation for $\alpha=-2uR\tilde{\beta}_{c}y_{jet}$.
Eastward and westward jets described in the previous section are still
critical points of entropy maximization.\\

\subsubsection{With a negative effective beta effect, eastward jets are statistical
equilibria}

We first consider the case $\widetilde{\beta}_{c}<0$. This occurs
when the zonal flow in the lower layer is eastward and sufficiently
strong ($U_{d}>R^{2}\beta_{c}$). If we compute the first order free
energy (\ref{eq:Energy_Libre_Jet_Effet_Beta}) for both the eastward
and the westward mid-latitude jet, it is easy to see that in order
to minimize $\mathcal{F}$, the domain $A_{+}$ has to be located
at the lower latitudes: taking $y=0$ at the interface, the term $-2u\int_{A_{+}}d^{2}{\bf r}\,\tilde{\beta}_{c}y=u\tilde{\beta}_{c}L_{x}L_{y}/4$
gives a negative contribution when the phase with $\phi=u$ (and $q=-u$)
is on the southern part of the domain ($A_{+}=(0,L_{x})\times(-\frac{L_{y}}{2},0)$).
This term would give the opposite contribution if the phase $\phi=u$
would occupy the northern part of the domain. Thus the statistical
equilibria is the one with negative streamfunction $\phi$ (corresponding
to positive potential vorticity $q$) on the northern part of the
domain. As discussed in the end of section \ref{sub:Eastward-jets-without_topography}
and illustrated on figure \ref{fig_pvfronts}, panel (b), this is
the case of an eastward jet.

Thus, we conclude that taking into account an effective negative beta-effect
term at first order breaks the westward-eastward jet symmetry. When
$\widetilde{\beta}_{c}<0$, statistical equilibria are flows with
mid-basin eastward jets. \\

\subsubsection{With a positive effective beta effect, westward jets are statistical
equilibria}

Let us now assume that the effective beta coefficient is positive.
This is the case when $U_{d}<R^{2}\beta_{c}$, i.e. when the lower
layer is either flowing westward, or eastward with a sufficiently
low velocity. The argument of the previous paragraph can then be used
to show that the statistical equilibrium is the solution presenting
a westward jet.%

\subsubsection{With a sufficiently small effective beta coefficient, eastward jets
are local statistical equilibria}

We have just proved that mid-basin eastward jets are not global equilibria
in the case of positive effective beta effect. They are however critical
points of entropy maximization. They still could be local entropy
maxima. We now consider this question: are mid-basin strong eastward
jets local equilibria for a positive effective beta coefficient?
In order to answer, we perturb the interface between the two phases,
while keeping constant the area they occupy, and compute the free
energy perturbation.

The unperturbed interface equation is $y=0$, the perturbed one is
$y=l(x)$, see figure \ref{fig_pvfronts}. Qualitatively, the contributions
to the free energy $\mathcal{F}$ (\ref{eq:Energy_Libre_Jet_Effet_Beta}),
of the jet on one hand and of the topography on the other hand, are
competing with each other. Any perturbation increases the jet length
$L=\int dx\ \sqrt{1+\left(\frac{dl}{dx}\right)^{2}}$ and then increases
the second term in equation (\ref{eq:Energy_Libre_Jet_Effet_Beta})
by $\delta\mathcal{F}_{1}=Re\int dx\,\left(dl/dx\right)^{2}$. Any
perturbation decreases the third term in equation (\ref{eq:Energy_Libre_Jet_Effet_Beta})
by $\delta\mathcal{F}_{2}=-2Ru\tilde{\beta}_{c}\int dx\,\ l^{2}$.

We suppose that $l=l_{k}\sin\frac{k\pi}{L_{x}}x$ where $k\ge1$ is
an integer. Then \[
\delta\mathcal{F}=\delta\mathcal{F}_{1}+\delta\mathcal{F}_{2}=-2u\tilde{\beta}_{c}+e\left(\frac{k\pi}{L_{x}}\right)^{2}.\]
 Because we minimize $\mathcal{F}$, we want to know if any perturbation
leads to positive variations of the free energy. The most unfavorable
case is for the smallest value of $k^{2}$, i.e. $k^{2}=1.$ Then
we conclude that eastward jets are local entropy maxima when \[
\widetilde{\beta}_{c}<\widetilde{\beta}_{c,cr}=\frac{1}{2}\frac{e}{u}\frac{\pi^{2}}{L_{x}^{2}}\ .\]
We thus conclude that eastward zonal jets are local equilibria for
sufficiently small values of $\widetilde{\beta}_{c}$.\\

The previous result can also be interpreted in terms of the domain
geometry, for a fixed value of $\widetilde{\beta}_{c}$. Eastward
jets are local entropy maxima if
\[ L_{x}<L_{x,cr}=\pi\sqrt{\frac{e}{2u\widetilde{\beta}_{c,cr}}}.\]
 Let us evaluate an order of magnitude for $L_{x,cr}$ for the ocean
case, first assuming there is no deep flow ($U_{d}=0$). Then $R\tilde{\beta}_{c}$ is the real coefficient of the beta plane approximation. Remembering
that a typical velocity of the jet is $U\sim uR$, and using
$e\sim u^{2}$  (see \cite{VenailleBouchetJPO} for more details).
Then $L_{x,cr}\approx\pi\sqrt{\frac{U}{\beta_{c,cr}}}$. This length
is proportional to the Rhine's' scale of geophysical fluid dynamics
\cite{VallisBook}. For jets like the Gulf Stream, typical jet velocity
is $1\ m.s^{-1}$ and $\beta_{c}\sim10^{-11}\ m^{-1}.s^{-1}$ at mid-latitude.
Then $L_{cr}\sim300\ km$. This length is much smaller than the typical
zonal extension of the inertial part of the Kuroshio or Gulf Stream
currents. We thus conclude that in a model with a quiescent lower
layer and the beta plane approximation, currents like the Gulf Stream
or the Kuroshio are not statistical equilibria, and they are not neither
close to local statistical equilibria. \\

Taking the oceanic parameters ($\beta_{c}=\ 10^{-11}\ m^{-1}s^{-1}$,
$R\sim50\ km$), we can estimate the critical eastward velocity in
the lower layer $U_{d,cr}=5\ cm\ s^{-1}$ above which the strong eastward
jet in the upper layer is a statistical equilibria. It is difficult
to make further conclusions about real mid-latitude jets; we conjecture
that their are marginally stable. This hypothesis of marginal stability
is in agreement with the observed instabilities of the Gulf-Stream
and Kuroshio current, but overall stability of the global structure
of the flow. A further discussion of these points will be the object
of future works.\\

In all of the preceding considerations, we have assumed that the term
$R\widetilde{\beta}_{c}$ was of order $R$ in dimensionless units.
This is self-consistent to compute the unstable states. To show that
a solution is effectively a statistical equilibria when $R\widetilde{\beta}_{c}$
in of order one, one has to use much less straightforward considerations
than in the preceding paragraphs, but the conclusions would be exactly
the same. \\

Notice that the description of an inertial solution presenting an
eastward jet in a closed domain constitutes in itself an important
step toward theoretical studies of oceanic mid-latitude jets, beside
the application to statistical mechanics. It can be for instance the
starting point of stability studies, by applying classical methods
to describe the evolution of perturbations around this mean state.

\subsection{Conclusion}

We have shown that when there is a sufficiently strong eastward flow
in the deep layer (i.e. when $U_{d}>U_{d,cr}$ with $U_{d,cr}=R^{2}\beta_{c,cr}$),
ocean mid-latitude eastward jets are statistical equilibria, even in
presence of a beta plane. When the flow in the deep layer is lower
than the critical value $U_{d,cr}$ but still almost compensate the
beta plane ($0<\beta_{c}-\frac{U_{d}}{R^{2}}<\frac{1}{2}\frac{e}{u}\frac{\pi^{2}}{L_{x}^{2}}R$),
the solutions with the eastward jets are local equilibria (metastable
states). When $\beta_{c}-\frac{U_{d}}{R^{2}}>\frac{1}{2}\frac{e}{u}\frac{\pi^{2}}{L_{x}^{2}}R$
the solution with an eastward jet are unstable.

We have also concluded that the inertial part of the real Gulf-Stream
or of the Kuroshio extension are likely to be marginally stable from
a statistical mechanics point of view.\\

The statistical equilibria that we have described in this section have a flow structure that differs notably from the celebrated Fofonoff solution \cite{Fofonoff:1954_steady_flow_frictionless}.

\begin{figure}[t!]

\begin{centering}
\includegraphics[width=\textwidth]{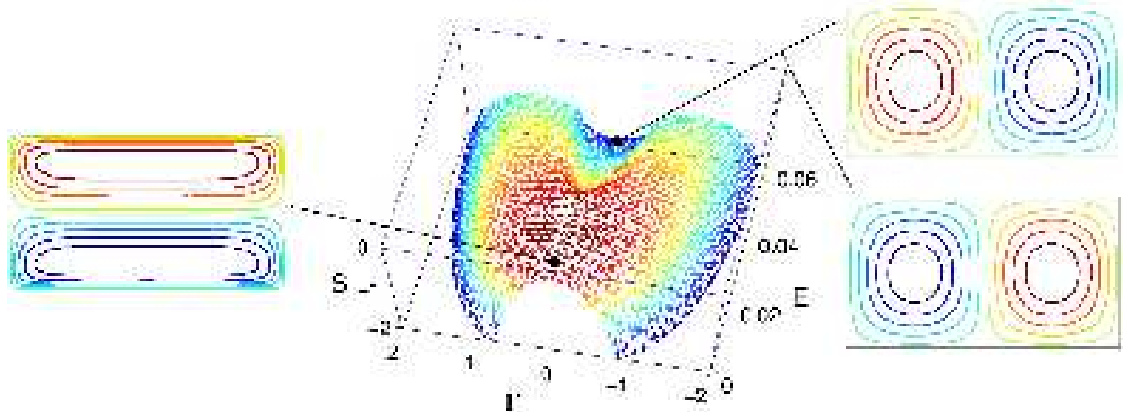}
\par\end{centering}

\caption{{\footnotesize  Phase diagrams of RSM statistical equilibrium states of the 1.5 layer quasi-geostrophic model,  characterized by a linear $q-\psi$ relationship, in a rectangular domain elongated in the $x$ direction.  $S(E,\Gamma)$ is the equilibrium entropy, $E$ is the energy and $\Gamma$ the circulation. Low energy states are the celebrated Fofonoff solutions  \cite{Fofonoff:1954_steady_flow_frictionless}, presenting a weak westward flow in the domain bulk.  High energy states have a very different structure (a dipole). Please note that at high energy the entropy is non-concave. This is related to ensemble inequivalence (see \ref{sub:Long-range-interactions} page \pageref{sub:Long-range-interactions}), which explain why such states were not  computed in previous studies. The method to compute explicitly this phase diagram is the same as the one presented in subsection \ref{sub:The-example-doubly-periodic} page \pageref{sub:The-example-doubly-periodic}.  See \cite{Venaille_Bouchet_PRL_2009} for more details.}}
\label{fig:Fofonoff}
\end{figure}

The Fofonoff solution is a stationary state of the quasi-geostrophic equations (\ref{QG}-\ref{dir}-\ref{u}) on a beta plane ($\eta_d = \beta_c y$) obtained by assuming a linear relationship between potential vorticity and streamfunction ($q=a \psi$), in the limit $a+R^{-2} \gg L^{-2}$, where $L$ is the domain size. In this limit, the Laplacian term in  (\ref{dir}) is negligible in the domain bulk. Then  $\psi \approx \beta_c / (a+R^{-2}) y$, which corresponds to a weak westward flow, as illustrated figure \ref{fig:Fofonoff}. Strong recirculating eastward jets occur at northern and southern boundaries, where the Laplacian term is no more negligible.

The original work of Fofonoff was carried independently from statistical mechanics considerations. The linear $q-\psi$ relationship was chosen as a starting point to compute analytically the flow structure. Because both the Salmon--Holloway--Hendershott statistical theory \cite{SalmonHollowayHendershott:1976_JFM_stat_mech_QG} (which is the extension of the Kraichnan energy-enstrophy theory in presence of topography) and the Bretherton--Haidvoguel minimum enstrophy principle \cite{BrethertonHaidvogel}  did predict a linear relationship between vorticity and streamfunction, it has been argued that statistical equilibrium theory predicts the emergence of the classical Fofonoff flows, which had effectively been reported in numerical simulations of freely decaying barotropic flows on a beta plane for some range of parameters \cite{WangVallis}.

We have seen in the last paragraph of subsection  \ref{sub:3D-Stat-mech-and-Energy-Enstrophy}  page \pageref{sub:3D-Stat-mech-and-Energy-Enstrophy}  and at the end of subsection  \ref{sub:Canonical-and-Grand-Canonical} page   \pageref{sub:Canonical-and-Grand-Canonical} that all those theories are particular cases of the RSM statistical mechanics theory.
On the one hand it has been actually proven that the classical Fofonoff solutions are  indeed RSM statistical equilibria in the limit of low energies \cite{Venaille_Bouchet_PRL_2009}. On the other hand, as illustrated by the results of this section, there exists a much richer variety of RSM equilibrium states than the sole classical Fofonoff solution. Even in the case of a linear $q-\psi$ relation, high energy statistical equilibrium states are characterized by a flow structure that differs notably from the original Fofonoff solution , as illustrated figure \ref{fig:Fofonoff}. These high energy states correpond  actually to the RSM equilibrium states of the Euler equation, originally computed by \cite{ChavanisSommeria:1996_JFM_Classification}. The transition from classical Fofonoff solutions to those high energy states  has been related the the occurrence of ensemble inequivalence \cite{Venaille_Bouchet_PRL_2009}. This explains also why such high energy states have not been reported in earlier studies, where computations were always performed in the (unconstrained) canonical ensemble (see the discussion at the end of subsection \ref{sub:3D-Stat-mech-and-Energy-Enstrophy}  page \pageref{sub:3D-Stat-mech-and-Energy-Enstrophy}).\\

The early work of Fofonoff \cite{Fofonoff:1954_steady_flow_frictionless} and  the equilibrium statistical mechanics of geophysical flows presented in this review are often referred to as the inertial approach of oceanic circulation, meaning that the effect of the forcing and the dissipation are neglected.

Ocean dynamics is actually much influenced by the forcing and the dissipation.
For instance the mass flux of a current like the Gulf Stream is mainly
explained by the Sverdrup transport. Indeed in the bulk of the ocean,
a balance between wind stress forcing and beta effect (the Sverdrup
balance) lead to a meridional global mass flux (for instance toward
the south on the southern part of the Atlantic ocean. This fluxes
is then oriented westward and explain a large part of the Gulf Stream
mass transport. This mechanism is at the base of the classical theories
for ocean dynamics \cite{Pedlosky:1998_OceanCirculationTheory}. Because
it is not an conservative process, the inertial approach does not
take this essential aspect into account. Conversely, the traditional
theory explains the Sverdrup transport, the westward intensification
and boundary current, but gives no clear explanation of the structure
of the inertial part of the current: the strongly eastward jets.

Each of the classical ocean theory \cite{Pedlosky:1998_OceanCirculationTheory}
or of the equilibrium statistical mechanics point of view give an
incomplete picture, and complement each other. Another interesting
approach consider the dynamics from the point of view of bifurcation
theory when the Reynolds number (or some other controlled parameters)
are increased. These three types of approaches seem complimentary
and we hope they may be combined in the future in a more comprehensive
non-equilibrium theory.

\newpage

\section{Non-equilibrium statistical mechanics of two-dimensional and geophysical
flows \label{sec:Out of equilibrium}}

In the previous chapters, we dealt with equilibrium statistical mechanics
for two-dimensional and geophysical flows. Assuming ergodicity, equilibrium
statistical mechanics describes long time outcome of the evolution
of the 2D Euler equations or the quasi-geostrophic  equations. Ergodicity
was then our only assumption, and all the presented results can be
derived rigorously.

In laboratory experiments or geophysical situations, most flows are
however subjected to dissipative processes. Very often such flows
are in statistically steady states, where forcing balance dissipation
on average, and where fluxes of energy and other conserved quantities
characterize the system. This is a situation of Non Equilibrium Steady
States (NESS), following the terminology of statistical mechanics.

In many situations of interest the action of forces and dissipation
mechanisms are weak compared to the inertial (Hamiltonian) part of
the dynamics. For instance, the turnover time scale can be small compared
to forcing time scale (i.e. a typical time needed to create the structure
starting from rest) or to a dissipation time scale (i.e. a typical
time needed to dissipate the structure if the force would be switched
off). In such situations of weak forces and dissipation, at leading
order one recovers the inertial dynamics: the Euler equations or the
quasi-geostrophic dynamics. Then a natural question is to know whether
we are close or not to some statistical equilibria, and if statistical
equilibrium could learn us something for these non-equilibrium situations.

A further objective is to make an non-equilibrium theory that could
predict the invariant measure and to predict the properties of this
NESS directly from the dynamics, for instance using a kinetic theory
approach. \\

In order to discuss these issues more precisely, we consider in the
following the 2D Navier-Stokes equations with viscous dissipation
$\nu$, linear friction $\alpha$ and stochastic forces $\eta$:\begin{equation}
\partial_{t}\omega+(\mathbf{v}\cdot\nabla)\omega=\nu\Delta\omega-\alpha\omega+\sqrt{\sigma}\eta(t,{\bf x})\,;\label{eq:2D-Stochastic-NavierStokes}\end{equation}
where $\sigma$ is the average energy injection rate by the stochastic
force $\eta$ ($\eta$ will be defined precisely latter on; $\eta$
is actually the curl of a force, but without ambiguity we call it
a stochastic force in the following). We recall that $\omega=\Delta\psi$
is the vorticity, and $\mathbf{v}=\mathbf{e}_{z}\times\nabla\psi$
the two-dimensional velocity field.

This is the most simple model for discussing the statistics of the
large scales of 2D and geophysical flows, in a statistically steady
regime. The type of reasoning presented in the following can be easily
generalized to other models.\\

In the case $\alpha=0$, many interesting mathematical results have
been recently obtained for the stochastic Navier-Stokes equations
(\ref{eq:2D-Stochastic-NavierStokes}): the existence of an invariant
measure, its properties in the Euler limit $\nu\rightarrow0$, the
validity of the law of large numbers, central limit theorems, ergodicity
(see \cite{Kuksin_2004_JStatPhys_EulerianLimit,Kuksin_Penrose_2005_JPhysStat_BalanceRelations,Kuksin_Shirikyan_2000_CMaPh,Bricmont_Kupianen_2001_Comm_Math_Phys_Ergodicity2DNavierStokes,Weinam_Mattingly_2001_Comm_Pure_Appl_Math_Ergodicity_NS,Mattingly_Sinai_1999math_3042M}
and references therein). We do not describe these results, but only
cite them when they are related to the more physical studies bellow.

This chapter is organized as follows. In section \ref{sub:NESS Two regimes}
we explain that two different regimes exist for the NESS of the 2D
Navier-Stokes equations, depending on the values of the forcing parameter
$\sigma$ and the linear friction parameter $\alpha$. The first one
is the classical regime of the self similar direct cascade of enstrophy
and inverse cascade of energy, first predicted by Kraichnan and studied
thoroughly during the last three decades. The second one is the regime
dominated by the largest scales of the flow. This turbulent large
scale regime is the interesting one as soon as one is interested in
predicting the statistics of the largest scales of geophysical flows.
We explain that it is natural to guess that this regime has some relations
with equilibrium statistical mechanics of the Euler equations, even
if the microcanonical measure does not describe its statistics.

In section \ref{sub:Phase Transition-1}, we explain that we can predict
many properties of this turbulent large scale regime from the equilibrium
statistical mechanics, for instance the topology of the average velocity
field. For instance, we show that we can predict non-equilibrium phase
transitions: situations of bistability between two different topologies
of the velocity fields. We also explain the strong limitations to
the use of equilibrium theory for such non-equilibrium situations.

In section \ref{sub:Kinetic_Theory}, we explain how a kinetic theory
could be developed to describe the turbulent large scales of turbulent
flows in a non-equilibrium steady state. We explain what would be
the minimal requirements for such a theory, and the associated difficulties.
In section \ref{sub:Relaxation-towards-equilibrium} we describe recent
progresses in this direction.

\subsection{Non-Equilibrium Steady States (NESS) for forced and dissipated turbulence
\label{sub:NESS Two regimes}}

In this subsection, we show that depending on the values of the friction
parameter $\alpha$ and the forcing parameter $\sigma$, there are
two main regimes for the stochastic Navier-Stokes equations (\ref{eq:2D-Stochastic-NavierStokes}).
We begin by some general considerations on the balance of energy and
other conserved quantities.

\subsubsection{Stochastic forces }

We first define the stochastic force $\eta(t,{\bf x})$. It is a sum
of random noises:\[
\eta(t,{\bf x})=\sum_{k}f_{k}{\bf e}_{k}\left(x,y\right)\eta_{k}(t),\]
 where $\left\{ e_{k}\right\} $ is the orthonormal basis of the Laplacian
eigenvectors with Dirichlet boundary conditions for the domain $\mathcal{D}$:
$-\Delta{\bf e}_{k}=\lambda_{k}{\bf e}_{k}$ with $\int_{\mathcal{D}}\mathrm{d}\mathbf{r}{\bf e}_{k}{\bf e}_{k'}=\delta_{kk'}$.
For a doubly periodic domain of size $\left(L_{x},L_{y}\right)$ we
have $\mathbf{k}=2\pi\left(n_{x}/L_{x},n_{y}/L_{y}\right)$ with integers
$n_{x}$ and $n_{y}$, $e_{\mathbf{k}}\left(\mathbf{r}\right)=\exp\left(i\mathbf{k}.\mathbf{r}\right)/L_{x}L_{y}$
and $\lambda_{k}=\mathbf{\left|k\right|}^{2}$. The terms $\eta_{k}$
are independent white noises $\left\langle \eta_{k}\left(t\right)\eta_{k'}\left(t\right)\right\rangle =\delta_{kk'}\delta\left(t-t'\right)$,
$f_{k}$ is the force spectrum and $\sigma$ is the force amplitude
that will be related to the energy and enstrophy injection rate later
on. In all the following we assume that $f_{k}$ decays rapidly for
large $k$: the stochastic force is white in time and smooth in space.

We rewrite (\ref{eq:2D-Stochastic-NavierStokes}) in the usual stochastic
form: \begin{equation}
d\omega=\left[-(\mathbf{u}\cdot\nabla)\omega+\nu\Delta\omega-\alpha\omega\right]dt+\sqrt{\sigma}\sum_{k}f_{k}{\bf e}_{k}dW_{k},\label{SQG}\end{equation}
 where $dW_{k}$ are the Wiener processes associated with $\eta_{k}$.\\

\subsubsection{Energy balance}

For the deterministic dynamics (i.e. without stochastic forcing, $\sigma=0$),
the energy balance reads \[
\frac{dE}{dt}=-\nu Z-2\alpha E,\]
 where $E=\frac{1}{2}\int_{\mathcal{D}}\mathrm{d}\mathbf{r}\ \mathbf{v}^{2}$
is the energy (\ref{eq:Energy}) and $Z=\int_{\mathcal{D}}\mathrm{d}\mathbf{r}\ \omega^{2}$
is the enstrophy (\ref{eq:Enstrophy}). For the stochastic dynamics,
application of the Ito formula to the energy evolution, starting from
(\ref{SQG}), leads to \begin{equation}
\frac{d\left\langle E\right\rangle }{dt}=-\nu\left\langle Z\right\rangle -2\alpha\left\langle E_{c}\right\rangle +\sigma,\label{eq:Energy_Balance}\end{equation}
 where the brackets are averages over the white noises realizations
and where we assume \begin{equation}
B_{0}\equiv\frac{1}{2}\sum_{k}\frac{|f_{k}|^{2}}{\lambda_{k}}=1.\label{eq:Bo}\end{equation}
 We see from (\ref{eq:Energy_Balance}) that $\sigma$ is the average
energy injection rate; the assumption $B_{0}=1$ is just equivalent
to defining $\sigma$. The energy $E$ is on units of $\mbox{m}{}^{4}\,\mbox{s}^{-2}$,
so the units for $\sigma$ are $\mbox{m}^{4}\,\mbox{s}^{-3}$ and
the units for $f_{k}$ are $m^{-1}$.

Similar balance relations can be easily derived for all the other
conserved quantities. We do not discuss these relations here as we
will not need them in the following. More details can be found in
\cite{Kuksin_Penrose_2005_JPhysStat_BalanceRelations} and \cite{Morita_Simonnet_Bouchet_2010_NS_Stochastic_Long}.
\\

\subsection{First regime: the Kraichnan self-similar cascades\label{sub:Kraichnan}}

We first consider the case where the force spectrum is peaked around
a given wave number $k_{f}$. Following Kolmogorov type of reasoning,
if $\alpha$ and $\nu$ are small enough, it is then possible to define
inertial ranges in which the effects of forcing and dissipation will
be negligible. Such inertial ranges will then be characterized by
fluxes of conserved quantities, for instance energy and enstrophy.
When the fluxes are dominated by dynamical processes which are local
in Fourier space, this is called a cascade regime. Such cascades,
in two-dimensional turbulence, have first been studied by Kraichnan
using ideas similar to Kolmogorov ideas for 3D turbulence. We give
here a very rapid account, mainly based on dimensional arguments.
More details can be found in \cite{Kraichnan_Motgommery_1980_Reports_Progress_Physics}
and in more recent reviews or lectures. Interesting recent results in this domain
include study of statistics of zero vorticity lines in relation with
the stochastic Loewner equation (SLE) \cite{Bernard_Boffetta_Celani_Falkovich_2006Nature}
and precise conditions for locality of turbulent cascades with applications
also to two-dimensional turbulence \textbf{\cite{Eyink_Locality_2005PhyD..207...91E,Eyink_Aluie_2009PhFl...21k5107E,Eyink_Aluie_II_2009PhFl...21k5108A}}.\\

The system is forced at wave number $k_{f}$ with an energy injection
rate per unit surface $\varepsilon_{f}$. Following the notations
of the previous subsection, we have $\varepsilon=\sigma/L^{2}$.
This corresponds to an enstrophy production rate $\eta\sim k_{f}^{^{2}}\varepsilon$.
We suppose that the system has reached a statistically steady state
and that the energy is limited to scales $k^{-1}$ much smaller that
the domain size $L$. A more precise statement of this hypothesis,
and a condition for its validity will be given at the end of this
section. In the following of this section, we assume cascade regimes
(for inertial scales $k\gg k_{f}$ and $k\ll k_{f}$, we assume constant
fluxes of conserved quantities due to local dynamical processes in
Fourier space).

In the limit where the large scale separation $Lk_{f}\gg1$, it is
then relevant to write the total energy per unit surface $E_{s}=E/L^{2}$
on the form of a continuous spectrum: $E_{s}=\int E(k)dk$ (see section
\ref{sub:Fjortof_Argument} for the definition of $E\left(k\right)$).
We consider the energy fluxes at scale $k$ (the energy going from
modes with wave numbers larger than $k$ to wave numbers smaller than
$k$, for a precise definition see \cite{Frisch_1996_Book}). One
could imagine a situation with both an upscale energy flux ($k<k_{f}$)
and a downscale one ($k>k_{f}$), both of the order of $\varepsilon$.
However, at small scale (large $k$), this would imply an enstrophy
flux of order $\varepsilon k^{2}$. But because $\varepsilon k^{2}\gg\epsilon k_{f}^{^{2}}\sim\eta$,
this would contradict the hypothesis of a steady regime, since there
would be much more enstrophy going downscale than the enstrophy injected.
We conclude that the energy injected at $k_{f}$ goes mostly toward
large scales, at a fixed rate $\varepsilon$. Using a similar argument
where the energy and enstrophy play a reverse role, we conclude that
in the limit or large scale separation $Lk_{f}\gg1$, enstrophy injected
at scale $k_{f}$ flows mostly toward small scales at a rate $\eta$.

We thus conclude that for statistically steady states, energy flows
mainly upscale and enstrophy mainly downscale; more precisely in
the limit of infinite inertial ranges all the energy flows upscale
and all enstrophy downscale. In section \ref{sub:Fjortof_Argument},
using Fjortoft argument, we obtained a qualitatively similar result
for the dynamics of decaying turbulence. In sections \ref{sub:3D-Stat-mech-and-Energy-Enstrophy}
and \ref{sub:Mean_Field} we explained that for the Euler equation,
statistical mechanics predicts that all energy is concentrated in
the largest scale and the all the enstrophy or other invariant excess
(a precise meaning being given by the entropy) flows towards smaller
and smaller scales. We see that these three precise results, for three
different situations, give a precise meaning to the statement that
energy goes towards large scales and enstrophy towards small scales
in two dimensional turbulence.\\

By using a dimensional analysis, Kraichnan gave a prediction for the
slope of the energy spectrum $E_{s}(k)$ in the inertial ranges, in
the cascade regimes. Let us first consider the inertial range for
scales above the injection scale ($k<k_{f}$). This is the inverse
energy cascade, as energy goes upscale in this region, with a given
flux $\varepsilon$ (with unit $m^{2}.s{}^{-3}$), and enstrophy flux
is negligible. Because of the locality hypothesis, at scale $k^{-1}$
(unit $L$), the energy spectrum $E_{s}(k)$ (unit $m^{3}.s^{-2}$)
can then depend only on $\epsilon$ and $k$. Dimensional analysis
then gives

\[
E_{s}(k)\sim\varepsilon^{2/3}k^{-5/3}\ \quad\text{for}\quad k\ll k_{f}\ .\]

Let us now consider the inertial region for scales below the injection
scale. This the enstrophy cascade inertial range as the enstrophy
goes downscale in this region, with a given flux $\eta$ (with unit
$s^{-2}$), and the energy flux is negligible. The energy spectrum
at scale $k^{-1}$ can then depend only on $k$ and $\eta$. Dimensional
analysis then gives

\[
E_{s}(k)\sim\eta^{2/3}k^{-3}\ \quad\text{for}\quad k\gg k_{f}.\]
 \\

\begin{figure}[t!]
\includegraphics[width=1\textwidth]{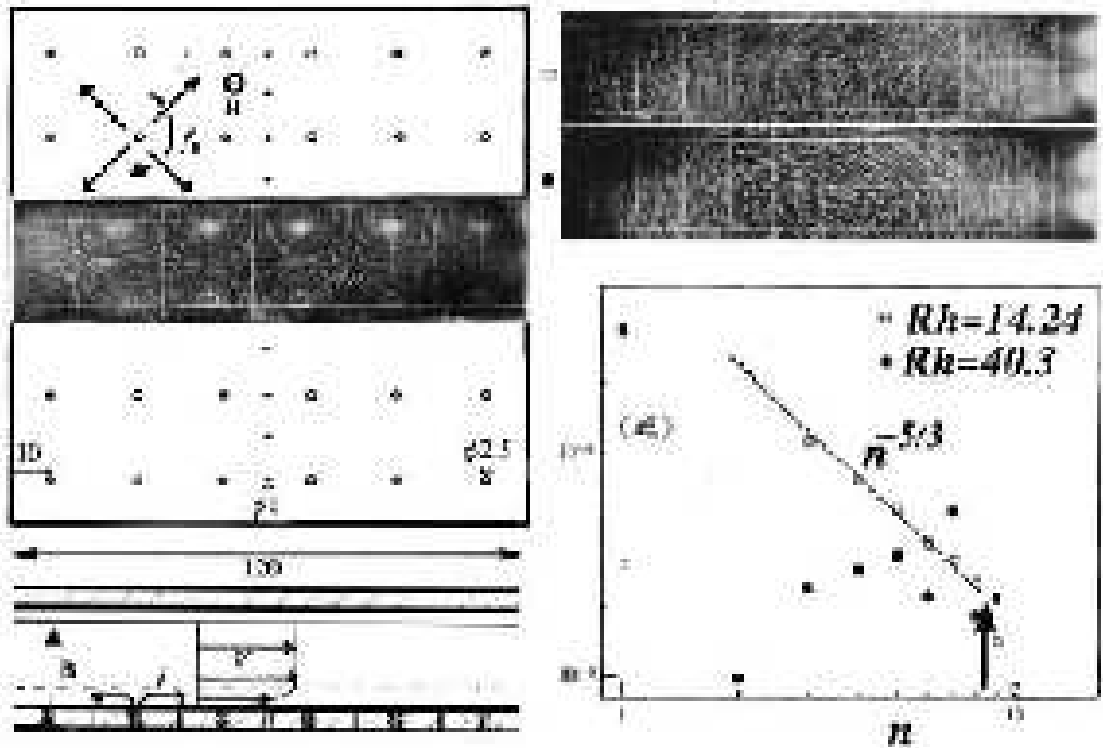}

\caption{\footnotesize First experimental observation of the inverse energy cascade
and the associated $k^{-5/3}$ spectrum, from \cite{Sommeria_1986_JFM_2Dinverscascade_MHD}.
The 2D turbulent flow is approached here by a thin layer of mercury
and a further ordering from a transverse magnetic field. The flow
is forced by an array of electrodes at the bottom, with an oscillating electric field. The parameter Rh is the ratio between inertial to bottom friction
terms. At low $Rh$ the flow has the structure of the forcing (left
panel). At sufficiently high $Rh$ the prediction of the self similar
cascade theory is well observed (right panel, bottom), and at even
higher $Rh$, the break up of the self similar theory along with the
organization of the flow into a coherent large scale flow is observed (see right
panel above).
}

\label{Fig:CascadeJoel}
\end{figure}

\begin{figure}[t!]
\begin{center}
\includegraphics[width=0.95\textwidth]{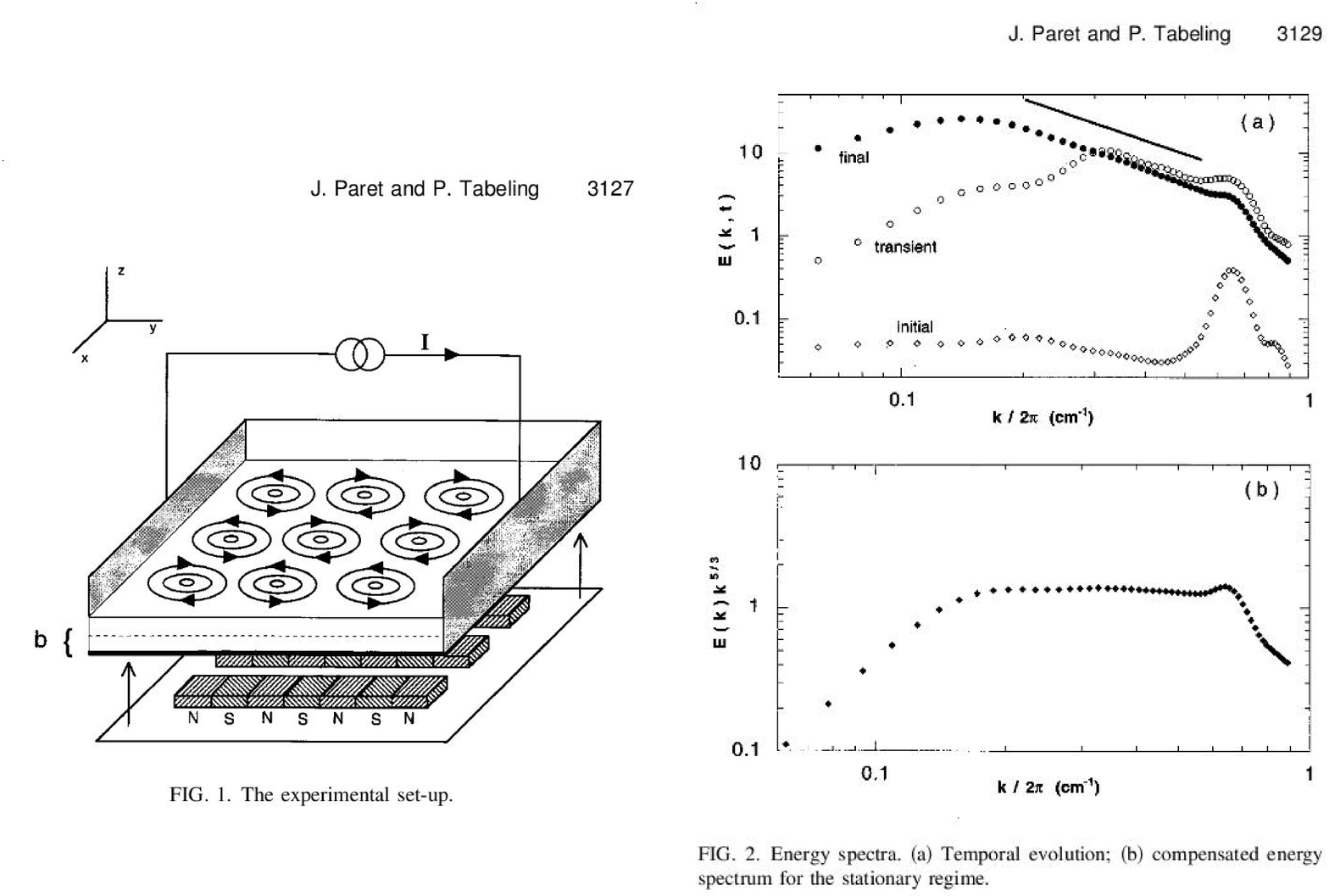}
\end{center}

\caption{\footnotesize Another experimental observation of the inverse energy cascade
and the associated $k^{-5/3}$, from \cite{Paret_Tabeling_1998_PhysFluids}.
The 2D turbulent flow is approached by a thin layer of (light) salty water lying above another thin layer of (dense) salty water. The stable stratification provides a further ordering. The flow is stirred at small scales by the interaction between an horizontal electric field imposed across the experimental cell and a vertical magnetic field imposed by an array of magnets located below the experimental cell. We see on the right panel that predictions of self-similar inverse energy cascade (the $k^{-5/3}$ spectrum) are well observed. Please note also the interesting transient evolution of the energy spectrum.}

\label{Fig:CascadeTabeling}
\end{figure}

Predictions of the self-similar cascade theory have been followed
by numerous experimental studies in many different settings over the
last three decades (see \cite{Bernard_Boffetta_Celani_Falkovich_2006Nature,Bernard_Boffetta_Celani_Falkovich_2007PRL}
and references therein), and this has led to beautiful experimental
results \cite{Sommeria_2001_CoursLesHouches,Paret_Tabeling_1998_PhysFluids}.
Among others, there have been measurements of the $-5/3$ slope of
the backward energy cascade part of the spectrum \cite{Sommeria_1986_JFM_2Dinverscascade_MHD,Paret_Tabeling_1998_PhysFluids}
(see also figures \ref{Fig:CascadeJoel} and \ref{Fig:CascadeTabeling}), as well as of the $-3$
slope of the forward enstrophy cascade part of the spectrum \cite{ParetJullienTabeling99}.
\\

In the previous paragraph, we dealt with the inertial range energy
spectrum. For large enough scales, the friction $-\alpha\omega$ is
no more be negligible and the inertial range hypothesis no more valid.
The energy is then be dissipated and no more flow towards larger and
larger scales. In experiments, this is visible as a maximum in the
energy spectrum $E_{s}(k)$. Using a classical argument based on dimensional
analysis  \cite{Lilly_1972GApFD...3..289L} (see also \cite{Danilov_Gurarie_2001PhRvE..63b0203D} for a critical discussion, or \cite{VallisBook}), we predict the scale $L_{I}$ at which the cascade stops. This scale can depend only on $\varepsilon$ ($m^{2}.s^{-3}$)
and on the friction $\alpha$ ($s^{-1}$): \begin{equation}
L_{I}=c\frac{\varepsilon^{1/2}}{\alpha^{3/2}},\label{eq:LI}\end{equation}
 where $c$ is a non-dimensional constant.

If the dynamics takes place in a finite box of size $L$, the energy
flux towards largest scales can be stopped by the box before it is
dissipated by the friction. In such a case the energy pile up at the
largest scale. The self-similar hypothesis for the spectrum then break
down, and we actually observe that energy fluxes are no more local.
The Kraichnan picture then break down. In a finite box, the Kraichnan
picture can be valid only if the scale $L_{I}$ is much smaller than
the box size $L$. Recalling that $\sigma=\varepsilon L^{2}$, we
thus conclude a necessary condition to observe a universal inverse
energy cascade is%
\footnote{The $\sqrt{2}$ is unimportant and is added for convenience in latter
computations.%
} \begin{equation}
R_{\alpha}=\sqrt{2}\frac{\sigma^{1/2}}{L^{2}\alpha^{3/2}}\ll1.\label{eq:Condition_Condensation}\end{equation}
In the opposite case, $R_{\alpha}\gg1$, for instance for too strong
energy injection or too weak dissipation for a given box size, the
energy cascade will not be arrested by friction, but will begin to
pile up at the largest scales. Then the largest scales self-organize
and create coherent vortices and jets. This is the second regime of
two dimensional turbulence, to be discussed in the next sections.

\subsection{Second regime: the coherent large scale flow}\label{sub:Second-regime:Large-Scales}

We are now interested in the regime where the flow self organize at
the largest scale $L$. Because the energy will flow towards the largest
scales; it is natural to neglect the energy dissipation by viscous
effect. We will give a more precise criterion for this to be valid
later on. From the energy balance (\ref{eq:Energy_Balance}), neglecting
viscous effect, we have \[
\left\langle E_{c}\right\rangle \simeq\frac{\sigma}{2\alpha}.\]
 A typical velocity for the large scale flow is thus $U=\sqrt{\left\langle E_{c}\right\rangle }/L=\frac{1}{L}\sqrt{\frac{\sigma}{2\alpha}}$.
A typical time scale for the largest scales (a turnover time) is then
$\tau=L/U=L^{2}\sqrt{\frac{2\alpha}{\sigma}}$. A natural non-dimensional
parameter is $R_{\alpha}$ the ratio of the dissipation time scale
$1/\alpha$ over the turnover time scale $\tau$ \[
R_{\alpha}=\sqrt{2}\frac{\sigma^{1/2}}{\alpha^{3/2}L^{2}}.\]
 $R_{\alpha}$ is indeed a Reynolds number based on the linear friction
$\alpha$ and the large scale flow velocities and length scale. We
note that the criteria for observing self-organization of the energy
at the largest scale (\ref{eq:Condition_Condensation}) is $R_{\alpha}>1$,
as could have been expected. The limit of large time scale separation,
$R_{\alpha}\gg1$, is particularly interesting.\\

It is natural to write non-dimensional dynamical equations using
as a length unit the domain size $L$ and as a time unit a typical
large scale turnover time $\tau=L^{2}\sqrt{\frac{2\alpha}{\sigma}}$:
$t=\tau t'$ and $(x,y)=L(x',y')$. In these non-dimensional units
the dynamical equations are \[
\begin{array}{l}
\partial_{t'}\omega'+(\mathbf{u}'\cdot\nabla')\omega'=\frac{1}{Re}\Delta'\omega'-\frac{1}{R_{\alpha}}\omega'+\sqrt{\frac{2}{R_{\alpha}}}\eta',\\
\mbox{with\ensuremath{\,\,\,}}\omega'=\Delta'\psi',\end{array}\]
 where \[
Re=UL/\nu=\sigma^{1/2}/\left(2\alpha\right)^{1/2}\nu\]
 is the Reynolds number based on the large scale velocity and domain
size. \\

We rewrite the non-dimensional equations dropping the primes and identifying
$\alpha$ to $1/R_{\alpha}$, $A$ to $a$ and $\nu$ to $1/Re$.
We then obtain \begin{equation}
\begin{array}{l}
\partial_{t}\omega+(\mathbf{u}\cdot\nabla)\omega=v\Delta\omega-\alpha\omega+\sqrt{2\alpha}\eta,\\
\mbox{with}\,\,\,\omega=\Delta\psi\end{array}\label{eq:SNavier-Stokes-Adimensionalized}\end{equation}
 and with\begin{equation}
\eta(t,{\bf x})=\sum_{k}f_{k}{\bf e}_{k}\left(x,y\right)\eta_{k}(t),\label{eq:Stochastic_Forces}\end{equation}
 where $\left\{ f_{k}\right\} $ verifies the constraint (\ref{eq:Bo}).

We are interested in the limit where the viscous Reynolds number $Re$
is much larger than the Reynolds number based on the linear friction
$R_{\alpha}$ (meaning that, as far as the large scales are concerned,
viscous dissipation is negligible compared to linear friction dissipation).
In these non-dimensional units, this condition reads $\nu\ll\alpha$.
The regime of large scale self organization is $\alpha\ll1$. We will
thus study the limit $\nu\ll\alpha\ll1$. We call this the limit of
weak forces and dissipation.\\

In the non-dimensional units, the energy balance (\ref{eq:Energy_Balance})
is \begin{equation}
\frac{d\left\langle E\right\rangle }{dt}=-2\nu\left\langle Z\right\rangle +2\alpha\left(1-\left\langle E\right\rangle \right),\label{eq:Energy_Balance_Adimensionalized}\end{equation}
 giving the stationary balance \begin{equation}
\left\langle E\right\rangle _{S}=1-\frac{\nu}{\alpha}\left\langle Z\right\rangle {}_{S}\leq1.\label{eq:Energy_Balance_Stationnary}\end{equation}
 We study the dynamics of the coherent large scale flow regime in
the next sections.

\subsection{Equilibrium statistical mechanics and NESS; prediction of non-equilibrium
phase transitions \label{sub:Phase Transition-1}}

\subsubsection{Are the largest scales of the 2D Navier-Stokes equations close to
statistical equilibrium, in the limit of weak forces and dissipation?}

We have stressed in the introduction that the Non Equilibrium Steady
States (NESS) of the two-dimensional equation break detailed balance,
and are the place of fluxes of conserved quantities. The microcanonical
measure we built from the Liouville theorem in section \ref{sub:microcanonical-measure}
and the equilibrium states we have studied in sections \ref{sec:Equilibrium-statistical-mechanics},
\ref{sec:First Order GRS and rings} and \ref{sec:Gulf Stream and Kuroshio}
have no such fluxes. Then the microcanonical measure can not describe
the details of the statistics of these NESS. There is however the
possibility that the stationary measure be close to the microcanonical
equilibrium one in the limit of weak forces and dissipation. We discuss
this subtle issue now.\\

In the limit of weak forces and dissipation, $\nu\ll\alpha\ll1$,
the non-Hamiltonian terms in the stochastic Navier-Stokes equation
(\ref{eq:SNavier-Stokes-Adimensionalized}) are vanishingly small.
The associated fluxes are also vanishingly small. Because of these
small parameters, it is then natural to assume that the flows will
be concentrated near to statistical equilibria. As the equilibrium
statistical mechanics predicts that the flow is concentrated close
to stationary solutions to the 2D Euler equations, a natural conjecture
is that in the limit $\nu\ll\alpha\ll1$, the stationary measure of
the stochastic Navier-Stokes equations will be also concentrated close
to ensembles of stationary solutions to the 2D Euler equations.

That such a behavior is plausible is actually supported by several
theorems. At the dynamical level, the Navier-Stokes equation is actually
well behaved in the limit $\nu\ll\alpha\ll1$: for arbitrary large
but finite times, its solutions remain close to the solutions of the
Euler equation. This is also true at a statistical level, as recently
proved by S. Kuksin for the case $\alpha=0$ \cite{Kuksin_2004_JStatPhys_EulerianLimit}:
in the limit $\nu\rightarrow0$ the invariant measure for the stochastic
Navier-Stokes equation is described by solutions to the Euler equation.
These mathematical theorems support the idea that the invariant measure
will be related to the statistics of Euler dynamics, but they do not
prove that the measure is concentrated close to ensemble of stationary solutions to the Euler equations. This remains a challenge for further
mathematical results.\\

We note that the situation is completely different in three dimensional
turbulence: as explained in section \ref{sub:3D-Stat-mech-and-Energy-Enstrophy},
equilibrium statistical mechanics of the 3D Euler equations predicts
a trivial measure with no flow. Then 3D turbulence is intrinsically
a non-equilibrium problem.\\

As we will see in the following, the conjecture that the invariant
measure is concentrated near stationary solutions to the
2D Euler equations is supported both by numerical simulations and experiments.
Even in this non-equilibrium context equilibrium statistical mechanics
is useful. This is a common situation in statistical physics. For
instance in systems with short range interactions driven non-equilibrium,
one expects local thermodynamic equilibrium to hold, when the temperature
gradient is small enough. In our case, interactions are non local,
but in the limit $\nu\ll\alpha\ll1$, we expect to be close to statistical
equilibria. We see this as a zeroth order prediction from equilibrium
statistical mechanics in a non-equilibrium situation.

This zeroth order prediction already gives us strong qualitative results
about the non-equilibrium dynamics. However the predictive range of
these arguments is very limited. For instance the energy distribution
and Casimir distributions will be determined by non Hamiltonian processes.
They can not be derived from equilibrium processes. As energy and
Casimirs are the control parameters of the equilibrium statistical
mechanics, we thus conclude that we can guess from equilibrium statistical
mechanics that we should be close to some ensemble of stationary solutions
to the 2D Euler equations, but that non-equilibrium theory is needed
to predict which ones and with which probability.

Moreover as soon as statistics of fluctuations is concerned, it is
meaningless to try to make predictions based on equilibrium statistical
mechanics, as fluctuations statistics will have to be consistent with
non-equilibrium fluxes.\\

We thus conclude that we expect the stochastic Navier-Stokes invariant
measure to be concentrated close to ensemble of statistical equilibria
(which are also stationary solutions of the 2D Euler equations). In section
\ref{sub:Close_To_Equilibrium_NavierStokes} we show that this is
confirmed by numerical simulations and experiments. In section \ref{sub:Phase Transition 2},
we show that this allows to predict non-equilibrium phase transitions.
To know which of the dynamical equilibrium states of the Euler equations
is selected by forcing and dissipation, and to predict the fluctuation
statistics, one needs an non-equilibrium theory. Possible candidates
for such theories will be discussed in section \ref{sub:Kinetic_Theory}.

\subsubsection{Non-equilibrium flows are close to statistical equilibria \label{sub:Close_To_Equilibrium_NavierStokes} }

In order to illustrate this discussion, we discuss the case of a doubly
periodic domain. Whereas this case has no experimental counterpart,
it is extremely interesting from an academic point of view. Indeed
because of the absence of boundaries, boundary layers are absent and
make the dynamical situation much more simple. Moreover, pseudo-spectral
codes allows for much more precise numerical simulations than in any
other geometries.

Numerical simulations of the 2D Navier-Stokes equations in a self-similar
transient regime, have been presented in \cite{Chertkov_Connaughton_andco_2007_PRL_EnergyCondesation},
for a square periodic box. This paper also show interesting power
laws for the vortex profiles, in this regime. In the following, we
concentrate on the statistically steady regime and discuss relation
with equilibrium statistical mechanics, and non-equilibrium phase transitions
\cite{Bouchet_Simonnet_2008,Morita_Simonnet_Bouchet_2010_NS_Stochastic_Long}.

We have described the equilibrium statistical mechanics of the 2D
Euler equations in doubly periodic domains in section \ref{sub:The-example-doubly-periodic}.
Figure \ref{fig:Equilibre}, page \pageref{fig:Equilibre}, shows
an equilibrium phase diagram. It shows that two types of flow topologies
may be expected: dipole flows or parallel flows, a crucial parameter
being the aspect ratio of the domain $\delta$.

\begin{figure}[t!]
\begin{center}
\includegraphics[width=\textwidth]{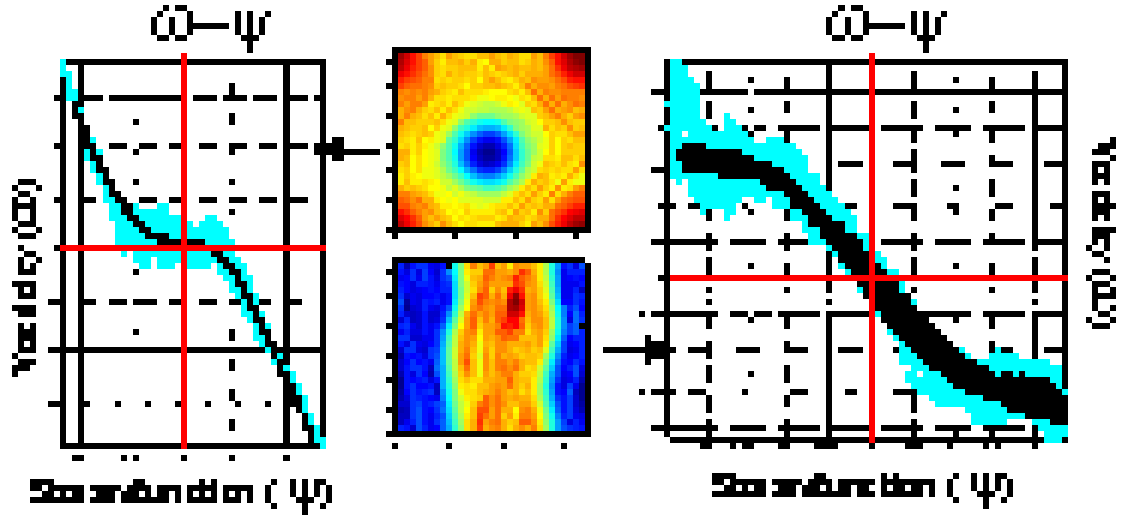}
\end{center}
\caption{\label{fig:Omega-Psi}  \footnotesize $\omega-\psi$ scatter-plots (cyan) (see color
figure on the .pdf version). In black the same after time averaging
(averaging windows $1\ll\tau\ll1/\nu$, the drift due to translational
invariance has been removed)\textbf{. }Left: dipole case with $\delta=1.03$.
Right: unidirectional case $\delta=1.10$. }

\end{figure}

Direct numerical simulations of the stochastic Navier-Stokes equations
in a square domain $\delta=1$ exhibit a statistically stationary
flow $\omega$ with a dipole structure (Fig. \ref{fig:Omega-Psi}
\textbf{a)}), whereas for $\delta\geq1.1$, nearly unidirectional
flows are observed (Fig. \ref{fig:Omega-Psi} \textbf{b)}). This result
has been confirmed both for $\alpha=0$ and $\alpha\neq0$, and for
different values of $\nu$ and force spectra. The structure of statistical
equilibria is thus observed also for non-equilibrium steady states.\\

As seen in section \ref{sec:Equilibrium-statistical-mechanics}, statistical
equilibria are characterized by a functional relationship $\omega=f\left(\psi\right)$
between vorticity and streamfunction. One observes in Fig. \ref{fig:Omega-Psi}
a $\omega-\psi$ scatter-plot (light blue or gray), for the two cases
of a dipole and a unidirectional flow. In the dipole case, the $\omega-\psi$
relationship is well observed for the larger values of $\left|\omega\right|$,
which corresponds to the core of the vortices. In the area in between
the vortices, the relationship between $\omega$ and $\psi$ is more
scattered. This correspond to the small scale filaments visible on
the vorticity picture (see figure \ref{fig:Omega-Psi}).

When we average the vorticity fields and the stream-functions over
several turnover times, we obtain the black curves which are very
nice $\omega-\psi$ relationships. The $\omega-\psi$ relationship
has the same convexity as a sinh (but is different from a $\sinh$),
in the dipole case and the same convexity as a tan-h in the unidirectional
case. We thus conclude that the observed flows are composed of average
quasi-stationary large scale structures, dipoles or unidirectional
flows, over which are superimposed fluctuations corresponding to small
scale filamentation.

In this sense, the structures are close to equilibrium. We note that
the parallel flows seems to be farther from equilibrium than the dipole
as the average relationship is a quite thick line. In this case this
is due to the presence of intermediate scale vortices as seen on the
vorticity picture.

This confirms the usefulness of the predictions of equilibrium statistical
mechanics in this non-equilibrium context. %

More details about the analysis of the proximity of the flow with
stationary solutions to the 2D Euler equation for doubly periodic conditions
are given in \cite{Bouchet_Simonnet_2008,Morita_Simonnet_Bouchet_2010_NS_Stochastic_Long}.
A similar conclusion can also be drawn from the numerical results
\cite{Chertkov_Connaughton_andco_2007_PRL_EnergyCondesation}, even
if the notion of stationary solution to the 2D Euler equation is not
used in this work.

In the limit of weak forces and dissipation, it is recognized for
a long time that the flow should be close to stationary solutions.
For instance, the quasi-geostrophic flows on a beta plane are known
to form zonal jets, which are dynamical equilibrium states of the
quasi-geostrophic equations, and this have been studied a lot recently.
In laboratory experiments, the importance of the formation of large
scale coherent structure close to stationary solutions has also been
recognized for a long time \cite{Sommeria_1986_JFM_2Dinverscascade_MHD}.

\subsubsection{Non-equilibrium phase transitions in the 2D-Stochastic Navier Stokes
equations \label{sub:Phase Transition 2}}

Phase transitions are situations where the qualitative properties
of the system change drastically. They are thus especially important
from a physical and a dynamical point of view.

We have stressed that we are not able to predict the probability of
the energy and of the vorticity moments for non-equilibrium situations.
However these are the main control parameters of the equilibrium properties.
For this reason the use of equilibrium theory for slightly non-equilibrium
situations give only qualitative results and does not provide a precise
prediction. However phase transitions give such drastic changes that
they should also be clearly identified even in non-equilibrium steady
states.

We illustrate this idea in the case of 2D Navier-Stokes equations
in the doubly periodic domain. Figure \ref{fig:Equilibre}, page \pageref{fig:Equilibre}
shows that a phase transition occur between dipoles and parallel flows
at equilibrium. A natural order parameter is $\left|z_{1}\right|$,
where $z_{1}=\frac{1}{\left(2\pi\right)^{2}}\int_{\mathcal{D}}\mathrm{d}\mathbf{r}\, \omega(x,y)\exp(iy)$.
Indeed, for unidirectional flow $\omega=ae_{1}$, $z_{1}=0$, whereas
for a dipole $\omega=a\left(e_{1}+e_{2}\right)$, $\left|z_{1}\right|=a$.

\begin{figure}[t!]
\begin{center}
\includegraphics[width=\textwidth]{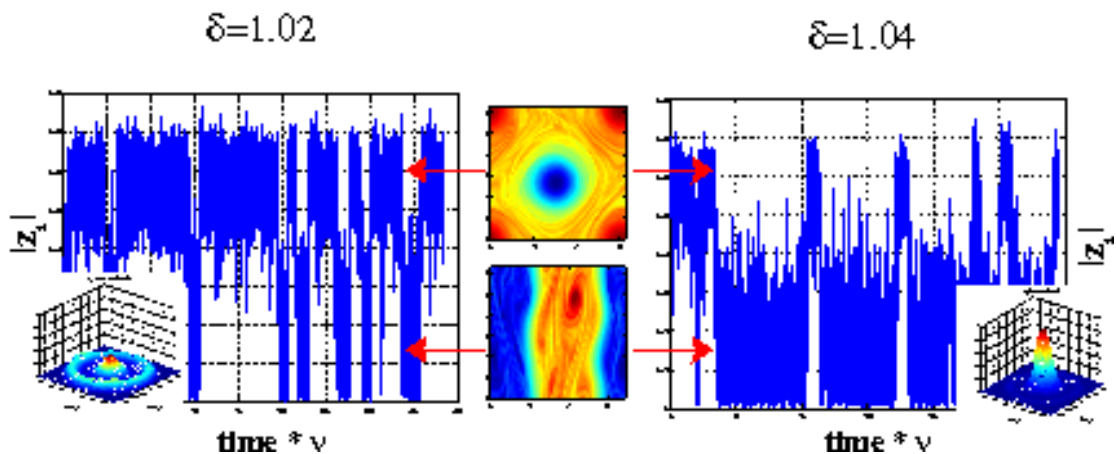}
\end{center}
\caption{ \footnotesize Dynamics of the 2D Navier--Stokes equations with stochastic forces in a doubly periodic domain of aspect ratio $\delta$, in a non-equilibrium phase transition regime. The two main plots are the time series and probability density functions (PDFs) of the modulus of the Fourier component $z_{1}=\frac{1}{\left(2\pi\right)^{2}}\int_{\mathcal{D}}\mathrm{d}\mathbf{r}\, \omega(x,y)\exp(iy)$ illustrating random changes between dipoles ($|z_1| \simeq 0.55$) and unidirectional flows ($|z_1| \simeq 0.55$). As discussed in section \ref{sub:Phase Transition 2}, the existence of such a non-equilibrium phase transition can be guessed from equilibrium phase diagrams (see figure \ref{fig:Equilibre}) \label{ts_f}}

\end{figure}

We have thus empirically looked for a non-equilibrium phase transition
(for the Navier-Stokes equations with random forces) that is the trace
of the equilibrium phase transition. A crucial control parameter is
the aspect ratio of the domain. We have made numerical simulations
for different values of $\delta$. Figure \ref{ts_f} shows $\left|z_{1}\right|$
time series for $\delta=1.02$ and $\delta=1.04$. The remarkable
observation is the bimodal behavior in this transition range. The
switches from $\left|z_{1}\right|$ values close to zero to values
of order of $0.6$ correspond to genuine transitions between unidirectional
and dipole flows. The probability distribution function (PDF) of the
complex variable $z_{1}$ (Fig. \ref{ts_f}) exhibits a circle corresponding
to the dipole state (a slow dipole random translation corresponds
into to a phase drift for $z_{1}$, explaining the circular symmetry).
The parallel flow state corresponds to the central peak. As $\delta$
increases, one observes less occurrences of the dipole. \\

\begin{figure}[t!]
\begin{centering}
\begin{tabular}{>{\centering}p{0.5\textwidth}>{\centering}p{0.5\textwidth}}
\centering{}\includegraphics[width=0.5\textwidth]{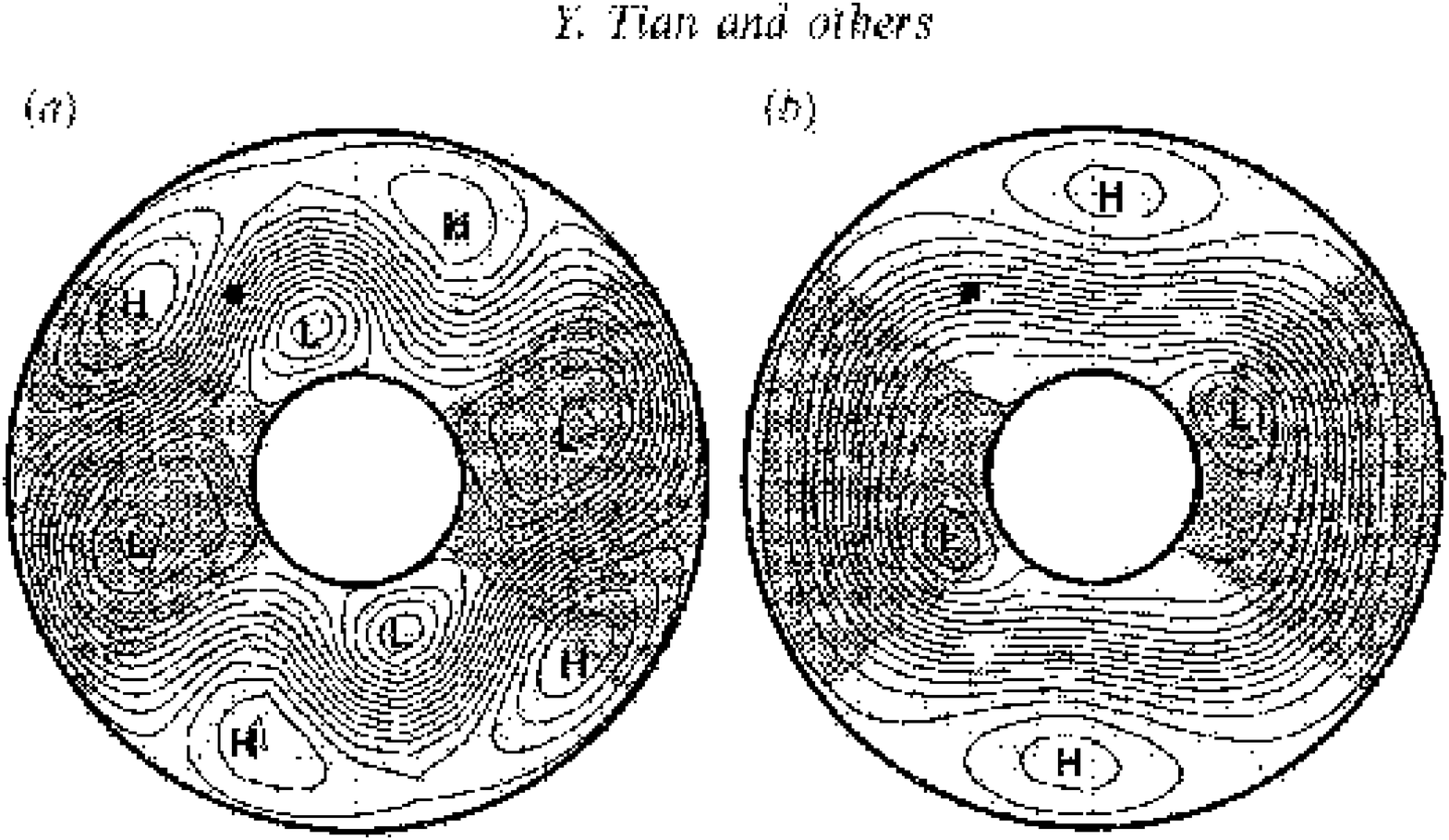}  & \centering{}\includegraphics[width=0.5\textwidth]{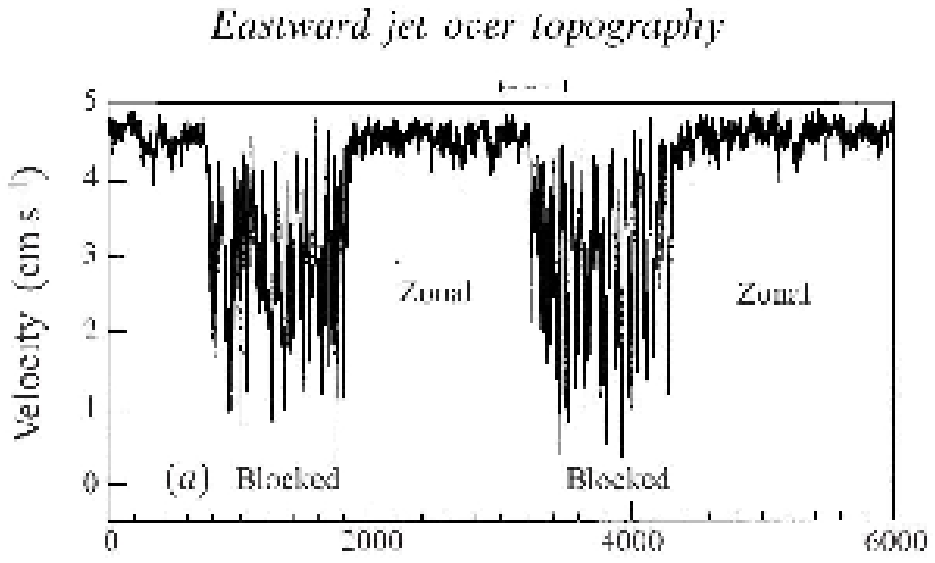} \tabularnewline
\end{tabular}
\par\end{centering}

\caption{  \footnotesize Bistability in a rotating tank experiment with topography (shaded area)\cite{Tian_Weeks_etc_Ghil_Swinney_2001_JFM_JetTopography,Weeks_Tian_etc_Swinney_Ghil_Science_1997}. The dynamics in this experiment would be well modelled by a 2D barotropic model with topography (the quasi-geostrophic model with $R=\infty$). The flow is alternatively close to two very distinct states, with random switches from one state to the other. Left: the streamfunction of each of these two states. Right: the time series of the velocity measured at the location of the black square on the left figure, illustrating clearly the bistable behavior. The similar theoretical structures for the 2D Euler equations on one hand and the quasi-geostrophic model on the other hand, suggest that the bistability in this experiment can be explained as a non equilibrium phase transition, as done in section \ref{sub:Phase Transition 2} (see also figure \ref{ts_f}) \label{fig:Swinney}}
\end{figure}

We thus conclude that situations of phase transitions are extremely
important. Prediction of equilibrium phase transition help at locating
non-equilibrium phase transitions in slightly non-equilibrium situations.
The ideas developed here in the context of the 2D Navier-Stokes equation
can be applied in a much broader context, for quasi-geostrophic or
shallow-water dynamics. Indeed we conjecture that this would explain
the observed bistability in recent quasi-geostrophic experiments \cite{Tian_Weeks_etc_Ghil_Swinney_2001_JFM_JetTopography,Weeks_Tian_etc_Swinney_Ghil_Science_1997}
(see figure \ref{fig:Swinney}).

\begin{figure}[t!]
\begin{centering}
\includegraphics[width=\textwidth]{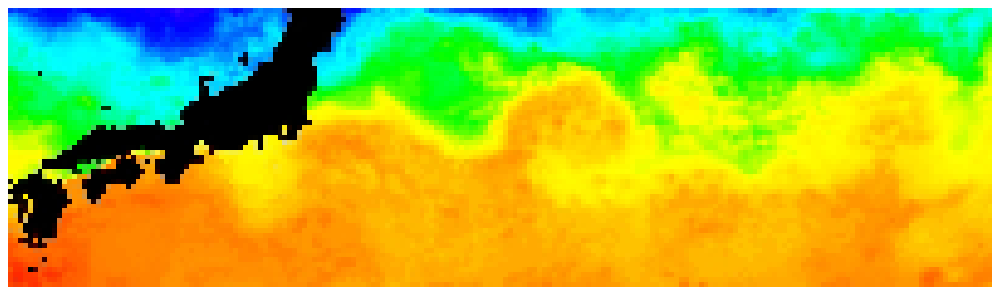}
\par\end{centering}

\caption{\footnotesize Kuroshio: sea surface temperature of the pacific ocean east of Japan,
February 18, 2009, infra-red radiometer from satellite (AVHRR, MODIS)
(New Generation Sea Surface Temperature (NGSST), data from JAXA (Japan
Aerospace Exploration Agency)).\protect \\
 The Kuroshio is a very strong current flowing along the coast,
south of Japan, before penetrating into the Pacific ocean. It is similar
to the Gulf Stream in the North Atlantic. In the picture, The strong
meandering color gradient (transition from yellow to green) delineates
the path of the strong jet (the Kuroshio extension) flowing eastward
from the coast of Japan into the Pacific ocean.\protect \\
 South of Japan, the yellowish area is the sign that, at the time
of this picture, the path of the Kuroshio had detached from the Japan
coast and was in a meandering state, like in the 1959-1962 period
(see figure \ref{fig:Kuroshio-Path}) \label{fig:SST-Kuroshio}}

\end{figure}

\begin{figure}[h!]
\begin{center}
\includegraphics[width=\textwidth]{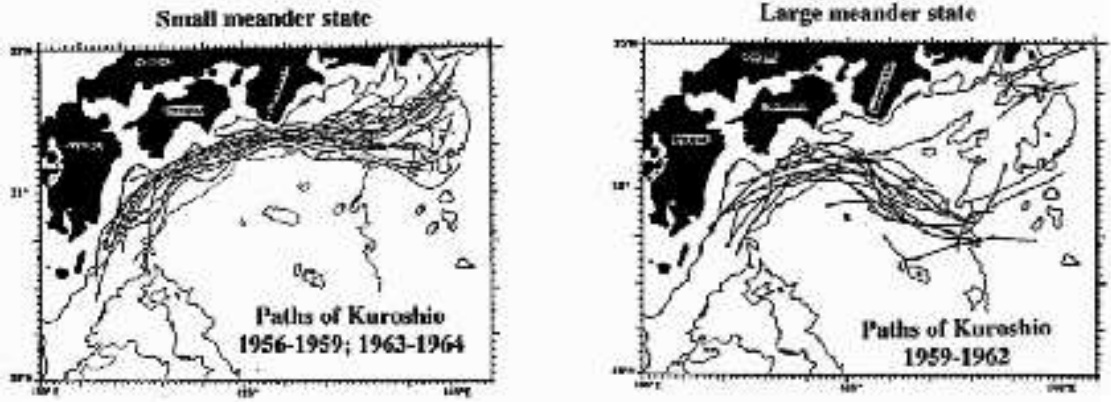}
\end{center}

\caption{\footnotesize Bistability of the paths of the Kuroshio during the 1956-1962 period
: paths of the Kuroshio in (left) its small meander state and (right)
its large meander state. The 1000-m (solid) and 4000-m (dotted) contours
are also shown. (figure from Schmeits and Dijkstraa \cite{Schmeits_Dijkstraa_2001_JPO_BimodaliteGulfStream},
adapted from Taft 1972). \label{fig:Kuroshio-Path}}

\end{figure}

\begin{figure}[h!]
\begin{center}
\includegraphics[height=35mm]{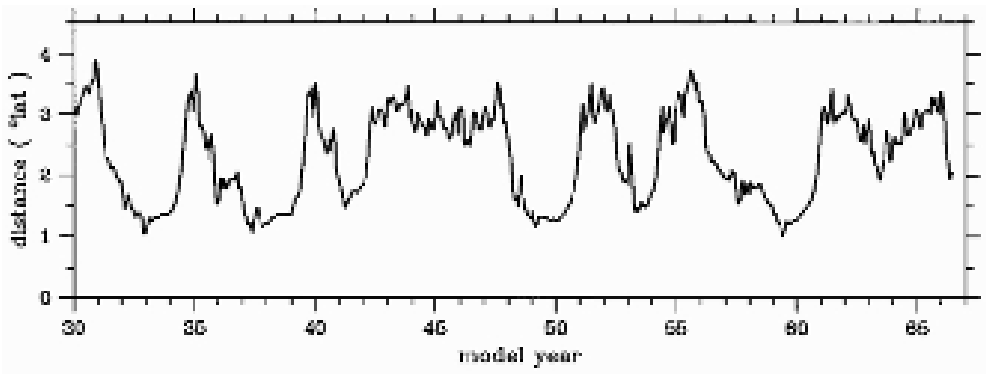}
\end{center}

\caption{\footnotesize Bistability of the paths of the Kuroshio, from Qiu and Miao \cite{Qiu_Miao_2000JPO....30.2124Q}:
time series of the distance of the Kuroshio jet axes from the coast,
averaged other the part of the coast between 132 degree and 140 degrees East,
from a numerical simulation using a two layer primitive equation model.
\label{fig:Kuroshio-Time-Series}}

\end{figure}

We also conjecture that this is an explanation of the bistability
of the Kuroshio current (Pacific ocean, east of Japan) (see figures
\ref{fig:SST-Kuroshio}, \ref{fig:Kuroshio-Path} and \ref{fig:Kuroshio-Time-Series}).

\subsection{Towards a kinetic theory of NESS \label{sub:Kinetic_Theory}}

We have explained in the previous sections that in the limit of weak
forces and dissipation, we expect to be close to some statistical
equilibria. This allows to predict qualitative properties of the flow
and non-equilibrium phase transitions. However, in order to be able
to predict which of these equilibria will be selected and to be able
to make predictions about the statistics of fluctuations, we cannot
rely on the equilibrium theory and we have to develop a non-equilibrium
theory. A way to proceed is to make a kinetic theory of these Non-Equilibrium Steady States.

Such a kinetic theory approach, as any kinetic theory, will be based
on an asymptotic expansion. Usually the small parameter is the ratio
of the typical time scale for the small scale fluctuations over the
typical time scale for the evolution of kinetic variables. For the
2D-stochastic Navier-Stokes equations, in the limit of small forces
and dissipation $\nu\ll\alpha\ll1$, the natural small parameter is
the friction coefficient $\alpha$. It is indeed the ratio of the
turnover time scale (the timescale at which fluctuation are advected)
over the time scale over which energy and other invariant of the inertial
dynamics evolve. \\

In the limit of weak forces and dissipation, the flow remains close
to statistically quasi stationary states (evolving on a time of
order $1/\alpha$, for instance dipoles or parallel flows in the case
of doubly periodic conditions, discussed in section \ref{sub:Close_To_Equilibrium_NavierStokes}),
with vorticity $\Omega_{0}\left(\mathbf{r},t\right)$ and velocity
$\mathbf{V}_{0}\left(\mathbf{r},t\right)$. At leading order in the
theory, we naturally obtain that $\Omega_{0}$ must be a dynamical
equilibrium state of the Euler equations $\mathbf{V}_{0}.\nabla\Omega_{0}=0$.
For it to be quasi-stationary, it also has to be dynamically stable.

The fluctuations evolve rapidly. The velocity fluctuations are expected
to be much smaller that the velocity $V_{0}$. It is then tempting
to try a perturbative expansion using this time scale separation.
We thus decompose the fields as\begin{equation}
\omega=\Omega_{0}+\delta\omega\,\,\,\mbox{and}\,\,\,\mathbf{v}=\mathbf{\mathbf{V}}_{0}+\delta\mathbf{v}.\label{eq:Avergae_Fluctuations}\end{equation}
In the more general case, $\Omega_{0}$ evolves slowly over time.
It may also exist cases where $\Omega_{0}$ is actually stationary.
For technical reasons, it will be simpler to discuss in the following
the case where $\Omega_{0}$ is stationary; however the generalization
to the quasi-stationary case or to situations with self-similar growth
is straightforward. We define $\left\langle .\right\rangle $ as an
average over the noise realization. Then $\Omega_{0}=\left\langle \omega\right\rangle $.

We start from the stochastic Navier-Stokes equations
\begin{equation}
\partial_{t}\omega+(\mathbf{v}\cdot\nabla)\omega=\nu\Delta\omega-\alpha\omega+\sqrt{2\alpha}\eta(t,{\bf x}).\label{eq:2D-Stochastic-Euler}
\end{equation}
We will need along the discussion the linearized Navier-Stokes equation
close to the base vorticity profile $\Omega_{0}$
\begin{equation}
\partial_{t}\delta\omega+L\left[\delta\omega\right]=\,\,\,\mbox{with}\,\,\, L\left[\delta\omega\right]=\mathbf{\mathbf{V}}_{0}.\nabla\delta\omega+\delta\mathbf{v} \cdot \nabla\Omega_{0}-\nu\Delta\omega+\alpha\omega.\label{eq:Euler2D_Linearized}\end{equation}
We decompose the fields as average plus fluctuations (\ref{eq:Avergae_Fluctuations}) ; The 2D Navier-Stokes equations  \ref{eq:2D-Stochastic-Euler} are then equivalent to the dynamics of the fluctuations, given by
\begin{equation}
\partial_{t}\delta\omega+L\left[\delta\omega\right]=\sqrt{2\alpha}\eta(t,{\bf x})-\mathbf{\delta v}\cdot\nabla\delta\omega-\alpha\Omega_{0} +\nu \Delta \Omega_{0} \label{eq:2D-NavierStokes-Fluctuations}
\end{equation}
Taking the average of (\ref{eq:2D-NavierStokes-Fluctuations}) gives
\[ \left\langle \delta\mathbf{v} \cdot \nabla\delta\omega\right\rangle =-\alpha\Omega_{0}+\nu\Delta\Omega_{0}.\]
This important equation just expresses that the mean vorticity profile
is determined by a balance between the average of the nonlinear contributions
of the fluctuations (Reynolds stresses) on one hand and the dissipation
on the other hand. The challenge is then to find a theory to compute
these Reynolds stresses.  \\

The Reynolds stress $\left\langle \delta\mathbf{v}.\nabla\delta\omega\right\rangle $
is a quadratic quantity; it can thus be evaluated from the two point
correlation function $\phi_{2}\left(\mathbf{r}_{1},\mathbf{r}_{2},t\right)=\left\langle \delta\omega\left(\mathbf{r}_{1},t\right)\delta\omega\left(\mathbf{r}_{2},t\right)\right\rangle $.
The equation for the time evolution of the two-point correlation function
is easily obtained from (\ref{eq:2D-Stochastic-Euler}), using the
Ito formula and averaging. We obtain \begin{equation}
\partial_{t}\phi_{2}+L_{1}\phi_{2}+L_{2}\phi_{2}=2\alpha F_{2}+NL_{2},\label{eq:TwoPoints_Correlation}\end{equation}
where $L_{1}$ (resp $L_{2}$) is the linearized Euler operator $L$
(\ref{eq:Euler2D_Linearized}) acting on the variable $\mathbf{r}_{1}$
(resp $\mathbf{r_{2}}$), $NL_{2}\left(\mathbf{r}_{1},\mathbf{r}_{2}\right)=-\left\langle \delta\mathbf{v}\left(\mathbf{r}_{1}\right).\nabla\delta\omega\left(\mathbf{r}_{1}\right)\delta\omega\left(\mathbf{r}_{2}\right)\right\rangle -\left\langle \delta\mathbf{v}\left(\mathbf{r}_{2}\right).\nabla\delta\omega\left(\mathbf{r}_{2}\right)\delta\omega\left(\mathbf{r}_{1}\right)\right\rangle $
is the contribution of the nonlinear term and $2\alpha F_{2}$ is
average effect of the stochastic force on the two-point correlation
function (with the stochastic force (\ref{eq:Stochastic_Forces}),
page \pageref{eq:Stochastic_Forces}, we have $F_{2}\left(\mathbf{r}_{1},\mathbf{r}_{2}\right)=\sum_{k}f_{k}^{2}e_{k}\left(\mathbf{r}_{1}\right)e_{k}\left(\mathbf{r}_{2}\right)$).

Due to the nonlinearity, the equation for the two points correlation
function (\ref{eq:TwoPoints_Correlation}) involves a three point
quantity $NL_{2}$. One could easily write the whole hierarchy for
the $n$-point correlators. Any truncation of such a hierarchy is
arbitrary, except in cases where a small parameter allows to neglect
the nonlinear terms in a self-consistent way. Such a situation occurs
for instance in kinetic theory, more specifically in the kinetic theory
of systems with long range interactions \cite{Bouchet_Dauxois:2005_PRE}
that share deep analogies with the present problem, one example being the kinetic theory of the point vortex model \cite{Chavanis_houches_2002} (an application of similar idea to the relaxation towards equilibrium of the 2D Euler equation as also been discussed, see \cite{Chavanis_Quasilinear_2000PhRvL} and further discussion in section \ref{sub:Relaxation-towards-equilibrium}). We then call such
an approach a kinetic approach.\\

Such a kinetic theory approach is a classical one. Similar ideas have
been discussed back in the seventies and eighties in other contexts
and are still studied currently (quasi-normal closures, rapid distortion
theories, second order cumulant truncations, and other related approaches).
Very few of these works however consider inhomogeneous flows dominated
by the large scales, as is our interest here. Some exceptions are
a series of theoretical and numerical works made during last decade
\cite{Dubrulle_Nazarenko_1997PhyD,Laval_Dubrulle_Nazarenko_2000PhyD,Nazarenko_Laval_JFM_2000,Nazarenko_PhysicsLetterA_2000,Chavanis_Quasilinear_2000PhRvL},
among them a very interesting model of 2D wall turbulence \cite{Nazarenko_PhysicsLetterA_2000}.
In the case of the large scales of geophysical flows, recent interesting
works have used numerical simulations, for instance to study the limits
of second order cumulants expansion \cite{Marston_Conover_Schneider_JAS2008},
or to study numerically a self consistent closure describing the coupling
of the mean flow and of the second order cumulant \cite{Farrell_Ioannou_JAS_2007}. Another line of research, on related issues, has been to search for crude closures \cite{Grote_Majda_CrudeClosure_1997PhFl,Grote_Majda_2000Nonli}, or more precise mathematical results \cite{Majda_Wang_Bombardement_2006_Comm}, when the system is subjected to random bombardments

In all of the previous works, some hypothesis of a phenomenological
nature are made in order to simplify the problem at some point (closure
without small parameter, assumption of scale separations, Markovianization),
that allows interesting studies to be pushed forward. However, there
still remains a lot of work to assess either numerically or theoretically
the validity or not of these hypothesis, and thus to be really able
to propose a clear theory of the large scales of two dimensional and
geophysical flows. Our belief is that any progress in this direction
requires a better theoretical understanding of the basic objects appearing
in the theory.

For instance any progress in the kinetic theory requires the understanding
of equation (\ref{eq:TwoPoints_Correlation}), and thus requires a
theoretical understanding of the two-point linear operator on the
rhs: $\partial_{t}\phi_{2}+L_{1}\phi_{2}+L_{2}\phi_{2}=F$. Similar
n-point linear operators, implying the linearized operator $L$, appear
at each level of the hierarchy of the equations for the $n$ point
correlation function. A prerequisite for any understanding of this
linear operator is a detailed understanding of the linearized Euler
equation and of its asymptotic behavior. The current theoretical understanding
of $L$ is readily not sufficient to go forward with the kinetic theory.

The results for the behavior of $L$ can not be universal. They depend
a lot on the boundary conditions, on the topology of the streamlines
and on the specific model (Euler, quasi-geostrophic, etc...). For
instance, any theory that would not depend explicitly on boundary
conditions would be promised to failure. The theoretical analysis
of the linear operator $L$ is one of the aims of next section.

\subsection{Relaxation towards equilibrium and asymptotic behavior of the 2D
Euler and linearized Euler equation \label{sub:Relaxation-towards-equilibrium}}

\subsubsection{Irreversibility of reversible dynamical systems}

The 2D Euler equations \begin{equation}
\partial_{t}\omega+(\mathbf{v}\cdot\nabla)\omega=0\label{eq:Euler_Relaxation}\end{equation}
is time reversible: it is invariant over the time reversal symmetry
$t\rightarrow-t$, $\omega\rightarrow-\omega$ (or equivalently $\mathbf{v}\rightarrow-\mathbf{v}$).
However, it has anyway an irreversible behavior. Indeed, as explained
in section \ref{sec:2D-Geostrophic-Turbulence}, for large times,
enstrophy and other Casimir invariant cascade towards lower and lower
scales and the largest scales of the flow converge towards a stationary solution to the 2D Euler equations. Such an apparent paradox
between the time reversal symmetry of the microscopic dynamics (here
the Euler equations) and the irreversible evolution of macroscopic
variables (here the largest scales of the flow) is a classic problem
of statistical mechanics.

This reversibility paradox is usually satisfactorily explained by
introducing in the discussion, the discussion of relative probabilities
of types of initial conditions (see for instance the classical discussion
\cite{Ritz_Einstein_1909PhyZ} about irreversible behavior in electromagnetism).
In the statistical mechanics this idea is formalized using the concepts
of microscopic versus macroscopic variables and by introducing the
notion of a probability for the macroscopic states. The entropy of
a macrostate quantifies the number of microstates corresponding to
a given macrostate. Then for a sample of rare initial microscopic
conditions (corresponding to a low entropy macrostate), an overwhelming
number of the trajectories evolve towards microscopic configurations
corresponding to a more probable (higher entropy) macrostate \cite{Goldstein_Lebowitz_2004PhyD..193...53G}.

For this classical explanation of the reversibility paradox to be
relevant, a clear distinction between microscopic and macroscopic
variable is essential; this requires to consider a limit with a large
number of degrees of freedom (usually the thermodynamic limit in
classical physical systems). The Euler equation is different from
those classical systems in the sense that it is a partial differential
equation that has from the beginning an infinite number of degrees
of freedom. \\

Beside the general qualitative explanation of the reversibility paradox,
there is only a few examples where one can prove mathematically the
irreversible evolution of the macroscopic variable directly from the
microscopic dynamics. The most famous example is probably Landford's
proof of the validity of the Boltzmann equation (and thus macroscopic
irreversibility), for a system of dilute particles (Grad limit), with
hard core interactions (see \cite{Spohn_1991} for a very clear presentation).
We want to stress that the Euler equation may be another example where
an irreversible behavior can be proved for a reversible equation (see
\cite{Bouchet_Morita_2010PhyD}).

The aim of the following discussion is to present the results in \cite{Bouchet_Morita_2010PhyD}
related to the irreversibility problem. These results moreover include
a detailed study of the linearized Euler equation which is directly
related to the discussion of section \ref{sub:Kinetic_Theory} about
the kinetic theory of the 2D Navier-Stokes equation. The general discussion
in \cite{Bouchet_Morita_2010PhyD} is however rather technical, and in
this section we only derive the main results for the special case
of a constant shear, for which explicit computation are straightforward
\cite{Case_1960_Phys_Fluids} and state the more general results.

\subsubsection{Irreversible relaxation of the linearized Euler equation}

We consider in the following the linearized 2D Euler equations close
to a stable parallel flows, in a doubly periodic domain or in a channel.
We stress however that all the following results should be valid for
a stable circular vortex in a disc geometry%
\footnote{For these results to be valid, some further conditions on the behavior
of the vorticity at the core of the vortex may be required; this remains
to be studied.%
}

Any parallel flow $\mathbf{v}_{0}$=$U\left(y\right)\mathbf{e}_{x}$
is a stationary solution to the 2D Euler equations (\ref{eq:Euler_Relaxation})
in a doubly periodic domain or in a channel. We consider the Euler
equations with initial conditions close to this base flow: $\Omega=\omega_{0}+\omega$
and $\mathbf{V}=\mathbf{v}+\mathbf{v}_{0}$, where $\omega_{0}\left(y\right)=-U'\left(y\right)$
is the base flow vorticity and $\omega$ and $\mathbf{v}$ are the
perturbation vorticity and velocity, respectively. It reads

\begin{equation}
\partial_t \omega +U\left(y\right) \partial_x \omega - v_{y}U''\left(y\right)=0,\label{eq:Linearized_Euler_Fourier}\end{equation}
 where $v_{y}$ is the transverse velocity component.

We assume that the base flow $U\left(y\right)$ is linearly stable
(there is no unstable mode to the linear equation (\ref{eq:Linearized_Euler_Fourier})).
We note that any $\omega\left(y\right)$ independent of $x$ is a
trivial neutral mode of (\ref{eq:Linearized_Euler_Fourier}). If we
decompose $\omega$ in Fourier modes along the longitudinal direction
$\omega(x,y)=\sum_{k}\omega_{k}\left(y\right)e^{ikx}$, the linearized
Euler equation for $\omega_{k}$ is\begin{equation}
\partial_t \omega_{k} + ikU\left(y\right)\omega_{k}-ik\psi_{k}U''\left(y\right)=0,\label{eq:Linearized_Euler_Fourier_Vraiment}\end{equation}
where $\psi_{k}$ is the Fourier transform of the streamfunction.
We assume that for all $k\neq0$, equations (\ref{eq:Linearized_Euler_Fourier_Vraiment})
have no neutral modes (this situation of a linear operator with no
modes may seem strange, it is however not unusual for a non-normal%
\footnote{A linear operator $L$ is said to be normal if it commutes with its
adjoint $LL^{*}=L^{*}L$. In finite dimensional spaces, a normal operator
can always be diagonalized on an orthogonal base. This result often
generalize to infinite dimensional space, for instance in the case
of bounded self-adjoint operators typical of quantum dynamics. By
contrast, non-normal operator may not be diagonalizable, and may not
have any mode as illustrated by many examples in fluid mechanics for
instance. %
} linear operator; for instance one can prove that that (\ref{eq:Linearized_Euler_Fourier_Vraiment})
has no mode as soon as $U$ is monotonic in a channel geometry \cite{Drazin_Reid_1981},
this is also true for the Kolmogorov flow in doubly periodic domains
\cite{Bouchet_Morita_2010PhyD}).

\paragraph*{The linear shear}

Because of its third l.h.s. term, a general discussion of (\ref{eq:Linearized_Euler_Fourier})
is rather complex and requires the use of complex mathematical tools
\cite{Bouchet_Morita_2010PhyD}. It is often argued that this third term
can be neglected, but this is usually a very bad approximation (see
\cite{Bouchet_Morita_2010PhyD}). However in the special case of a linear
shear flow $U(y)=\sigma y$, the third term vanishes and the equation
is then very simple \[
\partial_t \omega + \sigma y \partial_x \omega =0.\]

This equation can be easily solved: \[
\omega\left(x,y,t\right)=\omega(x-\sigma yt,y,0).\]
It is even more simple if we consider perturbation on the form $\omega(x,y,t)=\omega_{k}\left(y,t\right)\exp\left(ikx\right)$,
then \begin{equation}
\omega_{k}\left(y,t\right)=\omega_{k}(y,0)\exp\left(-ik\sigma yt\right).\label{eq:vorticite_cisaillementlineaire_Fourrier}\end{equation}

The velocity can be expressed from the vorticity using using a Green
function formalism. We have

\begin{equation}
\mathbf{v}_{k}(y,t)=\int\mathrm{d}y'\,\mathbf{G}_{k}(y,y')\omega_{k}(y',t),\label{eq:Evolution_velocity}\end{equation}
 where, using $\omega=\Delta\psi$, $v_{x}=-\frac{d\psi}{dy}\,\,\,\mbox{and}\,\,\, v_{y}=\frac{d\psi}{dx}$,
$\mathbf{G}_{k}$ is defined by \begin{equation}
\mathbf{G}_{k}(y,y')=\left(-\frac{\partial H_{k}}{\partial y},ikH_{k}\right)\left(y,y'\right)\,\,\,\mbox{with}\,\,\,\frac{\partial^{2}H_{k}}{\partial y^{2}}-k^{2}H_{k}=\delta\left(y-y'\right),\label{eq:Green Function}\end{equation}
with for instance a channel boundary conditions: $y\in(-L,L)$ with
$\psi\left(L\right)=\psi\left(-L\right)=H_{k}(L)=H_{k}(-L)=0$. Using
(\ref{eq:vorticite_cisaillementlineaire_Fourrier}), we have \begin{equation}
\mathbf{v}_{k}(y,t)=\int dy'\,\mathbf{G}_{k}\left(y,y'\right)\omega_{k}(y',0)\exp\left(-ik\sigma y't\right).\label{eq:velocity_ocillating_integral}\end{equation}

We consider the asymptotic behavior, for large times $t$, of the
oscillating integral (\ref{eq:velocity_ocillating_integral}). Since
Kelvin, very classical results do exist for the asymptotic behavior
of such integrals, the most well known results being the stationary
phase approximation. In our case there is no stationary phase, and
the asymptotic behavior of the velocity field is obtained by successive
integrations by parts, which lead to

\begin{equation}
v_{k,x}(y,t)\underset{t\rightarrow\infty}{\sim}\frac{\omega_{k}\left(y,0\right)}{ik}\frac{\exp\left(-iky\sigma t\right)}{\sigma t}\,\,\,\mbox{and}\label{eq:Vitesse-x-algebraic-linear}\end{equation}
 \begin{equation}
v_{k,y}(y,t)\underset{t\rightarrow\infty}{\sim}\frac{\omega_{k}\left(y\right)}{ik}\frac{\exp\left(-iky\sigma t\right)}{\sigma^{2}t^{2}}\ .\label{eq:Vitesse-y-algebraic-linear}\end{equation}
The exponents of the algebraic laws $1/t$ for $v_{x}$ and $1/t^{2}$
for $v_{y}$ are related to the singularities of the Green function
$\mathbf{G}_{k}$.

This shows that the velocity field decays algebraically for large
time. As illustrated by figure (\ref{fig:decroissance_vitesse}) in
the case of the Kolmogorov flow, the velocity actually decays much
faster (exponentially) for times of order $1/\sigma$ and then the
decrease has algebraic tails. This irreversible behavior of the velocity
field for a reversible equation (the linearized 2D Euler equations
are time reversible (symmetry $t\rightarrow-t$, $\omega\rightarrow-\omega$,
$\mathbf{v}\rightarrow-\mathbf{v}$, $U\rightarrow-U$ and $\Omega_{0}\rightarrow-\Omega_{0}$)
is a striking result.

\begin{figure}[t!]
\begin{centering}
\includegraphics[width=0.9\textwidth]{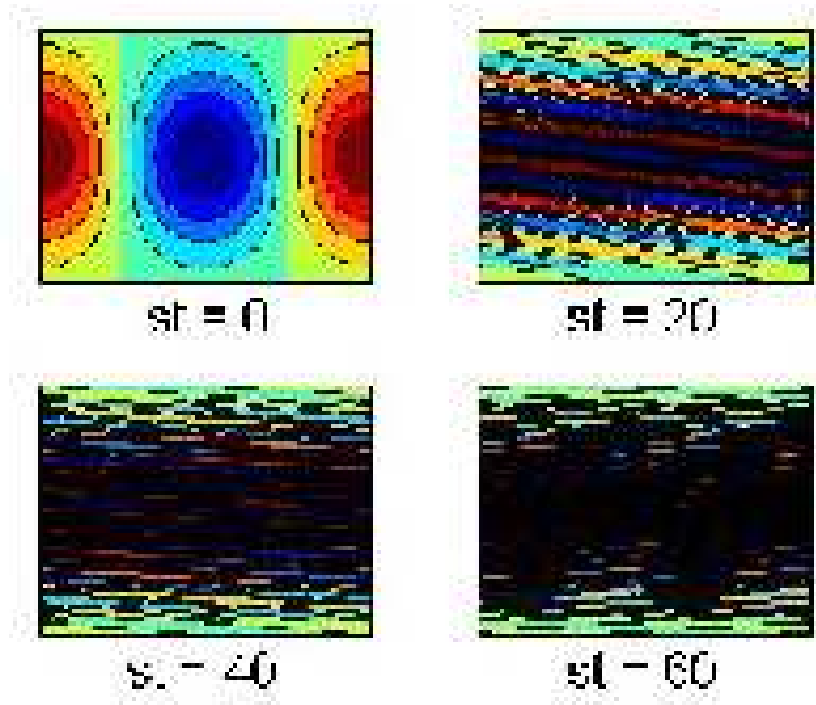}
\par\end{centering}

\caption{\footnotesize Evolution of $\omega(x,y,t)$ from an initial vorticity perturbation
$\omega(x,y,0)=\omega_{1}\left(y,0\right)\cos\left(x\right)$, by
the linearized 2D Euler equations close to a shear flow $U\left(y\right)=\sigma y$
. (colors in the .PDF document) }

\label{fig:shear_vorticity}
\end{figure}

Heuristically, the vorticity field is strongly sheared and produce
filaments at finer and finer scales as illustrated by figure (\ref{fig:shear_vorticity}).
The computation of the velocity field from the vorticity field involves
an integration, the contribution of these fine scale filaments is
then weaker and weaker.\\

\paragraph*{General parallel flow}

The algebraic decay of the velocity field for the linearized 2D Euler
equations close to a linear shear has been first obtained by Case
\cite{Case_1960_Phys_Fluids}, using an explicit computation rather
than the oscillating integral explanation given above. For more general
base flows with strictly monotonic profiles $U\left(y\right)$ (without
stationary streamline $U'\left(y_{0}\right)=0$), from classical arguments
\cite{Rosencrans_Sattinger_1966_J_Math_Phys,Briggs_BDL_1970_Phys_Fluids}
using the Laplace transform, one expects an asymptotic algebraic decrease
of the velocity field with the same $1/t$ and $1/t^{2}$ laws (see
also an ansatz for large time asymptotic in \cite{Brown_Stewartson_1980_JFM}).

\begin{figure}[t!]
\begin{centering}
\includegraphics[width=0.9\textwidth]{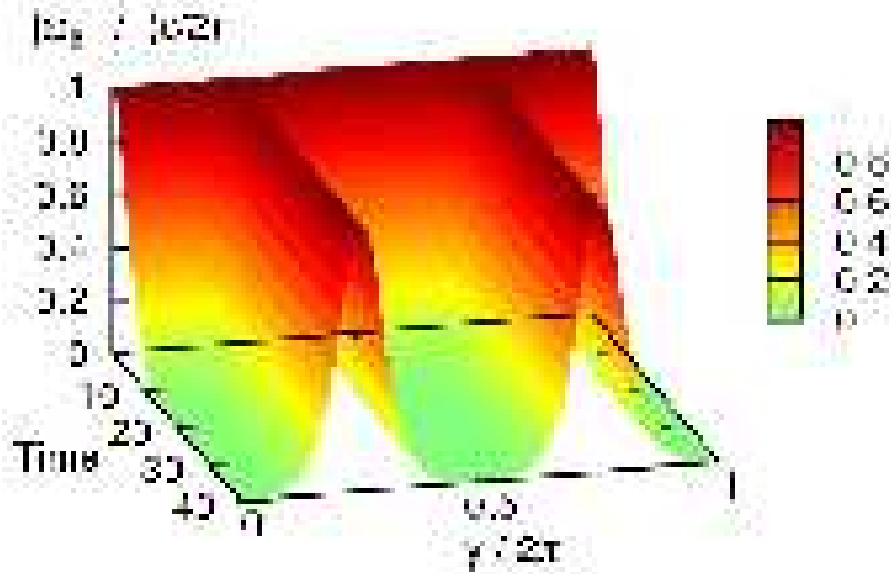}
\par\end{centering}

\caption{\footnotesize  \label{fig:Vorticity_Depletion}Evolution of the vorticity perturbation
$\omega(x,y,t)=\omega\left(y,t\right)\exp\left(ikx\right)$, close
to a parallel flow $\mathbf{v}_{0}(x,y)=U(y)\mathbf{e}_{x}$ with
$U\left(y\right)=\cos\left(y\right)$, in a doubly periodic domain
with aspect ratio $\delta$. The figure shows the modulus of the perturbation
$\left|\omega\left(y,t\right)\right|$ as a function of time and $y$.
One clearly sees that the vorticity perturbation rapidly converges
to zero close to the points where the velocity profile $U\left(y\right)$
has extrema ($U'(y_{0})=0$, with $y_{o}=0$ and $\pi$). This\emph{
depletion of the perturbation vorticity} at the stationary streamlines
$y_{0}$ is a new generic self-consistent mechanism, understood mathematically
as the regularization of the critical layer singularities at the edge
of the continuous spectrum (see \cite{Bouchet_Morita_2010PhyD}). (colors
in the .pdf document)}

\end{figure}

In the case of base flows $U(y)$, the oscillating phase in the integral
(\ref{eq:velocity_ocillating_integral}) is $ikU\left(y\right)t$.
Then for base flows with stationary streamlines $U'(y_{0})=0$, the
oscillating integral (\ref{eq:velocity_ocillating_integral}) has
a stationary phase and one expects other algebraic laws for the asymptotic
velocity fields (for instance $1/\sqrt{t}$) (see discussions by \cite{Brown_Stewartson_1980_JFM,Lundgren_1982_PhFl}).
It has however been proved recently that, unexpectedly, the same power
laws occur \cite{Bouchet_Morita_2010PhyD}. This is associated with a very
surprising non-local mechanism of vorticity depletion at the stationary
streamlines, not described before (see \cite{Bouchet_Morita_2010PhyD}
and figure \ref{fig:Vorticity_Depletion}).

\begin{figure}[t!]
\begin{centering}
(a) \includegraphics[width=0.9\textwidth]{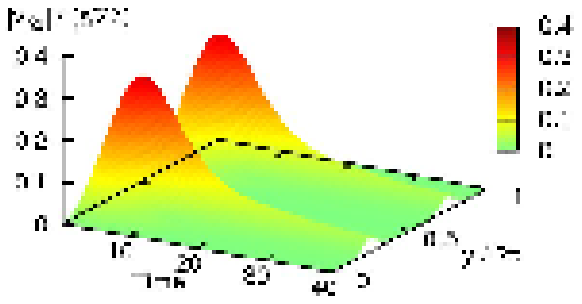}
\\
 (b) \includegraphics[width=0.9\textwidth]{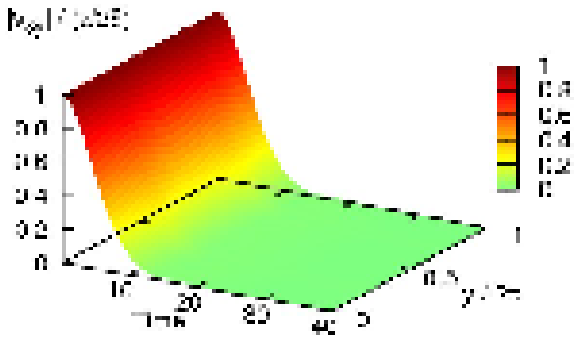}
\par\end{centering}

\caption{ \footnotesize The space-time series of perturbation velocity components, $|v_{\delta,x}(y,t)|$
(a) and $|v_{\delta,y}(y,t)|$ (b), for the initial perturbation profile
$\cos\left(x/\delta\right)$ in a doubly periodic domain with aspect
ratio $\delta=1.1$. Both the components relax toward zero, showing
the asymptotic stability of the Euler equations. (colors in the .pdf
document)}

\label{fig:v1_stseri}
\end{figure}

\begin{figure}[t]
\begin{centering}
(a) \includegraphics[width=0.7\textwidth]{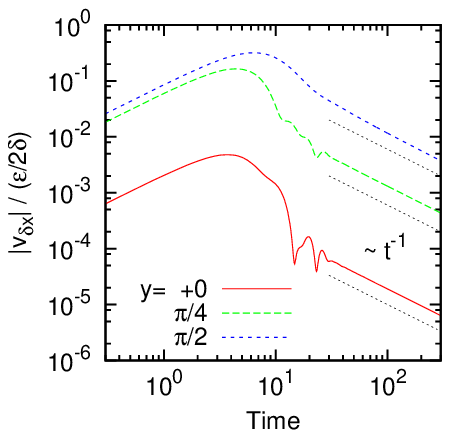}
\\
 (b) \includegraphics[width=0.7\textwidth]{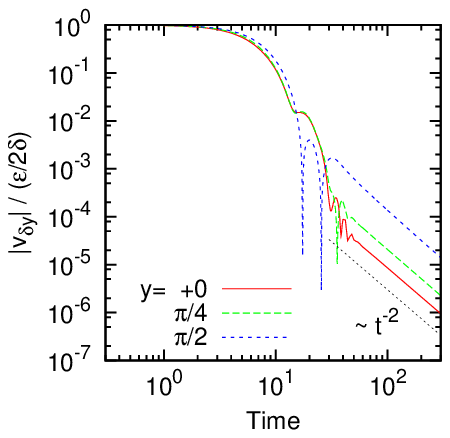}
\par\end{centering}

\caption{ \footnotesize The time series of perturbation velocity components $|v_{\delta,x}(y,t)|$
(a) and $|v_{\delta,y}(y,t)|$ (b) at three locations, $y=0$ (vicinity
of the stationary streamline) (red), $y=\pi/4$ (green), and $y=\pi/2$
(blue), for the initial perturbation profile $A(y)=1$ and the aspect
ratio $\delta=1.1$. We observe the asymptotic forms $|v_{\delta,x}(y,t)|\sim t^{-\alpha}$,
with $\alpha=1$, and $|v_{\delta,y}(y,t)|\sim t^{-\beta}$, with
$\beta=2$, in accordance with the theory for the asymptotic behavior
of the velocity (equations (\ref{eq:Vitesse-x-algebraic-linear})
and (\ref{eq:Vitesse-y-algebraic-linear})). The initial perturbation
profile is $\cos\left(x/\delta\right)$ in a doubly periodic domain
with aspect ratio $\delta=1.1$. (colors in the .pdf document)}

\label{fig:decroissance_vitesse}
\end{figure}

The general result \cite{Bouchet_Morita_2010PhyD}, valid for any stable
flow $U\left(y\right)$ without any mode for (\ref{eq:Linearized_Euler_Fourier_Vraiment}),
is then

\begin{equation}
\omega\left(y,t\right)\underset{t\rightarrow\infty}{\sim}\omega_{\infty}\left(y\right)\exp\left(-ikU(y)t\right)+\mathcal{O}\left(\frac{1}{t^{\gamma}}\right),\label{eq:Lungren}\end{equation}
with an algebraically decaying velocity for large times \begin{equation}
v_{x}(y,t)\underset{t\rightarrow\infty}{\sim}\frac{\omega_{\infty}\left(y\right)}{ik}\frac{\exp\left(-ikU(y)t\right)}{U'(y)t}\,\,\,\mbox{and}\label{eq:Vitesse-x-algebraic}\end{equation}
 \begin{equation}
v_{y}(y,t)\underset{t\rightarrow\infty}{\sim}\frac{\omega_{\infty}\left(y\right)}{ik}\frac{\exp\left(-ikU(y)t\right)}{\left(U'(y)t\right)^{2}};\label{eq:Vitesse-y-algebraic}\end{equation}
 where the asymptotic vorticity profile $\omega_{\infty}$ (see (\ref{eq:Lungren}),
(\ref{eq:Vitesse-x-algebraic}) and (\ref{eq:Vitesse-y-algebraic}))
can be computed from the Laplace transform of the linearized equation
(\ref{eq:Linearized_Euler_Fourier_Vraiment}) (see \cite{Bouchet_Morita_2010PhyD}).
The algebraic decay of the velocity field is illustrated on figures
\ref{fig:v1_stseri} and \ref{fig:decroissance_vitesse}, in the case
of the Kolmogorov flow $U(y)=\cos\left(y\right)$ on doubly periodic
domains.

\subsubsection{Relaxation and asymptotic stability of parallel flows for the 2D Euler
equations\label{sec:Asymptotic-stability}}

In the previous section, we have obtained results for the asymptotic
behavior of the linearized 2D Euler equations, with initial conditions
close to some parallel flows $\mathbf{v}_{0}\left(\mathbf{r}\right)=U\left(y\right)\mathbf{e}_{x}$.
We now address the evolution of the same initial conditions by the
nonlinear Euler equation (\ref{eq:Euler_Relaxation}).

The asymptotic stability of an ensemble of parallel flows means that
for any small perturbation of a parallel flow, the velocity converges
for large times towards another parallel flow close to the initial
one. The aim of this section is to explain why the linearized dynamics
is a good approximation for the non-linear dynamics for any time $t$,
and to explain why the flow velocity is asymptotically stable (in
kinetic energy norm), for small initial perturbation of the vorticity
(in the enstrophy norm). Such an irreversible convergence is a striking
phenomena for a reversible equation like the 2D Euler equations.\\

We consider the initial vorticity $\Omega\left(x,y,0\right)=-U'\left(y\right)+\epsilon\omega\left(x,y,0\right)$,
where $\epsilon$ is small. We suppose, without loss of generality,
that $\int dx\,\omega=0$. The perturbation $\omega$ can be decomposed
in Fourier modes along the $x$ direction \[
\omega\left(x,y,t\right)=\sum_{k}\omega_{k}\left(y,t\right)\exp\left(ikx\right).\]
 From the Euler equations (\ref{eq:Euler_Relaxation}), the equation
for $\omega_{k}$ is \begin{equation}
\partial_t \omega_{k} + ikU\left(y\right)\omega_{k}-ik\psi_{k}U''\left(y\right)=-\epsilon NL\,\,\,
\nonumber
\end{equation}
\begin{equation} \mbox{with}\,\,\, NL=\sum_{l}\left\{ -ik\frac{\partial\psi_{l}}{\partial y}\left(y,t\right)\omega_{k-l}\left(y,t\right)+\frac{\partial}{\partial y}\left[il\psi_{l}\left(y,t\right)\omega_{k-l}\left(y,t\right)\right]\right\} .\label{eq:Euler_k}\end{equation}
 The left hand side is the linearized Euler equation, whereas the
right hand side are the nonlinear corrections. We want to prove that,
for sufficiently small $\epsilon$, neglecting the nonlinear terms
is self-consistent.

For this we have to prove that the nonlinear terms remain uniformly
negligible for large times. We then use the asymptotic results for
the linearized equation (\ref{eq:Lungren}-\ref{eq:Vitesse-y-algebraic})
and $\omega_{k}=d^{2}\psi_{k}/dy^{2}-k^{2}\psi_{k}$. We have, for
any $k$,\begin{equation}
\psi_{k,L}\left(y,t\right)\underset{t\rightarrow\infty}{\sim}\frac{\omega_{k,L,\infty}\left(y\right)}{\left(ikU'\left(y\right)\right)^{2}}\frac{\exp\left(-ikU(y)t\right)}{t^{2}}\,\,\,\nonumber
\end{equation}
\begin{equation}
\mbox{and}\,\,\,\omega_{k,L}\left(y,t\right)\underset{t\rightarrow\infty}{\sim}\omega_{k,L,\infty}\left(y\right)\exp\left(-ikU(y)t\right),\label{eq:Asymptotics_k}\end{equation}
 where the subscript $L$ refers to the evolution according to the
linearized dynamics. We call a quasilinear approximation for the right
hand side of equation (\ref{eq:Euler_k}), the approximation where
$\psi_{k}$ and $\omega_{k}$ would be evaluated according to their
linearized evolution close to the base flow $U\left(y\right)$. From
(\ref{eq:Asymptotics_k}), one would expect at first sight that this
quasilinear approximation of the nonlinear term $NL_{QL}$, would
give contributions of order $O\left(1/t\right)$. The detailed computation,
easily performed from (\ref{eq:Asymptotics_k}), actually shows that
the contributions of order $O\left(1/t\right)$ identically vanish
for large times. This cancellation of terms is a remarkable property
with important consequences. Then \[
\epsilon NL_{k,QL}\underset{t\rightarrow\infty}{=}O\left(\frac{\epsilon}{t^{2}}\right).\]
This important remark proves that within a quasilinear approximation,
the contribution of the nonlinear terms $NL_{QL}$ remains uniformly
bounded, and more importantly it is integrable with respect to time.

Then we conjecture that the contribution of the nonlinear terms remains
always negligible. More precisely, we conjecture that within the fully
nonlinear equation, for sufficiently small $\epsilon$:\[
\psi_{k}\left(y,t\right)\underset{t\rightarrow\infty}{\sim}\frac{\omega_{k,\infty}\left(y\right)}{\left(ikU'\left(y\right)\right)^{2}}\frac{\exp\left(-ikU(y)t\right)}{t^{2}}\,\,\,\]\[ \mbox{and}\,\,\,\omega_{k}\left(y,t\right)\underset{t\rightarrow\infty}{\sim}\omega_{k,\infty}\left(y\right)\exp\left(-ikU(y)t\right),\]
 with \[
\omega_{k,\infty}\left(y\right)=\omega_{k,L,\infty}\left(y\right)+O\left(\epsilon\right)\]

A similar reasoning in order to evaluate the nonlinear evolution for
the profile $U\left(y\right)$ would lead to the conclusion that for
large times \[
\Omega_{0}\left(y,t\right)\underset{t\rightarrow\infty}{\sim}-U_{\infty}'\left(y\right)\,\,\,\mbox{with}\,\,\, U'_{\infty}\left(y\right)=U\left(y\right)+\delta U\left(y\right),\]
 where $\delta U=O\left(\epsilon^{2}\right)$.

This means that the parallel flow quickly stabilizes again towards
another parallel flow which is close to the initial one. This stabilization
is very rapid, it occurs on times scales of order $1/\sigma$ where
$\sigma$ is a typical shear rate.

We thus conclude that the relaxation towards stationary solutions to the 2D Euler equations is a very fast and simple process, leading to a stationary state on time scales given by the linearized dynamics. The velocity fluctuations are weakened extremely fast by the dynamics, such that their effect becomes soon negligible. This is by contrast with the phenomenology of particle models, like the point vortex model, where fluctuations are constantly produced due to the singularities of the vorticity field, related to the discrete point particles.

The long term evolution of point vortex models close to quasi-stationary states of the 2D Euler equations is thought to be described by a kinetic equation \cite{Chavanis_houches_2002,IUTAM_Symposium08} analogous of the Lenard-Balescu equation of plasma physics. A very natural hypothesis is that a similar kinetic equation could describe the long term evolution of initial conditions close to stationary states of the 2D Euler equations, as interestingly proposed by  \cite{Chavanis_Quasilinear_2000PhRvL}. Whether this is justified or not, and for which class of solutions, is a very complex issue, that has not been settled yet, neither from a theoretical nor from an empirical point of view. The results described in this section suggest that this is not the case for analytical initial conditions close to parallel flows, as then the fluctuations decays very quickly and the flow settles to a stationary states before a regime of long term quasi-stationary evolution could appear. As far as larger classes of initial conditions are concerned (close to other type of base flow than parallel flows, or with non analytic classes of initial conditions) the answer is unclear yet.\\

One might then want to compute the modified profile. The preceding
analysis leads to the quasilinear expression \begin{equation}
\delta U\left(y\right)=-\epsilon^{2}\int_{0}^{\infty}dt\, NL_{0,QL}\left(t\right)+o\left(\epsilon^{2}\right).\label{eq:Quasilinear_Profile}\end{equation}
 This expression involves integrals over times of the linearized Euler
equation. It is not amenable to any simple explicit expression, but
it can be evaluated using the Laplace transform of (\ref{eq:Linearized_Euler_Fourier_Vraiment}).

\begin{figure}[t!]
\begin{centering}
\includegraphics[width=0.9\textwidth]{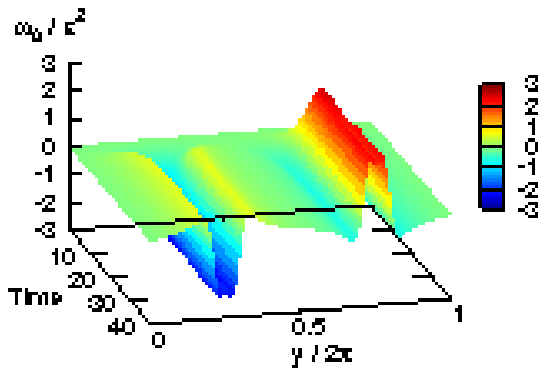}
\par\end{centering}

\caption{ \footnotesize The space-time series of the $x$-averaged perturbation vorticity,
$\omega_{0}(y,t)=\omega(y,t$)-$\Omega_{0}\left(y,0\right)$. The
initial condition is $\omega\left(y,t\right)=\Omega_{0}\left(y,0\right)+\epsilon\cos\left(x\right)$,
in a doubly periodic domain with aspect ratio $\delta=1.1$ (colors
in the .pdf document).}

\label{fig:w0_stseri}
\end{figure}

We conclude that for any profile $U$ with no unstable nor neutral
modes for (\ref{eq:Linearized_Euler_Fourier_Vraiment}), any perturbation
corresponding to a small vorticity, the assumption that the dynamics
can be treated with a quasi-linear approximation is a self consistent
hypothesis. Then the velocity converges for large times towards a
new parallel velocity profile which is close to the initial profile
$U$. Figures \ref{fig:v1_stseri} and \ref{fig:w0_stseri}, on page  \pageref{fig:v1_stseri},
show that numerical computations confirm this conjecture. %

From this discussion, we conclude that is natural to conjecture that
any profile $U$ verifying the hypothesis of this work (no unstable
and no neutral modes for (\ref{eq:Linearized_Euler_Fourier_Vraiment})),
any perturbation corresponding to a small vorticity will converge
at large times towards a new parallel velocity profile which is close
to the initial profile $U$. A possible theorem expressing this more
precisely this would require a detailed analysis of subsequent terms
in an asymptotic expansion for small $\epsilon$, in a similar way
to the results recently obtained by Mouhot and Villani \cite{Mouhot_Villani:2009},
for the Landau damping in the very close setup of the Vlasov equation.
A proof of such a theorem for the Euler equations is not known yet,
even in the simplest case of a profile $U$ without stationary points.

On the basis of the previous discussion, a further conjecture would
be that the ensemble of shear flows without unstable nor neutral modes
for (\ref{eq:Linearized_Euler_Fourier_Vraiment}) is asymptotically
stable %
\footnote{We think here to the notion of asymptotic stability of an ensemble
of stationary solutions of an infinite
dimensional Hamiltonian equations, see for example the work \cite{Pego_Weinstein_1994CMaPh.164..305P}
where stable solutions slightly perturbed are proved to converge for
large times towards another slightly different solution. Asymptotic
stability has been proved for other solutions of infinite dimensional
Hamiltonian systems.%
} in the sense given previously (initial perturbation controlled by
a vorticity norm, for instance the enstrophy and large time perturbation
controlled in kinetic energy norm) %
\footnote{A classical argument, presented in a rigorous framework by Caglioti
and Maffei \cite{Caglioti_Maffei_1998} in the context of the Vlasov
equation implies that stationary solutions to the Vlasov
equation for which Landau damping would occur, would be unstable in
a weak norm.\emph{ }At the core of the argument lies the time reversal
symmetry of the equations. These arguments would easily generalize
to the Euler equations. This may seem in contradiction with the notion
of asymptotic stability. However the notion of stability discussed
by Caglioti and Maffei involves weak topology for both the initial
conditions and final state. There is no contradiction with our definition
of asymptotic stability, as we control here the initial perturbation
in a vorticity norm and control the convergence in a velocity norm.%
}.

\newpage

\section{Conclusion and perspectives}

Statistical mechanics of two dimensional and geophysical flows brings
a new perspective to the study of the self-organization of turbulent
flows. It is complimentary to other studies, based on fluid dynamics,
non linear dynamics and numerical computations. The successes in modeling
Jupiter's troposphere or some aspects of the ocean vortices using
the drastic simplification provided by statistical mechanics concepts
is very encouraging. One of the current aim is to develop the theory
in order extend the range of validity and relevance of the approach,
for instance in order to address further problems in geophysical turbulence.
\\

The equilibrium statistical mechanics theory of the 2D Euler and quasi-geostrophic equations is still a actively
developing field. Recent results, not presented in this review,
include a classification of phase transitions and of ensemble inequivalence
\cite{Bouchet_Barre:2005_JSP}, which of important practical interest,
and extension of the results of stability of stationary solutions
\cite{EllisHavenTurkington:2002_Nonlinearity_Stability}. A complete
theory of phase transitions, specifically addressing the specificity
of two dimensional and quasi-geostrophic turbulence is however still
lacking.

Equilibrium statistical mechanics has also been extended recently
to magneto-hydrodynamics equations \cite{Jordan_Turkington_1997,LeprovostDubrulleChavanis05}, non-linear Schroedinger equations \cite{1999chao.dyn..5023J,2004CMaPh.244..187E}, the Shallow-Water model \cite{Chavanis_Sommeria_2002_Shallow_Water_PhRvE}, or models of axisymmetric turbulence \cite{Leprovost_Dubrule_2006PhRvE..73d6308L,Monchaux_Ravelet_Dubrulle_etal_2006PhRvL..96l4502M,NasoMonchauxChavanisDubrulle}. In experimental statistically stationary forced and dissipated turbulence, comparison of the PDF of velocity or vorticity fluctuations, with prediction from equilibrium statistical mechanics, is discussed in \cite{Jung_Morrison_2006JFM...554..433J} or in \cite{Leprovost_Dubrule_2006PhRvE..73d6308L,Monchaux_Ravelet_Dubrulle_etal_2006PhRvL..96l4502M,NasoMonchauxChavanisDubrulle}.

From a theoretical point of view, these new applications usually have
not the same level of rigor as the statistical mechanics of the 2D
Euler and quasi-geostrophic equations. More precisely, the classical
program of equilibrium statistical mechanics: building from the Liouville
theorem the natural invariant measure of the dynamical equations --- the
microcanonical measure ---, and being able to compute the real entropy
corresponding to the phase space volume, is not achieved for these
models. In some of the works cited above, some very natural hypothesis are made,
that will probably be proved to be true in the future; in some others
ad-hoc fixes are proposed the logic of which seems sometimes not to be based on
clear principles. Still some of the results are quite interesting
and lead to very appealing applications. It is thus a very exiting
challenge to try to develop the theory for the equilibrium statistical
mechanics of these models, in order to understand the validity or
not of the previous approaches, and to obtain more physical insights.
It is also essential to assess the limits of the range of validity
of equilibrium statistical mechanics, also for other models of interest
for geophysical flows.

As we explained is this review, equilibrium statistical mechanics
may give, in some specific circumstances, interesting results for
actual non-equilibrium flows. We discussed the examples of Jupiter's
troposphere where it exists a large separation between the time scales
for the inertial (Hamiltonian) and non-inertial aspects of the dynamics
(forces and dissipation) or for instance for ocean rings where the
dynamical process of their formation is extremely rapid. A large part
of the range of interest of equilibrium statistical mechanics, in the
laboratory or for geophysical flows, has still to be studied, and many
progresses shall be made in this direction in future works.\\

For many applications, a non-equilibrium statistical mechanics is
required. We discussed in the last section recent progresses for the
study of the relaxation towards stationary solutions of the 2D Euler
equations and recent progresses towards a kinetic description
of the 2D Navier Stokes equations with weak forces and dissipation.
This is a promising field of research, where theoretical physics and
mathematical results are foreseen in a near future. This type of works
is essential to explain the large scale organization of geophysical
turbulence.

Other approaches for non-equilibrium statistical mechanics, like linear
response theory, large deviations or path integral representations of stochastic processes
will probably be part of future theories for turbulent flows.

\section*{Acknowledgments}

We warmly thank M. Corvellec and S. Griffies for their comments about the manuscript.

This work was supported through the ANR program STATFLOW (ANR-06-JCJC-0037-01) and through the ANR program STATOCEAN (ANR-09-SYSC-014).

Freddy Bouchet thanks CNLS and LANL in Los Alamos for hosting him during most of the writing of this review.

\newpage

\bibliographystyle{elsart-num-sort}
\bibliography{./bib/FBouchet,./bib/Long_Range,./bib/Meca_Stat_Euler,./bib/Ocean,./bib/Experimental_2D_Flows,./bib/Euler_Stability,./bib/Jupiter,./bib/Turbulence_2D,./bib/Quasilinear,./bib/Cascade,./bib/Euler2D-Linearized,./bib/rings,./bib/Statistical-Mechanics,./bib/NS-Stochastic,./bib/Kinetic-Theories-Turbulence,./bib/FBouchet-Proceedings,./bib/FBouchet-Books,./bib/Kuroshio,./bib/Jets-QuasiGeostrophic}

\begin{thebibliography}{100}
\expandafter\ifx\csname url\endcsname\relax
  \def\url#1{\texttt{#1}}\fi
\expandafter\ifx\csname urlprefix\endcsname\relax\def\urlprefix{URL }\fi

\bibitem{berloff}
http://www.whoi.edu/science/PO/people/pberloff/.

\bibitem{Abramov_Majda_2003_PNAS}
R.~V. {Abramov}, A.~J. {Majda}, {Statistically relevant conserved quantities
  for truncated quasigeostrophic flow}, Proceedings of the National Academy of
  Science 100 (2003) 3841--3846.

\bibitem{Eyink_Aluie_II_2009PhFl...21k5108A}
H.~{Aluie}, G.~L. {Eyink}, {Localness of energy cascade in hydrodynamic
  turbulence. II. Sharp spectral filter}, Physics of Fluids 21~(11) (2009)
  115108.

\bibitem{IUTAM_Symposium08}
H.~{Aref}, {150 Years of vortex dynamics}, Theoretical and Computational Fluid
  Dynamics 24 (2010) 1--7.

\bibitem{Arnold_1966}
V.~I. {Arnold}, {On an a-priori estimate in the theory of hydrodynamic
  stability}, Izv. Vyssh. Uchebbn. Zaved. Matematika; Engl. transl.: Am. Math.
  Soc. Trans. 79~(2) (1966) 267--269.

\bibitem{Barnier06}
B.~{Barnier}, G.~{Madec}, T.~{Penduff}, J.~{Molines}, A.~{Treguier}, J.~{Le
  Sommer}, A.~{Beckmann}, A.~{Biastoch}, C.~{B{\"o}ning}, J.~{Dengg},
  C.~{Derval}, E.~{Durand}, S.~{Gulev}, E.~{Remy}, C.~{Talandier},
  S.~{Theetten}, M.~{Maltrud}, J.~{McClean}, B.~{de Cuevas}, {Impact of partial
  steps and momentum advection schemes in a global ocean circulation model at
  eddy-permitting resolution}, Ocean Dynamics 56 (2006) 543--567.

\bibitem{BerloffHogg}
P.~{Berloff}, A.~M. {Hogg}, W.~{Dewar}, {The Turbulent Oscillator: A Mechanism
  of Low-Frequency Variability of the Wind-Driven Ocean Gyres}, Journal of
  Physical Oceanography 37 (2007) 2363.

\bibitem{Bernard_Boffetta_Celani_Falkovich_2006Nature}
D.~{Bernard}, G.~{Boffetta}, A.~{Celani}, G.~{Falkovich}, {Conformal invariance
  in two-dimensional turbulence}, Nature Physics 2 (2006) 124--128.

\bibitem{Bernard_Boffetta_Celani_Falkovich_2007PRL}
D.~{Bernard}, G.~{Boffetta}, A.~{Celani}, G.~{Falkovich}, {Inverse Turbulent
  Cascades and Conformally Invariant Curves}, Physi. Rev. Lett. 98~(2) (2007)
  024501.

\bibitem{Binney_Tremaine_1987_Galactic_Dynamics}
J.~{Binney}, S.~{Tremaine}, {Galactic dynamics}, Princeton, NJ, Princeton
  University Press, 1987, 747 p., 1987.

\bibitem{Biryuk06}
A.~{Biryuk}, {On Invariant Measures of the 2D Euler Equation}, Journal of
  Statistical Physics 122 (2006) 597--616.

\bibitem{Boucher_Ellis_1999_AP}
C.~{Boucher}, R.~S. {Ellis}, B.~{Turkington}, { Spatializing Random Measures:
  Doubly Indexed Processes and the Large Deviation Principle }, Annals Prob. 27
  (1999) 297--324.

\bibitem{Bouchet_These}
F.~Bouchet, M\'ecanique statistique des \'ecoulements g\'eophysiques, PHD,
  Universit\'e Joseph Fourier-Grenoble, 2001.

\bibitem{Bouchet_2003_condmat}
F.~Bouchet, Parameterization of two dimensionnal turbulence using an
  anisotropic maximum entropy principle, cond-mat/0305205.

\bibitem{Bouchet:2008_Physica_D}
F.~{Bouchet}, Simpler variational problems for statistical equilibria of the 2d
  euler equation and other systems with long range interactions, Physica D
  Nonlinear Phenomena 237 (2008) 1976--1981.

\bibitem{Bouchet_Barre:2005_JSP}
F.~{Bouchet}, J.~{Barr{\'e}}, {Classification of Phase Transitions and Ensemble
  Inequivalence, in Systems with Long Range Interactions}, Journal of
  Statistical Physics 118 (2005) 1073--1105.

\bibitem{Bouchet_Barre_Venaille_2008_Proceeding_Assise}
F.~{Bouchet}, J.~{Barre}, A.~{Venaille}, Equilibrium and out of equilibrium
  phase transitions in systems with long range interactions and in 2d flows,
  in: A.~{Campa}, A.~{Giansanti}, G.~{Morigi}, F.~S. {Labini} (eds.), Dynamics
  and Thermodynamics of Systems with Long Range Interactions: Theory and
  Experiments, vol. 970 of American Institute of Physics Conference Series,
  2008, pp. 117--152.

\bibitem{Bouchet_Chavanis_Sommeria_2010_SW}
F.~{Bouchet}, P.~H. {Chavanis}, J.~{Sommeria}, {Statistical mechanics of
  Jupiter's Great Red Spot in the shallow water model}, Preprint, to be
  submitted.

\bibitem{Bouchet_Corvellec_JSTAT_2010}
F.~{Bouchet}, M.~{Corvellec}, {Invariant measures of the 2D Euler and Vlasov
  equations}, Journal of Statistical Mechanics: Theory and Experiment 8 (2010)
  P08021.

\bibitem{Bouchet_Dauxois:2005_PRE}
F.~{Bouchet}, T.~{Dauxois}, {Prediction of anomalous diffusion and algebraic
  relaxations for long-range interacting systems, using classical statistical
  mechanics}, Phys. Rev. E 72~(4) (2005) 045103.

\bibitem{Bouchet_Dumont_2003_condmat}
F.~Bouchet, T.~Dumont, Emergence of the great red spot of jupiter from random
  initial conditions, cond-mat/0305206.

\bibitem{Bouchet_Gupta_Mukamel_PRL_2009}
F.~{Bouchet}, S.~{Gupta}, D.~{Mukamel}, {Thermodynamics and dynamics of systems
  with long-range interactions}, Physica A (2010) 4389--4405.

\bibitem{Bouchet_Morita_2010PhyD}
F.~{Bouchet}, H.~{Morita}, {Large time behavior and asymptotic stability of the
  2D Euler and linearized Euler equations}, Physica D Nonlinear Phenomena 239
  (2010) 948--966.

\bibitem{Bouchet_Simonnet_2008}
F.~{Bouchet}, E.~{Simonnet}, {Random Changes of Flow Topology in
  Two-Dimensional and Geophysical Turbulence}, Physical Review Letters 102~(9)
  (2009) 094504.

\bibitem{Bouchet_Sommeria:2002_JFM}
F.~{Bouchet}, J.~{Sommeria}, {Emergence of intense jets and Jupiter's Great Red
  Spot as maximum-entropy structures}, Journal of Fluid Mechanics 464 (2002)
  165--207.

\bibitem{Brands_Chavanis_etc_1999PhFl...11.3465B}
H.~{Brands}, P.~H. {Chavanis}, R.~{Pasmanter}, J.~{Sommeria}, {Maximum entropy
  versus minimum enstrophy vortices}, Physics of Fluids 11 (1999) 3465--3477.

\bibitem{Brands_Stulemeyer_Pasmanter_1997PhFl....9.2465B}
H.~{Brands}, J.~{Stulemeyer}, R.~A. {Pasmanter}, T.~J. {Schep}, {A mean field
  prediction of the asymptotic state of decaying 2D turbulence}, Physics of
  Fluids 9 (1997) 2465--2467.

\bibitem{BrethertonHaidvogel}
F.~P. {Bretherton}, D.~B. {Haidvogel}, {Two-dimensional turbulence above
  topography}, Journal of Fluid Mechanics 78 (1976) 129--154.

\bibitem{Bricmont_Kupianen_2001_Comm_Math_Phys_Ergodicity2DNavierStokes}
J.~{Bricmont}, A.~{Kupiainen}, R.~{Lefevere}, {Ergodicity of the 2D
  Navier-Stokes Equations with Random Forcing}, Com.. Math. Phys. 224 (2001)
  65--81.

\bibitem{Briggs_BDL_1970_Phys_Fluids}
R.~J. {Briggs}, J.~D. {Daugherty}, R.~H. {Levy}, {Role of Landau Damping in
  Crossed-Field Electron Beams and Inviscid Shear Flow}, Physics of Fluids 13
  (1970) 421--432.

\bibitem{Brown_Stewartson_1980_JFM}
S.~N. {Brown}, K.~{Stewartson}, {On the algebraic decay of disturbances in a
  stratified linear shear flow}, Journal of Fluid Mechanics 100 (1980)
  811--816.

\bibitem{Bruneau_Kellay_2005PhRvE}
C.~H. {Bruneau}, H.~{Kellay}, {Experiments and direct numerical simulations of
  two-dimensional turbulence}, \pre 71~(4) (2005) 046305.

\bibitem{CagliotiLMP:1995_CMP_II(Inequivalence)}
E.~{Caglioti}, P.~L. {Lions}, C.~{Marchioro}, M.~{Pulvirenti}, {A special class
  of stationary flows for two-dimensional euler equations: A statistical
  mechanics description. Part II}, Commun. Math. Phys. 174 (1995) 229--260.

\bibitem{Caglioti_Maffei_1998}
E.~Caglioti, C.~Maffei, Time asymptotics for solutions of vlasov poisson
  equation in a circle., J. Stat. Phys. 92~(1) (1998) 301--323.

\bibitem{Caglioti_Rousset_2007_JStatPhys_QSS}
E.~{Caglioti}, F.~{Rousset}, {Quasi-Stationary States for Particle Systems in
  the Mean-Field Limit}, J. Stat. Phys. 129~(2) (2007) 241--263.

\bibitem{Callen_Thermodynamics_1985tait.book.....C}
H.~B. {Callen}, {Thermodynamics and an Introduction to Thermostatistics, 2nd
  Edition}, 1985.

\bibitem{Campa_Dauxois_Ruffo_Revues_2009_PhR...480...57C}
A.~{Campa}, T.~{Dauxois}, S.~{Ruffo}, {Statistical mechanics and dynamics of
  solvable models with long-range interactions}, Phys. Rep. 480 (2009) 57--159.

\bibitem{Campa_Giansanti_Morigi_Labini_2008AIPC..970.....C}
A.~{Campa}, A.~{Giansanti}, G.~{Morigi}, F.~S. {Labini} (eds.), {Dynamics and
  Thermodynamics of Systems with Long Range Interactions: Theory and
  Experiments}, vol. 970 of American Institute of Physics Conference Series,
  2008 (2008).

\bibitem{Capel_Pasmanter_2000PhFl...12.2514C}
H.~W. {Capel}, R.~A. {Pasmanter}, {Evolution of the vorticity-area density
  during the formation of coherent structures in two-dimensional flows},
  Physics of Fluids 12 (2000) 2514--2521.

\bibitem{Carnevale_Frederiksen_NLstab_statmech_topog_1987JFM}
G.~F. {Carnevale}, J.~S. {Frederiksen}, {Nonlinear stability and statistical
  mechanics of flow over topography}, Journal of Fluid Mechanics 175 (1987)
  157--181.

\bibitem{Case_1960_Phys_Fluids}
K.~M. {Case}, {Stability of Inviscid Plane Couette Flow}, Physics of Fluids 3
  (1960) 143--148.

\bibitem{Chavanis_Quasilinear_2000PhRvL}
P.~H. {Chavanis}, {Quasilinear Theory of the 2D Euler Equation}, Physical
  Review Letters 84 (2000) 5512--5515.

\bibitem{Chavanis_houches_2002}
P.~H. {Chavanis}, Statistical mechanis of two-dimensional vortices and stellar
  systems, in: T.~{Dauxois}, S.~{Ruffo}, E.~{Arimondo}, M.~{Wilkens} (eds.),
  Dynamics and Thermodynamics of Systems With Long Range Interactions, vol. 602
  of Lecture Notes in Physics, Springer-Verlag, 2002, pp. 208--289.

\bibitem{Chavanis_2006IJMPB_Revue_Auto_Gravitant}
P.~H. {Chavanis}, {Phase Transitions in Self-Gravitating Systems},
  International Journal of Modern Physics B 20 (2006) 3113--3198.

\bibitem{ChavanisEPJB2009}
P.~H. {Chavanis}, {Dynamical and thermodynamical stability of two-dimensional
  flows: variational principles and relaxation equations}, European Physical
  Journal B 70 (2009) 73--105.

\bibitem{Chavanis_Naso_Dubrulle_2010EPJB_Relaxation}
P.~H. {Chavanis}, A.~{Naso}, B.~{Dubrulle}, {Relaxation equations for
  two-dimensional turbulent flows with a prior vorticity distribution},
  European Physical Journal B 77 (2010) 167--186.

\bibitem{ChavanisSommeria:1996_JFM_Classification}
P.~H. {Chavanis}, J.~{Sommeria}, {Classification of self-organized vortices in
  two-dimensional turbulence: the case of a bounded domain}, J. Fluid Mech. 314
  (1996) 267--297.

\bibitem{ChavanisPRL97}
P.~H. {Chavanis}, J.~{Sommeria}, {Thermodynamical Approach for Small-Scale
  Parametrization in 2D Turbulence}, Physical Review Letters 78 (1997)
  3302--3305.

\bibitem{Chavanis_Sommeria_1998JFM_LocalizedEquilibria...356..259C}
P.~H. {Chavanis}, J.~{Sommeria}, {Classification of robust isolated vortices in
  two-dimensional hydrodynamics}, Journal of Fluid Mechanics 356 (1998)
  259--296.

\bibitem{Chavanis_Sommeria_2002_Shallow_Water_PhRvE}
P.~H. {Chavanis}, J.~{Sommeria}, {Statistical mechanics of the shallow water
  system}, Phys. Rev. E 65~(2) (2002) 026302.

\bibitem{Chavanis_etal_APJ_1996}
P.~H. {Chavanis}, J.~{Sommeria}, R.~{Robert}, {Statistical Mechanics of
  Two-dimensional Vortices and Collisionless Stellar Systems}, Astro. Phys.
  Jour. 471 (1996) 385.

\bibitem{Chelton07}
D.~B. {Chelton}, M.~G. {Schlax}, R.~M. {Samelson}, R.~A. {de Szoeke}, {Global
  observations of large oceanic eddies}, \grl 34 (2007) 15606.

\bibitem{Chen_Cross_1994PhRvE}
P.~{Chen}, M.~C. {Cross}, {Phase diagram for coherent vortex formation in the
  two-dimensional inviscid fluid in circular geometries}, \pre 50 (1994)
  2022--2029.

\bibitem{Chen_Cross_1996PhRv}
P.~{Chen}, M.~C. {Cross}, {Mean field equilibria of single coherent vortices},
  \pre 54 (1996) 6356--6363.

\bibitem{Chen_Cross_1996_PhRvL}
P.~{Chen}, M.~C. {Cross}, {Mixing and Thermal Equilibrium in the Dynamical
  Relaxation of a Vortex Ring}, Physical Review Letters 77 (1996) 4174--4177.

\bibitem{Chertkov_Connaughton_andco_2007_PRL_EnergyCondesation}
M.~{Chertkov}, C.~{Connaughton}, I.~{Kolokolov}, V.~{Lebedev}, {Dynamics of
  Energy Condensation in Two-Dimensional Turbulence}, Phys. Rev. Lett. 99
  (2007) 084501.

\bibitem{Clercx_Massen_VanHeijst_1999PhFl...11..611C}
H.~J.~H. {Clercx}, S.~R. {Maassen}, G.~J.~F. {van Heijst}, {Decaying
  two-dimensional turbulence in square containers with no-slip or stress-free
  boundaries}, Physics of Fluids 11 (1999) 611--626.

\bibitem{Danilov_Gurarie_2001PhRvE..63b0203D}
S.~{Danilov}, D.~{Gurarie}, {Nonuniversal features of forced two-dimensional
  turbulence in the energy range}, \pre 63~(2) (2001) 020203.

\bibitem{Dauxois_Ruffo_Arimondo_Wilkens_2002LNP...602....1D}
T.~{Dauxois}, S.~{Ruffo}, E.~{Arimondo}, M.~{Wilkens}, Dynamics and
  Thermodynamics of Systems with Long-Range Interactions, vol. 602 of Lecture
  Notes in Physics, Berlin Springer Verlag, 2002.

\bibitem{DibattistaMajda00}
M.~T. {Dibattista}, A.~J. {Majda}, {An Equilibrium Statistical Theory for
  Large-Scale Features of Open-Ocean Convection}, Journal of Physical
  Oceanography 30 (2000) 1325--1353.

\bibitem{DibattistaMajda02}
M.~T. {Dibattista}, A.~J. {Majda}, J.~{Marshall}, {A Statistical Theory for the
  ``Patchiness'' of Open-Ocean Deep Convection: The Effect of Preconditioning},
  Journal of Physical Oceanography 32 (2002) 599--626.

\bibitem{Dowling_Review_1995AnRFM..27..293D}
T.~E. {Dowling}, {Dynamics of jovian atmospheres}, Annual Review of Fluid
  Mechanics 27 (1995) 293--334.

\bibitem{Dowling_Ingersoll_1988JAtS...45.1380D}
T.~E. {Dowling}, A.~P. {Ingersoll}, {Potential vorticity and layer thickness
  variations in the flow around Jupiter's Great Red SPOT and White Oval BC},
  Journal of Atmospheric Sciences 45 (1988) 1380--1396.

\bibitem{Drazin_Reid_1981}
P.~G. {Drazin}, W.~H. {Reid}, {Hydrodynamic stability}, Cambridge university
  press, 2004, second edition.

\bibitem{Dritschel_McIntyre_2008JAtS}
D.~G. {Dritschel}, M.~E. {McIntyre}, {Multiple Jets as PV Staircases: The
  Phillips Effect and the Resilience of Eddy-Transport Barriers}, Journal of
  Atmospheric Sciences 65 (2008) 855.

\bibitem{Dubin_ONeil_1988_PhysRevLett_Kinetic_Point_Vortex}
D.~H.~E. Dubin, T.~M. O\char39{}Neil, Two-dimensional guiding-center transport
  of a pure electron plasma, Phys. Rev. Lett. 60~(13) (1988) 1286--1289.

\bibitem{Dubinkina_Frank_2010JCoPh}
S.~{Dubinkina}, J.~{Frank}, {Statistical relevance of vorticity conservation in
  the Hamiltonian particle-mesh method}, Journal of Computational Physics 229
  (2010) 2634--2648.

\bibitem{Dubrulle_Nazarenko_1997PhyD}
B.~{Dubrulle}, S.~{Nazarenko}, {Interaction of turbulence and large-scale
  vortices in incompressible 2D fluids}, Physica D 110 (1997) 123--138.

\bibitem{EllisHavenTurkington:2000_Inequivalence}
R.~S. {Ellis}, K.~{Haven}, B.~{Turkington}, {Large Deviation Principles and
  Complete Equivalence and Nonequivalence Results for Pure and Mixed
  Ensembles}, J. Stat. Phys. 101 (2000) 999.

\bibitem{EllisHavenTurkington:2002_Nonlinearity_Stability}
R.~S. {Ellis}, K.~{Haven}, B.~{Turkington}, {Nonequivalent statistical
  equilibrium ensembles and refined stability theorems for most probable flows
  }, Nonlinearity 15 (2002) 239--255.

\bibitem{2004CMaPh.244..187E}
R.~S. {Ellis}, R.~{Jordan}, P.~{Otto}, B.~{Turkington}, {A Statistical Approach
  to the Asymptotic Behavior of a Class of Generalized Nonlinear
  Schr{\"o}dinger Equations}, Communications in Mathematical Physics 244 (2004)
  187--208.

\bibitem{Esler08}
J.~G. {Esler}, {The turbulent equilibration of an unstable baroclinic jet},
  Journal of Fluid Mechanics 599 (2008) 241--268.

\bibitem{Eyink_Locality_2005PhyD..207...91E}
G.~L. {Eyink}, {Locality of turbulent cascades}, Physica D Nonlinear Phenomena
  207 (2005) 91--116.

\bibitem{Eyink_Aluie_2009PhFl...21k5107E}
G.~L. {Eyink}, H.~{Aluie}, {Localness of energy cascade in hydrodynamic
  turbulence. I. Smooth coarse graining}, Physics of Fluids 21~(11) (2009)
  115107.

\bibitem{Eyink_Spohn_1993_JSP....70..833E}
G.~L. {Eyink}, H.~{Spohn}, {Negative-temperature states and large-scale,
  long-lived vortices in two-dimensional turbulence}, Journal of Statistical
  Physics 70 (1993) 833--886.

\bibitem{Eyink_Sreenivasan_2006_Rev_Modern_Physics}
G.~L. {Eyink}, K.~R. {Sreenivasan}, {Onsager and the theory of hydrodynamic
  turbulence}, Rev. Mod. Phys. 78 (2006) 87--135.

\bibitem{Farrell_Ioannou_JAS_2007}
B.~F. {Farrell}, P.~J. {Ioannou}, {Structure and Spacing of Jets in Barotropic
  Turbulence}, Journal of Atmospheric Sciences 64 (2007) 3652.

\bibitem{Farrel_Ioannou_2009JAtS}
B.~F. {Farrell}, P.~J. {Ioannou}, {A Theory of Baroclinic Turbulence}, Journal
  of Atmospheric Sciences 66 (2009) 2444.

\bibitem{Fjortoft53}
R.~{Fjortoft}, {On the changes in the spectral distribution of kinetic energy
  for two-dimensional nondivergent flow}, Tellus 5 (1953) 225--230.

\bibitem{Fofonoff:1954_steady_flow_frictionless}
N.~P. Fofonoff, Steady flow in a frictionless homogeneous ocean., J. Mar. Res.
  13 (1954) 254--262.

\bibitem{fredericksen91GAFDa}
J.~S. {Frederiksen}, {Nonlinear studies on the effects of topography on
  baroclinic zonal flows}, Geophysical and Astrophysical Fluid Dynamics 59
  (1991) 57--82.

\bibitem{FrederiksenOKane08}
J.~S. {Frederiksen}, T.~J. {O'Kane}, {Entropy, Closures and Subgrid Modeling},
  Entropy 10 (2008) 635--683.

\bibitem{Frisch_1996_Book}
U.~{Frisch}, {Turbulence}, Turbulence, by Uriel Frisch, pp.~310.~ISBN
  0521457130.~Cambridge, UK: Cambridge University Press, January 1996., 1996.

\bibitem{FrishKurien08}
U.~Frisch, S.~Kurien, R.~Pandit, W.~Pauls, S.~S. Ray, A.~Wirth, J.-Z. Zhu,
  Hyperviscosity, galerkin truncation, and bottlenecks in turbulence, Phys.
  Rev. Lett. 101~(14) (2008) 144501.

\bibitem{GillBook}
A.~E. {Gill}, {Atmosphere-Ocean Dynamics}, 1982.

\bibitem{GGS74}
A.~E. {Gill}, J.~S.~A. {Green}, A.~{Simmons}, {Energy partition in the
  large-scale ocean circulation and the production of mid-ocean eddies},
  Deep-Sea Research 21 (1974) 499--528.

\bibitem{Goldstein_Lebowitz_2004PhyD..193...53G}
S.~{Goldstein}, J.~L. {Lebowitz}, {On the (Boltzmann) entropy of
  non-equilibrium systems}, Physica D 193 (2004) 53.

\bibitem{Grote_Majda_CrudeClosure_1997PhFl}
M.~J. {Grote}, A.~J. {Majda}, {Crude closure dynamics through large scale
  statistical theories}, Physics of Fluids 9 (1997) 3431--3442.

\bibitem{Grote_Majda_2000Nonli}
M.~J. {Grote}, A.~J. {Majda}, {Crude closure for flow with topography through
  large-scale statistical theory}, Nonlinearity 13 (2000) 569--600.

\bibitem{HallbergMESO_06}
R.~{Hallberg}, A.~{Gnanadesikan}, {The Role of Eddies in Determining the
  Structure and Response of the Wind-Driven Southern Hemisphere Overturning:
  Results from the Modeling Eddies in the Southern Ocean (MESO) Project},
  Journal of Physical Oceanography 36 (2006) 2232.

\bibitem{Held_Larichev_1996JAtS...53..946H}
I.~M. {Held}, V.~D. {Larichev}, {A Scaling Theory for Horizontally Homogeneous,
  Baroclinically Unstable Flow on a Beta Plane.}, Journal of Atmospheric
  Sciences 53 (1996) 946--952.

\bibitem{Hertel_Thirring_1971_AnnPhys}
W.~{Hertel}, P.~{Thirring}, {Soluble model for a system with negative specific
  heat}, Annals Phys. 63 (1971) 520--533.

\bibitem{HoggStommel85}
N.~G. {Hogg}, H.~M. {Stommel}, {The Heton, an Elementary Interaction Between
  Discrete Baroclinic Geostrophic Vortices, and Its Implications Concerning
  Eddy Heat-Flow}, Royal Society of London Proceedings Series A 397.

\bibitem{HollowayReview04}
G.~{Holloway}, {From Classical To Statistical Ocean Dynamics}, Surveys in
  Geophysics 25 (2004) 203--219.

\bibitem{Holm_etal_PhysRep_1985}
D.~D. {Holm}, J.~E. {Marsden}, T.~{Ratiu}, A.~{Weinstein}, {Nonlinear stability
  of fluid and plasma equilibria}, Phys. Rep. 123 (1985) 1--2.

\bibitem{Holm_Marsden_Ratiu_1998_EulerPoincare}
D.~D. {Holm}, J.~E. {Marsden}, T.~S. {Ratiu}, {The Euler-Poincare Equations and
  Semidirect Products with Applications to Continuum Theories}, in: eprint
  arXiv:chao-dyn/9801015, 1998, p. 1015.

\bibitem{Ingersoll_Vasavada_1998IAUSS...1.1042I}
A.~P. {Ingersoll}, A.~R. {Vasavada}, {Dynamics of Jupiter's atmosphere.}, IAU
  Special Session 1 (1998) 1042--1049.

\bibitem{Isichenko_1997_PhRvL}
M.~B. {Isichenko}, {Nonlinear Landau Damping in Collisionless Plasma and
  Inviscid Fluid}, Physical Review Letters 78 (1997) 2369--2372.

\bibitem{JinDubin98}
D.~Z. {Jin}, D.~H.~E. {Dubin}, {Regional Maximum Entropy Theory of Vortex
  Crystal Formation}, Physical Review Letters 80 (1998) 4434--4437.

\bibitem{Jin_Dubin_2000PhRvL}
D.~Z. {Jin}, D.~H.~E. {Dubin}, {Characteristics of Two-Dimensional Turbulence
  That Self-Organizes into Vortex Crystals}, Phys. Rev. Lett. 84 (2000)
  1443--1446.

\bibitem{1999chao.dyn..5023J}
R.~{Jordan}, C.~{Josserand}, {Self-organization in nonlinear wave turbulence},
  in: eprint arXiv:chao-dyn/9905023, 1999, p. 5023.

\bibitem{Jordan_Turkington_1997}
R.~{Jordan}, B.~{Turkington}, {Statistical mechanics of organized structures in
  two-dimensional magnetofluid turbulence}, vol.~30, 1997, pp. 3629--3636.

\bibitem{Joyce_Montgommery_1973}
G.~{Joyce}, D.~{Montgomery}, {Negative temperature states for the
  two-dimensional guiding-centre plasma}, Journal of Plasma Physics 10 (1973)
  107.

\bibitem{Jung_Morrison_2006JFM...554..433J}
S.~{Jung}, P.~J. {Morrison}, H.~L. {Swinney}, {Statistical mechanics of
  two-dimensional turbulence}, Journal of Fluid Mechanics 554 (2006) 433--456.

\bibitem{Jutner_Thess_Sommeria_1995PhFl}
B.~{J{\"u}ttner}, A.~{Thess}, J.~{Sommeria}, {On the symmetry of self-organized
  structures in two-dimensional turbulence}, Physics of Fluids 7 (1995)
  2108--2110.

\bibitem{Kazantsev_Sommeria_Verron_1998JPO....28.1017K}
E.~{Kazantsev}, J.~{Sommeria}, J.~{Verron}, {Subgrid-Scale Eddy
  Parameterization by Statistical Mechanics in a Barotropic Ocean Model},
  Journal of Physical Oceanography 28 (1998) 1017--1042.

\bibitem{Kellay_Glodburg_2002_Rep_Prog_Phsyics}
H.~{Kellay}, W.~I. {Goldburg}, {Two-dimensional turbulence: a review of some
  recent experiments }, Rep. Prog. Phys. 65 (2002) 845--894.

\bibitem{Kiessling_2008AIPC}
M.~{Kiessling}, {Statistical equilibrium dynamics}, in: {A.~Campa,
  A.~Giansanti, G.~Morigi, \& F.~S.~Labini} (ed.), Dynamics and Thermodynamics
  of Systems with Long Range Interactions: Theory and Experiments, vol. 970 of
  American Institute of Physics Conference Series, 2008, pp. 91--108.

\bibitem{Kiessling_Lebowitz_1997_PointVortex_Inequivalence_LMathPhys}
M.~K.~H. {Kiessling}, J.~L. {Lebowitz}, {The Micro-Canonical Point Vortex
  Ensemble: Beyond Equivalence}, Lett. Math. Phys. 42~(1) (1997) 43--56.

\bibitem{KossinSchubert}
J.~P. {Kossin}, W.~H. {Schubert}, {Mesovortices, Polygonal Flow Patterns, and
  Rapid Pressure Falls in Hurricane-Like Vortices.}, Journal of Atmospheric
  Sciences 58 (2001) 2196--2209.

\bibitem{Kraichnan_1967PhFl...10.1417K}
R.~H. {Kraichnan}, {Inertial Ranges in Two-Dimensional Turbulence}, Physics of
  Fluids 10 (1967) 1417--1423.

\bibitem{Kraichnan_Phys_Fluid_1967_2Dturbulence}
R.~H. {Kraichnan}, {Inertial Ranges in Two-Dimensional Turbulence}, Phys.
  Fluids 10 (1967) 1417.

\bibitem{Kraichnan_Motgommery_1980_Reports_Progress_Physics}
R.~H. {Kraichnan}, D.~{Montgomery}, {Two-dimensional turbulence}, Reports on
  Progress in Physics 43 (1980) 547--619.

\bibitem{Kuksin_Penrose_2005_JPhysStat_BalanceRelations}
S.~{Kuksin}, .~{Penrose}, {A family of balance relations for the
  two-dimensional Navier-Stokes equations with random forcing}, J. Stat. Phys.
  118~(3-4) (2005) 437--449.

\bibitem{Kuksin_Shirikyan_2000_CMaPh}
S.~{Kuksin}, A.~{Shirikyan}, {Stochastic Dissipative PDE's and Gibbs Measures},
  Communications in Mathematical Physics 213 (2000) 291--330.

\bibitem{Kuksin_2004_JStatPhys_EulerianLimit}
S.~B. {Kuksin}, {The eulerian limit for 2D statistical hydrodynamics}, J. Stat.
  Phys. 115 (2004) 469--492.

\bibitem{Landau_Lifshitz_1996_Book}
L.~D. {Landau}, E.~M. {Lifshitz}, {Statistical Physics. Vol. 5 of the Course of
  Theoretical Physics}, Pergamon Press, 1980.

\bibitem{Lapeyre_Held_2003JAtS}
G.~{Lapeyre}, I.~M. {Held}, {Diffusivity, Kinetic Energy Dissipation, and
  Closure Theories for the Poleward Eddy Heat Flux.}, Journal of Atmospheric
  Sciences 60 (2003) 2907--2916.

\bibitem{Laval_Dubrulle_Nazarenko_2000PhyD}
J.~{Laval}, B.~{Dubrulle}, S.~{Nazarenko}, {Dynamical modeling of sub-grid
  scales in 2D turbulence}, Physica D Nonlinear Phenomena 142 (2000) 231--253.

\bibitem{Lee52}
Lee, On some statitical properties of hydrodynamical and magnetohydrodynamical
  fields, Quart. Appl. Math 10~(69).

\bibitem{LeprovostDubrulleChavanis05}
N.~{Leprovost}, B.~{Dubrulle}, P.~H. {Chavanis}, {Thermodynamics of
  magnetohydrodynamic flows with axial symmetry}, Physical Review E 71~(3)
  (2005) 036311.

\bibitem{Leprovost_Dubrule_2006PhRvE..73d6308L}
N.~{Leprovost}, B.~{Dubrulle}, P.~H. {Chavanis}, {Dynamics and thermodynamics
  of axisymmetric flows: Theory}, \pre 73~(4) (2006) 046308.

\bibitem{Lifshitz_Pitaevskii_1981_Physical_Kinetics}
E.~M. {Lifshitz}, L.~P. {Pitaevskii}, {Physical kinetics}, Course of
  theoretical physics, Oxford: Pergamon Press, 1981, 1981.

\bibitem{Lilly_1972GApFD...3..289L}
D.~K. {Lilly}, {Numerical simulation studies of two-Dimensional turbulence: I.
  Models of statistically steady turbulence}, Geophysical and Astrophysical
  Fluid Dynamics 3 (1972) 289--319.

\bibitem{Lim_2001PhFl...13.1961L}
C.~C. {Lim}, {A long range spherical model and exact solutions of an energy
  enstrophy theory for two-dimensional turbulence}, Physics of Fluids 13 (2001)
  1961--1973.

\bibitem{Lim_SinghMavi_2007PhyA}
C.~C. {Lim}, R.~{Singh Mavi}, {Phase transitions of barotropic flow coupled to
  a massive rotating sphere. Derivation of a fixed point equation by the Bragg
  method}, Physica A Statistical Mechanics and its Applications 380 (2007)
  43--60.

\bibitem{Lundgren_1982_PhFl}
T.~S. {Lundgren}, {Strained spiral vortex model for turbulent fine structure},
  Physics of Fluids 25 (1982) 2193--2203.

\bibitem{LyndenBell:1968_MNRAS}
D.~{Lynden-Bell}, R.~{Wood}, {The gravo-thermal catastrophe in isothermal
  spheres and the onset of red-giant structure for stellar systems}, Mon. Not.
  R. Astron. Soc. 138 (1968) 495.

\bibitem{Majda_Wang_Book_Geophysique_Stat}
A.~J. {Majda}, X.~{Wang}, {Nonlinear Dynamics and Statistical Theories for
  Basic Geophysical Flows}, Cambridge University Press, 2006.

\bibitem{Majda_Wang_Bombardement_2006_Comm}
A.~J. {Majda}, X.~{Wang}, {The emergence of large-scale coherent structure
  under small-scale random bombardments}, Comm. Pure App. Maths 59~(4) (2006)
  467--500.

\bibitem{1993ARA&A..31..523M}
P.~S. {Marcus}, {Jupiter's Great Red SPOT and other vortices}, Ann. Rev.
  Astron. Astrophys. 31 (1993) 523--573.

\bibitem{Marston_Conover_Schneider_JAS2008}
J.~B. {Marston}, E.~{Conover}, T.~{Schneider}, {Statistics of an Unstable
  Barotropic Jet from a Cumulant Expansion}, Journal of Atmospheric Sciences 65
  (2008) 1955.

\bibitem{Marteau_Cardoso_Tabeling_1995PhRvE}
D.~{Marteau}, O.~{Cardoso}, P.~{Tabeling}, {Equilibrium states of
  two-dimensional turbulence: An experimental study}, \pre 51 (1995)
  5124--5127.

\bibitem{Mattingly_Sinai_1999math_3042M}
J.~C. {Mattingly}, Y.~G. {Sinai}, {An Elementary Proof of the Existence and
  Uniqueness Theorem for the Navier-Stokes Equations}, ArXiv Mathematics
  e-prints:math/9903042.

\bibitem{Merryfield98JFM}
W.~J. {Merryfield}, {Effects of stratification on quasi-geostrophic inviscid
  equilibria}, Journal of Fluid Mechanics 354 (1998) 345--356.

\bibitem{Merryfield_Cummins_Holloway_2001JPO....31.1880M}
W.~J. {Merryfield}, P.~F. {Cummins}, G.~{Holloway}, {Equilibrium Statistical
  Mechanics of Barotropic Flow over Finite Topography}, Journal of Physical
  Oceanography 31 (2001) 1880--1890.

\bibitem{Michel_Robert_LargeDeviations1994CMaPh.159..195M}
J.~{Michel}, R.~{Robert}, {Large deviations for young measures and statistical
  mechanics of infinite dimensional dynamical systems with conservation law},
  Communications in Mathematical Physics 159 (1994) 195--215.

\bibitem{Michel_Robert_1994_JSP_GRS}
J.~{Michel}, R.~{Robert}, {Statistical mechanical theory of the great red spot
  of jupiter}, Journal of Statistical Physics 77 (1994) 645--666.

\bibitem{Miller:1990_PRL_Meca_Stat}
J.~Miller, Statistical mechanics of euler equations in two dimensions, Phys.
  Rev. Lett. 65~(17) (1990) 2137--2140.

\bibitem{Miller_Weichman_Cross_1992PhRvA}
J.~{Miller}, P.~B. {Weichman}, M.~C. {Cross}, {Statistical mechanics, Euler's
  equation, and Jupiter's Red Spot}, \pra 45 (1992) 2328--2359.

\bibitem{1987_Modica_ArchRatMechAna}
L.~{Modica}, {The gradient theory of phase transitions and the minimal
  interface criterion}, Archive for Rational Mechanics and Analysis 98 (1987)
  123--142.

\bibitem{Monchaux_Ravelet_Dubrulle_etal_2006PhRvL..96l4502M}
R.~{Monchaux}, F.~{Ravelet}, B.~{Dubrulle}, A.~{Chiffaudel}, F.~{Daviaud},
  {Properties of Steady States in Turbulent Axisymmetric Flows}, Physical
  Review Letters 96~(12) (2006) 124502.

\bibitem{Morita_Simonnet_Bouchet_2010_NS_Stochastic_Long}
H.~{Morita}, E.~{Simonnet}, F.~{Bouchet}, {Statistical properties of out of
  equilibrium phase transitions for the 2D Navier-Stokes equations}, Submitted
  to Phys. Rev. E.

\bibitem{Morrison_1998_HamiltonianFluid_RvMP}
P.~J. {Morrison}, {Hamiltonian description of the ideal fluid}, Reviews of
  Modern Physics 70 (1998) 467--521.

\bibitem{Mouhot_Villani:2009}
C.~{Mouhot}, C.~{Villani}, On the landau damping, arXiv 0904.2760.

\bibitem{NasoChavanisDubrulle2}
A.~{Naso}, P.~H. {Chavanis}, B.~{Dubrulle}, {Statistical mechanics of Fofonoff
  flows in an oceanic basin}, ArXiv e-prints:1007.0164.

\bibitem{NasoChavanisDubrulle}
A.~{Naso}, P.~H. {Chavanis}, B.~{Dubrulle}, {Statistical mechanics of
  two-dimensional Euler flows and minimum enstrophy states}, European Physical
  Journal B 77 (2010) 187--212.

\bibitem{NasoMonchauxChavanisDubrulle}
A.~{Naso}, R.~{Monchaux}, P.~H. {Chavanis}, B.~{Dubrulle}, {Statistical
  mechanics of Beltrami flows in axisymmetric geometry: Theory reexamined},
  Physical Review E 81~(6) (2010) 066318.

\bibitem{Nazarenko_PhysicsLetterA_2000}
S.~{Nazarenko}, {Exact solutions for near-wall turbulence theory}, Physics
  Letters A 264 (2000) 444--448.

\bibitem{Nazarenko_Laval_JFM_2000}
S.~{Nazarenko}, J.-P. {Laval}, {Non-local two-dimensional turbulence and
  Batchelor's regime for passive scalars}, J. Fluid Mech. 408 (2000) 301--321.

\bibitem{Nazarenko_Quinn_PRL_2009}
S.~{Nazarenko}, B.~{Quinn}, {Triple Cascade Behavior in Quasigeostrophic and
  Drift Turbulence and Generation of Zonal Jets}, Physical Review Letters
  103~(11) (2009) 118501.

\bibitem{Nicholson_1991}
D.~{Nicholson}, {Introduction to plasma theory}, {Wiley, New-York}, 1983.

\bibitem{Onsager:1949_Meca_Stat_Points_Vortex}
L.~{Onsager}, {Statistical hydrodynamics}, Nuovo Cimento 6 (No. 2 (Suppl.))
  (1949) 249--286.

\bibitem{ParetJullienTabeling99}
J.~{Paret}, M.~{Jullien}, P.~{Tabeling}, {Vorticity Statistics in the
  Two-Dimensional Enstrophy Cascade}, Physical Review Letters 83 (1999)
  3418--3421.

\bibitem{Paret_Tabeling_1998_PhysFluids}
J.~{Paret}, P.~{Tabeling}, {Intermittency in the two-dimensional inverse
  cascade of energy: Experimental observations}, Phys. Fluids 10 (1998)
  3126--3136.

\bibitem{PedloskyBook}
J.~{Pedlosky}, {Geophysical fluid dynamics}, 1982.

\bibitem{Pedlosky:1998_OceanCirculationTheory}
J.~{Pedlosky}, {Ocean Circulation Theory}, New York and Berlin,
  Springer-Verlag, 1998.

\bibitem{Pego_Weinstein_1994CMaPh.164..305P}
R.~L. {Pego}, M.~I. {Weinstein}, {Asymptotic stability of solitary waves},
  Communications in Mathematical Physics 164 (1994) 305--349.

\bibitem{Pomeau_Cargese_1995}
Y.~{Pomeau}, {Statistical approach (to 2D turbulence)}, in: P.~Tabeling,
  O.~Cardoso (eds.), Turbulence: A Tentative Dictionary, Plenum Press, New
  York, 1995, pp. 117--123.

\bibitem{Prieto_Schubert_2001JAtS}
R.~{Prieto}, W.~H. {Schubert}, {Analytical Predictions for Zonally Symmetric
  Equilibrium States of the Stratospheric Polar Vortex.}, Journal of
  Atmospheric Sciences 58 (2001) 2709--2728.

\bibitem{Qiu_Miao_2000JPO....30.2124Q}
B.~{Qiu}, W.~{Miao}, {Kuroshio Path Variations South of Japan: Bimodality as a
  Self-Sustained Internal Oscillation}, Journal of Physical Oceanography 30
  (2000) 2124--2137.

\bibitem{Ritz_Einstein_1909PhyZ}
W.~{Ritz}, A.~{Einstein}, {Zum gegenw{\"a}rtigen Stand des Strahlungsproblems},
  Physikalische Zeitschrift 10 (1909) 323--324.

\bibitem{Robert:1990_CRAS}
R.~{Robert}, {Etats d'\'equilibre statistique pour l'\'ecoulement
  bidimensionnel d'un fluide parfait}, C. R. Acad. Sci. 1 (1990) 311:575--578.

\bibitem{Robert:1991_JSP_Meca_Stat}
R.~{Robert}, {A maximum-entropy principle for two-dimensional perfect fluid
  dynamics}, J. Stat. Phys. 65 (1991) 531--553.

\bibitem{Robert_2000_CommMathPhys-TruncationEuler}
R.~{Robert}, {On the Statistical Mechanics of 2D Euler Equation},
  Communications in Mathematical Physics 212 (2000) 245--256.

\bibitem{Robert_Rosier_1997JSP....86..481R}
R.~{Robert}, C.~{Rosier}, {The modeling of small scales in two-dimensional
  turbulent flows: A statistical mechanics approach}, Journal of Statistical
  Physics 86 (1997) 481--515.

\bibitem{Robert_Rosier_2001NPGeo...8...55R}
R.~{Robert}, C.~{Rosier}, {Long range predictability of atmospheric flows},
  Nonlinear Processes in Geophysics 8 (2001) 55--67.

\bibitem{SommeriaRobert:1991_JFM_meca_Stat}
R.~{Robert}, J.~{Sommeria}, {Statistical equilibrium states for two-dimensional
  flows}, J. Fluid Mech. 229 (1991) 291--310.

\bibitem{RobertSommeria:1992_PRL_Relaxation_Meca_Stat}
R.~Robert, J.~Sommeria, Relaxation towards a statistical equilibrium state in
  two-dimensional perfect fluid dynamics, Phys. Rev. Lett. 69~(19) (1992)
  2776--2779.

\bibitem{Rosencrans_Sattinger_1966_J_Math_Phys}
S.~I. {Rosencrans}, D.~H. {Sattinger}, {On the spectrum of an operator occuring
  in the theory of hydrodynamic stability}, J. Math. Phys. 45 (1966) 289--300.

\bibitem{Salmon_1998_Book}
R.~{Salmon}, {Lectures on Geophysical Fluid Dynamics}, Oxford University Press,
  1998.

\bibitem{SalmonHollowayHendershott:1976_JFM_stat_mech_QG}
R.~{Salmon}, G.~{Holloway}, M.~C. {Hendershott}, {The equilibrium statistical
  mechanics of simple quasi-geostrophic models}, Journal of Fluid Mechanics 75
  (1976) 691--703.

\bibitem{Scecter_etal_2000_PhysicsFluids}
D.~A. {Schecter}, D.~H.~E. {Dubin}, A.~C. {Cass}, C.~F. {Driscoll}, I.~M.
  {Lansky}, T.~M. {O'Neil}, {Inviscid damping of asymmetries on a
  two-dimensional vortex}, Physics of Fluids 12 (2000) 2397--2412.

\bibitem{Schecter_Dubin_etc_Vortex_Crystals_2DEuler1999PhFl}
D.~A. {Schecter}, D.~H.~E. {Dubin}, K.~S. {Fine}, C.~F. {Driscoll}, {Vortex
  crystals from 2D Euler flow: Experiment and simulation}, Phys. Fluids 11
  (1999) 905--914.

\bibitem{Schmeits_Dijkstraa_2001_JPO_BimodaliteGulfStream}
M.~J. Schmeits, H.~A. Dijkstraa, Bimodal behavior of the kuroshio and the gulf
  stream., J. Phys. Oceanogr. 31 (2001) 3425--3456.

\bibitem{Segre_Kida_1997chao.dyn..9020S}
E.~{Segre}, S.~{Kida}, {Late states of incompressible 2D decaying vorticity
  fields}, in: eprint arXiv:chao-dyn/9709020, 1997, p. 9020.

\bibitem{kellayPRL08}
F.~{Seychelles}, Y.~{Amarouchene}, M.~{Bessafi}, H.~{Kellay}, {Thermal
  Convection and Emergence of Isolated Vortices in Soap Bubbles}, Physical
  Review Letters 100~(14) (2008) 144501.

\bibitem{Smith_ONeil_Physics_Fluids_1990_Inequivalence}
R.~A. {Smith}, T.~M. {O'Neil}, {Nonaxisymmetric thermal equilibria of a
  cylindrically bounded guiding-center plasma or discrete vortex system}, Phys.
  Fluids B 2 (1990) 2961--2975.

\bibitem{Sommeria_1986_JFM_2Dinverscascade_MHD}
J.~{Sommeria}, {Experimental study of the two dimensional inverse energy
  cascade in a square box}, J. Fluid. Mech. 170 (1986) 139--168.

\bibitem{Sommeria_2001_CoursLesHouches}
J.~{Sommeria}, {Two-Dimensional Turbulence}, in: S.~Berlin (ed.), New trends in
  turbulence, vol.~74 of Les Houches, 2001, pp. 385--447.

\bibitem{Sommeria_Nore_Dumont_Robert_1991CRASB.312..999S}
J.~{Sommeria}, C.~{Nore}, T.~{Dumont}, R.~{Robert}, {Statistical theory of the
  Great Red SPOT of Jupiter}, Academie des Science Paris Comptes Rendus Serie B
  Sciences Physiques 312 (1991) 999--1005.

\bibitem{Staquet}
J.~{Sommeria}, C.~{Staquet}, R.~{Robert}, {Final equilibrium state of a
  two-dimensional shear layer}, Journal of Fluid Mechanics 233 (1991) 661--689.

\bibitem{Spitzer_1991}
L.~{Spitzer}, {Dynamical evolution of Globular Clusters}, {Princeton University
  Press}, 1991.

\bibitem{Spohn_1991}
H.~{Spohn}, {Large Scale Dynamics of Interacting Particles}, {Springer,
  New-York}, 2002.

\bibitem{Stammer97}
D.~{Stammer}, {Global Characteristics of Ocean Variability Estimated from
  Regional TOPEX/POSEIDON Altimeter Measurements}, Journal of Physical
  Oceanography 27 (1997) 1743--1769.

\bibitem{Tabeling02}
P.~{Tabeling}, {Two-dimensional turbulence: a physicist approach}, \physrep 362
  (2002) 1--62.

\bibitem{thess94}
A.~{Thess}, J.~{Sommeria}, B.~{J{\"u}ttner}, {Inertial organization of a
  two-dimensional turbulent vortex street}, Physics of Fluids 6 (1994)
  2417--2429.

\bibitem{Thomson_Young_2007JAts}
A.~F. {Thompson}, W.~R. {Young}, {Two-Layer Baroclinic Eddy Heat Fluxes: Zonal
  Flows and Energy Balance}, Journal of Atmospheric Sciences 64 (2007) 3214.

\bibitem{Kelvin_1887_bis}
W.~{Thomson}, {On the stability of steady and of periodic fluid motion},
  Philos. Mag. 24 (1887) 188--196.

\bibitem{Tian_Weeks_etc_Ghil_Swinney_2001_JFM_JetTopography}
Y.~{Tian}, E.~R. {Weeks}, K.~{Ide}, J.~S. {Urbach}, C.~N. {Baroud}, M.~{Ghil},
  H.~L. {Swinney}, {Experimental and numerical studies of an eastward jet over
  topography}, J. Fluid Mech. 438 (2001) 129--157.

\bibitem{1983_Turkington_CommMathPhys1}
B.~{Turkington}, {On steady vortex flow in two dimensions, I}, Communications
  in Partial Differential Equations 8 (9) (1983) 999--1030.

\bibitem{1983_Turkington_CommMathPhys2}
B.~{Turkington}, {On steady vortex flow in two dimensions, II}, Communications
  in Partial Differential Equations 8 (9) (1983) 999--1030.

\bibitem{TurkingtonMHD:2001_PNAS_GRS}
B.~{Turkington}, A.~{Majda}, K.~{Haven}, M.~{Dibattista}, {Statistical
  equilibrium predictions of jets and spots on Jupiter}, PNAS 98 (2001)
  12346--12350.

\bibitem{244830}
B.~Turkington, N.~Whitaker, Statistical equilibrium computations of coherent
  structures in turbulent shear layers, SIAM J. Sci. Comput. 17~(6) (1996)
  1414--1433.

\bibitem{VallisBook}
G.~K. {Vallis}, {Atmospheric and Oceanic Fluid Dynamics}, 2006.

\bibitem{VallisYoungCarnevaleJFM}
G.~K. {Vallis}, G.~F. {Carnevale}, W.~R. {Young}, {Extremal energy properties
  and construction of stable solutions of the Euler equations}, Journal of
  Fluid Mechanics 207 (1989) 133--152.

\bibitem{Venaille_Bouchet_PRL_2009}
A.~{Venaille}, F.~{Bouchet}, {Statistical Ensemble Inequivalence and Bicritical
  Points for Two-Dimensional Flows and Geophysical Flows}, Physical Review
  Letters 102~(10) (2009) 104501.

\bibitem{VenailleBouchetJPO}
A.~{Venaille}, F.~{Bouchet}, {Oceanic rings and jets as statistical equilibrium
  states}, ArXiv e-prints:1011.2556.

\bibitem{VenailleBouchetJSP}
A.~{Venaille}, F.~{Bouchet}, {Solvable phase diagrams and ensemble
  inequivalence for two-dimensional and geophysical turbulent flows}, ArXiv
  e-prints:1011.2309.

\bibitem{WangVallis}
J.~{Wang}, G.~K. {Vallis}, {Emergence of Fofonoff states in inviscid and
  viscous ocean circulation models}, Journal of Marine Research 52 (1994)
  83--127.

\bibitem{Weeks_Tian_etc_Swinney_Ghil_Science_1997}
E.~R. {Weeks}, Y.~{Tian}, J.~S. {Urbach}, K.~{Ide}, H.~L. {Swinney}, M.~{Ghil},
  {Transitions Between Blocked and Zonal Flows in a Rotating Annulus}, Science
  278 (1997) 1598.

\bibitem{Weichman_2006PhRvE}
P.~B. {Weichman}, {Equilibrium theory of coherent vortex and zonal jet
  formation in a system of nonlinear Rossby waves}, \pre 73~(3) (2006) 036313.

\bibitem{Weinam_Mattingly_2001_Comm_Pure_Appl_Math_Ergodicity_NS}
E.~{Weinan}, J.~C. {Mattingly}, {Ergodicity for the Navier-Stokes equation with
  degenerate random forcing: Finite-dimensional approximation}, Comm. Pure
  Appl. Math. 54 (2001) 1386--1402.

\bibitem{1994PhFl....6.3963W}
N.~{Whitaker}, B.~{Turkington}, {Maximum entropy states for rotating vortex
  patches}, Physics of Fluids 6 (1994) 3963--3973.

\bibitem{Wolansky_Ghil_1998_CMaPh}
G.~{Wolansky}, M.~{Ghil}, {Nonlinear Stability for Saddle Solutions of Ideal
  Flows and Symmetry Breaking}, Commun. Math. Phys. 193 (1998) 713--736.

\bibitem{Yin_Montgomery_Clercx_2003PhFluids}
Z.~{Yin}, D.~C. {Montgomery}, H.~J.~H. {Clercx}, {Alternative
  statistical-mechanical descriptions of decaying two-dimensional turbulence in
  terms of ``patches'' and ``points''}, Phys. Fluids 15 (2003) 1937--1953.

\bibitem{Zeitlin_1991_HamiltonianTruncations}
V.~{Zeitlin}, {Finite-mode analogs of 2D ideal hydrodynamics: Coadjoint orbits
  and local canonical structure}, Physica D Nonlinear Phenomena 49 (1991)
  353--362.

\bibitem{Zou_Holloway_1994JFM...263..361Z}
J.~{Zou}, G.~{Holloway}, {Entropy maximization tendency in topographic
  turbulence}, Journal of Fluid Mechanics 263 (1994) 361--374.

\end{thebibliography}

\end{document}